\documentclass[12pt]{report}
\pdfoutput=1 

\usepackage[T1]{fontenc} 
\usepackage{color}
\usepackage{fancybox}
\usepackage{amsfonts}
\usepackage{amsmath}
\usepackage{mathrsfs}
\usepackage{slashed}
\usepackage{amssymb}
\usepackage{stmaryrd}
\usepackage{indentfirst}
\usepackage{subcaption}
\usepackage{textcomp}
\usepackage{mathtools}
\usepackage{graphicx}
\usepackage[font={small,it}]{caption}
\usepackage[justification=centering]{caption}
\usepackage{braket}

\RequirePackage[colorlinks=true
,urlcolor=blue
,anchorcolor=blue
,citecolor=blue
,filecolor=blue
,linkcolor=blue
,menucolor=blue
,pagecolor=blue
,linktocpage=true
,pdfproducer=medialab
,pdfa=true
]{hyperref}

\allowdisplaybreaks

\newcommand{\be}{\begin{equation}}
\newcommand{\bea}{\begin{eqnarray}}

\newcommand{\ee}{\end{equation}}
\newcommand{\eea}{\end{eqnarray}}

\DeclareMathAlphabet{\mathpzc}{OT1}{pzc}{m}{it}

\renewcommand{\baselinestretch}{1.2}
\begin{document}
\pagestyle{plain}
\def\baselinestretch{1.2}
\hoffset=-1.0 true cm
\voffset=-2 true cm
\topmargin=1.0cm
\thispagestyle{empty}


\thicklines
\begin{picture}(370,60)(0,0)
\setlength{\unitlength}{1pt}
\put(40,53){\line(2,3){15}}
\put(40,53){\line(5,6){19}}
\put(40,53){\line(1,1){27}}
\put(40,53){\line(6,5){33}}
\put(40,53){\line(3,2){25}}
\put(40,53){\line(2,1){19}}
\put(40,53){\line(5,-6){17}}
\put(40,53){\line(1,-1){22}}
\put(40,53){\line(6,-5){30}}
\put(40,53){\line(3,-2){22}}
\put(40,53){\line(-2,1){15}}
\put(40,53){\line(-3,1){23}}
\put(40,53){\line(-4,1){26}}
\put(40,53){\line(-6,1){36}}
\put(40,53){\line(-1,0){40}}
\put(40,53){\line(-6,-1){32}}
\put(40,53){\line(-3,-1){20}}
\put(40,53){\line(-2,-1){10}}
\put(75,45){\Huge \bf IFT}
\put(180,56){\small \bf Instituto de F\'\i sica Te\'orica}
\put(165,42){\small \bf Universidade Estadual Paulista}
\put(-25,2){\line(1,0){433}}
\put(-25,-2){\line(1,0){433}}
\end{picture}


\vskip .3cm

\noindent
{TESE DE DOUTORAMENTO}
\hfill    IFT-T.012/19\\

\vspace{3cm}
\begin{center}
{\large \bf Pure spinors and $D=11$ supergravity}

\vspace{1.2cm}
Max Guillen
\end{center}

\vskip 3cm
\hfill Orientador
\vskip 0.4cm
\hfill {\em Nathan Berkovits}
\vskip 3.5cm
\vfill
\begin{center}
Dezembro de 2019
\end{center}

\newpage
\pagenumbering{roman}
\begin{center}
\null

\vfill
 {\it To my family}
\end{center}

\newpage
\begin{center}
\null 
\vfill
{\it "If I have seen further it is by standing on the shoulders of Giants"}\\

\null \hfill -Issac Newton {\it } 

\end{center}

\newpage

\begin{center}
{\Large \bf Acknowledgements}
\end{center}
I would like to thank Nathan Berkovits for his supervision, encouragement, support, patience and collaboration during the last six years. It has been a remarkable and enjoyable experience to be his student and learn from him a lot about physics and research. His high ability to select open problems as well as efficient methods for their possible resolutions was one of the things which impressed me and influenced me the most. I will always keep his extraordinary style of doing physics as a example to follow.

\vspace{2mm}
I am also grateful to my co-supervisor in Oxford, Lionel Mason. I enjoyed each weekly discussion we had during my visit to the Mathematical Institute. His particular mathematical view of physics and great sense of humor made our meetings very interesting, productive and dynamical. His hospitality, kindness, patience and time made my stay in Oxford an unforgettable professional and personal experience.

\vspace{2mm}
I also want to acknowledge my co-supervisor at Perimeter Institute, Freddy Cachazo. His gentleness, patience, generosity and hospitality made my visit to Perimeter Institute very comfortable and productive. Discussing with him several ideas on amplitudes, quantum field theory, string theory, twistor strings, F-theory, etc, was an unprecedented experience to me. His authentic style of doing science have influenced me in several ways, and not only in physics. I will certainly carry with me all of the lessons and suggestions I received from him.

\vspace{2mm}
I would also like to thank my family for his constant support and encouragement during all these years. Each time I traveled back to Brazil from Peru, I felt completely renewed because of the unique hospitality each member of my family gave me, especially my adorable parents Nancy Quiroz and Luis Guillen. Even being outside Peru, I still could feel their concerns and huge love through messages, calls, e-mails, etc. This was very important for my emotional wellness and helped me pretty much to face tough moments in life. Thanks family!

\vspace{2mm}
I also want to thank my girlfriend Caroline for all her love, support, encouragement and companionship. Our several discussions on science and life have been more than useful for me and helped me to mature and grow as a person. She has played a fundamental role in this stage of my carreer, and I feel happy and fortunate I had an extraordinary and beautiful person as her on my side during all of this process.

\vspace{2mm}
I also wish to express my thanks to Egnaldo Costa and Michele Fernandes for all their support, love and hospitality during my visits to Brasilia, and for all our conversations about Politics, Science, Economics, History, etc, in and after the so tasty lunches and dinners which this lovely family used to prepare for us. I will keep all these nice memories with deep gratitude.

\vspace{2mm}
Moreover, I would also like to thank my IFT friends: Luis Alejo, Segundo Parra, Daniel Reyes, Vilson Fabricio, Carlos Bautista, Carlisson Miller, Fernando Serna, Johan Malpartida, Enzo Leon, Victor Cesar, Matheus Lize, Daniel Wagner, Heliudson Bernardo, Prieslei Goulart, Henrique Flores, Lucas Martins, Ana Lucia Retore, Dennis Zavaleta, Dean Valois, Gabriel Caro, Andres Vasquez and Luis Alberto Ypanaque for all our conversations on physics, politics, economics, etc., and for all the social meetings we shared and all the good times we had at the historic Ademir's bar. Likewise, I want to thank my ICTP friends Oscar Chacaltana, Jose Luis Herrera, Andrea Guerrieri, Antonino Troja, Vasco Goncalves and Georgios Itsios for all of our conversations on academic and non-academic life and all the fun moments we shared. Also, I want to express my gratitude to Ivana Azuaje, Luiza Victoretti and Rosane Mascarello for their special friendship and company in each \emph{oficial call} I used to do many Fridays of the year. 

\vspace{2mm}
I am also thankful to my Oxford friends: Eduardo Casali, Stefan Nekovar, Diego Berdeja and Matteo Parisi for several discussions on physics and mathematics and their kind hospitality during my visit to Oxford.

\vspace{2mm} 
I also want to thank my PI friends: Frank Coronado, Alfredo Guevara, Ignacio Reyes, Yilber Bautista, Diego Garcia, Andre Pereira and Jairo Martinez for all of the interesting discussions we had on physics during my stay in Waterloo, and for all of the enjoyable moments we shared in and outside Perimeter.

\vspace{2mm}
I also wish to thank Ashoke Sen, Edward Witten, Barton Zwiebach, Cumrun Vafa, Kumar Narain, John Schwarz, Rajesh Gopakumar, Nima Arkani-Hamed, Martin Schnabl, Ted Erler, Sergio Ferrara, Pietro Grassi, Ellis Ye Yuan, Carlos Mafra, Yvonne Geyer, Oliver Schlotterer, Renann Jusinskas and Thales Azevedo for all their illuminating explanations on several topics in physics and their extraordinary suggestions which ended up being very useful for my research.

\vspace{2mm}
Last but not least, I want to thank FAPESP for financial support through grants 2015/23732-2, 2018/10159-0, 2018/16785-0.

\newpage

\begin{center}
{\Large \bf Resumo}
\end{center}
\vskip 0.005cm

\noindent Nessa tese estudamos as abordagens de primeira e segunda quantiza\c{c}\~ao que descrevem supergravidade em $D=11$ dimens\~oes utilizando vari\'aveis de espinores puros. Introduzimos a chamada superpart\'icula de espinores puros atrav\'es de argumentos de cohomologia BRST come\c{c}ando a partir da superpart\'icula de Brink-Schwarz em $D=11$ dimens\~oes no calibre do semi-cone de luz. Ap\'os realizar uma an\'alise no cone de luz da cohomologia BRST de espinor puro com n\'umero de fantasma tr\^es, encontramos as equa\c{c}\~oes de movimento linearizadas de supergravidade em $D=11$ dimens\~oes no superespa\c{c}o em $D=9$ dimen\~oes. Al\'em disso, constru\'imos um operador de v\'ertice BRST fechado de n\'umero de fantasma um, feito de campos na linha de mundo e supercampos de supergravidade em $D=11$ dimens\~oes, e encontramos uma inconsist\^encia ao construir um operador de v\'ertice de n\'umero de fantasma zero satisfazendo uma descent-equation usual. Em seguida, introduzimos a superpart\'icula de espinor puro em $D=11$ dimens\~oes na vers\~ao n\~ao m\'inima, na qual um fantasma $b$ satisfazendo $\{Q,b\} = P^{2}$ pode ser constru\'ido. Entretanto, sua express\~ao complicada torna a demonstra\c{c}\~ao de sua nilpot\^encia dif\'icil. Ap\'os introduzir um vetor fermi\^onico de $SO(1,10)$ $\bar{\Sigma}^{a}$, n\'os iremos notavelmente simplificar sua forma e mostrar que de fato $\{Q,b\} = P^{2}$ e $\{b,b\}=$ BRST-exato. Usando esse fantasma $b$, n\'os propomos um operador de v\'ertice de n\'umero de fantasma zero alternativo satisfazendo uma descent-equation usual. Entretanto, esse operador depender\'a de vari\'aveis de espinores puros n\~ao m\'inimos de maneira bastante complicada. Depois de discutir essa abordagem de primeira quantiza\c{c}\~ao para supergravidade em $D=11$ dimens\~oes, n\'os prosseguimos com a discuss\~ao das a\c{c}\~oes maestras de espinores puros introduzidas por Cederwall para estudar teorias de calibre supersim\'etricas. N\'os mostramos que essas a\c{c}\~oes de fato descrevem super-Yang-Mills e super-Born-Infeld em $D=10$ dimens\~oes e supergravidade em $D=11$ dimens\~oes extraindo as equa\c{c}\~oes de movimento no superespa\c{c}o ordin\'ario para cada uma dessas teorias.
\vskip 0.3cm
\noindent
{\bf Palavras Chaves}: Superpart\'icula; Supermembrana; Supergravidade; Espinores Puros.
\vskip 0.3cm
\noindent
{\bf \'{A}reas do conhecimento}: Teoria de Supercordas; Teoria M; Supergravidade.

\newpage

\begin{center}
{\Large \bf Abstract}
\end{center}
\vskip 2.0cm

\noindent  In this Thesis we study first- and second-quantized approaches describing $D=11$ supergravity using pure spinor variables. We introduce the so-called $D=11$ pure spinor superparticle through BRST cohomology arguments starting from the semi-light-cone gauge $D=11$ Brink-Schwarz-like superparticle. After performing a light-cone gauge analysis of the pure spinor BRST cohomology at ghost number three, we find the linearized equations of motion of $D=11$ supergravity in $D=9$ superspace. In addition, we construct a BRST-closed, ghost number one vertex operator made out of worldline fields and $D=11$ supergravity superfields, and we run into an inconsistency when constructing a ghost number zero vertex operator satisfying a standard descent equation. We then introduce the non-minimal version of the $D=11$ pure spinor superparticle, in which a composite $b$-ghost can be constructed satisfying $\{Q,b\} = P^{2}$. However, its complicated expression makes it difficult to check its nilpotency.  We show that introducing an $SO(1,10)$ fermionic vector $\bar{\Sigma}^{a}$ simplifies the form of the $b$-ghost considerably, which allows us to verify that $\{Q,b\} = P^{2}$ and $\{b,b\}=$ BRST-exact. Using this $b$-ghost we propose an alternative ghost number zero vertex operator satisfying a standard descent equation. However, its expression will depend on non-minimal pure spinor variables in a very complicated fashion. After discussing this first-quantized approach for $D=11$ supergravity, we move on to discussing the pure spinor master actions introduced by Cederwall for studying maximally supersymmetric gauge theories. We show that these actions indeed describe $D=10$ super-Yang-Mills, $D=10$ super-Born-Infeld and $D=11$ supergravity by extracting the equations of motion in ordinary superspace for each one of these theories.

\vskip 0.5cm
\noindent
{\bf Areas}: Superstring Theory; M-Theory; Supergravity.

\newpage
\centerline {\bf List of publications}

\vspace{1.5cm}
\noindent{\bf [1]} Nathan Berkovits, Max Guillen, ``Simplified $D=11$ $b$-ghost'', {\bf JHEP 1707 (2017) 115}, arxiv: 1703.05116 [hep-th].\\

\noindent{\bf [2]} Max Guillen, ``Equivalence of the 11D pure spinor and Brink-Schwarz-like superparticle cohomologies'', {\bf Phys. Rev. D 97}, 066002.\\

\noindent{\bf [3]} Nathan Berkovits, Max Guillen, ``Equations of motion from Cederwall\'s pure spinor superspace actions'', {\bf JHEP 1808 (2018) 033}, arXiv: 1804.06979 [hep-th].\\

\noindent{\bf [4]} Nathan Berkovits, Eduardo Casali, Max Guillen, Lionel Mason, ``Notes on the $D=11$ pure spinor superparticle'', {\bf JHEP 1908 (2019) 178}, arXiv: 1905.03737 [hep-th].\\

\noindent{\bf [5]} Nathan Berkovits, Max Guillen, Lionel Mason, ``Supertwistor description of ambitwistor strings'', arXiv: 1908.06899 [hep-th]\\

\noindent Chapter 2 is based on paper number [3]. Chapter 3 is based on paper number [4]. Chapter 4 is based on paper number [1]. Chapter 5 is based on paper number [3].

\vfill \eject

\tableofcontents

\chapter{Introduction}
\pagenumbering{arabic}

$D=11$ supergravity was introduced in \cite{Cremmer:1978km} as a gauge theory with the maximal number of supersymmetries describing physically consistent interactions. It contains a simple spectrum consisting of a spin-2 field, a spin-$\frac{3}{2}$ field and a 3-form gauge field. Its uniqueness comes from the fact that all their interactions are completely fixed by demanding its action to be diffeomorphism-, Lorentz- and supersymmetry-invariant. This theory has been shown to be related to type IIA supergravity after compactifying on a circle, and just as Type IIA supergravity possesses a UV completion described by a type IIA superstring, $D=11$ supergravity has been conjectured to be UV completed by the so-called M-theory \cite{Hull:1994ys,Townsend:1995kk,Witten:1995ex}. In this sense, $D=11$ supergravity is said to be the low-energy limit of M-theory. However, although there are some hints to suspect that $D=11$ supergravity will exhibit UV divergences at some loop order, there is not any concrete evidence so far which demonstrates this hypothesis. Therefore the finiteness of the ultraviolet behavior of $D=11$ supergravity still remains as an open problem.

\vspace{2mm}
One way to address this problem is by calculating several Feynman diagrams by using standard quantum field theory techniques. Although in principle this should be possible, it is a highly time-consuming task. Moreover, since the number of Feynman diagrams grows exponentially with the number of external particles and loops in a certain given scattering process, the computational cost of it is extremely large. These technical issues force us to look for alternative approaches where one has enough control to answer these types of questions. 

\vspace{2mm}
A first step towards this direction was presented in \cite{Green:1999by}. There, the $D=11$ Brink-Schwarz-like superparticle was first-quantized in the light-cone gauge and its physical spectrum was shown to contain an $SO(9)$ traceless symmetric tensor, an $SO(9)$ $\Gamma$-traceless vector-spinor and an $SO(9)$ 3-form which are exactly the $D=11$ supergravity physical degrees of freedom. Using light-cone gauge techniques inspired in string theory, they were able to reproduce the Chern-Simons term appearing in the $D=11$ supergravity action from a worldline correlator of light-cone gauge vertex operators. Although similar results were already known for ten-dimensional maximally supersymmetric gauge theories from a stringy point of view, the application of these ideas for $D=11$ dimensions was completely new and so gave rise to a new framework for computing scattering processes of $D=11$ supergravity states.

\vspace{2mm}
The main disadvantage of this superparticle formalism was the lack of Lorentz covariance when fixing the light-cone gauge. This problem had already been found in other space-time dimensions, and it was shown to be related to the constrained nature of the superparticle models. More specifically, the first- and second-class constraints present in this model could not be separated out in a Lorentz covariant way. Eventually this problem was solved from different perspectives but they all exhibited an intrinsic complexity for practical purposes \cite{Siegel:1985xj,Mikovic:1988bj,Essler:1990az,Essler:1990yq,Berkovits:1990yc}.

\vspace{2mm}
On the other hand, the search for an off-shell description of maximally supersymmetric gauge theories \cite{Nilsson:1985cm}, a geometric description of these in larger spaces \cite{HOWE1991141,Howe:1991bx} and a covariant quantization approach for superstrings \cite{Berkovits:2000fe} led to the introduction of ten-dimensional pure spinors, that is, complex bosonic Weyl spinors satisfying $\lambda\gamma^{m}\lambda = 0$\footnote{The formal definition of a pure spinor is due to Cartan, who defines an $SO(2n)$ pure spinor as a chiral spinor $\lambda^{\alpha}$ for $\alpha=1,\ldots, 2^{n-1}$, satisfying the constraints $\lambda\Gamma^{m_{1}\ldots m_{n-4}}\lambda =\lambda\Gamma^{m_{1}\ldots m_{n-8}}\lambda = \lambda\Gamma^{m_{1}\ldots m_{n-12}}\lambda = \ldots =0$, where $\Gamma^{m_{1}\ldots m_{k}}$ is the totally antisymmetrized product of $k$ $SO(2n)$ gamma matrices. Note that this definition assumes an even number of spacetime dimensions.}, as a main ingredient for studying theories possessing maximal supersymmetry. Remarkably, the covariant program gave rise to a totally new formalism for studying string theory which harnesses the power of BRST methods and pure spinor identities. In this manner, a covariant quantization scheme straightforwardly follows from the simplicity of the pure spinor BRST charge. Furthermore, the pure spinor superstring formalism has also been shown to be more efficient than the other existing superstring formalisms for computing multiloop scattering amplitudes. This feature is a direct consequence of manifest supersymmetry in the pure spinor framework.

\vspace{2mm}
This pure spinor construction for superstrings was later applied to the $D=10$ superparticle which gave rise to what is now known as the $D=10$ pure spinor superparticle \cite{Berkovits:2001rb,Berkovits:2002zk}. The problem of covariant quantization thus translated into the problem of finding the non-trivial cohomology of the ten-dimensional pure spinor BRST charge. As a result, physical states turned out to appear at different (up to three) ghost numbers obeying the same equations of motion and gauge invariances as those dictated by the Batalin-Vilkovisky (BV) description of $N=1$ $D=10$ super Yang-Mills. These ideas were also applied to the $N=2$ superparticle and the pure spinor quantization gave an elegant covariant description of Type IIA/IIB supergravity in their BV formulations \cite{Berkovits:2002uc}. 

\vspace{2mm} 
Some years later, a worldline approach for computing scattering amplitudes in $N=1$ $D=10$ super Yang-Mills and Type IIA/IIB supergravity was introduced in \cite{Bjornsson:2010wm,Bjornsson:2010wu} which is based on the pure spinor superparticle models in their non-minimal versions, that is, models resulting from adding a couple of constrained variables to the original theory though the quartet mechanism \cite{Kugo:1979gm}. In this manner, several properties of the 4-point functions up to 5-loops in $N=4$ $D=4$ super Yang-Mills and $N=8$ $D=4$ supergravity were predicted from simple power counting arguments. In particular, it was argued that the 4-point super Yang-Mills amplitude is UV finite at all loop orders and that the 4-point $N=8$ $D=4$ supergravity amplitude should present a logarithmic divergence at $L=7$ loops.

\vspace{2mm}
These notable features of the pure spinor formalism in $D=10$ dimensions led one to wonder if there is any eleven-dimensional pure spinor construction for $D=11$ supergravity. As discussed in \cite{Berkovits:2002uc}, the answer to this question turned out to be in the affirmative. In this work, Berkovits proposes a pure spinor action for the $D=11$ superparticle and supermembrane together with a standard pure spinor BRST charge. The quantization of this system is described by the BRST cohomology of an eleven-dimensional pure spinor, namely an eleven-dimensional bosonic Majorana spinor satisfying $\lambda\Gamma^{m}\lambda = 0$\footnote{This definition is somewhat different from the one given by Howe in \cite{Howe:1991bx}. 
For this reason some authors refer to spinors used in this thesis as semi-pure spinors. However we will ignore these subtleties and simply call them pure spinors.
}. 
Strikingly, physical states were found at different (up to seven) ghost numbers obeying the same equations of motion and gauge invariances as dictated by the BV description of $D=11$ supergravity. Although a prescription for computing N-point correlation functions involving vertex operators with different ghost numbers was conjectured to exist there, there has not been any significant progress in this direction.

\vspace{2mm}
One decade later, a second-quantization approach was proposed by Cederwall \cite{Cederwall:2009ez,Cederwall:2010tn,Cederwall:2011vy,Cederwall:2013vba} for studying maximally supersymmetric gauge theories such as $N=1$ $D=10$ super Yang-Mills, $N=1$ $D=10$ super Born-Infeld and $D=11$ supergravity from a pure spinor antifield formalism. This framework uses the power of pure spinor superfields encoding the full spectrum coming from a BV description of a certain theory for constructing pure spinor master actions, that is, actions made out of pure spinor superfields satisfying a master equation defined through a pure spinor superfield antibracket. In this manner, pure spinor master actions were found for each of these theories and they turned out to be extremely simple as compared to their component actions. So for instance, $N=1$ $D=10$ super Born-Infeld ($N=1$ $D=11$ supergravity) possesses a pure spinor action which is cubic (quartic) in a scalar ten-dimensional (eleven-dimensional) pure spinor superfield. In addition, these actions are manifestly suppersymmetric which make them attractive for computing scattering amplitudes. However, despite these notable features of the formalism, an explicit procedure for extracting dynamical information on ordinary superspace was not found.

\vspace{2mm} 
In this thesis we will study two different approaches for studying $D=11$ supergravity in a manifestly Lorentz covariant and supersymmetric way using pure spinor variables. The former will be through a worldline framework using the $D=11$ pure spinor superparticle. This will be motivated and introduced from the $D=11$ Brink-Schwarz-like superparticle in semi-light-cone gauge in chapter \ref{newchapter2}. We will review there the standard superparticle model in detail and discuss its advantages and disadvantages for describing $D=11$ linearized supergravity. After introducing a new couple of canonical conjugate variables together with a set of first-class constraints, which allow us to recover the original gauge-fixed action, and performing two sucessive similarity transformations, the $D=11$ pure spinor superparticle model will be found. Furthermore, after performing a light-cone gauge analysis on the ghost number three BRST cohomology, we will find the equations of motion of $D=11$ linearized supergravity in $D=9$ superspace.

\vspace{2mm}
In order to compute $D=11$ supergravity scattering amplitudes, an $N$-point correlation function prescription was given in \cite{Anguelova:2004pg}. This uses three type of vertex operators of ghost numbers three, one and zero. We will discuss the construction of each one of them in chapter \ref{newchapter3}. In particular, we will find a BRST-closed, ghost number one vertex operator which differs from the one proposed in \cite{Anguelova:2004pg}, and we will show an inconsistency in attempting to construct a ghost number zero vertex operator depending on pure spinor variables and $D=11$ supergravity superfields satisfying a standard descent equation. The way in which this affects the results found in \cite{Anguelova:2004pg} can be viewed through the 3-point correlation function. We will find a formula for it differing from the one in \cite{Anguelova:2004pg}, but it takes a similar form to the 3-point coupling proposed by Cederwall in \cite{Cederwall:2009ez,Cederwall:2010tn}.

\vspace{2mm}  
In chapter \ref{newchapter4}, we will introduce the $D=11$ non-minimal pure spinor superparticle by adding a bosonic pure spinor $\bar{\Lambda}^{\alpha}$ satisfying $\bar{\Lambda}\Gamma^{a}\bar{\Lambda} = 0$ and a fermionic vector $R_{\alpha}$ satisfying $\bar{\Lambda}\Gamma^{a}R = 0$ to the original $D=11$ pure spinor superparticle through the quartet argument. As a result, a composite $b$-ghost satisfying $\{Q , b\} = P^{a}P_{a}$ can be shown to exist. However, its very intricate form makes its use difficult for practical purposes. In particular, its nilpotency property has not been checked yet. Using an $SO(1,10)$ fermionic vector $\bar{\Sigma}^{a}$, we will notably simplify its form and verify that it satisfies $\{Q,b\} = P^{a}P_{a}$ and $\{b,b\} =$ BRST-exact. Likewise, we will propose a ghost number zero vertex operator obeying $\{b, U^{(1)}\} = V^{(0)}$ with $U^{(1)}$ the ghost number one vertex operator. It will trivially satisfy a standard descent equation, but depend on non-minimal pure spinor variables in a complicated way.

\vspace{2mm}
After having studied the first-quantized approach for $D=11$ supergravity, we move on to studying a second-quantized approach through pure spinor master actions in chapter \ref{newchapter5}. This framework makes use of the field-antifield structure encoded by pure spinor superfields in $D=10$ and $D=11$ dimensions for describing interacting theories through a BV-like formalism. We will review the connection between the $D=10$ and $D=11$ non-minimal pure spinor superparticles with $D=10$ super-Maxwell and $D=11$ linearized supergravity, respectively. This will be useful to motivate the construction of the pure spinor integration measures in $D=10$ and $D=11$ dimensions, which will later enter the construction of pure spinor master actions. We then study the $D=10$ super-Yang-Mills, $D=10$ Abelian supersymmmetric-Born-Infeld and $D=11$ supergravity pure spinor actions proposed in \cite{Cederwall:2011vy,Cederwall:2010tn} and deduce the equations of motion and gauge symmetries coming from each one of them. Due to the presence of non-minimal pure spinor variables in each one of these equations, it is a non-trivial task to obtain physical information from them. We will explain in detail a systematic procedure for this purpose, and obtain the equations of motion at first order in the coupling constant in ordinary superspace for each one of these theories.

\vspace{2mm}
We have added some Appendices at the end of this thesis in order to make it as self contained as possible. Thus, explicit realizations of $D=10$ and $D=11$ gamma matrices and a brief review of octonions can be found in Appendix \ref{appendix1}. Several ten-dimensional discussions illustrating the ideas used for the $D=11$ case in a simpler way are found in Appendices \ref{appendix2} and \ref{appendix3}. A review of superspace formulations of $D=10$ super-Yang-Mills and $D=11$ supergravity are discussed in Appendices \ref{appendix4} and \ref{appendixB}. A set of $D=11$ pure spinor identities and heavy computations involving $\bar{\Sigma}^{a}$ can be found in Appendix \ref{appendix5}. Finally, Appendix \ref{appendix6} discusses several useful $D=10$ and $D=11$ gamma matrix identities, a brief review of the pure spinor superfield formalism and an explicit computation of $R^{a}\Psi$.

\chapter{$D=11$ Pure spinor superparticle}\label{newchapter2}
The $D=11$ Brink-Schwarz-like superparticle \cite{Green:1999by} possesses first-class and second-class constraints which do not allow a manifestly covariant quantization of the theory. However, it is possible to quantize the theory in the light-cone gauge and it can be shown that the spectrum is described by an $SO(9)$ traceless symmetric tensor, an $SO(9)$ $\Gamma$-traceless vector-spinor and an $SO(9)$ 3-form which describe $D=11$ linearized supergravity.  However, light-cone gauge breaks the manifest covariance of the theory.

\vspace{2mm}
It is interesting and useful to look for covariant descriptions which manifestly preserve as many symmetries as possible. One candidate that addresses this point is the pure spinor version of the $D=11$ Brink-Schwarz-like superparticle, known as the \emph{$D=11$ pure spinor superparticle} \cite{Berkovits:2002uc}. This description preserves supersymmetry and Lorentz symmetry in a manifestly covariant way. The physical states of this pure spinor version are defined as elements in the cohomology of the BRST operator $Q = \Lambda^{\alpha}D_{\alpha}$, where $\Lambda^{\alpha}$ is a $D=11$ pure spinor and $D_{\alpha}$ are the fermionic constraints of the $D=11$ Brink-Schwarz-like superparticle. The elements of this $Q$-cohomology describe the BV version of $D=11$ linearized supergravity \cite{Berkovits:2002uc}. 
In this chapter we will introduce the $D=11$ pure spinor superparticle starting from the $D=11$ Brink-Schwarz-like superparticle by using BRST cohomology arguments and two different group decompositions\footnote{In \cite{Bandos:2007mi,Bandos:2007wm} I. Bandos relates these two models by using the Lorentz harmonics approach. We will address the problem in a different way, by focusing on the $D=11$ light-cone Brink-Schwarz-like superparticle.}. A direct and immediate consequence of this approach, it is the physical equivalence of both models.


\vspace{2mm}

Furthermore we will deduce the light-cone gauge equations of motion satisfied by the physical fields of $D=11$ linearized supergravity \cite{Green:1999by} from the light-cone gauge analysis of the BRST cohomology of the $D=11$ pure spinor superparticle. We will focus on the ghost number 3 vertex operator $V = \Lambda^{\alpha}\Lambda^{\beta}\Lambda^{\delta}C_{\alpha\beta\delta}$ which has been previously shown to contain the $D=11$ supergravity fields after imposing on it the pure spinor physical state condition \cite{Berkovits:2002uc}.

\vspace{2mm}
The chapter is organized as follows: In section \ref{s2.2} we review the $D=11$ Brink-Schwarz-like superparticle. In section \ref{s2.3} we present the $D=11$ pure spinor superparticle and show the equivalence between the cohomologies of this theory and the previous one by decomposing $D=11$ objects into their $SO(1,1)\times SO(9)$ and $SO(3,1)\times SO(7)$ components. In section \ref{s2.4} we study the light-cone gauge pure spinor cohomology and show that it is described by the usual $SO(9)$ irreducible representations that describe $D=11$ supergravity and satisfy linearized equations of motion in $D=9$ superspace.

\vspace{2mm}
All the ideas developed in this chapter can be properly applied to the ten-dimensional case. Since our main goal is to understand the role of $D=11$ pure spinors in the study of $D=11$ supergravity, we will leave the ten-dimensional analysis for Appendix \ref{appendix2}. The untrained reader might find it useful to first read this Appendix before tackling the eleven-dimensional case.

\section{Review of the $D=11$ Brink-Schwarz-like superparticle}\label{s2.2}
The $D=11$ Brink-Schwarz-like superparticle is defined by the action \cite{Berkovits:2002uc,Green:1999by}:
\begin{equation}\label{eq60}
S = \int d\tau(P^{m}\Pi_{m} + eP^{m}P_{m})
\end{equation}
where $\Pi_{m} = \partial_{\tau}X_{m} - \partial_{\tau}\Theta^{\alpha}(\Gamma_{m})_{\alpha\beta}\Theta^{\beta}$, and $\Theta^{\alpha}$ is a Majorana spinor. Let us now fix conventions. We will denote $SO(10,1)$ vector indices by $m, n, p, \ldots$, and spinor indices by $\alpha,\beta, \ldots$ ($m = 0, \ldots, 10$ and $\alpha = 1, \ldots, 32$). The $D=11$ gamma matrices $\Gamma^{m}$ are $32\times 32$ symmetric matrices which satisfy $\Gamma^{m}_{\alpha\beta}\Gamma^{n\,\beta\gamma} + \Gamma^{n}_{\alpha\beta}\Gamma^{m\,\beta\gamma} = 2\eta^{mn}\delta_{\alpha}^{\gamma}$ and $\eta_{mn}\Gamma^{m}_{(\alpha\beta}\Gamma^{np}_{\gamma\delta)} = 0$. In contrast to the $D=10$ case, in $D=11$ there exists an antisymmetric metric tensor $C_{\alpha\beta}$ (and its inverse $(C^{-1})^{\alpha\beta}$) which will allow us to lower (and raise) indices (for instance $\Gamma^{m\,\alpha\beta} = C^{\alpha\delta}\Gamma^{m\,\beta}_{\delta}$, etc). We also note that any $D=11$ antisymmetric bispinor can be decomposed into a scalar, three-form, and four form as $f^{[\alpha\beta]} = C^{\alpha\beta}f + (\Gamma_{mnp})^{\alpha\beta}f^{mnp} + (\Gamma_{mnpq})^{\alpha\beta}f^{mnpq}$, and that any $D=11$ symmetric bispinor can be written in terms of a one-form, two-form and five-form as $g^{(\alpha\beta)} = \Gamma^{\alpha\beta}_{m}g^{m} + (\Gamma_{mn})^{\alpha\beta}g^{mn} + (\Gamma_{mnpqr})^{\alpha\beta}g^{mnpqr}$.\\

The action \eqref{eq60} is invariant under reparametrizations, SUSY transformations and $\kappa$-transformations which are defined by the following equations:
\begin{eqnarray*}
\mbox{Reparametrizations} &\rightarrow & d\tau\ensuremath{'} = \frac{d\tau\ensuremath{'}}{d\tau}d\tau \hspace{2mm} \mbox{,} \hspace{2mm} e\ensuremath{'}(\tau)=\frac{d\tau}{d\tau\ensuremath{'}}d\tau\\
\mbox{SUSY transformations} &\rightarrow& \delta\Theta^{\alpha} = \epsilon^{\alpha} \hspace{2mm} \mbox{,} \hspace{2mm} \delta X^{m} = \Theta^{\alpha}\Gamma^{m}_{\alpha\beta}\epsilon^{\beta} \hspace{2mm} \mbox{,} \hspace{2mm} \delta P_{m} = \delta e = 0\\
\mbox{$\kappa$ (local) transformations} &\rightarrow& \delta\Theta^{\alpha} = iP^{m}\Gamma^{\alpha\beta}_{m}\kappa_{\beta} \hspace{2mm} \mbox{,} \hspace{2mm} \delta X^{m} = -\Theta^{\alpha}\Gamma^{m}_{\alpha\beta}\delta\Theta^{\beta} \hspace{2mm} \mbox{,} \hspace{2mm} \delta P_{m} = 0 ,\\
&& \delta e = 2i(\partial_{\tau}\Theta^{\beta})\kappa_{\beta}
\end{eqnarray*}
The conjugate momentum to $\Theta^{\alpha}$ is
\begin{equation}
P_{\alpha} = \frac{\partial L}{\partial (\partial_{\tau}\Theta^{\alpha})} = -(\Gamma^{m})_{\alpha\beta}\Theta^{\beta}P_{m}
\end{equation}
Therefore, this system possesses constraints,
\begin{equation}\label{eq501}
D_{\alpha} = P_{\alpha} + (\Gamma^{m})_{\alpha\beta}\Theta^{\beta}P_{m}
\end{equation}
and considering that 
 $\{\Theta^{\alpha}, P_{\beta}\}_{P.B} = i\delta^{\alpha}_{\beta}$, we get the constraint algebra
\begin{equation}\label{eq61}
\{D_{\alpha}, D_{\beta}\} = 2i(\Gamma^{m})_{\alpha\beta}P_{m},
\end{equation}
where $\{\cdot ,\cdot \}$ denotes a Poisson bracket. One can show that $K^{\alpha} = P^{m}(\Gamma_{m})^{\alpha\beta}D_{\beta}$ are the first-class constraints that generate the $\kappa$-symmetry.
From \eqref{eq61}, we realize that we have 16 first-class constraints and 16 second-class constraints, and there is no simple way to covariantly separate them out. However, the physical spectrum can be easily found by using the semi light-cone gauge, which is defined by:
\begin{eqnarray}
X^{+} = \frac{1}{\sqrt{2}}(X^{0} + X^{9}) \hspace{4mm}&,&\hspace{4mm} \Gamma^{+} = \frac{1}{\sqrt{2}}(\Gamma^{0} + \Gamma^{9})\\
X^{-} = \frac{1}{\sqrt{2}}(X^{0} - X^{9}) \hspace{4mm}&,&\hspace{4mm} \Gamma^{-} = \frac{1}{\sqrt{2}}(\Gamma^{0} - \Gamma^{9})
\end{eqnarray}
In these light-cone gauge coordinates one can use the $\kappa$-transformation to choose a gauge where $(\Gamma^{+}\Theta)_{\alpha} = 0$\footnote{An easy way to see this is to choose a frame where $P^{m}=(P,0,\ldots,P,0)$. The $\kappa$-transformation takes the form $\delta\Theta^{\alpha} = -iP^{+}\Gamma^{-\,\alpha\beta}\kappa_{\beta}$, and thus it follows immediately that $(\Gamma^{+}\Theta)_{\alpha} = 0$.}. With this choice we can rewrite the action as follows
\begin{equation}\label{section2eq72}
S = \int d\tau[P^{m}\partial_{\tau}X_{m} + \frac{i}{2}S^{A}\partial_{\tau}S_{A} + e P^{m}P_{m}]
\end{equation}

where $S_{A}$ is an $SO(9)$ Majorana spinor, which can be written in terms of $SO(9)$ component of $\Theta^{\alpha}$. The conjugate momentum to $S^{A}$ is:
\begin{equation}
p_{A} = \frac{\partial L}{\partial (\partial_{\tau}S^{A})} = -\frac{i}{2}S_{A}
\end{equation}
So, the constraints for this gauge-fixed system are:
\begin{equation}
\tilde{D}_{A} = p_{A} + \frac{i}{2}S_{A}
\end{equation}
Considering that $\{S_{A}, p_{B}\} = -i\delta_{AB}$, we obtain
\begin{eqnarray}
\{\tilde{D}_{A}, \tilde{D}_{B}\} &=& \delta_{AB}
\end{eqnarray}
Hence, the constraint matrix is $C_{AB} = \delta_{AB}$, and its corresponding inverse is $(C^{-1})^{AB} = \delta^{AB}$. This allows us to compute the following Dirac Bracket:
\begin{eqnarray}
\{S_{A}, S_{B}\}_{D} &=& \{S_{A}, S_{B}\}_{P} - \sum_{E,F}\{S_{A}, \tilde{D}_{E}\}_{P}(C^{-1})^{EF}\{\tilde{D}_{F}, S_{B}\}_{P}\nonumber\\
&=& 0 - \sum_{E,F}(-i\delta_{AE})(\delta^{EF})(-i\delta_{FB})\nonumber\\
&=& \delta_{AB} \label{eq71}
\end{eqnarray}
As well known, the representation of the algebra \eqref{eq71} defines the space of physical states. These states will be denoted $\vert IJ\rangle$, $\vert BI\rangle$ and $\vert LMN\rangle$, where we represent $SO(9)$ vector indices by $I, J, K, L, \ldots$, and spinor indices by $A, B, C, D, \ldots$. These states correspond to an $SO(9)$ traceless symmetric tensor, an $SO(9)$ $\Gamma$-traceless vectorspinor and an $SO(9)$ 3-form, which, together, form the field content of $D=11$ supergravity. The action of the operators $S_{A}$ on the physical states is defined by
\begin{eqnarray}
S_{A}\vert IJ\rangle &=& \Gamma^{I}_{AB}\vert BJ\rangle + \Gamma^{J}_{AB}\vert BI\rangle\\
S_{A}\vert BI\rangle &=& \frac{1}{4}\Gamma^{J}_{AB}\vert IJ\rangle + \frac{1}{72}(\Gamma^{ILMN}_{AB} + 6\delta^{IL}\Gamma^{MN}_{AB})\vert LMN\rangle\\
S_{A}\vert LMN\rangle &=& \Gamma^{LM}_{AB}\vert BN\rangle + \Gamma^{MN}_{AB}\vert BL\rangle + \Gamma^{NL}_{AB}\vert BM\rangle
\end{eqnarray}
We can check that these definitions indeed reproduce the desired algebra. Let us check the statement explicitly for the graviton $\vert IJ\rangle$:
\begin{eqnarray*}
S_{A}S_{B}\vert IJ\rangle &=& \Gamma^{I}_{BC}S_{A}\vert CJ\rangle + \Gamma^{J}_{BC}S_{A}\vert CI\rangle\\
&=& \Gamma^{I}_{BC}[\frac{1}{4}\Gamma^{K}_{AC}\vert JK\rangle + \frac{1}{72}(\Gamma^{JLMN}_{AC} + 6\delta^{JL}\Gamma^{MN}_{AC})\vert LMN\rangle]\\ &+& \Gamma^{J}_{BC}[\frac{1}{4}\Gamma^{K}_{AC}\vert IK\rangle + \frac{1}{72}(\Gamma^{ILMN}_{AC} + 6\delta^{JL}\Gamma^{MN}_{AC})\vert LMN\rangle]
\end{eqnarray*}
Analogously,
\begin{eqnarray*}
S_{B}S_{A}\vert IJ\rangle &=& \Gamma^{I}_{AC}S_{B}\vert CJ\rangle + \Gamma^{J}_{AC}S_{B}\vert CI\rangle\\
&=& \Gamma^{I}_{AC}[\frac{1}{4}\Gamma^{K}_{BC}\vert JK\rangle + \frac{1}{72}(\Gamma^{JLMN}_{BC} + 6\delta^{JL}\Gamma^{MN}_{BC})\vert LMN\rangle]\\ &+& \Gamma^{J}_{AC}[\frac{1}{4}\Gamma^{K}_{BC}\vert IK\rangle + \frac{1}{72}(\Gamma^{ILMN}_{BC} + 6\delta^{JL}\Gamma^{MN}_{BC})\vert LMN\rangle]
\end{eqnarray*}
Thus, the anticommutator is
\begin{eqnarray*}
\{S_{A}, S_{B}\}\vert IJ\rangle &=& \frac{1}{4}[\Gamma^{I}_{BC}\Gamma^{K}_{AC} + \Gamma^{I}_{AC}\Gamma^{K}_{BC}]\vert JK\rangle + \frac{1}{4}[\Gamma^{J}_{BC}\Gamma^{K}_{AC} + \Gamma^{J}_{AC}\Gamma^{K}_{BC}]\vert IK\rangle\\
&+& \frac{1}{72}[(\Gamma^{I}_{BC}\Gamma^{JLMN}_{AC} + \Gamma^{J}_{BC}\Gamma^{ILMN}_{AC} + \Gamma^{I}_{AC}\Gamma^{JLMN}_{BC} + \Gamma^{J}_{AC}\Gamma^{ILMN}_{BC})\\
&+& 6(\delta^{JL}\Gamma^{I}_{BC}\Gamma^{MN}_{AC} + \delta^{IL}\Gamma^{J}_{BC}\Gamma^{MN}_{AC} + \delta^{JL}\Gamma^{I}_{AC}\Gamma^{MN}_{BC} + \delta^{IL}\Gamma^{J}_{AC}\Gamma^{MN}_{BC})]\vert LMN\rangle\\
&=& \frac{1}{4}(2\delta^{IK}\delta_{AB}\vert JK\rangle + 2\delta^{JK}\delta_{AB}\vert IK\rangle) + \frac{1}{72}[4!(\delta^{I[J}\Gamma^{LMN]}_{BA} + \delta^{J[I}\Gamma^{LMN]}_{BA} \\&+& \delta^{I[J}\Gamma^{LMN]}_{AB} + \delta^{J[I}\Gamma^{LMN]}_{AB}) + 6(\delta^{JL}\Gamma^{I}_{BC}\Gamma^{MN}_{AC} + \delta^{IL}\Gamma^{J}_{BC}\Gamma^{MN}_{AC}\\ &+& \delta^{JL}\Gamma^{I}_{AC}\Gamma^{MN}_{BC} + \delta^{IL}\Gamma^{J}_{AC}\Gamma^{MN}_{BC})]\vert LMN\rangle
\end{eqnarray*}
Now, let us consider the symmetry properties of the $SO(9)$ $\Gamma$-matrices. The 1-form and 4-form are symmetric in their spinor indices, and the 2-form and 3-form are antisymmetric in their spinor indices. Therefore,
\begin{eqnarray}
\{S_{A}, S_{B}\}\vert IJ\rangle &=& \delta_{AB}\vert IJ\rangle + \frac{1}{12}(\delta^{JL}\Gamma^{I}_{BC}\Gamma^{MN}_{AC} + \delta^{IL}\Gamma^{J}_{BC}\Gamma^{MN}_{AC}\nonumber\\
&+& \delta^{JL}\Gamma^{I}_{AC}\Gamma^{MN}_{BC} + \delta^{IL}\Gamma^{J}_{AC}\Gamma^{MN}_{BC})\vert LMN\rangle \nonumber\\
&=& \delta_{AB}\vert IJ\rangle + \frac{1}{12}[\delta^{JL}(\Gamma^{IMN}_{BA} + \delta^{I[M}\Gamma^{N]} + \Gamma^{IMN}_{AB} + \delta^{I[M}\Gamma^{N]}_{AB}) \nonumber\\
&+& \delta^{IL}(\Gamma^{JMN}_{BA} + \delta^{J[M}\Gamma^{N]} + \Gamma^{JMN}_{AB}  + \delta^{J[M}\Gamma^{N}_{AB})]\vert LMN\rangle\nonumber\\
&=& \delta_{AB}\vert IJ\rangle + \frac{1}{12}[\delta^{JL}\delta^{IM}\Gamma^{N} - \delta^{JL}\delta^{IN}\Gamma^{M} + \delta^{JL}\delta^{IM}\Gamma^{N} - \delta^{IN}\delta^{JL}\Gamma^{M}\nonumber\\
&+& \delta^{IL}\delta^{JM}\Gamma^{N} - \delta^{IL}\delta^{JN}\Gamma^{M} + \delta^{IL}\delta^{JM}\Gamma^{N} - \delta^{IL}\delta^{JN}\Gamma^{M}]\vert LMN\rangle \nonumber\\
&=& \delta_{AB}\vert IJ\rangle + \frac{1}{12}[2\Gamma^{N}\vert JIN\rangle - 2\Gamma^{M}\vert JMI\rangle + 2\Gamma^{N}\vert IJN\rangle - 2\Gamma^{M}\vert IMJ\rangle]\nonumber\\
&=& \delta_{AB}\vert IJ\rangle
\end{eqnarray}
as expected. One can similarly show that this algebra is satisfied for the action of $S_{A}$ on the other two fields. Therefore, we have shown that the $D=11$ superparticle spectrum describes the physical degrees of freedom of $D=11$ supergravity.

\section{D=11 pure spinor superparticle}\label{s2.3}
As for the $D=10$ case \cite{Berkovits:2002zk}, we will obtain the $D=11$ pure spinor superparticle from the semi-light-cone gauge Brink-Schwarz-like superparticle \eqref{section2eq72} by introducing a new set of variables $(\Theta^{\alpha}, P_{\alpha})$ and a new symmetry coming from the following first-class constraints:
\begin{equation}
\hat{D}_{\alpha} = D_{\alpha} + \frac{1}{\sqrt{\sqrt{2}P^{+}}}(\Gamma^{m}\Gamma^{+}S)_{\alpha}P_{m}
\end{equation}
where $\{S_{A}, S_{B}\} = \delta_{AB}$ and $D_{\alpha} = P_{\alpha} + (\Gamma^{m})_{\alpha\beta}\Theta^{\beta}P_{m}$. Using the relation $\{\Theta^{\alpha}, P_{\beta}\} = i\delta^{\alpha}_{\beta}$, one can show that $\{D_{\alpha}, D_{\beta}\} = 2i(\Gamma^{m})_{\alpha\beta}P_{m}$. Let us check that these ones are indeed first-class constraints:
\begin{eqnarray}
\{\hat{D}_{\alpha}, \hat{D}_{\beta}\} &=& \{D_{\alpha}, D_{\beta}\} + \frac{1}{\sqrt{2}P^{+}}(\Gamma^{m})_{\alpha\lambda}(\Gamma^{+})^{\lambda A}(\Gamma^{n})_{\beta\delta}(\Gamma^{+})^{\delta A}P_{m}P_{n}\nonumber \\
&=& 2i(\Gamma^{m})_{\alpha\beta}P_{m} - \frac{\sqrt{2}i}{\sqrt{2}P^{+}}\Gamma^{m}_{\alpha\lambda}\Gamma^{n}_{\beta\delta}\Gamma^{+\,\lambda\delta}P_{m}P_{n}\nonumber \\
&=& 2i(\Gamma^{m})_{\alpha\beta}P_{m} - \frac{i}{P^{+}}(\Gamma^{m}\Gamma^{+}\Gamma^{n})_{\alpha\beta}P_{m}P_{n}
\end{eqnarray}
Since $\Gamma^{m}\Gamma^{+} = -\Gamma^{+}\Gamma^{m} + 2\eta^{m+}$, we obtain
\begin{eqnarray}
\{\hat{D}_{\alpha}, \hat{D}_{\beta}\} &=& 2i(\Gamma^{m})_{\alpha\beta}P_{m} - \frac{i}{P^{+}}(\Gamma^{m}\Gamma^{+}\Gamma^{n})_{\alpha\beta}P_{m}P_{n}\nonumber\\
&=& 2i(\Gamma^{m})_{\alpha\beta}P_{m} - \frac{i}{P^{+}}(2\eta^{+m}(\Gamma^{n})_{\alpha\beta})P_{m}P_{n} + \frac{i}{P^{+}}(\Gamma^{+})_{\alpha\beta}P^{2}\nonumber\\
&=& 2i(\Gamma^{m})_{\alpha\beta}P_{m} - 2i(\Gamma^{m})_{\alpha\beta}P_{m} + \frac{i}{P^{+}}(\Gamma^{+})_{\alpha\beta}P^{2}\nonumber\\
&=& \frac{i}{P^{+}}(\Gamma^{+})_{\alpha\beta}P^{2} \label{eq35}
\end{eqnarray}
Thus, the modified Brink-Schwarz-like action will be:
\begin{equation}
S = \int d\tau (P_{m}\partial_{\tau}X^{m} + \frac{i}{2}S^{A}\partial_{\tau}S_{A} + P_{\alpha}\partial_{\tau}\Theta^{\alpha} + f^{\alpha}\hat{D}_{\alpha} + e P^{m}P_{m})
\end{equation}
where we have added the usual kinetic term for the variables $(\Theta^{\alpha}, P_{\alpha})$ and the last term takes into account the new constraint through the fermionic Lagrange multiplier $f^{\alpha}$. The standard BRST method gives us the following gauge-fixed action: 
\begin{equation}\label{eq36}
S = \int d\tau(P_{m}\dot{X}^{m} + \frac{i}{2}S^{A}\partial_{\tau}S_{A} + P_{\alpha}\partial_{\tau}\Theta^{\alpha} + b\partial_{\tau}c + \hat{W}_{\alpha}\partial_{\tau}\hat{\Lambda}^{\alpha} - \frac{1}{2}P^{m}P_{m})
\end{equation}
and the BRST operator
\begin{equation}
\hat{Q} = \hat{\Lambda}^{\alpha}\hat{D}_{\alpha} + cP^{m}P_{m} - \frac{i}{2P^{+}}(\hat{\Lambda}\Gamma^{+}\hat{\Lambda})b,
\end{equation}
once we choose the gauge $e=-\frac{1}{2}$ and $f^{\alpha}=0$. The ghosts $c$, $\hat{\Lambda}^{\alpha}$ come from gauge-fixing the reparametrization symmetry and the new fermionic symmetry, respectively.

\vspace{2mm}
Now we will show that the cohomology of the BRST operator $\hat{Q}$ is equivalent to the cohomology of a BRST operator $Q = \Lambda^{\alpha}D_{\alpha}$, where $\Lambda^{\alpha}$ is a pure spinor. We will show this claim in two steps. First, we show that the $\hat{Q}$-cohomology is equivalent to a $Q\ensuremath{'}$-cohomology, where $Q\ensuremath{'} = \Lambda\ensuremath{'}^{\alpha}\hat{D}_{\alpha}$ and $\Lambda\ensuremath{'}\Gamma^{+}\Lambda\ensuremath{'} = 0$. Finally, we will prove that the $Q\ensuremath{'}$-cohomology is equivalent to the $Q$-cohomology.

\vspace{2mm}
Let us start by defining the operator $Q_{0} = \Lambda_{0}^{\alpha}\hat{D}_{\alpha}$. Notice that when $\Lambda_{0}^{\alpha}$ is equal to $\hat{\Lambda}^{\alpha}$ or $\Lambda\ensuremath{'}^{\alpha}$, $Q_{0}$ becomes the first term of $\hat{Q}$ or $Q\ensuremath{'}$, respectively. Now, let $V$ be a state such that $Q_{0}V = (\Lambda_{0}\Gamma^{+}\Lambda_{0})W$, for some W. Because of the property that $\Lambda\ensuremath{'}^{\alpha}$ satisfies, V is annihilated by $Q\ensuremath{'}$. Also, using \eqref{eq35}, we find that $(Q_{0})^{2} = \frac{i}{2P^{+}}P^{m}P_{m}(\Lambda_{0}\Gamma^{+}\Lambda_{0})$. So, we conclude that $Q_{0}W = \frac{i}{2P^{+}}P^{m}P_{m}V$. We can then show that the state $\hat{V} = V - 2iP^{+}cW$ is annihilated by $\hat{Q}$:
\begin{eqnarray}
\hat{Q}\hat{V} &=& \hat{Q}(V - 2iP^{+}cW)\nonumber\\
 &=& \hat{Q}V - 2iP^{+}(\hat{Q}c)W + 2iP^{+}c(\hat{Q}W) \nonumber\\
 &=& (\hat{\Lambda}\Gamma^{+}\hat{\Lambda})W + c P^{m}P_{m}V - 2iP^{+}(-\frac{i}{2P^{+}})(\hat{\Lambda}\Gamma^{+}\hat{\Lambda})W + 2iP^{+}c(\frac{i}{2P^{+}})P^{m}P_{m}V \nonumber\\
 &=& (\hat{\Lambda}\Gamma^{+}\hat{\Lambda})W + c P^{m}P_{m}V - (\hat{\Lambda}\Gamma^{+}\hat{\Lambda})W - c P^{m}P_{m}V\nonumber\\
 &=& 0
\end{eqnarray}
where we have assumed that $b$ annihilates physical states. Now, let us show that if a state $V$ is BRST-trivial (in the $Q\ensuremath{'}$-cohomology), we can find a state $\hat{V} = V - 2iP^{+}cW$ which is also BRST-trivial (in the $\hat{Q}$-cohomology). Let V be a state which satisfies $V = Q_{0}\Omega + (\Lambda_{0}\Gamma^{+}\Lambda_{0})Y$, for some $Y$. It is clear that if $\Lambda_{0}^{\alpha} = \Lambda\ensuremath{'}^{\alpha}$, we have that $V$ is $Q\ensuremath{'}$-exact and if $\Lambda_{0}^{\alpha} = \hat{\Lambda}^{\alpha}$, we have that the first term of $\hat{Q}\Omega$ is equal to $V - (\hat{\Lambda}\Gamma^{+}\hat{\Lambda})Y$. So we see that
\begin{eqnarray}
\hat{Q}(\Omega + 2iP^{+}cY) &=& \hat{Q}\Omega + 2iP^{+}(\hat{Q}c)Y - 2iP^{+}c(\hat{Q}Y)\nonumber\\
 &=& V - (\hat{\Lambda}\Gamma^{+}\hat{\Lambda})Y + cP^{m}P_{m}\Omega + 2iP^{+}(-\frac{i}{2P^{+}})(\hat{\Lambda}\Gamma^{+}\hat{\Lambda})Y \nonumber\\
 && -
2iP^{+}c(W - \frac{i}{2P^{+}}P^{m}P_{m}\Omega)
\end{eqnarray}
where we used the fact that $b$ annihilates $\Omega$ as well as the result $\hat{Q}_{0}Y =
W - \frac{i}{2P^{+}}P^{m}P_{m}$, which follows from the definition of $V$. Hence, we obtain
\begin{eqnarray}
\hat{Q}(\Omega + 2iP^{+}cY) &=& V - (\hat{\Lambda}\Gamma^{+}\hat{\Lambda})Y + cP^{m}P_{m}\Omega + (\hat{\Lambda}\Gamma^{+}\hat{\Lambda})Y - 2iP^{+}cW - cP^{m}P_{m}\Omega\nonumber\\
 &=& V - 2iP^{+}cW\nonumber\\
 &=& \hat{V}
\end{eqnarray}
Therefore, we have proven that for each state $V$ in the $Q\ensuremath{'}$-cohomology, we can find a state $\hat{V}$ in the $\hat{Q}$-cohomology. If we reverse the arguments given above we can show that any state in the $\hat{Q}$-cohomology corresponds to a state in the $Q\ensuremath{'}$-cohomology.

\vspace{2mm}
The last step is to show that the $Q\ensuremath{'}$-cohomology is equivalent to the $Q$-cohomology. We will do this by using two different approaches.

\subsection{Group decomposition $SO(9) \rightarrow SU(2)\times SU(4)$}
The $SO(10,1)$ spinors $\Lambda^{\alpha}$ and $D_{\alpha}$ can be expressed in terms of their $SO(8)$ components in the following way:
\begin{equation}\label{section2eq500}
\begin{aligned}
\Lambda\ensuremath{'}^{\alpha} &= \begin{pmatrix}
\lambda\ensuremath{'}^{a}\\
\lambda\ensuremath{'}^{\dot{a}}\\
\tilde{\lambda}\ensuremath{'}^{a}\\
\tilde{\lambda}\ensuremath{'}^{\dot{a}}
\end{pmatrix}
\end{aligned}, \hspace{4mm}
\begin{aligned}
D_{\alpha} &= \begin{pmatrix}
\tilde{d}^{a}\\
\tilde{d}^{\dot{a}}\\
-d^{a}\\
-d^{\dot{a}}
\end{pmatrix}
\end{aligned},
\end{equation}
where $a, \dot{a} = 1,\ldots, 8$. The constraint $\Lambda\ensuremath{'}\Gamma^{+}\Lambda\ensuremath{'} = 0$ can be written in terms of these $SO(8)$ components as follows 
\begin{equation}
\lambda\ensuremath{'}^{\dot{a}}\lambda\ensuremath{'}^{\dot{a}} + \tilde{\lambda}\ensuremath{'}^{a}\tilde{\lambda}\ensuremath{'}^{a} = 0
\end{equation}
The particular representation for $SO(10,1)$ $\Gamma$-matrices used in this section is studied in detail in Appendix \ref{appA}. Now, we find it useful to break $SO(9)$ into $SU(2)\times SU(4)$. The branching rule for the spinor representation is  $16 \rightarrow (2,4)+(2,\bar{4})$. Explicit expressions for the $SU(2)\times SU(4)$ components corresponding to $S^{a}, \bar{S}^{\dot{a}}, d^{\dot{a}}, \tilde{d}^{a}, \lambda\ensuremath{'}^{a}, \tilde{\lambda}\ensuremath{'}^{\dot{a}}$ are given below:
\begin{equation}\label{section2eq400}
\begin{aligned}
S_{\hat{A}} &= \frac{1}{\sqrt{2}}(S^{2a} + iS^{2a-1})\\
S_{\bar{\hat{A}}} &= \frac{1}{\sqrt{2}}(S^{2a} - iS^{2a-1})\\
\tilde{S}_{\hat{A}} &= \frac{1}{\sqrt{2}}(\bar{S}^{2\dot{a}} + i\bar{S}^{2\dot{a}-1})\\
\tilde{S}_{\bar{\hat{A}}} &= \frac{1}{\sqrt{2}}(\bar{S}^{2\dot{a}} - i\bar{S}^{2\dot{a}-1})
\end{aligned}\hspace{4mm}
\begin{aligned}
d_{\hat{A}} &= \frac{1}{\sqrt{2}}(d^{2\dot{a}} + id^{2\dot{a}-1})\\
d_{\bar{\hat{A}}} &= \frac{1}{\sqrt{2}}(d^{2\dot{a}} - id^{2\dot{a}-1})\\
\tilde{d}_{\hat{A}} &= \frac{1}{\sqrt{2}}(\tilde{d}^{2a} + i\tilde{d}^{2a-1})\\
\tilde{d}_{\bar{\hat{A}}} &= \frac{1}{\sqrt{2}}(\tilde{d}^{2a} - i\tilde{d}^{2a-1})
\end{aligned}\hspace{4mm}
\begin{aligned}
\lambda\ensuremath{'}_{\hat{A}} &= \frac{1}{\sqrt{2}}(\lambda\ensuremath{'}^{2a} + i\lambda\ensuremath{'}^{2a-1})\\
\lambda\ensuremath{'}_{\bar{\hat{A}}} &= \frac{1}{\sqrt{2}}(\lambda\ensuremath{'}^{2a} - i\lambda\ensuremath{'}^{2a-1})\\
\tilde{\lambda}\ensuremath{'}_{\hat{A}} &= \frac{1}{\sqrt{2}}(\tilde{\lambda}\ensuremath{'}^{2\dot{a}} + i\tilde{\lambda}\ensuremath{'}^{2\dot{a}-1})\\
\tilde{\lambda}\ensuremath{'}_{\bar{\hat{A}}} &= \frac{1}{\sqrt{2}}(\tilde{\lambda}\ensuremath{'}^{2\dot{a}} - i\tilde{\lambda}\ensuremath{'}^{2\dot{a}-1})
\end{aligned}
\end{equation}
where the $SO(9)$ spinor $S_{A}$ has been expressed in terms of its $SO(8)$ components:
\begin{equation}
S_{A} = \begin{pmatrix}
S^{a}\\
\bar{S}^{\dot{a}}
\end{pmatrix}
\end{equation}
and $\hat{A}, \bar{\hat{A}} = 1, \ldots , 4$. It should be clear in \eqref{section2eq400} that fields in the same representation of $SU(4)$ ($4$ or $\bar{4}$) form $SU(2)$ doublets. So, for instance, $\begin{pmatrix}
d_{\hat{A}}\\
\tilde{d}_{\hat{A}}
\end{pmatrix}$ transforms under $(2,4)$, $\begin{pmatrix}
\lambda\ensuremath{'}_{\bar{\hat{A}}}\\
\tilde{\lambda}\ensuremath{'}_{\bar{\hat{A}}}
\end{pmatrix}$ transforms under $(2,\bar{4})$, etc. Notice that the representations $4$ and $\bar{4}$ are defined by the null spinor $(\Gamma^{+}\Lambda\ensuremath{'})^{A}$ by using the fact that one can always choose an $SU(4)$ subgroup under which this spinor is invariant. Therefore we define the antifundamental representation ($\bar{4}$) in such a way that $(\Gamma^{J})_{(\Upsilon\bar{\hat{A}})A}(\Gamma^{+}\Lambda\ensuremath{'})^{A} = 0$, where $J=1,\ldots ,9$, $\Upsilon$ is an $SU(2)$ vector index and $A$ is an $SO(9)$ spinor index. After making the following shifts:
\begin{eqnarray}
S_{\hat{A}} &\rightarrow & S_{\hat{A}} - (\frac{\sqrt{\sqrt{2}}}{2\sqrt{P^{+}}})\tilde{d}_{\hat{A}}\\
\tilde{S}_{\hat{A}} &\rightarrow & \tilde{S}_{\hat{A}} + (\frac{\sqrt{\sqrt{2}}}{2\sqrt{P^{+}}})d_{\hat{A}}
\end{eqnarray}
the operator $Q\ensuremath{'}$ will change by the similarity transformation:
\begin{equation}
Q\ensuremath{'} \rightarrow e^{-[K(S_{\bar{\hat{A}}}\tilde{d}_{\hat{A}} - \tilde{S}_{\bar{\hat{A}}}d_{\hat{A}})]}Q\ensuremath{'}e^{[K(S_{\bar{\hat{A}}}\tilde{d}_{\hat{A}} - \tilde{S}_{\bar{\hat{A}}}d_{\hat{A}})]}
\end{equation}
where $K = -\frac{\sqrt{\sqrt{2}}}{2\sqrt{P{{+}}}}$. This result can be expanded by using the BCH formula:
\begin{equation}\label{eq402}
e^{-Z}Xe^{Z} = X + [X, Z] + \frac{1}{2}[[X, Z], Z] + \ldots
\end{equation}
where $X = Q\ensuremath{'} = \Lambda\ensuremath{'}^{\alpha}\hat{D}_{\alpha}$ and $Z = K(S_{\bar{\hat{A}}}\tilde{d}_{\hat{A}} - \tilde{S}_{\bar{\hat{A}}}d_{\hat{A}})$. The first term is just $Q\ensuremath{'}$,  which can be cast as
\begin{eqnarray}
Q\ensuremath{'} &=& \Lambda\ensuremath{'}^{\alpha}D_{\alpha} + \frac{1}{\sqrt{\sqrt{2}P^{+}}}(\Lambda\ensuremath{'}\Gamma^{m}\Gamma^{+}S) P_{m} \nonumber\\
&=& \lambda\ensuremath{'}_{\dot{a}}\tilde{d}_{\dot{a}} + \lambda\ensuremath{'}_{\bar{\hat{A}}}\tilde{d}_{\hat{A}} + \lambda\ensuremath{'}_{\hat{A}}\tilde{d}_{\bar{\hat{A}}} - \tilde{\lambda}\ensuremath{'}_{a}d_{a} - \tilde{\lambda}\ensuremath{'}_{\bar{\hat{A}}}d_{\hat{A}} - \tilde{\lambda}\ensuremath{'}_{\hat{A}}d_{\bar{\hat{A}}} + \sqrt{2\sqrt{2}P^{+}}\lambda\ensuremath{'}_{\bar{\hat{A}}}S_{\hat{A}} + \sqrt{2\sqrt{2}P^{+}}\lambda\ensuremath{'}_{\hat{A}}S_{\bar{\hat{A}}}\nonumber\\
&& + \sqrt{2\sqrt{2}P^{+}}\tilde{\lambda}\ensuremath{'}_{\bar{\hat{A}}}\tilde{S}_{\hat{A}} + \sqrt{2\sqrt{2}P^{+}}\tilde{\lambda}\ensuremath{'}_{\hat{A}}\tilde{S}_{\bar{\hat{A}}}+ \sqrt{\frac{\sqrt{2}}{P^{+}}}[\lambda\ensuremath{'}^{\dot{a}}(\sigma^{\hat{i}})_{\dot{a}\hat{A}}S_{\bar{\hat{A}}}P_{\hat{i}} - \tilde{\lambda}\ensuremath{'}^{a}(\sigma^{\hat{i}})_{\hat{A}a}\tilde{S}_{\bar{\hat{A}}}P_{\hat{i}}]\nonumber\\
&& + \sqrt{\frac{\sqrt{2}}{P^{+}}}[\tilde{\lambda}\ensuremath{'}_{\hat{A}}S_{\bar{\hat{A}}} + \lambda\ensuremath{'}_{\hat{A}}\tilde{S}_{\bar{\hat{A}}}]P^{11}\nonumber\\\label{eq107}
\end{eqnarray}
To find the second term in \eqref{eq402}, it is necessary to compute the $SU(4)$ (anti)commutation relations, which can be obtained from the $SO(8)$ relations:
\begin{equation}
\begin{aligned}
\{\tilde{d}^{a}, \tilde{d}^{b}\} &= -2\sqrt{2}\delta^{ab}P^{+} \hspace{4mm}, \\ 
\{d^{\dot{a}}, d^{\dot{b}}\} &= -2\sqrt{2}\delta^{\dot{a}\dot{b}}P^{+} \hspace{4mm} ,\\
\{d^{a}, d^{b}\} &= -2\sqrt{2}\delta^{ab}P^{-} \hspace{4mm}, \\ 
\{\tilde{d}^{\dot{a}}, \tilde{d}^{\dot{b}}\} &= -2\sqrt{2}\delta^{\dot{a}\dot{b}}P^{-} \hspace{4mm} ,
\end{aligned}\hspace{4mm}
\begin{aligned}
\{d^{a}, \tilde{d}^{b}\} &= 2\delta^{ab}P^{11}\\ 
\{\tilde{d}^{\dot{a}}, d^{\dot{b}}\} &= 2\delta^{\dot{a}\dot{b}}P^{11} \\
\{d^{a}, d^{\dot{b}}\} &= 2(\sigma^{\hat{i}})^{a\dot{b}}P^{\hat{i}}  \\ 
\{\tilde{d}^{\dot{a}}, d^{b}\} &= -2(\sigma^{\hat{i}})^{\dot{a}b}P^{\hat{i}} 
\end{aligned}
\end{equation}
Using these, together with \eqref{section2eq400}, leads us to the following $SU(4)$ relations:
\begin{equation}
\begin{aligned}
\{S_{\hat{A}}, S_{\bar{\hat{A}}}\} &= \eta_{\hat{A}\bar{\hat{A}}} \hspace{16.3mm}, \\ 
\{\tilde{S}_{\hat{A}}, \tilde{S}_{\bar{\hat{A}}}\} &= \eta_{\hat{A}\bar{\hat{A}}} \hspace{16.3mm} ,\\
\{d_{\hat{A}}, d_{\bar{\hat{A}}}\} &= -2\sqrt{2}\eta_{\hat{A}\bar{\hat{A}}}P^{+}\hspace{1mm}, \\
\{\tilde{d}_{\hat{A}}, \tilde{d}_{\bar{\hat{A}}}\} &= -2\sqrt{2}\eta_{\hat{A}\bar{\hat{A}}}P^{+}\hspace{1mm} ,
\end{aligned}
\begin{aligned}
\{ \tilde{d}_{\dot{a}}, d_{\hat{A}}\} &= 2\delta_{\dot{a}A}P^{11} \hspace{4mm} ,\\
\{ \tilde{d}_{\dot{a}}, d_{\bar{\hat{A}}}\} &= 2\delta_{\dot{a}\bar{A}}P^{11}\hspace{4mm} ,\\
\{ d_{a}, \tilde{d}_{\hat{A}}\} &= 2\delta_{a A}P^{11} \hspace{4mm} ,\\
\{ d_{a}, \tilde{d}_{\bar{\hat{A}}}\} &= 2\delta_{a \bar{A}}P^{11}\hspace{4mm} ,
\end{aligned}
\begin{aligned}
\{d_{a}, d_{\hat{A}}\} &=2(\sigma^{\hat{i}})_{a \hat{A}}P^{\hat{i}}\\
\{d_{a}, d_{\bar{\hat{A}}}\} &=2(\sigma^{\hat{i}})_{a \bar{\hat{A}}}P^{\hat{i}}\\
\{\tilde{d}_{\dot{a}}, \tilde{d}_{\hat{A}}\} &=-2(\sigma^{\hat{i}})_{\dot{a} \hat{A}}P^{\hat{i}}\\
\{\tilde{d}_{\dot{a}}, \tilde{d}_{\bar{\hat{A}}}\} &=-2(\sigma^{\hat{i}})_{\dot{a} \bar{\hat{A}}}P^{\hat{i}}
\end{aligned}
\end{equation}

Hence, we get
\begin{eqnarray}
K\left[Q\ensuremath{'}, S_{\bar{\hat{A}}}\tilde{d}_{\hat{A}} - \tilde{S}_{\bar{\hat{A}}}d_{\hat{A}}\right] &=& 
 -\sqrt{2\sqrt{2}P^{+}}\lambda\ensuremath{'}_{\hat{A}}S_{\bar{\hat{A}}} -\sqrt{2\sqrt{2}P^{+}}\tilde{\lambda}\ensuremath{'}_{\hat{A}}\tilde{S}_{\bar{\hat{A}}} - \lambda\ensuremath{'}_{\bar{\hat{A}}}\tilde{d}_{\hat{A}} + \tilde{\lambda}\ensuremath{'}_{\bar{\hat{A}}} d_{\hat{A}}\nonumber\\
 && -\sqrt{\frac{\sqrt{2}}{P^{+}}}\tilde{\lambda}\ensuremath{'}_{\hat{A}}S_{\bar{\hat{A}}}P^{11} -\sqrt{\frac{\sqrt{2}}{P^{+}}}\lambda\ensuremath{'}_{\hat{A}}\tilde{S}_{\bar{\hat{A}}}P^{11} -\sqrt{\frac{\sqrt{2}}{P^{+}}}\lambda\ensuremath{'}^{\dot{a}}(\sigma^{\hat{i}})_{\dot{a}\hat{A}}S_{\bar{\hat{A}}}P_{\hat{i}}\nonumber\\
 && + \sqrt{\frac{\sqrt{2}}{P^{+}}}\tilde{\lambda}\ensuremath{'}^{a}(\sigma^{\hat{i}})_{\hat{A} a}\tilde{S}_{\bar{A}}P_{\hat{i}}\label{eq106}
\end{eqnarray}
From this expression it is easy to see that:
\begin{equation}
[[Q\ensuremath{'}, Z], Z] = 0
\end{equation}
and so the third term and all of the other ones in \eqref{eq402} (which were represented by $\ldots$) vanish.

\vspace{2mm}
Therefore, we have arrived at the following result:
\begin{equation}
Q\ensuremath{'} \rightarrow \lambda\ensuremath{'}_{\dot{a}}\tilde{d}_{\dot{a}} + \lambda\ensuremath{'}_{\hat{A}}\tilde{d}_{\bar{\hat{A}}} - \tilde{\lambda}\ensuremath{'}_{a}d_{a} - \tilde{\lambda}\ensuremath{'}_{\hat{A}}d_{\bar{\hat{A}}} + \sqrt{2\sqrt{2}P^{+}}\lambda\ensuremath{'}_{\bar{\hat{A}}}S_{\hat{A}} + \sqrt{2\sqrt{2}P^{+}}\tilde{\lambda}\ensuremath{'}_{\bar{\hat{A}}}\tilde{S}_{\hat{A}}
\end{equation}
where $\lambda\ensuremath{'}^{\dot{a}}$ and $\tilde{\lambda}\ensuremath{'}^{a}$ satisfy the relation $\lambda\ensuremath^{\dot{a}}\lambda\ensuremath^{\dot{a}} + \tilde{\lambda}\ensuremath{'}^{a}\tilde{\lambda}\ensuremath{'}^{a} = 0$. If we define a spinor $\Lambda^{\alpha} = [\lambda_{\hat{A}}, \lambda_{\bar{\hat{A}}}, \lambda_{\dot{a}}, \tilde{\lambda}_{a}, \tilde{\lambda}_{\hat{A}}, \tilde{\lambda}_{\bar{\hat{A}}}] = [\lambda\ensuremath{'}_{\hat{A}}, 0, \lambda\ensuremath{'}_{\dot{a}}, \tilde{\lambda}\ensuremath{'}_{a}, \tilde{\lambda}\ensuremath{'}_{\hat{A}}, 0]$, the previous expression can be written as
\begin{equation}
Q\ensuremath{'} \rightarrow \Lambda^{\alpha}D_{\alpha} + \sqrt{2\sqrt{2}P^{+}}\lambda\ensuremath{'}_{\bar{\hat{A}}}S_{\hat{A}} + \sqrt{2\sqrt{2}P^{+}}\tilde{\lambda}\ensuremath{'}_{\bar{\hat{A}}}\tilde{S}_{\hat{A}}
\end{equation} 
Furthermore, after using the quartet argument \cite{Kugo:1979gm}, it is clear that the $Q\ensuremath{'}$-cohomology is equivalent to the $Q$-cohomology\footnote{That is, the states in the Hilbert space will be independent of $\lambda\ensuremath{'}_{\bar{\hat{A}}}$, $S_{\hat{A}}$, $\tilde{\lambda}\ensuremath{'}_{\bar{\hat{A}}}$, $\tilde{S}_{\hat{A}}$, and their respective conjugate momenta $w\ensuremath{'}_{\hat{A}}$, $S_{\bar{\hat{A}}}$, $\tilde{w}\ensuremath{'}_{\hat{A}}$, $\tilde{S}_{\bar{\hat{A}}}$.}:
\begin{equation}
Q\ensuremath{'} \rightarrow Q = \Lambda^{\alpha}D_{\alpha}
\end{equation}
where $\Lambda^{\alpha}$ is a pure spinor.

\subsection{Group decomposition $SO(9) \rightarrow U(1) \times SO(7)$}
We will express $SO(10,1)$ spinors in terms of their $SO(3,1) \times SO(7)$ components:
\begin{equation}
\chi^{\alpha} = \begin{pmatrix}
\chi^{\pm\pm 0}\\
\chi^{\pm\pm i}
\end{pmatrix}
\end{equation}
where $i = 1, \ldots, 7$. The notation $\pm$ and the representation of the $SO(10,1)$ gamma matrices used here are explained in detail in Appendix \ref{appB}. Using this notation, we can express the anticommutation relations studied above in the $SO(3,1) \times SO(7)$ language:
\begin{equation}\label{eq405}
\begin{aligned}
\{D^{--0}, D^{-+0}\} &= 2\sqrt{2}P^{+}\hspace{9.2mm},\\ 
\{D^{--i}, D^{-+j}\} &= -2\sqrt{2}P^{+}\delta^{ij}\hspace{2mm},\\
\{D^{++0}, D^{+-0}\} &= 2\sqrt{2}P^{-}\hspace{9.4mm},\\ 
\{D^{++i}, D^{+-j}\} &= -2\sqrt{2}P^{-}\delta^{ij}\hspace{2.1mm},
\end{aligned}
\begin{aligned}
\{D^{--0}, D^{+-0}\} &= 2\sqrt{2}P^{2+3i}\hspace{9.2mm},\\
\{D^{--i}, D^{+-j}\} &= -2\sqrt{2}P^{2+3i}\delta^{ij}\hspace{2mm},\\
\{D^{++0}, D^{-+0}\} &= 2\sqrt{2}P^{2-3i}\hspace{9.4mm},\\
\{D^{++i}, D^{-+j}\} &= -2\sqrt{2}P^{2-3i}\delta^{ij}\hspace{2.1mm},
\end{aligned}
\begin{aligned}
\{D^{--i}, D^{++0}\} &= -2P^{i}\\
\{D^{++i}, D^{--0}\} &= 2P^{i}\\
\{D^{-+i}, D^{+-0}\} &= 2P^{i}\\
\{D^{+-i}, D^{-+0}\} &= -2P^{i}\\
\end{aligned}
\end{equation}
and also
\begin{equation}
\begin{aligned}
\{S^{--0}, S^{-+0}\} &= -1 \\
\{S^{--i}, S^{-+j}\} &= \delta^{ij} 
\end{aligned}
\end{equation}
and any other anticommutator vanishes. Under a certain subgroup $U(1)\times SO(7) \subset SO(9)$, the null spinor $(\Gamma^{+}\Lambda\ensuremath{'})^{A}$ will be invariant up to rescaling. This subgroup is chosen in such a way that $(\Gamma^{2+3i})_{(-0)A}(\Gamma^{+}\Lambda\ensuremath{'})^{A}$ 
$ = (\Gamma^{j})_{(-0)A}(\Gamma^{+}\Lambda\ensuremath{'})^{A} = 0$ 
$= 0$, where we have dropped out the minus sign associated to the first $U(1)$ charge, and $j=1,\ldots ,7$.
\vspace{2mm}
The BRST operator $Q\ensuremath{'}$ can be expressed in terms of $SO(3,1)\times SO(7)$ variables:
\begin{eqnarray}
Q\ensuremath{'} &=& \Lambda\ensuremath{'}^{\alpha}D_{\alpha} + \frac{1}{\sqrt{\sqrt{2}P^{+}}}[-(\Lambda\ensuremath{'}\Gamma^{-}\Gamma^{+}S)P^{+} +  (\Lambda\ensuremath{'}\Gamma^{2-3i}\Gamma^{+}S)P^{2+3i} +  (\Lambda\ensuremath{'}\Gamma^{2+3i}\Gamma^{+}S)P^{2-3i}\nonumber\\
&& + (\Lambda\ensuremath{'}\Gamma^{j}\Gamma^{+}S)P^{j}]\nonumber\\
&=& \Lambda\ensuremath{'}^{\alpha}D_{\alpha} - \frac{2\sqrt{P^{+}}}{\sqrt{\sqrt{2}}}(\Lambda\ensuremath{'}^{+-0}S^{-+0} - \Lambda\ensuremath{'}^{+-i}S^{-+i} + \Lambda\ensuremath{'}^{++0}S^{--0} - \Lambda\ensuremath{'}^{++i}S^{--i})\nonumber\\
&& +\sqrt{\frac{2\sqrt{2}}{P^{+}}}(\Lambda^{-+0}S^{-+0})P^{2+3i} - \sqrt{\frac{2\sqrt{2}}{P^{+}}}(\Lambda^{--j}S^{--j})P^{2-3i}-\sqrt{\frac{\sqrt{2}}{P^{+}}}(\Lambda^{--j}S^{-+0})P^{j}\nonumber\\
&& -\sqrt{\frac{\sqrt{2}}{P^{+}}}(\Lambda^{-+0}S^{--j})P^{j}\nonumber
\end{eqnarray}

After performing the following shifts:
\begin{eqnarray}
S^{--0} \rightarrow S^{--0} - \frac{\sqrt{\sqrt{2}}}{2\sqrt{P^{+}}}D^{--0}\\
S^{-+i} \rightarrow S^{-+i} - \frac{\sqrt{\sqrt{2}}}{2\sqrt{P^{+}}}D^{-+i}
\end{eqnarray}
the BRST operator will change by
\begin{equation}
Q\ensuremath{'} \rightarrow e^{-Z}Q\ensuremath{'}e^{Z},
\end{equation}
where $Z = \frac{\sqrt{\sqrt{2}}}{2\sqrt{P^{+}}}(S^{-+0}D^{--0} + S^{--i}D^{-+i})$. The BCH formula \eqref{eq402} gives us the result
\begin{eqnarray}
Q\ensuremath{'} &\rightarrow & Q\ensuremath{'} + [Q\ensuremath{'}, Z] + \frac{1}{2}[[Q\ensuremath{'}, Z], Z] + \ldots\nonumber\\
&\rightarrow & -\Lambda\ensuremath{'}^{++0}D^{--0} + \Lambda\ensuremath{'}^{++i}D^{--i} + \Lambda\ensuremath{'}^{--0}D^{++0} - \Lambda\ensuremath{'}^{--i}D^{++i} -\Lambda\ensuremath{'}^{+-0}D^{-+0} + \Lambda\ensuremath{'}^{+-i}D^{-+i} \nonumber\\
&&+ \Lambda\ensuremath{'}^{-+0}D^{+-0} - \Lambda\ensuremath{'}^{-+i}D^{+-i} - \frac{2\sqrt{P^{+}}}{\sqrt{\sqrt{2}}}(\Lambda\ensuremath{'}^{+-0}S^{-+0} - \Lambda\ensuremath{'}^{++i}S^{--i} + \Lambda\ensuremath{'}^{++0}S^{--0} \nonumber\\
&&- \Lambda\ensuremath{'}^{+-i}S^{-+i}) +\sqrt{\frac{2\sqrt{2}}{P^{+}}}(\Lambda^{-+0}S^{-+0})P^{2+3i} - \sqrt{\frac{2\sqrt{2}}{P^{+}}}(\Lambda^{--j}S^{--j})P^{2-3i}\nonumber\\
&&-\sqrt{\frac{\sqrt{2}}{P^{+}}}(\Lambda^{--j}S^{-+0})P^{j}-\sqrt{\frac{\sqrt{2}}{P^{+}}}(\Lambda^{-+0}S^{--j})P^{j}+ \frac{\sqrt{\sqrt{2}}}{2\sqrt{P^{+}}}[2\sqrt{2}\Lambda\ensuremath{'}^{+-0}S^{-+0}P^{+}\nonumber\\
&& + 2\Lambda\ensuremath{'}^{--i}S^{-+0}P^{i} - 2\sqrt{2}\Lambda\ensuremath{'}^{-+0}S^{-+0}P^{2+3i}  - 2\sqrt{2}P^{+}\Lambda\ensuremath{'}^{++i}S^{--i} + 2\sqrt{2}P^{2-3i}\Lambda\ensuremath{'}^{--i}S^{--i} \nonumber\\
&&+ 2\Lambda\ensuremath{'}^{-+0}S^{--i}P^{i} - \Lambda\ensuremath{'}^{++0}D^{--0} +\Lambda\ensuremath{'}^{+-i}D^{-+i}]+\ldots
\end{eqnarray}
where the ellipsis represents $\frac{1}{2!}[[Q\ensuremath{'}, Z], Z] + \frac{1}{3!}[[[Q\ensuremath{'}, Z], Z], Z] + \ldots$. However, these terms vanish because $[[Q\ensuremath{'}, Z], Z] = 0$, as can be seen from eqn. \eqref{eq405}. Thus, we are left with
\begin{eqnarray}
Q\ensuremath{'}&\rightarrow & \Lambda\ensuremath{'}^{++i}D^{--i} + \Lambda\ensuremath{'}^{--0}D^{++0} - \Lambda\ensuremath{'}^{--i}D^{++i} - \Lambda\ensuremath{'}^{+-0}D^{-+0} + \Lambda\ensuremath{'}^{-+0}D^{+-0} - \Lambda\ensuremath{'}^{-+i}D^{+-i}\nonumber\\
&& -\sqrt{2\sqrt{2}P^{+}}(\Lambda\ensuremath{'}^{++0}S^{--0} - \Lambda\ensuremath{'}^{+-i}S^{-+i})
\end{eqnarray}
If we define a spinor $\Lambda^{\alpha} = [\Lambda^{++0}, \Lambda^{++i}, \Lambda^{--0}, \Lambda^{--i}, \Lambda^{+-0}, \Lambda^{+-i}, \Lambda^{-+0}, \Lambda^{-+i}]$ $=$ $[0$, $\Lambda\ensuremath{'}^{++i}$, $\Lambda\ensuremath{'}^{--0}$, $\Lambda\ensuremath{'}^{--i}$, $\Lambda\ensuremath{'}^{+-0}$ , $0$, $\Lambda\ensuremath{'}^{-+0}$, $\Lambda\ensuremath{'}^{-+i}]$ where $\Lambda\ensuremath{'}\Gamma^{+}\Lambda\ensuremath{'} = 0$, the resulting BRST operator can be written as
\begin{equation}\label{eq410}
Q\ensuremath{'} \rightarrow \Lambda^{\alpha}D_{\alpha} - \sqrt{2\sqrt{2}P^{+}}(\Lambda\ensuremath{'}^{++0}S^{-+0} - \Lambda\ensuremath{'}^{+-i}S^{--i})
\end{equation}
From this last expression, we can conclude that the space of physical states will not depend on the canonical variables $S^{-+0}$, $S^{--i}$, $\Lambda\ensuremath{'}^{++0}$, $\Lambda\ensuremath{'}^{+-i}$, and their respective conjugate momenta $S^{--0}$, $S^{-+i}$, $W\ensuremath{'}^{--0}$, $W\ensuremath{'}^{-+i}$. Therefore the BRST operator takes the simple form
\begin{equation}
Q\ensuremath{'} \rightarrow Q = \Lambda^{\alpha}D_{\alpha}
\end{equation}
where $\Lambda^{\alpha}$ is a $D=11$ pure spinor. Therefore, we have proved that the modified Brink-Schwarz-like superparticle \eqref{eq36} is equivalent to the theory described by the manifestly Lorentz covariant action
\begin{equation}
S = \int d\tau (P_{m}\partial_{\tau}X^{m} + P_{\alpha}\partial_{\tau}\Theta^{\alpha} + W_{\alpha}\partial_{\tau}\Lambda^{\alpha} - \frac{1}{2}P^{m}P_{m})
\end{equation}
and the BRST operator $Q = \Lambda^{\alpha}D_{\alpha}$, where $\Lambda\Gamma^{m}\Lambda = 0$. This theory is the $D=11$ pure spinor superparticle.

\section{Light-cone gauge analysis of the pure spinor cohomology}\label{s2.4}
In this section it will be shown that the pure spinor physical condition implies light-cone gauge equations of motion for $D=11$ linearized supergravity in $D=9$ superspace, which coincide with those found in \cite{Green:1999by}. To see this, let us write $Q$ in $SO(9)$ notation (see Appendix \ref{appA}):
\begin{equation}
Q = \Lambda^{A}D_{A} + \bar{\Lambda}^{A}\bar{D}_{A}
\end{equation}
and define the operator
\begin{equation}
R = \frac{P^{I}\bar{N}_{I}}{\sqrt{2}P^{+}}
\end{equation}
where $I=1,\ldots,8,11$ and $\bar{N}^{I} = \Lambda^{A}\Gamma^{I}_{AB}\bar{W}^{B}$. The corresponding similarity transformation generated by this operator is
\begin{eqnarray}
\tilde{Q} &=& e^{-R}Qe^{R} \nonumber\\
&=& Q + [Q,R] + \frac{1}{2}[[Q,R],R] + \ldots \nonumber\\
&=& \Lambda^{A}D_{A} + \bar{\Lambda}^{A}\bar{D}_{A} + \frac{i}{\sqrt{2}P^{+}}P_{I}(\Lambda^{A}\Gamma^{I}_{AB}\bar{D}^{B})\nonumber\\
&=& \Lambda^{A}[D_{A} + \frac{i}{\sqrt{2}P^{+}}P_{I}(\Gamma^{I}\bar{D})_{A}] + \bar{\Lambda}^{A}\bar{D}_{A}\nonumber\\
&=& \Lambda^{A}G_{A} + \bar{\Lambda}^{A}\bar{D}_{A}
\end{eqnarray}
where $G_{A}$ is defined by the relation
\begin{eqnarray}
G_{A} &=& D_{A} + \frac{i}{\sqrt{2}P^{+}}P_{I}(\Gamma^{I}\bar{D})_{A}\nonumber\\
&=& D_{A} + \frac{1}{\sqrt{2}P^{+}}P_{\hat{i}}(\gamma^{9}\gamma^{\hat{i}}\bar{D})_{A} - \frac{1}{\sqrt{2}P^{+}}P_{11}\bar{D}_{A}
\end{eqnarray}
where $\hat{i}$ is an $SO(8)$ vector index. This object can be written in the compact form
\begin{equation}\label{eq502}
G_{A} = \frac{1}{2P^{+}}P^{m}(\Gamma^{+}\Gamma_{m}D)_{A}
\end{equation}
It will be useful to keep in mind the following $SO(9)$ relations which can be deduced from \eqref{eq501}, \eqref{eq61}:
\begin{equation}\label{section2eq503}
\begin{aligned}
\{D_{A}, D_{B}\} &= -2\sqrt{2}\,\delta_{AB}P^{-}\hspace{2mm},\\
\{\bar{D}_{A}, \bar{D}_{B}\} &= -2\sqrt{2}\,\delta_{AB}P^{+}\hspace{2mm},
\end{aligned}
\begin{aligned}
\{D_{A}, \bar{D}_{B}\} &= 2[(\gamma^{9}\gamma^{\hat{i}})_{AB}P_{\hat{i}} - \delta_{AB}P_{11}]\\
\{\bar{D}_{A}, D_{B}\} &= 2[-(\gamma^{9}\gamma^{\hat{i}})_{AB}P_{\hat{i}} - \delta_{AB}P_{11}]
\end{aligned}
\end{equation}
where $D_{A}, \bar{D}_{A}$ are given by
\begin{eqnarray}
D_{A} &=& P_{A} + \sqrt{2}i\Theta_{A}P^{-} - i(\gamma^{9}\gamma^{\hat{i}}\bar{\Theta})_{A}P_{\hat{i}} + i\bar{\Theta}_{A}P_{11}\\
\bar{D}_{A} &=& \bar{P}_{A} + \sqrt{2}i\bar{\Theta}_{A}P^{+} + i(\gamma^{9}\gamma^{\hat{i}}\Theta)_{A}P_{\hat{i}} + i\Theta_{A}P_{11}
\end{eqnarray}
or in a more compact form
\begin{eqnarray}
D_{A} &=& P_{A} + \sqrt{2}i\Theta_{A}P^{-} + \Gamma^{I}_{AB}\bar{\Theta}^{B}P_{I}\\
\bar{D}_{A} &=& \bar{P}_{A} + \sqrt{2}i\bar{\Theta}_{A}P^{+} + \Gamma^{I}_{AB}\Theta^{B}P_{I}
\end{eqnarray}
where $\Gamma^{I}_{A\bar{B}} = (-i(\gamma^{9}\gamma^{\hat{i}})_{AB},i\delta_{AB})$, $\Gamma^{I}_{\bar{A}B} = (i(\gamma^{9}\gamma^{\hat{i}})_{AB},i\delta_{AB})$.
 Using eqns. \eqref{eq502}, \eqref{section2eq503} one can show that
\begin{eqnarray}
\{G_{A}, \bar{D}_{B}\} &=& 0\label{eq411}\\
\{G_{A}, G_{B}\} &=& \frac{\sqrt{2}}{P^{+}}(P^{m}P_{m})\delta_{AB}
\end{eqnarray}
Notice that the nilpotency of $\tilde{Q}$ no longer requires the validity of the $SO(9)$ pure spinor constraint $\Lambda^{A}\Gamma^{I}_{AB}\bar{\Lambda}^{B} = 0$ as can be seen from \eqref{eq411}. A further similarity transformation induced by the operator
\begin{equation}
\hat{R} = -\frac{1}{\sqrt{2}P^{+}}(\Theta^{A}\Gamma^{I}_{AB}\bar{P}^{B})P_{I}
\end{equation}
will transform the operators $\bar{D}_{A}$, $G_{A}$ into
\begin{eqnarray}
\hat{\bar{D}}_{A} &=& \bar{P}_{A} + \sqrt{2}i\bar{\Theta}_{A}P^{+}\\
\hat{G}_{A} &=& P_{A} - \frac{i}{\sqrt{2}P^{+}}(P^{m}P_{m})\Theta_{A}
\end{eqnarray}
Hence the pure spinor BRST operator will take the form
\begin{equation}
\tilde{\tilde{Q}} = \Lambda^{A}\hat{G}_{A} + \bar{\Lambda}^{A}\hat{\bar{D}}_{A}
\end{equation}
The supersymmetry invariance of this operator follows from the supersymmetry invariance of $\hat{G}_{A}$ and $\hat{\bar{D}}_{A}$ under the operators
\begin{eqnarray}
\hat{\bar{Q}}_{A} &=& \bar{P}_{A} - \sqrt{2}i\bar{\Theta}_{A}P^{+}\\
\hat{Q}_{A} &=& P_{A} + \frac{i}{\sqrt{2}P^{+}}(P^{m}P_{m})\Theta_{A} - \frac{i}{\sqrt{2}P^{+}}P_{I}\Gamma^{I}_{AD}\hat{\bar{Q}}_{D} 
\end{eqnarray}
which are the $\tilde{R}$-transformed versions of the supersymmetry generators
\begin{eqnarray}
Q_{A} &=& P_{A} - \sqrt{2}iP^{-}\Theta_{A} + i(\gamma^{9}\gamma^{\hat{i}}\bar{\Theta})_{A}P_{\hat{i}} - i\bar{\Theta}_{A}P_{11}\\
\bar{Q}_{A} &=& \bar{P}_{A} - \sqrt{2}iP^{+}\bar{\Theta}_{A} - (\gamma^{9}\gamma^{\hat{i}}\Theta)_{A}P_{\hat{i}} - i\Theta_{A}P_{11}
\end{eqnarray} 
\subsection{Light-cone gauge equations of motion}
The physical fields are contained in the ghost number 3 superfield $V = \Lambda^{\alpha}\Lambda^{\beta}\Lambda^{\sigma}C_{\alpha\beta\sigma}$ \cite{Berkovits:2002uc}. This superfield can be written in $SO(9)$ notation as
\begin{eqnarray}
V &=& \Lambda^{A}\Lambda^{B}\Lambda^{C}C_{(+A)(+B)(+C)} + 3\bar{\Lambda}^{A}\Lambda^{B}\Lambda^{C}C_{(-A)(+B)(+C)}\nonumber\\
&& + 3\bar{\Lambda}^{A}\bar{\Lambda}^{B}\Lambda^{C}C_{(-A)(-B)(+C)} + \bar{\Lambda}^{A}\bar{\Lambda}^{B}\bar{\Lambda}^{C}C_{(-A)(-B)(-C)},\label{eq504}
\end{eqnarray}
where the signs $\pm$ come from the splitting $SO(10,1)\rightarrow SO(1,1)\times SO(9)$. The use of the gauge transformation $\delta V = \tilde{\tilde{Q}}\Omega$, with $\Omega$ being an arbitrary ghost number 2 superfield, allows us to cancel out the last three terms in \eqref{eq504}:
\begin{eqnarray}
\tilde{\tilde{Q}}\Omega &=& \Lambda^{A}\Lambda^{B}\Lambda^{C}\hat{G}_{A}\Omega_{(+B)(+C)} + 2\Lambda^{A}\bar{\Lambda}^{B}\Lambda^{C}\hat{G}_{A}\Omega_{(-B)(+C)} + \Lambda^{A}\bar{\Lambda}^{B}\bar{\Lambda}^{C}\hat{G}_{A}\Omega_{(-B)(-C)}\nonumber\\
&& + \bar{\Lambda}^{A}\Lambda^{B}\Lambda^{C}\hat{\bar{D}}_{A}\Omega_{(+B)(+C)} + 2\bar{\Lambda}^{A}\bar{\Lambda}^{B}\Lambda^{C}\hat{\bar{D}}_{A}\Omega_{(-B)(+C)} + \bar{\Lambda}^{A}\bar{\Lambda}^{B}\bar{\Lambda}^{C}\hat{\bar{D}}_{A}\Omega_{(-B)(-C)},\nonumber
\end{eqnarray}
after conveniently choosing $\Omega_{(-B)(-C)}$, $\Omega_{(+B)(-C)}$, $\Omega_{(+B)(+C)}$. Therefore we are left with 
\begin{equation}\label{eq512}
V = \Lambda^{A}\Lambda^{B}\Lambda^{C}C_{ABC},
\end{equation}
where we have dropped the $SO(1,1)$ index for convenience. The $\tilde{\tilde{Q}}$-closedness condition for $V$ implies the following equations for $C_{BCD}$:
\begin{eqnarray}
\hat{\bar{D}}_{A}C_{BCD} &=& (\Gamma^{J})_{A(B}C_{|J|CD)} + \delta_{(BC}\chi_{D)A}\label{eq510}\\
\hat{G}_{(A}C_{BCD)} &=& \delta_{(AB}\xi_{CD)} + (\Gamma^{JK})_{A(B}C_{|JK|CD)} + (\Gamma^{JKL})_{A(B}C_{|JKL|CD)},\label{eq511}
\end{eqnarray}
where $\chi_{DA}$, $\xi_{CD}$, $C_{JCD}$, $C_{JKCD}$, $C_{JKLCD}$ are $SO(9)$ p-form-bispinors. Each of these possesses a certain symmetry determined by \eqref{eq510}, \eqref{eq511}. To find the physical spectrum and the corresponding equations of motion, we should solve these equations subject to the constraints:
\begin{eqnarray}
\{\hat{\bar{D}}_{A}, \hat{\bar{D}}_{B}\} &=& -2\sqrt{2}P^{+}\delta_{AB}\label{eq522}\\
\{\hat{G}_{A}, \hat{G}_{B}\} &=& \frac{\sqrt{2}}{P^{+}}(P^{m}P_{m})\delta_{AB}\label{eq521}
\end{eqnarray}

A way to solve this constrained system of equations is by using the supersymmetry algebra \eqref{eq522}.
Let us choose the only non-zero component of the spinor $\Lambda^{A}$ to be $\Lambda^{+0}$. This choice will imply $\bar{\Lambda}^{-i} = \bar{\Lambda}^{+0} = 0$, where $i$ is the usual $SO(7)$ vector index. With these constraints, the only $\hat{\bar{D}}_{A}$ that act non-trivially on $C_{(+0)(+0)(+0)}$ are $\hat{\bar{D}}_{-i}$ and $\hat{\bar{D}}_{+0}$. Therefore, we will have $2^{8}$ states in $C_{(+0)(+0)(+0)}$: 128 bosonic and 128 fermionic states. The other componens of $C_{ABC}$ can be shown to be related to $C_{(+0)(+0)(+0)}$ by $SO(9)$ rotations (see Appendix \ref{appC}) given by the operator
\begin{equation}
R^{IJ} = \frac{1}{\sqrt{8\sqrt{2}P^{+}}}(\hat{\bar{D}}\Gamma^{IJ}\hat{\bar{D}}),
\end{equation}
which satisfies the algebra
\begin{equation}
[R^{IJ}, R^{KL}] = \eta^{IK}R^{JL} - \eta^{JK}R^{IL} - \eta^{IL}R^{JK} + \eta^{JL}R^{IK}.
\end{equation} 
The 128 fermionic states can be adequately represented by the lowest order term in $\tilde{f}_{JD}$:
\begin{equation}\label{eq530}
C_{BCD} = (\Gamma^{J})_{(BC}\tilde{f}_{|J|D)},
\end{equation}
where $\tilde{f}_{JD}$ is $\Gamma$-traceless. The 128 bosonic states can be accommodated in the $SO(9)$ traceless symmetric tensor $g_{JK}$ and the 3-form $H^{LMN}$. Therefore we can write
\begin{equation}\label{eq531}
C_{JCD} = a(\Gamma^{K})_{CD}g_{JK} + b(\Gamma_{JKLM})_{CD}H^{KLM}
\end{equation}
After replacing \eqref{eq530}, \eqref{eq531} in \eqref{eq510} one obtains
\begin{eqnarray}
\Gamma^{J}_{(BC}\bar{D}_{|A}\tilde{f}_{J|D)} &=& a(\Gamma^{K})_{A(B}(\Gamma^{J})_{CD)}g_{JK} + b(\Gamma_{J})_{A(B}(\Gamma^{JKLM})_{CD)}H_{KLM}\nonumber\\
&& + \frac{2b}{3}\delta_{(BC}(\Gamma^{KLM})_{D)A}H_{KLM}
\end{eqnarray}
Next we use the $SO(9)$ Fierz identities
\begin{eqnarray}
\delta_{(BC}(\Gamma^{KLM})_{D)A} &=& 3(\Gamma^{[K})_{(BC}(\Gamma^{LM]})_{D)A} + (\Gamma_{J})_{(BC}(\Gamma^{JKLM})_{D)A},\\
(\Gamma^{JKLM})_{(BC}(\Gamma_{J})_{D)A} &=& -(\Gamma_{J})_{(BC}(\Gamma^{JKLM})_{D)A},
\end{eqnarray}
which can be found by using the Mathematica package GAMMA \cite{Gran:2001yh}, to obtain
\begin{eqnarray}
\Gamma^{J}_{(BC}\bar{D}_{|A}\tilde{f}_{J|D)} &=& a(\Gamma^{J})_{(BC}(\Gamma^{K})_{D)A}g_{JK} + 2b(\Gamma^{J})_{(BC}(\Gamma^{LM})_{D)A}H_{JLM}\nonumber\\
&& - \frac{b}{3}(\Gamma_{J})_{(BC}(\Gamma^{JKLM})_{D)A}H_{KLM},
\end{eqnarray}
which implies
\begin{equation}
\hat{\bar{D}}_{A}\tilde{f}_{JD} = a(\Gamma^{K})_{AD}g_{JK} - 2b(\Gamma^{LM})_{AD}H_{JLM} - \frac{b}{3}(\Gamma_{JKLM})_{AD}H^{KLM},
\end{equation}
where the constants $a$, $b$ will be determined from supersymmetry. To do this we should know how $\hat{\bar{D}}_{A}$ acts on $g_{JK}$ and $H_{KLM}$. An educated guess based on linearity and symmetry properties is
\begin{eqnarray}
\hat{\bar{D}}_{A}g_{JK} &=& -2\sqrt{2}P^{+}[(\Gamma_{J})_{AE}\tilde{f}_{K\,E} + (\Gamma_{K})_{AE}\tilde{f}_{J\,E}],\\
\hat{\bar{D}}_{A}H^{KLM} &=& -2\sqrt{2}P^{+}[(\Gamma^{KL})_{AE}\tilde{f}^{M}_{E} - (\Gamma^{KM})_{AE}\tilde{f}^{L}_{E} + (\Gamma^{LM})_{AE}\tilde{f}^{K}_{E}],
\end{eqnarray}
where the factor $-2\sqrt{2}P^{+}$ was chosen for convenience. These equations of motion should satisfy the supersymmetry algebra \eqref{eq522}. This requirement fixes the values of $a$, $b$ to be $a=\frac{1}{4}$, $b=\frac{1}{72}$. Therefore the whole set of light-cone gauge equations of motion is
\begin{eqnarray}
\hat{\bar{D}}_{A}g_{JK} &=& -2\sqrt{2}P^{+}[(\Gamma_{J})_{AE}\tilde{f}_{K\,E} + (\Gamma_{K})_{AE}\tilde{f}_{J\,E}],\\
\hat{\bar{D}}_{A}\tilde{f}_{JD} &=& \frac{1}{4}(\Gamma^{K})_{AD}g_{JK} + \frac{1}{72}[(\Gamma_{JKLM})_{AD} + 6\eta^{JK}(\Gamma^{LM})_{AD}]H^{KLM},\\
\hat{\bar{D}}_{A}H^{KLM} &=& -2\sqrt{2}P^{+}[(\Gamma^{KL})_{AE}\tilde{f}^{M}_{E} - (\Gamma^{KM})_{AE}\tilde{f}^{L}_{E} + (\Gamma^{LM})_{AE}\tilde{f}^{K}_{E}].
\end{eqnarray}
These expressions are the same equations of motion obtained for $D=11$ linearized supergravity from the $D=11$ Brink-Schwarz-like superparticle in light-cone gauge \cite{Green:1999by} studied in section 2.1 of this chapter. 

\vspace{2mm}
The mass-shell condition can be obtained from \eqref{eq521} after using the tracelessness condition for $C_{BCD}$, which is necessary to have a non-trivial vertex operator $V$. This condition gives rise to the equation:
\begin{equation}
\hat{G}_{A}C_{BCD} + \hat{G}_{B}C_{ACD} + \hat{G}_{C}C_{ABD} + \hat{G}_{D}C_{ABC} = 0
\end{equation}
which has solution only if $\hat{G}_{A}C_{BCD} = 0$. This result, together with \eqref{eq521}, implies that $k^{m}k_{m} = 0$,  where $k^{m}$ is the momentum. Consequently, $C_{BCD}$ depends only on $\bar{\Theta}$, $C_{BCD}=C_{BCD}(\bar{\Theta})$. To obtain the pure spinor vertex operator in the $Q$-cohomology one just performs the similarity transformation generated by $-(R+\hat{R})$. The result is
\begin{eqnarray}
V &=& V(\hat{\bar{\Theta}})e^{ik.X}
\end{eqnarray}
where $\hat{{\bar{\Theta}}}^{A} = \bar{\Theta}^{A} - \frac{i}{\sqrt{2}P^{+}}\Theta_{B}(\Gamma^{I})^{AB}k_{I}$.

\vspace{2mm}
In this chapter we have described in detail the procedure which leads us to the $D=11$ pure spinor superparticle from the gauge fixed $D=11$ Brink-Schwarz-like superparticle. This method gave us automatically a formal proof of equivalence between both cohomologies, which allows us to claim both theories are physically the same. Furthermore the equations of motion in $D=9$ superspace found by studying the light-cone gauge pure spinor cohomology, match the $SO(9)$ equations of motion obtained through the study of the $D=11$ Brink-Schwarz-like superparticle in light-cone gauge. 

\vspace{2mm}
Now that the $D=11$ pure spinor superparticle model has been consistently defined, we move on to the study of $D=11$ vertex operators, which should play a crucial role for computing supergravity correlation functions.

\chapter{$D=11$ Pure spinor superparticle vertex operators}\label{newchapter3}
\vspace{2mm}
Pure spinors were introduced in $D=10$ and $D=11$ supersymmetric field theories in \cite{HOWE1991141} and \cite{Howe:1991bx}, and were introduced in the context of superstring theory in \cite{Berkovits:2000fe} as extra dynamical variables on the worldsheet. These extra variables allowed super-Poincar\'e covariant quantization using a simple BRST operator and simplified the computation of multiloop scattering amplitudes as compared to the other superstring formalisms.
Pure spinors have also been  used for worldline field theory computations in quantum field theories \cite{Bjornsson:2010wm,Bjornsson:2010wu} where the ultraviolet behavior of the 4-point amplitude for ten-dimensional super Yang-Mills and Type II supergravity up to 5-loops was studied using power counting arguments. 

\vspace{2mm}
The eleven-dimensional analogs of pure spinors that are discussed here for the superparticle were introduced in \cite{Berkovits:2002uc} and used by \cite{Anguelova:2004pg} to set up a framework for computing $N$-point correlation functions at tree and loop level using a worldline field theory framework. Some higher-loop computations using the non-minimal $D=11$ pure spinor formalism of \cite{Cederwall:2010tn} have been performed in \cite{Cederwall:2012es,Karlsson:2014xva}. 

\vspace{2mm}
In this chapter we provide evidence that contradicts some of the assumptions made in \cite{Anguelova:2004pg}. We construct the ghost number one vertex operator as a perturbation of the BRST operator.  This will be BRST invariant only when the $D=11$ supergravity equations of motion are imposed. This vertex operator takes the form
\begin{eqnarray}
U^{(1)} &=& \Lambda^{\alpha}[h_{\alpha}{}^{a}P_{a} - h_{\alpha}{}^{\beta}D_{\beta} + \Omega_{\alpha ab}N^{ab}]
\end{eqnarray}
where $h_{\alpha}{}^{a}$, $h_{\alpha}{}^{\beta}$, $\Omega_{\alpha ab}$ come from small perturbations of the eleven-dimensional vielbeins and the structure equations of linearized $D=11$ supergravity. They satisfy equations of motion and gauge freedoms arising from the $D=11$ supergravity dynamical constraints.  These  determine their full $\theta$-expansions as explained in \cite{Tsimpis:2004gq} and these are required  in correlation function prescriptions involving $U^{(1)}$.

\vspace{2mm}
The eleven-dimensional pure spinor prescription for computing 
tree-level $N$-point correlation functions given in \cite{Anguelova:2004pg} requires the existence of a ghost number zero vertex operator satisfying the standard descent relation
\begin{eqnarray}\label{int1}
\{Q, V^{(0)}\} &=& [H, U^{(1)}]
\end{eqnarray}
where $H = P^{2}$ is the particle Hamiltonian. We will show that eqn. \eqref{int1} is incompatible with linearized $D=11$ supergravity and discuss some possible ways to fix this problem without going into details. This incompatibility seems to be an obstruction for computing $D=11$ supergravity scattering amplitudes from a pure spinor wordline framework\footnote{It is worth mentioning that this inconsistency also appears to be an obstacle for extending the 11-dimensional pure-spinor superparticle to an 11-dimensional pure-spinor ambitwistor-string following \cite{Mason:2013sva,Berkovits:2013xba}.}.

\vspace{2mm}
The chapter is organized as follows. In section \ref{s3.2} we review the $D=11$ pure spinor superparticle. In section \ref{sec3} we construct the ghost number one vertex operator by requiring that the pure spinor BRST operator be nilpotent at first order as an on-shell geometric deformation of the BRST charge. In section \ref{sec4}, we show the inconsistency between the descent equation \eqref{int1} relating ghost number one and zero vertex operators and the structure equations of $D=11$ supergravity. A self-contained review of the superspace formulation of $D=11$ supergravity is left for Appendix \ref{appendix4}.

\vspace{2mm}
As before, the ten-dimensional analog of the analysis presented here is left for Appendix \ref{appendix3}. The unexperienced reader might find it useful to first read this Appendix before moving on to the more complicated eleven-dimensional case.

\section{$D=11$ pure spinor superparticle revisited}\label{s3.2}
The eleven-dimensional pure spinor superparticle action in a flat background \cite{Berkovits:2002uc,Guillen:2017mte} was studied in the previous chapter and is given by 
\begin{equation}\label{eeq1}
S = \int d\tau [P_{m}\partial_{\tau}X^{m} + P_{\mu}\partial_{\tau}\Theta^{\mu} + W_{\alpha}\partial_{\tau}\Lambda^{\alpha} - \frac{1}{2}P^{m}P_{m}]\, .
\end{equation}
We will use lowercase letters from the beginning/middle of the Greek alphabet to denote $SO(10,1)$ tangent/curved-space spinor indices and we will let lowercase letters from the beginning/midle of the Latin alphabet denote $SO(10,1)$ tangent/curved-space vector indices. The superspace fermionic coordinate $\Theta^{\mu}$ is an $SO(10,1)$ Majorana spinor and $P_{\mu}$ is its respective canonical conjugate momentum, and $P_{m}$ is the momentum for $X^m$. The variable $\Lambda^{\alpha}$ is a $D=11$ pure spinor variable\footnote{Note that we do not require $\Lambda\Gamma^{ab}\Lambda=0$ which would also be imposed by Cartan's definition of purity. } satisfying $\Lambda\Gamma^{a}\Lambda = 0$, and $W_{\alpha}$ is its conjugate momentum which is defined up to the gauge transformation $\delta W_{\alpha} = (\Gamma^{a}\Lambda)_{\alpha}r_{m}$, for an arbitrary gauge parameter $r_{m}$. The $SO(10,1)$ gamma matrices denoted by $\Gamma^{a}$ satisfy the Clifford algebra $(\Gamma^{a})_{\alpha\beta}(\Gamma^{b})^{\beta\sigma} + (\Gamma^{b})_{\alpha\beta}(\Gamma^{a})^{\beta\sigma} = 2\eta^{ab}\delta^{\sigma}_{\alpha}$. As seen before, in $D=11$ dimensions there exist an antisymmetric spinor metric $C_{\alpha\beta}$ (and its inverse $(C^{-1})^{\alpha\beta}$) which allows us to lower (and raise) spinor indices. 

\vspace{2mm}
The BRST operator associated to this theory was found to be
\begin{eqnarray}
Q &=& \Lambda^{\alpha}D_{\alpha}
\end{eqnarray}
where $D_{\alpha} = P_{\alpha} - \frac{1}{2}(\Gamma^{a}\Theta)_{\alpha}P_{a}$ are the fermionic constraints of the $D=11$ Brink-Schwarz-like superparticle. The nilpotency of this operator follows immediately from the pureness of $\Lambda^{\alpha}$, and thus physical states can be defined as elements of its cohomology. As shown in \cite{Berkovits:2002uc}, this BRST cohomology turns out to describe linearized $D=11$ supergravity in its Batalin-Vilkovisky formulation. The $D=11$ supergravity physical fields are found in the ghost number three sector of the cohomology. To see this, one can write the most general ghost number three superfield
\begin{eqnarray}
U^{(3)} &=& \Lambda^{\alpha}\Lambda^{\beta}\Lambda^{\delta}C_{\alpha\beta\delta}(x,\theta)
\end{eqnarray}
The physical state conditions will constrain the functional form of $C_{\alpha\beta\delta}$ to be
\begin{eqnarray}
C_{\alpha\beta\delta} &=& (\Gamma^{a}\Theta)_{\alpha}(\Gamma^{b}\Theta)_{\beta}(\Gamma^{c}\Theta)_{\delta}C_{abc}(x) + (\Gamma^{(a}\Theta)_{\alpha}(\Gamma^{b)c}\Theta)_{\beta}(\Gamma_{c}\Theta)_{\delta}h_{ab}(x)\nonumber\\
&& + (\Gamma^{b}\Theta)_{\alpha}[(\Gamma^{c}\Theta)_{\beta}(\Gamma^{d}\Theta)_{\delta}(\Theta\Gamma_{cd})_{\epsilon} - (\Gamma_{cd}\Theta)_{\beta}(\Gamma_{c}\Theta)_{\delta}(\Gamma_{d}\Theta)_{\epsilon}]\chi_{b}^{\epsilon}(x) + \ldots
\end{eqnarray}
where the fields $C_{abc}(x)$, $h_{ab}(x)$, $\chi_{b}^{\alpha}$ satisfy the linearized $D=11$ supergravity equations of motion and gauge invariances
\begin{eqnarray}
\partial^{c}[\partial_{c}h_{ab} - 2\partial_{(a}h_{b)c}] - \partial_{a}\partial_{b}h^{c}{}_{c} = 0\hspace{2mm}&,&\hspace{2mm} \delta h_{ab} = \partial_{(b}\Omega_{c)}\nonumber\\
\partial^{d}\partial_{[a}C_{bcd]} = 0\hspace{2mm}&,&\hspace{2mm} \delta C_{abc} = \partial_{[a}\Omega_{bc]}\nonumber\\
(\gamma^{abc})_{\alpha\beta}\partial_{b}\chi_{c}^{\beta} = 0 \hspace{2mm}&,&\hspace{2mm} \delta \chi_{a}^{\beta} = \partial_{a}\Omega^{\beta}
\end{eqnarray}
where $\Omega_{a}$, $\Omega_{bc}$, $\Omega^{\alpha}$ are arbitrary gauge parameters. The other BV fields of linearized $D=11$ supergravity are placed into different ghost sectors up to ghost number 7. 

\vspace{2mm}
Following 
$D=10$ dimensions \cite{Bjornsson:2010wm, Bjornsson:2010wu}, one can attempt to 
define a pure spinor measure from the eleven-dimensional scalar top cohomology, namely $\langle \Lambda^{7}\Theta^{9} \rangle = 1$ in order to give a consistent prescription for computing $N$-point correlation functions. This measure is easily shown to be BRST-invariant and supersymmetric and has already been successfully used to get the kinetic terms of the $D=11$ supergravity action from a second-quantized point of view \cite{Berkovits:2002uc}. Using this one can then propose that the N-point amplitude should be given by a correlation function of the form  \cite{Anguelova:2004pg}
\begin{eqnarray}\label{eq2}
\mathcal{A}^{11D}_{N} &=& \langle U^{(3)}_{1}(\tau_{1})U^{(3)}_{2}(\tau_{2})U^{(1)}_{3}(\tau_{3})\int d\tau_{4}V^{(0)}_{4}(\tau_{4})\ldots \int d\tau_{N} V^{(0)}_{N}(\tau_{N})\rangle
\end{eqnarray}
In this expression $U^{(3)}$ is the ghost number three vertex operator described above, $U^{(1)}$ is a ghost number one vertex operator and $V^{(0)}$ is a vertex operator of ghost number zero. Although it is possible to write an alternative prescription involving the ghost number four vertex operator containing the antifields of the $D=11$ supergravity physical fields, the existence of $V^{(0)}$ clearly plays a crucial role for the computation of the N-point correlation functions beyond $N=3$ in this framework. 

\vspace{2mm}
Having established the importance of the ghost number one and zero vertex operators, we now discuss their construction.


\section{Ghost number one vertex operator}\label{sec3}
The ghost number one vertex operator will be constructed from a small perturbation of the pure spinor BRST operator whose nilpotency will follow from the $D=11$ linearized supergravity equations of motion and the pure spinor constraint. We give a detailed review of the superspace formulation of $D=11$ supergravity in Appendix \ref{appendixB}\footnote{The linearized description of it can be readily obtained by dropping out interacting terms in the equations of motion displayed in this Appendix.}. Let us write the eleven-dimensional vielbeins in their linearized form
\begin{eqnarray}
E^{A} = E_{0}^{A} + h^{A} = (\mathcal{D}x^{a} + h_{b}{}^{a}\mathcal{D}x^{b} + h_{\beta}{}^{a}d\Theta^{\beta}, d\Theta^{\alpha} + h_{\beta}{}^{\alpha}d\Theta^{\beta} + \psi_{b}^{\alpha}\mathcal{D}x^{b})
\end{eqnarray}
where
\begin{eqnarray}
\mathcal{D}x^{b} = dx^{b} + \frac{1}{2}(\Theta\Gamma^{b}d\Theta) \hspace{2mm}&,&\hspace{2mm} \mathcal{D}_{\alpha} = \partial_{\alpha} - \frac{1}{2}(\Gamma^{c}\Theta)_{\alpha}\partial_{c}
\end{eqnarray}
These give dually to first order
\begin{eqnarray}\label{new0002}
\hat{\mathcal{D}}_{\alpha} = \mathcal{D}_{\alpha} - h_{\alpha}{}^{\beta}\mathcal{D}_{\beta} - h_{\alpha}{}^{a}\partial_{a} \hspace{2mm}&,&\hspace{2mm} \hat{\mathcal{D}}_{a} = \partial_{a} - \psi_{a}^{\alpha}\mathcal{D}_{\alpha} - h_{a}{}^{b}\partial_{b}
\end{eqnarray}
On the other hand, using eqn. \eqref{section3eq12} one can show that at linear order
\begin{eqnarray}
[\hat{\mathcal{D}}_{C}, \hat{\mathcal{D}}_{D}\} &=& T_{CD}{}^{A}\hat{\mathcal{D}}_{A} - 2\Omega_{[CD\}}{}^{A}\hat{\mathcal{D}}_{A}\label{newweq5}
\end{eqnarray}
where $[\cdot,\cdot\}$ is the graded (anti)commutator. Using the $D=11$ supergravity constraints \eqref{eq18}, one then finds that
\begin{eqnarray}\label{eq3}
\{\hat{\mathcal{D}}_{\alpha}, \hat{\mathcal{D}}_{\beta}\} &=& (\Gamma^{a})_{\alpha\beta}\hat{\mathcal{D}}_{a} - 2\Omega_{(\alpha\beta)}{}^{\gamma}\hat{\mathcal{D}}_{\gamma}
\end{eqnarray}
Thus if one defines the BRST operator to be
\begin{eqnarray}\label{section3eq4}
Q &=& \Lambda^{\alpha}(\hat{\mathcal{D}}_{\alpha} + \Omega_{\alpha\beta}{}^{\gamma}\Lambda^{\beta}\frac{\partial}{\partial\Lambda_{\gamma}})
\end{eqnarray}
then its nilpotency property immediately follows from the e.o.m \eqref{eq22}
\begin{eqnarray}\label{eq8}
\{Q, Q\} = \Lambda^{\alpha}\Lambda^{\beta}\Lambda^{\delta}R_{(\alpha\beta\delta)}{}^{\epsilon}\frac{\partial}{\partial\Lambda^{\epsilon}} = 0
\end{eqnarray} 

After converting \eqref{section3eq4} into a worldline vector with ghost number 1 by replacing operators by corresponding worldline fields, one concludes that
\begin{eqnarray}
Q = Q_{0} + U^{(1)} + \ldots \hspace{2mm}&,&\hspace{2mm} Q_{0} = \Lambda^{\alpha}D_{\alpha}
\end{eqnarray}
where
\begin{eqnarray}\label{eq60}
U^{(1)} &=& \Lambda^{\alpha}(h_{\alpha}{}^{a}P_{a} - h_{\alpha}{}^{\beta}D_{\beta} - \Omega_{\alpha\beta}{}^{\gamma}N_{\gamma}{}^{\beta})
\end{eqnarray}
and $\dots$ means higher order terms. Thus $\{Q, Q\} = 0$  yields directly
\begin{eqnarray}
\{Q_{0} , U^{(1)}\} &=& 0
\end{eqnarray}
as desired.

\vspace{2mm}
The e.o.m satisfied by the superfields in \eqref{eq60} can be easily found by plugging \eqref{new0002} into \eqref{newweq5}. From the relation $\{\hat{\mathcal{D}}_{\alpha}, \hat{\mathcal{D}}_{\beta}\}$, one gets
\begin{eqnarray}
2 \mathcal{D}_{(\alpha}h_{\beta)}{}^{a} + 2 h_{(\alpha}{}^{\delta}(\Gamma^{a})_{\beta)\delta} - h_{b}{}^{a}(\Gamma^{b})_{\alpha\beta} &=& 0\label{eq70}\\
2\mathcal{D}_{(\alpha}h_{\beta)}{}^{\delta} - 2\Omega_{(\alpha\beta)}{}^{\delta} - (\Gamma^{a})_{\alpha\beta}\psi_{a}{}^{\delta} &=& 0\label{eq71}
\end{eqnarray}
From the relation $\{\hat{\mathcal{D}}_{a}, \hat{\mathcal{D}}_{\alpha}\}$ one finds
\begin{eqnarray}
\partial_{a}h_{\alpha}{}^{\beta} - \mathcal{D}_{\alpha}\psi_{a}{}^{\beta} + T_{a\alpha}{}^{\beta} - \Omega_{a\alpha}{}^{\beta} &=& 0\label{eq82}\\
\partial_{a}h_{\alpha}{}^{b} - \mathcal{D}_{\alpha}h_{a}{}^{b} + \psi_{a}{}^{\beta}(\Gamma^{b})_{\beta\alpha} + \Omega_{\alpha\,a}{}^{b} &=& 0\label{eq80}
\end{eqnarray}
From the relation $\{\hat{\mathcal{D}}_{a}, \hat{\mathcal{D}}_{b}\}$ one obtains
\begin{eqnarray}
\partial_{a}\psi_{b}{}^{\alpha} - \partial_{b}\psi_{a}{}^{\alpha} + T_{ab}{}^{\alpha} &=& 0\\
\partial_{a}h_{b}{}^{c} - \partial_{b}h_{a}{}^{c} - 2\Omega_{[ab]}{}^{c} &=& 0
\end{eqnarray}
Moreover, the linearized supercurvature components can be written in terms of the super spin-conection using eqn. \eqref{eq11}
\begin{eqnarray}
R_{\alpha\beta c}{}^{d} &=& 2 \mathcal{D}_{(\alpha}\Omega_{\beta)c}{}^{d} - (\Gamma^{a})_{\alpha\beta}\Omega_{a\,c}{}^{d}\label{eq72}\\
R_{a\alpha b}{}^{c} &=& \partial_{a}\Omega_{\alpha b}{}^{c} - \mathcal{D}_{\alpha}\Omega_{ab}{}^{c}\\
R_{abc}{}^{d} &=& 2\partial_{[a}\Omega_{b]c}{}^{d}
\end{eqnarray}
As a consistency check, one can verify that $\{Q_{0}, U^{(1)}\} = 0$ as a consequence of the e.o.m \eqref{eq70}, \eqref{eq71}, \eqref{eq72}.

\section{The 3-point function}
One can now use the prescription \eqref{eq2} to calculate the $D=11$ supergravity 3-point correlation function \cite{Anguelova:2004pg}. Let us locate the two $U^{(3)}$ at positions $\tau_{1}$, $\tau_{2}$ and $U^{(1)}$ at $\tau_{3}$. Then,
\begin{eqnarray}
\mathcal{A}^{11D}_{3-pt} &=& \langle U^{(3)}(\tau_{1})U^{(3)}(\tau_{2})U^{(1)}(\tau_{3})\rangle\nonumber\\
&=& \langle U^{(3)}(\tau_{1})U^{(3)}(\tau_{2})[\Lambda^{\alpha}(h_{\alpha}{}^{a}P_{a} - h_{\alpha}{}^{\beta}D_{\beta} - \frac{1}{2}\Omega_{\alpha ab}N^{ab})](\tau_{3})\rangle\ \nonumber\\
&=& \langle U^{(3)}(\tau_{1})U^{(3)}(\tau_{2})\Phi^{a}P_{a}(\tau_{3})\rangle - \langle U^{(3)}(\tau_{1})U^{(3)}(\tau_{2}) \Phi^{\beta}D_{\beta}(\tau_{3})\rangle\nonumber\\
&& - \frac{1}{2}\langle U^{(3)}(\tau_{1})U^{(3)}(\tau_{2})\Phi_{ab}N^{ab}(\tau_{3})\rangle\ \label{newnew1}
\end{eqnarray}
where we have defined $\Phi^{a} = \Lambda^{\alpha}h_{\alpha}{}^{a}$, $\Phi^{\beta} = \Lambda^{\alpha}h_{\alpha}{}^{\beta}$, $\Phi_{ab} = \Lambda^{\alpha}\Omega_{\alpha ab}$. We will now show that the last two terms cancel out on-shell. To see this, let us recall the standard tree-level correlators of the worldline formalism \cite{Strassler:1992zr}:
\begin{eqnarray}
\langle X^{m}(\tau_{1})X^{n}(\tau_{2})\rangle_{tree} &=& \eta^{mn}(\frac{|\tau_{1}-\tau_{2}|}{2} + A + B\tau_{2})\\
\langle P_{\mu}(\tau_{1})\Theta^{\nu}(\tau_{2})\rangle_{tree} &=& \delta_{\mu}^{\nu}\,sgn(\tau_{2}-\tau_{1})
\end{eqnarray}
and assume that $\tau_{1} < \tau_{3} < \tau_{2}$. The second term in \eqref{newnew1} then reads
\begin{eqnarray}
\mbox{2nd-term} &=& \langle U^{(3)}(\tau_{1})U^{(3)}(\tau_{2}) \Phi^{\beta}D_{\beta}(\tau_{3})\rangle\nonumber\\
&=& -\langle \mathcal{D}_{\beta}U^{(3)}(\tau_{1})U^{(3)}(\tau_{2}) \Phi^{\beta}(\tau_{3})\rangle - \langle U^{(3)}(\tau_{1})\mathcal{D}_{\beta} U^{(3)}(\tau_{2}) \Phi^{\beta}(\tau_{3})\rangle \nonumber\\
&=& -3\langle (\Lambda^{\delta}\Lambda^{\epsilon}C_{\beta\delta\epsilon}(\tau_{1}))U^{(3)}(\tau_{2}) Q_{0}\Phi^{\beta}(\tau_{3})\rangle\nonumber\\
&& + 3\langle (\Gamma^{a}\Lambda)_{\beta}\Lambda^{\alpha}\Lambda^{\delta}C_{a\alpha\delta}(\tau_{1})U^{(3)}(\tau_{2})\Phi^{\beta}(\tau_{3})\rangle + (1 \leftrightarrow 2)\nonumber\\
&=& 3\langle (\Lambda^{\delta}\Lambda^{\epsilon}C_{\beta\delta\epsilon}(\tau_{1}))U^{(3)}(\tau_{2}) \Lambda^{\gamma}\Lambda^{\alpha}\Omega_{\gamma\alpha}{}^{\beta}(\tau_{3})\rangle\nonumber\\
&& + 3\langle (\Gamma^{a}\Lambda)_{\beta}\Lambda^{\alpha}\Lambda^{\delta}C_{a\alpha\delta}(\tau_{1})U^{(3)}(\tau_{2})\Phi^{\beta}(\tau_{3})\rangle + (1 \leftrightarrow 2)
%
%
%
\label{newnew2}
\end{eqnarray}
where we used eqn. \eqref{eq71} in the last line of \eqref{newnew2}. On the other hand, the last term in \eqref{newnew1} can be written as
\begin{eqnarray}
\mbox{3rd-term} &=& \frac{1}{2}\langle U^{(3)}(\tau_{1})U^{(3)}(\tau_{2}) \Phi_{ab}N^{ab}(\tau_{3})\rangle\nonumber\\
&=& -\frac{3}{4}\langle(\Lambda\Gamma^{ab})^{\alpha}\Lambda^{\beta}\Lambda^{\delta}C_{\alpha\beta\delta}(\tau_{1})U^{(3)}(\tau_{2})\Lambda^{\epsilon}\Omega_{\epsilon ab}(\tau_{3})\rangle + (1 \leftrightarrow 2)\nonumber\\
&=& -3\langle(\Lambda^{\beta}\Lambda^{\delta}C_{\alpha\beta\delta}(\tau_{1}))U^{(3)}(\tau_{2})\Lambda^{\gamma}\Lambda^{\epsilon}\Omega_{\epsilon\gamma}{}^{\alpha}(\tau_{3})\rangle + (1 \leftrightarrow 2)\label{newnew3}
\end{eqnarray}
Plugging eqns. \eqref{newnew2}, \eqref{newnew3} into \eqref{newnew1}, one finds that
\begin{eqnarray}
\mathcal{A}^{11D}_{3-pt} &=& \langle (\partial_{a}U^{(3)}(\tau_{1}))U^{(3)}(\tau_{2})\Phi^{a}(\tau_{3})\rangle - 3\langle (\Gamma^{a}\Lambda)_{\beta}\Lambda^{\alpha}\Lambda^{\delta}C_{a\alpha\delta}(\tau_{1})U^{(3)}(\tau_{2})\Phi^{\beta}(\tau_{3})\rangle\nonumber\\
&& + (1 \leftrightarrow 2)
\end{eqnarray}
We shall now use the relationship between $\Phi^{a}$ and $U^{(3)}$ discussed in Appendix \ref{appendix6}. Using eqn. \eqref{eqeqe305}, one then learns that
\begin{eqnarray}
\mathcal{A}^{11D}_{3-pt} &\propto & \delta^{11}(\sum_{r=1}^{3}k_{r}^{m})\bigg[\frac{1}{4}\langle(\Lambda\Gamma^{ab}\Lambda)\Phi_{a}(\tau_{1})U^{(3)}(\tau_{2})\Phi_{b}(\tau_{3})\rangle\nonumber\\
&& + 3\langle Q_{0}(\Lambda^{\alpha}\Lambda^{\beta}C_{b\alpha\beta})(\tau_{1})U^{(3)}(\tau_{2})\Phi^{b}(\tau_{3})\rangle\nonumber\\
&& + 3\langle (\Gamma^{a}\Lambda)_{\beta}\Lambda^{\alpha}\Lambda^{\delta}C_{a\alpha\delta}(\tau_{1})U^{(3)}(\tau_{2})\Phi^{\beta}(\tau_{3})\rangle + (1 \leftrightarrow 2) \bigg]\nonumber\\
&\propto & \delta^{11}(\sum_{r=1}^{3}k_{r}^{m})\bigg[\frac{1}{4}\langle(\Lambda\Gamma^{ab}\Lambda)\Phi_{a}(\tau_{1})U^{(3)}(\tau_{2})\Phi_{b}(\tau_{3})\rangle\nonumber\\
&& + 3\langle(\Lambda^{\alpha}\Lambda^{\beta}C_{b\alpha\beta})(\tau_{1})U^{(3)}(\tau_{2})Q_{0}\Phi^{b}(\tau_{3})\rangle\nonumber\\
&& + 3\langle (\Gamma^{a}\Lambda)_{\beta}\Lambda^{\alpha}\Lambda^{\delta}C_{a\alpha\delta}(\tau_{1})U^{(3)}(\tau_{2})\Phi^{\beta}(\tau_{3})\rangle + (1 \leftrightarrow 2) \bigg]\nonumber\\
&\propto & \frac{1}{4}\delta^{11}(\sum_{r=1}^{3}k_{r}^{m})\bigg(\langle(\Lambda\Gamma^{ab}\Lambda)\Phi_{1\,a}\Phi_{3\,b}U_{2}^{(3)}\rangle + (1 \leftrightarrow 2)\bigg)\label{newnew5}
\end{eqnarray} 
where we used eqn. \eqref{eq70} in the last line of \eqref{newnew5} and integrated out the zero modes of $X^{m}$ to get the eleven-dimensional momentum conservation delta function. We have also written down numeric subscripts in \eqref{newnew5} to label the external particles, so $\Phi_{1\,a}$ is the ghost number one vector superfield whose polarization vectors and momenta correspond to the particle 1, etc. We will see later that eqn. \eqref{newnew5}, obtained here from a wordline perspective, can be used to construct a 3-point coupling for a second-quantized description of $D=11$ supergravity on a pure spinor superspace \cite{Cederwall:2009ez,Cederwall:2010tn}.

\section{Ghost number zero vertex operator}\label{sec4}
In order for a consistent standard equation to be satisfied, a ghost number zero vertex operator should exist and satisfy the relation
\begin{eqnarray}\label{eq75}
\{Q_{0}, V^{(0)}\} &=& P^{a}\partial_{a}U^{(1)}
\end{eqnarray}
where $U^{(1)}$ is the ghost number one vertex operator discussed above. To solve eqn. \eqref{eq75}, let us write first the most general ghost number zero vertex operator which is gauge invariant under the pure spinor constraint
\begin{eqnarray}\label{eq76}
V^{(0)} &=& P^{a}P^{b}\mathcal{G}_{ab} + P^{a}D_{\beta}\Psi_{a}^{\beta} + P^{a}N^{bc}\mathcal{W}_{abc}\nonumber\\
&& + D_{\alpha}D_{\beta}\mathcal{P}^{\alpha\beta} + D_{\alpha}N^{ab}\mathcal{T}_{ab}{}^{\alpha} + N^{ab}N^{cd}\mathcal{R}_{ab,cd}
\end{eqnarray}
One can now compute the e.o.m that the superfields in \eqref{eq76} should satisfy such that \eqref{eq75} holds. After some algebraic manipulations one finds that
\begin{eqnarray}
\Lambda^{\alpha}P^{a}P^{b}[\mathcal{D}_{\alpha}\mathcal{G}_{ab} - \Psi_{(a}^{\beta}(\Gamma_{b)})_{\alpha\beta} - \partial_{a}h_{\alpha b}] &=& 0\label{section3eq100}\\
\Lambda^{\alpha}P^{a}D_{\beta}[-\mathcal{D}_{\alpha}\Psi_{a}^{\beta} - \frac{1}{2}\mathcal{W}_{a}{}^{bc}(\Gamma_{bc})^{\beta}{}_{\alpha} - 2(\Gamma_{a})_{\alpha\gamma}\mathcal{P}^{\gamma\beta} + \partial_{a}h_{\alpha}{}^{\beta}] &=& 0\label{eq81}\\
\Lambda^{\alpha}P^{a}N^{bc}[\mathcal{D}_{\alpha}\mathcal{W}_{abc} - (\Gamma_{a})_{\alpha\beta}\mathcal{T}_{bc}{}^{\beta} - \partial_{a}\Omega_{\alpha bc}] &=& 0\label{004}\\
\Lambda^{\alpha}D_{\beta}D_{\gamma}[\mathcal{D}_{\alpha}\mathcal{P}^{\beta\gamma} + \frac{1}{2}(\Gamma^{ab})^{\gamma}_{\hspace{2mm}\alpha}\mathcal{T}_{ab}{}^{\beta}] &=& 0 \label{001}\\
\Lambda^{\beta}D_{\alpha}N^{ab}[\mathcal{D}_{\beta}\mathcal{T}_{ab}{}^{\alpha} + \frac{1}{2}(\Gamma^{cd})^{\alpha}_{\hspace{2mm}\beta}\mathcal{R}_{cdab} + \frac{1}{2}(\Gamma^{cd})^{\alpha}{}_{\beta}\mathcal{R}_{abcd}] &=& 0\label{003}\\
\Lambda^{\alpha}N^{ab}N^{cd}\mathcal{D}_{\alpha}\mathcal{R}_{abcd} &=& 0\label{007}
\end{eqnarray}
The first equation can be automatically solved if one identifies $\mathcal{G}_{ab} = h_{ab}$, $\Psi_{a}^{\alpha} = \psi_{a}^{\alpha}$ as can be seen from \eqref{eq80}. Replacing this into \eqref{eq81} one gets
\begin{eqnarray}
\Lambda^{\alpha}P^{a}D_{\beta}[-\mathcal{D}_{\alpha}\psi_{a}^{\beta} - \frac{1}{2}\mathcal{W}_{a}{}^{bc}(\Gamma_{bc})^{\beta}{}_{\alpha} - 2(\Gamma_{a})_{\alpha\gamma}\mathcal{P}^{\gamma\beta} + \partial_{a}h_{\alpha}{}^{\beta}] &=& 0
\end{eqnarray}
After taking a look at eqn. \eqref{eq82}, one concludes that this equation becomes an identity if one identifies $\mathcal{W}_{abc} = \frac{1}{2}\Omega_{abc}$, $-2(\Gamma_{a})_{\alpha\gamma}\mathcal{P}^{\gamma\beta} = T_{a\alpha}{}^{\beta}$. However this solution for $\mathcal{P}^{\alpha\beta}$ is inconsistent as will be shown now. If this identification were true, it would imply that
\begin{eqnarray}\label{eq85}
\mathcal{P}^{\alpha\beta} &=& \frac{5}{192.11}(\Gamma^{abcd})^{\alpha\beta}H_{abcd}
\end{eqnarray}
If one now tries to recover $T_{a\delta}{}^{\beta}$ by multiplying eqn. \eqref{eq85} by $-2(\Gamma_{a})_{\delta\alpha}$, one finds that
\begin{eqnarray}
-2(\Gamma_{a})_{\delta\alpha}P^{\alpha\beta} 
&=& -\frac{5}{24.11}[(\Gamma^{cde})_{\delta}{}^{\beta}H_{acde} + \frac{1}{4}(\Gamma_{a}{}^{bcde})_{\delta}{}^{\beta}H_{bcde}]
\end{eqnarray}
which is clearly an inconsistency because of the eleven-dimensional structure of maximal supergravity (see eqn. \eqref{eq24}). 

\vspace{2mm}
Further evidence that $D=11$ supergravity is inconsistent with eqns. \eqref{section3eq100}-\eqref{007} can be found when trying to solve eqn. \eqref{001}. To see this, let us identify $\mathcal{T}_{ab}{}^{\alpha}$ with one of the $D=11$ supergravity fields. Using dimensional analysis arguments one concludes that the most general expression for $\mathcal{T}_{ab}{}^{\alpha}$ should have the form
\begin{eqnarray}\label{eq91}
\mathcal{T}_{ab}{}^{\alpha} &=& T_{ab}{}^{\alpha} + a_{1}(\Gamma_{[a}{}^{c})^{\alpha}{}_{\delta}T_{b]c}{}^{\delta} + a_{2}(\Gamma_{ab}{}^{cd})^{\alpha}{}_{\delta}T_{cd}{}^{\delta}
\end{eqnarray}
where $a_{1}$, $a_{2}$ are numerical constants to be determined. Using eqn. \eqref{eq90}, one can relate the last two terms on the right hand side of \eqref{eq91} to the first one, since $(\Gamma_{[a}{}^{c})^{\alpha}{}_{\delta}T_{b]c}{}^{\delta} = T_{ab}{}^{\alpha}$ and $(\Gamma_{ab}{}^{cd})^{\alpha}{}_{\delta}T_{cd}{}^{\delta} = 2T_{ab}{}^{\alpha}$. This implies that $\mathcal{T}_{ab}{}^{\alpha} = b_{1}T_{ab}{}^{\alpha}$ where $b_{1}$ is a constant normalization factor. After plugging this and eqn. \eqref{eq85} into \eqref{001} one demonstrates that
\begin{eqnarray}
\Lambda^{\alpha}D_{\beta}D_{\gamma}[-\frac{5}{96}(\Gamma^{bcde})^{\beta\gamma}\mathcal{D}_{\alpha}H_{bcde} + b_{1}(\Gamma^{ab})^{\gamma}{}_{\alpha}T_{ab}{}^{\beta}] &=& 0
\end{eqnarray}
Since this equation is antisymmetric in $(\beta,\gamma)$, it should be true for the all antisymmetric gamma matrix projections of it, namely $C_{\beta\gamma}$, $(\Gamma^{fgh})_{\beta\gamma}$, $(\Gamma^{fghi})_{\beta\gamma}$. In particular, the 3-form projection requires 
\begin{eqnarray}
(\Gamma_{fgh})_{\beta\gamma}(\Gamma^{ab})_{\alpha}{}^{\gamma}T_{ab}{}^{\beta} &=& 0
\end{eqnarray}
However, the use of eqn. \eqref{eq90} allows one to show that 
\begin{eqnarray}
(\Gamma_{fgh})_{\beta\gamma}(\Gamma^{ab})_{\alpha}{}^{\gamma}T_{ab}{}^{\beta} &=& 24(\Gamma_{[h})_{\alpha\beta}T_{fg]}{}^{\beta}
\end{eqnarray}
which is non-zero and thus inconsistent with \eqref{001}. Thus it is not possible to obtain a ghost number zero vertex operator from the $D=11$ supergravity fields that satisfies the standard descent equation \eqref{eq75}.

\vspace{2mm}
In this chapter we have constructed a ghost number one vertex operator involving more terms in its definition compared to that presented in \cite{Anguelova:2004pg}. In principle, there is no physical reason to ignore them in the 3-point function computations. In fact, when they are consistently taken into account, one finds eqn. \eqref{newnew5} which differs from the one found in \cite{Anguelova:2004pg}, but it takes the same form as the 3-point coupling used to study $D=11$ supergravity from a second-quantized perspective \cite{Cederwall:2010tn,Cederwall:2012es,Berkovits:2018gbq}. 


\vspace{2mm}
On the other hand, we have shown that it is not possible to write a ghost number zero vertex operator made out of the $D=11$ supergravity superfields satisfying a standard descent equation. One possible resolution is to extend the present framework to its non-minimal version by introducing the standard non-minimal pure spinor variables. This is what we will do in next chapter.

\vspace{2mm}
Another approach is to impose additional constraints on the eleven-dimensional pure spinor in such a way that more terms in the ghost number zero vertex operator are allowed. For example, one could impose the full Cartan purity condition $\Lambda\Gamma^{ab}\Lambda=0$ as was considered in \cite{Howe:1991bx,Babalic:2008ga}, and it would be interesting to see if there is some relation of this constraint with our approach.


\chapter{$D=11$ Non-minimal pure spinor superparticle and the b-ghost}\label{newchapter4}
Motivated by the non-minimal version of the pure spinor superstring \cite{Berkovits:2005bt}, Cederwall formulated the $D=11$ non-minimal pure spinor superparticle by introducing a new set of variables $\bar{\Lambda}_{\alpha}$, $R_{\beta}$ and their respective momenta $\bar{W}^{\alpha}$, $S^{\beta}$, where $\bar{\Lambda}_{\alpha}$ is a $D=11$ bosonic spinor and $R_{\beta}$ is a $D=11$ fermionic spinor satisfying the constraints $\bar{\Lambda}\Gamma^{a}\bar{\Lambda} = 0$ and $\bar{\Lambda}\Gamma^{a}R = 0$ \cite{Cederwall:2009ez,Cederwall:2010tn}. In order for the new variables to not affect the physical spectrum, the BRST operator should be modified to $Q = \Lambda^{\alpha}D_{\alpha} + R_{\alpha}\bar{W}^{\alpha}$, as in the quartet argument of \cite{Kugo:1979gm}. In the non-minimal pure spinor formalism of superstring, one can formulate a consistent prescription to compute scattering amplitudes by constructing a non-fundamental $b$ ghost satisfying $\{Q, b\} = T$. Therefore, it is important to know if a similar $b$ ghost can be constructed
 in the $D=11$ superparticle case.


\vspace{2mm}
The $D=11$ $b$-ghost was first constructed in \cite{Cederwall:2012es} in terms of quantities which are not manifestly invariant under the gauge symmetries of $w_\alpha$ generated by $\Lambda \Gamma^a \Lambda=0$. This $b$-ghost was later shown in \cite{Karlsson:2014xva} to be $Q$-equivalent to one written in terms of the gauge-invariant quantities $N_{ab}$ and $J$, and we   will focus on this manifestly gauge-invariant version of the $b$-ghost. 

\vspace{2mm}
The complicated form of the $b$-ghost in \cite{Karlsson:2014xva} makes it difficult to treat, so for instance its nilpotency property $\{b,b\}$ has not yet been analyzed. A similar complication exists in $D=10$ dimensions, however, it was shown in \cite{Berkovits:2013pla} that the $D=10$ $b$-ghost could be simplified by defining new fermionic vector variables. In this chapter, a similar simplification involving fermionic vector variables will be found for the $D=11$ $b$-ghost which will simplify the computations of $\{Q,b\}=T$ and $\{b,b\}$.

\vspace{2mm}
The chapter is organized as follows: In section \ref{s4.2} we review the $D=10$ non-minimal pure spinor superparticle, constructing the corresponding pure spinor $b$-ghost and its simplification\footnote{Unlike the two previous chapters, we will include in this chapter a discussion on the ten-dimensional case. This will turn out to be useful for next chapters where a general framework for constructing pure spinor actions for $D=10$ and $D=11$ maximally supersymmetric gauge theories will be constructed.}. In section \ref{s4.3} we review the $D=11$ pure spinor superparticle, constructing the manifestly gauge-invariant $b$-ghost and explaining how to translate the simplification of the $D=10$ $b$-ghost to the $D=11$ $b$-ghost by defining the $SO(10,1)$ composite fermionic vector $\bar{\Sigma}^{j}$. We then construct the simplified $D=11$ $b$-ghost and show that it satisfies the relations $\{Q , b\} = T$ and $\{b , b\} =$ BRST-trivial. Finally, in section \ref{s4.3} we make use of the $D=11$ $b$-ghost to propose a ghost number zero vertex operator satisfying the standard descent equation \eqref{int1}. Some comments are given at the end of the chapter concerning the relation between the $b$-ghost found in \cite{Karlsson:2014xva} and this simplified $b$-ghost.

\section{$D=10$ non-minimal pure spinor superparticle}\label{s4.2}
The $D=10$ minimal pure spinor superparticle \cite{Berkovits:2001rb} is briefly reviewed in Appendices \ref{appendix2}, \ref{appendix3}. Its action is given by
\begin{equation}
S = \int d\tau (P_{m}\partial_{\tau}X^{m} + p_{\mu}\partial_{\tau}\theta^{\mu} + w_{\mu}\partial_{\tau}\lambda^{\mu} - \frac{1}{2}P^{m}P_{m})
\end{equation}
where $m$, $\mu$ are $SO(9,1)$ vector/spinor indices, $\theta^{\mu}$ is an $SO(9,1)$ Majorana-Weyl spinor, $p_{\mu}$ is its corresponding conjugate momentum and $P^{m}$ is the momentum. The variable $\lambda^{\mu}$ is a $D=10$ pure spinor satisfying the constraint $\lambda\gamma^{m}\lambda = 0$ where $m$ is an $SO(9,1)$ vector index, and $w_{\mu}$ is its corresponding conjugate momentum. Because of the pure spinor constraint this $SO(9,1)$ antichiral spinor is defined up to the gauge transformation $\delta w_{\mu} = (\gamma^{m}\lambda)_{\mu}f_{m}$, where $f_{m}$ is an arbitrary vector. The $SO(9,1)$ gamma matrices denoted by $\gamma^{m}$ satisfy the Clifford algebra $(\gamma^{m})_{\mu\nu}(\gamma^{n})^{\nu\rho} + (\gamma^{n})_{\mu\nu}(\gamma^{m})^{\nu\rho} = 2\eta^{mn}\delta^{\rho}_{\mu}$. The physical states are defined as elements of the cohomology of the BRST operator $Q = \lambda^{\mu}d_{\mu}$, where $d_{\mu}=p_\mu - P_m (\gamma^m \theta)_\mu$ are the first-class constraints of the $D=10$ Brink-Schwarz superparticle \cite{Brink:1981nb} (see Appendix \ref{appendix2}). The spectrum turns out to describe the BV version of $D=10$ (abelian) Super Yang-Mills \cite{Berkovits:2001rb,Berkovits:2002zk,Bedoya:2009np}. 

\vspace{2mm}
In the \emph{non-minimal} version of the pure spinor superparticle \cite{Berkovits:2005bt}\cite{Bjornsson:2010wm}, one  introduces a new pure anti-Weyl spinor $\bar{\lambda}_{\mu}$, and a fermionic field $r_{\mu}$ satisfying the constraint $\bar{\lambda}\gamma^{m}r = 0$, together with their respective conjugate momenta $\bar{w}^{\mu}$, $s^{\mu}$. In order to not affect the cohomology corresponding to $Q_{min}$, the \emph{non-minimal} BRST operator is defined as $Q_{non-min} = \lambda^{\mu}d_{\mu} + \bar{w}^{\mu}r_{\mu}$. Thus the $D=10$ non-minimal pure spinor superparticle is described by the action:
\begin{equation}
S = \int d\tau (P_{m}\partial_{\tau}X^{m} + p_{\mu}\partial_{\tau}\theta^{\mu} + w_{\mu}\partial_{\tau}\lambda^{\mu} + \bar{w}^{\mu}\partial_{\tau}\bar{\lambda}_{\mu} + s^{\mu}\partial_{\tau}r_{\mu} - \frac{1}{2}P^{m}P_{m}) 
\end{equation}
and the BRST operator $Q = \lambda^{\mu}d_{\mu} + \bar{w}^{\mu}r_{\mu}$. By construction, the physical spectrum also describes BV $D=10$ (abelian) Super Yang-Mills.


\subsection{$D=10$ b-ghost}
As discussed in \cite{Bjornsson:2010wm,Bjornsson:2010wu} a consistent scattering amplitude prescription can be defined using a composite $b$-ghost satisfying $\{Q, b\} = T$, where $Q$ is the non-minimal BRST operator and $T = -\frac{1}{2}P^{m}P_{m}$ is the stress-energy tensor. This superparticle $b$-ghost is obtained by dropping the worldsheet non-zero modes in the superstring $b$ ghost and is
\begin{eqnarray}\label{section4eeq5}
b &=& \frac{1}{2}\frac{(\bar{\lambda}\gamma_{m}d)}{\bar{\lambda}\lambda}P^{m} - \frac{1}{192}\frac{(\bar{\lambda}\gamma^{mnp}r)[(d\gamma_{mnp}d) + 24N_{mn}P_{p}]}{(\bar{\lambda}\lambda)^{2}} + \frac{1}{16}\frac{(r\gamma_{mnp}r)(\bar{\lambda}\gamma^{m}d)N^{np}}{(\bar{\lambda}\lambda)^{3}} \nonumber \\ 
&& - \frac{1}{128}\frac{(r\gamma_{mnp}r)(\bar{\lambda}\gamma^{pqr}r)N^{mn}N_{qr}}{(\bar{\lambda}\lambda)^{4}}
\end{eqnarray}
where $N_{mn} = \frac{1}{2}(\lambda \gamma_{mn} w)$.

\vspace{2mm}
The complicated nature of this expression makes it difficult to prove nilpotence \cite{Jusinskas:2013yca}, however it was shown in 
\cite{Berkovits:2013pla} that the $b$-ghost can be simplified by introducing an $SO(9,1)$ composite fermionic vector $\bar{\Gamma}^{m}$ satisfying the constraint $ (\gamma_{m}\bar{\lambda})^{\mu}\bar{\Gamma}^{m} = 0$. In the expression (\ref{section4eeq5}), the terms involving $d_{\mu}$ always appear in the combination
\begin{equation}\label{neweqnonew}
\bar{\Gamma}^{m} = \frac{1}{2}\frac{(\bar{\lambda}\gamma^{m}d)}{(\bar{\lambda}\lambda)} - \frac{1}{8}\frac{(\bar{\lambda}\gamma^{mnp}r)N_{np}}{(\bar{\lambda}\lambda)^{2}, }
\end{equation}
and using this $\bar{\Gamma}^{m}$, the $b$-ghost can be written in the simpler form:
\begin{equation}\label{neweeq6new}
b = P^{m}\bar{\Gamma}_{m} - \frac{1}{4}\frac{(\lambda\gamma^{mn}r)}{(\bar{\lambda}\lambda)}\bar{\Gamma}_{m}\bar{\Gamma}_{n}
\end{equation} 
This simplified $D=10$ $b$-ghost was shown to satisfy the property $\{Q , b\} = T$ in \cite{Berkovits:2014ama} and the nilpotence property $\{b,b\} =0$ easily follows from
$\{\bar\Gamma_m, \bar\Gamma_n\}=0$ and $[\bar\Gamma_m, \bar\lambda\lambda]=0$. Let us check this explicitly as a warm up exercise before moving on to the eleven-dimensional case. Let us start with the commutator $\{\bar{\Gamma}^{m}, \bar{\Gamma}^{n}\}$:
\begin{eqnarray}
\{\bar{\Gamma}^{m}, \bar{\Gamma}^{n}\} &=& \{\frac{1}{2}\frac{(\bar{\lambda}\gamma^{m}d)}{(\bar{\lambda}\lambda)} - \frac{1}{8}\frac{(\bar{\lambda}\gamma^{mrs}r)N_{rs}}{(\bar{\lambda}\lambda)^{2}}, \frac{1}{2}\frac{(\bar{\lambda}\gamma^{n}d)}{(\bar{\lambda}\lambda)} - \frac{1}{8}\frac{(\bar{\lambda}\gamma^{npq}r)N_{pq}}{(\bar{\lambda}\lambda)^{2}}\}\nonumber \\
&=& -\frac{1}{4(\bar{\lambda}\lambda)^{2}}(\bar{\lambda}\gamma^{m}\gamma^{k}\gamma^{n}\bar{\lambda})P_{k} +  \frac{1}{32(\bar{\lambda}\lambda)^{4}}(\bar{\lambda}\gamma^{m}d)(\bar{\lambda}\gamma^{npq}r)(\lambda\gamma_{pq}\bar{\lambda})\nonumber \\
&& + \frac{1}{32(\bar{\lambda}\lambda)^{4}}(\bar{\lambda}\gamma^{n}d)(\bar{\lambda}\gamma^{mrs}r)(\lambda\gamma_{rs}\bar{\lambda}) + \frac{1}{16(\bar{\lambda}\lambda)^{4}}(\bar{\lambda}\gamma^{mrs}r)(\bar{\lambda}\gamma^{nps}r)N_{pr} \nonumber \\
&& + \frac{1}{64(\bar{\lambda}\lambda)^{5}}(\bar{\lambda}\gamma^{mrs}r)(\bar{\lambda}\gamma^{npq}r)(\lambda\gamma_{rs}\bar{\lambda})N_{pq}\nonumber - \frac{1}{64(\bar{\lambda}\lambda)^{5}}(\bar{\lambda}\gamma^{mrs}r)N_{rs}(\bar{\lambda}\gamma^{npq}r)(\lambda\Gamma_{pq}\bar{\lambda})
\end{eqnarray}
where the constraint algebra $\{d_{\mu}, d_{\nu}\} = -\gamma^{m}_{\mu\nu}P_{m}$ was used. The term proportional to $P^{m}$ is zero because the pure spinor constraint and the symmetry property of the ten-dimensional 3-form gamma matrix. Furthermore, one can show that $(\bar{\lambda}\gamma^{npq}r)(\lambda\gamma_{pq}\bar{\lambda}) = 0$ from the identity \eqref{ap2}. In this way, one is left with
\begin{equation}
\{\bar{\Gamma}^{m}, \bar{\Gamma}^{n}\} = \frac{1}{16}(\bar{\lambda}\gamma^{mrs}r)(\bar{\lambda}\gamma^{nps}r)N_{pr}
\end{equation}
Using eqn. \eqref{ap4}, this expression can be put into the form
\begin{equation}
\{\bar{\Gamma}^{m}, \bar{\Gamma}^{n}\} = -\frac{1}{3!16^{2}}(\bar{\lambda}\gamma^{mrs}\gamma^{tuv}\gamma^{nps}\bar{\lambda})(r\gamma_{tuv}r)N_{pr}
\end{equation}
One can now use the GAMMA package \cite{Gran:2001yh} to do the gamma matrix manipulations. The final result is easily found to be
\begin{eqnarray}
\{\bar{\Gamma}^{m}, \bar{\Gamma}^{n}\} &=& -\frac{1}{3!16^{2}}[\eta^{np}(\bar{\lambda}\gamma^{tuvmr}\bar{\lambda}) + \eta^{mp}(\bar{\lambda}\gamma^{tuvnr}\bar{\lambda}) - \eta^{mn}(\bar{\lambda}\gamma^{tuvpr}\bar{\lambda})\nonumber \\
&& + 6\eta^{tn}(\bar{\lambda}\gamma^{uvmpr}\bar{\lambda}) + 6\eta^{tm}(\bar{\lambda}\gamma^{uvnpr}\bar{\lambda})](r\gamma_{tuv}r)N_{pr}
\end{eqnarray}
Using the identity \eqref{ap5} and the pure spinor constraints, one then demonstrates that
\begin{equation}\label{section4eq1}
\{\bar{\Gamma}^{m}, \bar{\Gamma}^{n}\} = 0
\end{equation}
The use of \eqref{section4eq1} allows us to calculate $\{b,b\}$ directly:
\begin{eqnarray}
\{b , b\} &=& \{P^{m}\bar{\Gamma}_{m} - \frac{1}{4}\frac{(\lambda\gamma^{mn}r)}{(\bar{\lambda}\lambda)}\bar{\Gamma}_{m}\bar{\Gamma}_{n}, P^{p}\bar{\Gamma}_{p} - \frac{1}{4}\frac{(\lambda\gamma^{pq}r)}{(\bar{\lambda}\lambda)}\bar{\Gamma}_{p}\bar{\Gamma}_{q}\}\nonumber \\
&=& 0
\end{eqnarray}
where we used that $[\bar{\Gamma}^{m}, \bar{\lambda}\lambda] = \frac{1}{16(\bar{\lambda}\lambda)^{2}}(\bar{\lambda}\gamma^{mnp}r)(\lambda\gamma_{np}\bar{\lambda}) = 0$ and $\{\bar{\Gamma}^{m}, \lambda\gamma^{rs}r\}\bar{\Gamma}_{r}\bar{\Gamma}_{s} = \frac{1}{8(\bar{\lambda}\lambda)}(\bar{\lambda}\gamma^{mnp}r)(\lambda\gamma_{np}\gamma^{rs}r)\bar{\Gamma}_{r}\bar{\Gamma}_{s} = \frac{1}{2(\bar{\lambda}\lambda)}[-(\bar{\lambda}\gamma^{mns}r)(\lambda\gamma_{n}^{\hspace{2mm}r}r) + (\bar{\lambda}\gamma^{mnr}r)(\lambda\gamma_{n}^{\hspace{2mm}s}r)]\bar{\Gamma}_{r}\bar{\Gamma}_{s} = 0$ because of the constraint $(\gamma_{m}\bar{\lambda})^{\mu}\bar{\Gamma}^{m} = 0$.

\vspace{2mm}
It is worthwhile mentioning one can also show the nilpotency of the $D=10$ $b$-ghost through a simple $U(5)$-covariant analysis. We will leave this non Lorentz-covariant proof for Appendix \ref{apI} for the interested reader and move on now to the study of the eleven-dimensional case.

\section{$D=11$ non-minimal pure spinor superparticle}\label{s4.3}
The $D=11$ non-minimal pure spinor superparticle action in a flat background is given by \cite{Berkovits:2002uc}
\begin{equation}\label{section4eeq1}
S = \int d\tau (P_{a}\partial_{\tau}X^{a} + P_{\alpha}\partial_{\tau}\Theta^{\alpha} + W_{\alpha}\partial_{\tau}\Lambda^{\alpha} + \bar{W}^{\alpha}\partial_{\tau}\bar{\Lambda}_{\alpha} + S^{\alpha}\partial_{\tau}R_{\alpha} - \frac{1}{2}P^{a}P_{a})
\end{equation}
We use letters of the beginning of the Greek alphabet ($\alpha, \beta, \ldots$) to denote $SO(10,1)$ spinor indices and henceforth we will use Latin letters ($a, b, \ldots, j, k$) to denote $SO(10,1)$ vector indices, unless otherwise stated. In \eqref{section4eeq1} $\Theta^{\alpha}$ is an $SO(10,1)$ Majorana spinor and $P_{\alpha}$ is its corresponding conjugate momentum, and $P_{a}$ is the momentum for $X^a$. The variables $\Lambda^{\alpha}$, $\bar{\Lambda}_{\alpha}$ are $D=11$ pure spinors and $W_{\alpha}$, $\bar{W}^{\alpha}$ are their respective conjugate momenta, $R_{\alpha}$ is an $SO(10,1)$ fermionic spinor satisfying $\bar{\Lambda}\Gamma^{a}R = 0$ and $S^{\alpha}$ is its corresponding conjugate momentum. The $SO(10,1)$ gamma matrices denoted by $\Gamma^{a}$ satisfy the Clifford algebra $(\Gamma^{a})_{\alpha\beta}(\Gamma^{b})^{\beta\sigma} + (\Gamma^{b})_{\alpha\beta}(\Gamma^{a})^{\beta\sigma} = 2\eta^{ab}\delta^{\sigma}_{\alpha}$. In $D=11$ dimensions there exist an antisymmetric spinor metric $C_{\alpha\beta}$ (and its inverse $(C^{-1})^{\alpha\beta}$) which allows us to lower (and raise) spinor indices (e.g. $(\Gamma^{a})^{\alpha\beta} = C^{\alpha\sigma}C^{\beta\delta}(\Gamma^{a})_{\sigma\delta}$, $(\Gamma^{a})^{\alpha}_{\hspace{2mm}\beta} = C^{\alpha\sigma}(\Gamma^{a})_{\sigma\beta}$, etc).

\vspace{2mm}
The physical states described by this theory are defined as elements of the cohomology of the BRST operator $Q = \Lambda^{\alpha}D_{\alpha} + R_{\alpha}\bar{W}^{\alpha}$ where $D_\alpha = P_\alpha - P_a (\Gamma^a \Theta)_\alpha$ and describe $D=11$ linearized supergravity.

\subsection{$D=11$ $b$-ghost and its simplification}
As in the $D=10$ case, a composite $D=11$ $b$-ghost can be constructed satisfying the properties $\{Q , b\} = T$ where $T = -P^a P_a$, and was found in \cite{Cederwall:2012es,Karlsson:2014xva,Karlsson:2015qda} to be:
\begin{align}\label{section4eq16}
b &= \frac{1}{2}\eta^{-1}(\bar{\Lambda}\Gamma_{ab}\bar{\Lambda})(\Lambda\Gamma^{ab}\Gamma^{i}D)P_{i} + \eta^{-2}L^{(1)}_{ab,cd}[(\Lambda\Gamma^{a}D)(\Lambda\Gamma^{bcd}D) + 2(\Lambda\Gamma^{abc}_{\hspace{4mm}ij}\Lambda)N^{di}P^{j} \notag \\
& + \frac{2}{3}(\eta^{b}_{\hspace{2mm}f}\eta^{d}_{\hspace{2mm}g} - \eta^{bd}\eta_{fg})(\Lambda\Gamma^{afcij}\Lambda)N_{ij}P^{g}] - \frac{1}{3}\eta^{-3}L^{(2)}_{ab,cd,ef}\{(\Lambda\Gamma^{abcij}\Lambda)(\Lambda\Gamma^{def}D)N_{ij} \notag \\
 & - 12[ (\Lambda\Gamma^{abcei}\Lambda)\eta^{fj} - \frac{2}{3}\eta^{f[a}(\Lambda\Gamma^{bce]ij}\Lambda](\Lambda\Gamma^{d}D)N_{ij}\} \notag \\
& + \frac{4}{3}\eta^{-4}L^{(3)}_{ab,cd,ef,gh}(\Lambda\Gamma^{abcij}\Lambda)[(\Lambda\Gamma^{defgk}\Lambda)\eta^{hl} - \frac{2}{3}\eta^{h[d}(\Lambda\Gamma^{efg]kl}\Lambda)]\{N_{ij},N_{kl}\}
\end{align}
where
\begin{eqnarray}
\eta &=& (\Lambda\Gamma^{ab}\Lambda)(\bar{\Lambda}\Gamma_{ab}\bar{\Lambda})\\
L^{(n)}_{a_{0}b_{0}, a_{1}b_{1}, \ldots, a_{n}b_{n}} &=& (\bar{\Lambda}\Gamma_{\llbracket a_{0}b_{0}}\bar{\Lambda})(\bar{\Lambda}\Gamma_{a_{1}b_{1}}R)\ldots (\bar{\Lambda}\Gamma_{a_{n}b_{n}\rrbracket}R)
\end{eqnarray}
and $\llbracket \rrbracket$ means antisymmetrization between each pair of indices. The $D=11$ ghost current is defined by $N_{ij} = \Lambda\Gamma_{ij}W$.


\vspace{2mm}
To simplify this complicated expression for the $D=11$ $b$-ghost, we shall
mimic the procedure explained above for the $D=10$ $b$-ghost and look for a similar object to $\bar{\Gamma}^{m}$. A hint comes from looking at the quantity multiplying the momentum $P^{i}$ in the expression for the $D=11$ $b$-ghost:
\begin{eqnarray}
b &=& P^{i}[\frac{1}{2}\eta^{-1}(\bar{\Lambda}\Gamma_{ab}\bar{\Lambda})(\Lambda\Gamma^{ab}\Gamma_{i}D) + \eta^{-2}L^{(1)}_{ab,cd}[2(\Lambda\Gamma^{abc}_{\hspace{4mm}ki}\Lambda)N^{dk}\nonumber \\
&& 
+ \frac{2}{3}(\eta^{b}_{\hspace{2mm}f}\eta^{d}_{\hspace{2mm}i} - \eta^{bd}\eta_{fi})(\Lambda\Gamma^{afcgj}\Lambda)N_{gj}]] + \ldots
\end{eqnarray}
Therefore our candidate to play the analog role to $\bar{\Gamma}^{m}$ is:
\begin{equation}\label{eq406}
\bar{\Sigma}^{i} = \bar{\Sigma}^{i}_{0} + \frac{2}{\eta^{2}}L^{(1)}_{ab,cd}(\Lambda\Gamma^{abcki}\Lambda)N^{d}_{\hspace{1mm}k} + \frac{2}{3\eta^{2}}L^{(1)\hspace{2mm}i}_{ab,c}(\Lambda\Gamma^{abcfj}\Lambda)N_{fj} - \frac{2}{3\eta^{2}}L^{(1)\hspace{2mm}d}_{ad,c}(\Lambda\Gamma^{aicfj}\Lambda)N_{fj}
\end{equation}
where $\bar{\Sigma}^{i}_{0} = \frac{1}{2}\eta^{-1}(\bar{\Lambda}\Gamma_{ab}\bar{\Lambda})(\Lambda\Gamma^{ab}\Gamma^{i}D)$ is the only term containing $D_{\alpha}$'s. 
Using the identities \eqref{newapp1}, \eqref{newapp2} in Appendix \ref{apapB}, one finds that
$\bar{\Sigma}^{j}$ satisfies the constraint:
\begin{equation}
(\bar{\Lambda}\Gamma_{ab}\bar{\Lambda})\bar{\Sigma}^{a} = 0.
\end{equation}

\vspace{2mm}
Furthermore, it will be shown in Appendix \ref{apapC} that the $D_{\alpha}$'s appearing in $\bar{\Sigma}_{0}^{i}$ are the same as those appearing in the $b$-ghost. Therefore a plausible assumption for the simplification of the $b$-ghost would be
$b = P^{i}\bar{\Sigma}_{i} + O(\bar{\Sigma}^{2})$. As will now be shown, the simplified form of 
the $b$-ghost satisfying $\{Q, b\} = T$ is indeed
\begin{align}
b & = P^{i}\bar{\Sigma}_{i} - \frac{2}{\eta}(\bar{\Lambda}\Gamma_{ac}R)(\Lambda\Gamma^{aj}\Lambda)\bar{\Sigma}_{j}\bar{\Sigma}^{c} - \frac{1}{\eta}(\bar{\Lambda}R)(\Lambda\Gamma^{jk}\Lambda)\bar{\Sigma}_{j}\bar{\Sigma}_{k} \label{section4eq28}
\end{align}

\subsection{Computation of $\{Q, \bar\Sigma^j\}$ }

To show that the $b$-ghost of \eqref{section4eq28} satisfies $\{Q, b\}=T$, it will be convenient to first compute
$\{Q,\bar\Sigma^i\}$ where, 
using the identities \eqref{newapp5}, \eqref{newapp10},
\begin{align}\label{section4eeq22}
\bar{\Sigma}^{i} & = \bar{\Sigma}_{0}^{i} + \frac{2}{\eta^{2}}(\bar{\Lambda}\Gamma_{ab}\bar{\Lambda})(\bar{\Lambda}\Gamma_{cd}R)(\Lambda\Gamma^{abcki}\Lambda)N^{d}_{\hspace{2mm}k} + \frac{2}{3\eta^{2}}(\bar{\Lambda}\Gamma_{ab}\bar{\Lambda})(\bar{\Lambda}\Gamma_{c}^{\hspace{2mm}i}R)(\Lambda\Gamma^{abcfj}\Lambda)N_{fj} \notag \\
&  - \frac{2}{3\eta^{2}}(\bar{\Lambda}\Gamma_{ac}\bar{\Lambda})(\bar{\Lambda}R)(\Lambda\Gamma^{aicfj}\Lambda)N_{fj}
\end{align}
Using eqn. \eqref{section4eeq22} and the identities \eqref{newapp19}, \eqref{newapp20}, \eqref{newapp21}, \eqref{newapp22}:
\begin{align}
\{Q, \bar{\Sigma}^{i}\} & = -P^{i} - \frac{2}{\eta}[(\bar{\Lambda}\Gamma^{fb}\bar{\Lambda})(\Lambda\Gamma_{b}^{\hspace{1mm}i}\Lambda) - (\bar{\Lambda}\Gamma^{ib}\bar{\Lambda})(\Lambda\Gamma_{b}^{\hspace{1mm}f}\Lambda)]P_{f} + \frac{2}{\eta}(\bar{\Lambda}\Gamma_{fg}R)(\Lambda\Gamma^{fg}\Lambda)\bar{\Sigma}^{i}_{0} \notag \\
& + \frac{4}{\eta}(\bar{\Lambda}\Gamma^{fg}R)(\Lambda\Gamma_{fg}\Lambda)(\bar{\Sigma}^{i} - \bar{\Sigma}^{i}_{0})-\frac{1}{\eta}(\bar{\Lambda}\Gamma_{ab}R)(\Lambda\Gamma^{ab}\Gamma^{i}D) \notag \\
& - \frac{2}{\eta^{2}}(\bar{\Lambda}\Gamma_{ab}\bar{\Lambda})(\bar{\Lambda}\Gamma_{cd}R)(\Lambda\Gamma^{abcki}\Lambda)(\Lambda\Gamma^{d}_{\hspace{1mm}k}D) - \frac{2}{3\eta^{2}}(\bar{\Lambda}\Gamma_{ab}\bar{\Lambda})(\bar{\Lambda}\Gamma_{c}^{\hspace{2mm}i}R)(\Lambda\Gamma^{abcdk}\Lambda)(\Lambda\Gamma_{dk}D)\notag \\
& - \frac{2}{3\eta^{2}}(\bar{\Lambda}\Gamma_{ab}\bar{\Lambda})(\bar{\Lambda}R)(\Lambda\Gamma^{iabdk}\Lambda)(\Lambda\Gamma_{dk}D) - \frac{4}{\eta^{2}}(\bar{\Lambda}\Gamma_{ab}R)(\bar{\Lambda}\Gamma_{cd}R)(\Lambda\Gamma^{abcki}\Lambda)N^{d}_{\hspace{1mm}k}\notag \\
& - \frac{4}{3\eta^{2}}(\bar{\Lambda}\Gamma_{ab}R)(\bar{\Lambda}\Gamma_{c}^{\hspace{2mm}i}R)(\Lambda\Gamma^{abcdk}\Lambda)N_{dk} - \frac{4}{3\eta^{2}}(\bar{\Lambda}\Gamma_{ab}R)(\bar{\Lambda}R)(\Lambda\Gamma^{iabdk}\Lambda)N_{dk} \notag \\
& - \frac{2}{3\eta^{2}}(\bar{\Lambda}\Gamma_{ab}\bar{\Lambda})(RR)(\Lambda\Gamma^{iabdk}\Lambda)N_{dk} \label{neweq26}
\end{align}

As shown in Appendix \ref{apD1}, this expression in invariant under the same gauge transformations under which $\bar{\Sigma}_{0}^{i}$ is invariant:
\begin{equation}\label{eq400}
 \delta D_{\alpha} = (\Gamma^{ij}\Lambda\ensuremath{'})_{\alpha}f_{ij}
\end{equation} 
where $(\Lambda\ensuremath{'})^{\alpha} = \frac{1}{2\eta}(\bar{\Lambda}\Gamma_{ab}\bar{\Lambda})(\Lambda\Gamma^{ab})^{\alpha}$ is a pure spinor, and $f_{ij}$ is an antisymmetric gauge parameter. Therefore we can write all $D_\alpha$'s in this object in terms of $\bar{\Sigma}^{i}_{0}$, and the result is (see Appendix \ref{apD1}):
\begin{align}
\{Q, \bar{\Sigma}^{i}\} & = -P^{i} - \frac{2}{\eta}[(\bar{\Lambda}\Gamma^{fb}\bar{\Lambda})(\Lambda\Gamma_{b}^{\hspace{1mm}i}\Lambda) - (\bar{\Lambda}\Gamma^{ib}\bar{\Lambda})(\Lambda\Gamma_{b}^{\hspace{1mm}f}\Lambda)]P_{f} + \frac{4}{\eta}(\bar{\Lambda}\Gamma_{fg}R)(\Lambda\Gamma^{fg}\Lambda)(\bar{\Sigma}^{i} - \bar{\Sigma}^{i}_{0}) \notag \\
& - \frac{2}{\eta}(\bar{\Lambda}\Gamma^{ci}R)(\Lambda\Gamma_{ck}\Lambda)\bar{\Sigma}_{0}^{k} + \frac{4}{\eta}(\bar{\Lambda}\Gamma_{cd}R)(\Lambda\Gamma^{ci}\Lambda)\bar{\Sigma}_{0}^{d} + \frac{2}{\eta}(\bar{\Lambda}R)(\Lambda\Gamma^{ik}\Lambda)\bar{\Sigma}_{0\,k} \notag \\
& - \frac{2}{\eta^{2}}(\bar{\Lambda}\Gamma_{cd}R)(\Lambda\Gamma^{cd}\Lambda)(\bar{\Lambda}\Gamma^{ig}\bar{\Lambda})(\Lambda\Gamma^{gk}\Lambda)\bar{\Sigma}_{0\,k} - \frac{4}{\eta^{2}}(\bar{\Lambda}\Gamma_{ab}R)(\bar{\Lambda}\Gamma_{cd}R)(\Lambda\Gamma^{abcki}\Lambda)N^{d}_{\hspace{1mm}k}\notag\\
& - \frac{4}{3\eta^{2}}(\bar{\Lambda}\Gamma_{ab}R)(\bar{\Lambda}\Gamma_{c}^{\hspace{2mm}i}R)(\Lambda\Gamma^{abcdk}\Lambda)N_{dk} - \frac{4}{3\eta^{2}}(\bar{\Lambda}\Gamma_{ab}R)(\bar{\Lambda}R)(\Lambda\Gamma^{iabdk}\Lambda)N_{dk}\notag \\
& - \frac{2}{3\eta^{2}}(\bar{\Lambda}\Gamma_{ab}\bar{\Lambda})(RR)(\Lambda\Gamma^{iabdk}\Lambda)N_{dk}\label{section4eq401}
\end{align}

After plugging \eqref{section4eeq22} into \eqref{section4eq401}, all of the terms explicitly depending on $N_{ab}$ are cancelled and we get (see appendix \ref{apD2}): 
\begin{align}\label{section4eq500}
\{Q, \bar{\Sigma}^{i}\} & = -P^{i} - \frac{2}{\eta}[(\bar{\Lambda}\Gamma^{fb}\bar{\Lambda})(\Lambda\Gamma_{b}^{\hspace{1mm}i}\Lambda) - (\bar{\Lambda}\Gamma^{ib}\bar{\Lambda})(\Lambda\Gamma_{b}^{\hspace{1mm}f}\Lambda)]P_{f} -\frac{2}{\eta}(\bar{\Lambda}\Gamma^{ci}R)(\Lambda\Gamma_{ck}\Lambda)\bar{\Sigma}^{k} \notag \\
& + \frac{4}{\eta}(\bar{\Lambda}\Gamma_{cd}R)(\Lambda\Gamma^{ci}\Lambda)\bar{\Sigma}^{d} \notag  + \frac{2}{\eta}(\bar{\Lambda}R)(\Lambda\Gamma^{ik}\Lambda)\bar{\Sigma}_{k} - \frac{2}{\eta^{2}}(\bar{\Lambda}\Gamma_{cd}R)(\Lambda\Gamma^{cd}\Lambda)(\bar{\Lambda}\Gamma^{ig}\bar{\Lambda})(\Lambda\Gamma^{gk}\Lambda)\bar{\Sigma}_{k}\\
\end{align}

\subsection{$\{Q, b\} = T$}

Using \eqref{section4eq500}  it is now straightforward to compute $\{Q , b\}$:
\begin{align}
\{Q , b\} & = P^{i}\{Q , \bar{\Sigma}_{i}\} - \frac{4}{\eta^{2}}(\Lambda\Gamma^{fg}\Lambda)(\bar{\Lambda}\Gamma_{fg}R)(\bar{\Lambda}\Gamma^{aj}R)(\Lambda\Gamma_{ak}\Lambda)\bar{\Sigma}^{k}\bar{\Sigma}_{j} \notag \\
& + \frac{2}{\eta}(\bar{\Lambda}\Gamma^{aj}R)(\Lambda\Gamma_{ak}\Lambda)(\{Q , \bar{\Sigma}^{k}\})\bar{\Sigma}_{j} - \frac{2}{\eta}(\bar{\Lambda}\Gamma^{aj}R)(\Lambda\Gamma_{ak}\Lambda)\bar{\Sigma}^{k}(\{Q , \bar{\Sigma}_{j}\}) \notag \\
& - \frac{2}{\eta^{2}}(\Lambda\Gamma^{fg}\Lambda)(\bar{\Lambda}\Gamma_{fg}R)(\bar{\Lambda}R)(\Lambda\Gamma^{jk}\Lambda)\bar{\Sigma}_{j}\bar{\Sigma}_{k} + \frac{1}{\eta}(RR)(\Lambda\Gamma^{jk}\Lambda)\bar{\Sigma}_{j}\bar{\Sigma}_{k} \notag \\
& + \frac{1}{\eta}(\bar{\Lambda}R)(\Lambda\Gamma^{jk}\Lambda)(\{Q , \bar{\Sigma}_{j}\})\bar{\Sigma}_{k} - \frac{1}{\eta}(\bar{\Lambda}R)(\Lambda\Gamma^{jk}\Lambda)\bar{\Sigma}_{j}(\{Q , \bar{\Sigma}_{k}\}) \label{section4eq27}
\end{align}
To make the computations transparent, each term in \eqref{section4eq27} involving $\{Q , \bar{\Sigma}_{i}\} $ will be simplified separately:
\begin{align}
M_{1} & = P^{i}\{Q , \bar{\Sigma}_{i}\} \notag \\
& = P_{i}\{-P^{i} - \frac{2}{\eta}[(\Lambda\Gamma^{ib}\Lambda)(\bar{\Lambda}\Gamma_{bf}\bar{\Lambda}) - (\Lambda\Gamma^{fb}\Lambda)(\bar{\Lambda}\Gamma^{bi}\bar{\Lambda})]P_{f} \notag \\
& - \frac{2}{\eta}(\bar{\Lambda}\Gamma^{ci}R)(\Lambda\Gamma_{ck}\Lambda)\bar{\Sigma}^{k} + \frac{4}{\eta}(\bar{\Lambda}\Gamma_{cd}R)(\Lambda\Gamma^{ci}\Lambda)\bar{\Sigma}^{d} \notag \\
& + \frac{2}{\eta}(\bar{\Lambda}R)(\Lambda\Gamma^{ik}\Lambda)\bar{\Sigma}_{k} - \frac{2}{\eta^{2}}(\bar{\Lambda}\Gamma^{cd}R)(\Lambda\Gamma_{cd}\Lambda)(\bar{\Lambda}\Gamma^{ig}\bar{\Lambda})(\Lambda\Gamma_{gk}\Lambda)\bar{\Sigma}^{k}\}\notag \\
& = -P^{2} - \frac{2}{\eta}(\bar{\Lambda}\Gamma^{ci}R)(\Lambda\Gamma_{ck}\Lambda)P_{i}\bar{\Sigma}^{k} + \frac{4}{\eta}(\bar{\Lambda}\Gamma_{cd}R)(\Lambda\Gamma^{ci}\Lambda)P_{i}\bar{\Sigma}^{d} \notag \\
& + \frac{2}{\eta}(\bar{\Lambda}R)(\Lambda\Gamma^{ik}\Lambda)P_{i}\bar{\Sigma}_{k} - \frac{2}{\eta^{2}}(\bar{\Lambda}\Gamma^{cd}R)(\Lambda\Gamma_{cd}\Lambda)(\bar{\Lambda}\Gamma^{ig}\bar{\Lambda})(\Lambda\Gamma_{gk}\Lambda)P_{i}\bar{\Sigma}^{k}
\end{align}
\begin{align}
M_{2} & = \frac{2}{\eta}(\bar{\Lambda}\Gamma^{aj}R)(\Lambda\Gamma_{ak}\Lambda)(\{Q , \bar{\Sigma}^{k}\})\bar{\Sigma}_{j}\notag \\
& = \frac{2}{\eta}(\bar{\Lambda}\Gamma^{aj}R)(\Lambda\Gamma_{ak}\Lambda)[-P^{k} + \frac{2}{\eta}(\Lambda\Gamma_{fb}\Lambda)(\bar{\Lambda}\Gamma^{bk}\bar{\Lambda})P^{f} \notag - \frac{2}{\eta}(\bar{\Lambda}\Gamma^{ck}R)(\Lambda\Gamma_{cf}\Lambda)\bar{\Sigma}^{f}\\
& - \frac{2}{\eta^{2}}(\bar{\Lambda}\Gamma^{cd}R)(\Lambda\Gamma_{cd}\Lambda)(\bar{\Lambda}\Gamma^{kg}\bar{\Lambda})(\Lambda\Gamma_{gf}\Lambda)\bar{\Sigma}^{f}]\bar{\Sigma}_{j}\notag \\
& = -\frac{2}{\eta}{\bar{\Lambda}\Gamma^{aj}R}(\Lambda\Gamma_{ak}\Lambda)P^{k}\bar{\Sigma}_{j} + \frac{4}{\eta^{2}}(\bar{\Lambda}\Gamma^{aj}R)(\Lambda\Gamma_{ak}\Lambda)(\Lambda\Gamma_{fb}\Lambda)(\bar{\Lambda}\Gamma^{bk}\bar{\Lambda})P^{f}\bar{\Sigma}_{j}\notag \\
& - \frac{4}{\eta^{2}}(\bar{\Lambda}\Gamma^{aj}R)(\Lambda\Gamma_{ak}\Lambda)(\bar{\Lambda}\Gamma^{ck}R)(\Lambda\Gamma_{cf}\Lambda)\bar{\Sigma}^{f}\bar{\Sigma}_{j} \notag \\
& - \frac{4}{\eta^{3}}(\bar{\Lambda}\Gamma^{aj}R)(\Lambda\Gamma_{ak}\Lambda)(\bar{\Lambda}\Gamma^{cd}R)(\Lambda\Gamma_{cd}\Lambda)(\bar{\Lambda}\Gamma^{kg}\bar{\Lambda})(\Lambda\Gamma_{gf}\Lambda)\bar{\Sigma}^{f}\bar{\Sigma}_{j} \notag
\end{align}
Using \eqref{newapp2}, we get
\begin{align}
M_{2} & = -\frac{2}{\eta}({\bar{\Lambda}\Gamma^{aj}R})(\Lambda\Gamma_{ak}\Lambda)P^{k}\bar{\Sigma}_{j} + \frac{2}{\eta}(\bar{\Lambda}\Gamma^{aj}R)(\Lambda\Gamma_{fa}\Lambda)P^{f}\bar{\Sigma}_{j} \notag \\ 
& - \frac{2}{\eta^{2}}(\bar{\Lambda}\Gamma^{aj}R)(\Lambda\Gamma_{af}\Lambda)(\bar{\Lambda}\Gamma^{ck}R)(\Lambda\Gamma_{ck}\Lambda)\bar{\Sigma}^{f}\bar{\Sigma}_{j} \notag \\
& - \frac{2}{\eta^{2}}(\bar{\Lambda}\Gamma^{aj}R)(\Lambda\Gamma_{fa}\Lambda)(\bar{\Lambda}\Gamma^{cd}R)(\Lambda\Gamma_{cd}\Lambda)\bar{\Sigma}^{f}\bar{\Sigma}_{j} \notag \\
& = -\frac{4}{\eta}({\bar{\Lambda}\Gamma^{aj}R})(\Lambda\Gamma_{ak}\Lambda)P^{k}\bar{\Sigma}_{j}
\end{align}

\begin{align}
M_{3} & = - \frac{2}{\eta}(\bar{\Lambda}\Gamma^{aj}R)(\Lambda\Gamma_{ak}\Lambda)\bar{\Sigma}^{k}(\{Q , \bar{\Sigma}_{j}\}) \notag \\
& = - \frac{2}{\eta}(\bar{\Lambda}\Gamma^{aj}R)(\Lambda\Gamma_{ak}\Lambda)\bar{\Sigma}^{k}\{-P_{j} - \frac{2}{\eta}[(\Lambda\Gamma_{jb}\Lambda)(\bar{\Lambda}\Gamma^{bf}\bar{\Lambda}) - (\Lambda\Gamma^{fb}\Lambda)(\bar{\Lambda}\Gamma_{bj}\bar{\Lambda})]P_{f} \notag \\
& - \frac{2}{\eta}(\bar{\Lambda}\Gamma_{cj}R))(\Lambda\Gamma^{cf}\Lambda)\bar{\Sigma}_{f} + \frac{4}{\eta}(\bar{\Lambda}\Gamma^{cd}R)(\Lambda\Gamma_{cj}\Lambda)\bar{\Sigma}_{d} + \frac{2}{\eta}(\bar{\Lambda}R)(\Lambda\Gamma_{jf}\Lambda)\bar{\Sigma}^{f} \notag \\ & 
- \frac{2}{\eta^{2}}(\bar{\Lambda}\Gamma^{cd}R)(\Lambda\Gamma_{cd}\Lambda)(\bar{\Lambda}\Gamma_{jg}\bar{\Lambda})(\Lambda\Gamma^{gf}\Lambda)\bar{\Sigma}_{f}\} \notag \\
& = \frac{2}{\eta}(\bar{\Lambda}\Gamma^{aj}R)(\Lambda\Gamma_{ak}\Lambda)\bar{\Sigma}^{k}P_{j} + \frac{4}{\eta^{2}}(\bar{\Lambda}\Gamma^{aj}R)(\Lambda\Gamma_{ak}\Lambda)\bar{\Sigma}^{k}(\Lambda\Gamma_{jb}\Lambda)(\bar{\Lambda}\Gamma^{bf}\bar{\Lambda})P_{f} \notag \\
& - \frac{4}{\eta^{2}}(\bar{\Lambda}\Gamma^{aj}R)(\Lambda\Gamma_{ak}\Lambda)\bar{\Sigma}^{k}(\Lambda\Gamma^{fb}\Lambda)(\bar{\Lambda}\Gamma_{bj}\bar{\Lambda})P_{f} + \frac{4}{\eta^{2}}(\bar{\Lambda}\Gamma^{aj}R)(\Lambda\Gamma_{ak}\Lambda)\bar{\Sigma}^{k}(\bar{\Lambda}\Gamma^{cj}R)(\Lambda\Gamma_{cf}\Lambda)\bar{\Sigma}^{f} \notag \\
& - \frac{8}{\eta^{2}}(\bar{\Lambda}\Gamma^{aj}R)(\Lambda\Gamma_{ak}\Lambda)\bar{\Sigma}^{k}(\bar{\Lambda}\Gamma^{cd}R)(\Lambda\Gamma_{cj}\Lambda)\bar{\Sigma}_{d} - \frac{4}{\eta^{2}}(\bar{\Lambda}\Gamma^{aj}R)(\Lambda\Gamma_{ak}R)\bar{\Sigma}^{k}(\bar{\Lambda}R)(\Lambda\Gamma_{jf}\Lambda)\bar{\Sigma}^{f} \notag \\
& + \frac{4}{\eta^{3}}(\bar{\Lambda}\Gamma^{aj}R)(\Lambda\Gamma_{ak}\Lambda)\bar{\Sigma}^{k}(\bar{\Lambda}\Gamma^{cd}R)(\Lambda\Gamma_{cd}\Lambda)(\bar{\Lambda}\Gamma_{jg}\bar{\Lambda})(\Lambda\Gamma^{gf}\Lambda)\bar{\Sigma}_{f}\notag
\end{align}
Using \eqref{newapp2}, \eqref{newapp12}, \eqref{newapp25}:
\begin{align}
M_{3} & = \frac{2}{\eta}(\bar{\Lambda}\Gamma^{aj}R)(\Lambda\Gamma_{ak}\Lambda)\bar{\Sigma}^{k}P_{j} + \frac{2}{\eta^{2}}(\bar{\Lambda}\Gamma^{aj}R)(\Lambda\Gamma_{aj}\Lambda)(\Lambda\Gamma^{kb}\Lambda)(\bar{\Lambda}\Gamma_{bf}\bar{\Lambda})\bar{\Sigma}_{k}P^{f} \notag \\
& + \frac{2}{\eta}(\bar{\Lambda}R)(\Lambda\Gamma^{fk}\Lambda)P_{f}\bar{\Sigma}_{k} - \frac{2}{\eta^{2}}(\bar{\Lambda}\Gamma_{ac}R)(\Lambda\Gamma^{ac}\Lambda)(\bar{\Lambda}R)(\Lambda\Gamma^{kf}\Lambda)\bar{\Sigma}_{k}\bar{\Sigma}_{f} \notag \\
& - \frac{1}{\eta}(RR)(\Lambda\Gamma^{kf}\Lambda)\bar{\Sigma}_{k}\bar{\Sigma}_{f} + \frac{4}{\eta^{2}}(\bar{\Lambda}\Gamma^{aj}R)(\Lambda\Gamma_{aj}\Lambda)(\bar{\Lambda}\Gamma^{cd}R)(\Lambda\Gamma_{ck}\Lambda)\bar{\Sigma}^{k}\bar{\Sigma}_{d} \notag \\
& + \frac{2}{\eta^{2}}(\bar{\Lambda}\Gamma^{aj}R)(\Lambda\Gamma_{aj}\Lambda)(\bar{\Lambda}R)(\Lambda\Gamma^{kf}\Lambda)\bar{\Sigma}_{k}\bar{\Sigma}_{f} - \frac{2}{\eta^{2}}(\bar{\Lambda}R)(\Lambda\Gamma^{kf}\Lambda)(\bar{\Lambda}\Gamma_{cd}R)(\Lambda\Gamma^{cd}\Lambda)\bar{\Sigma}_{k}\bar{\Sigma}_{f}\notag \\
& = \frac{2}{\eta}(\bar{\Lambda}\Gamma^{aj}R)(\Lambda\Gamma_{ak}\Lambda)\bar{\Sigma}^{k}P_{j} + \frac{2}{\eta^{2}}(\bar{\Lambda}\Gamma^{aj}R)(\Lambda\Gamma_{aj}\Lambda)(\Lambda\Gamma^{kb}\Lambda)(\bar{\Lambda}\Gamma_{bf}\bar{\Lambda})\bar{\Sigma}_{k}P^{f} \notag \\
& + \frac{2}{\eta}(\bar{\Lambda}R)(\Lambda\Gamma^{fk}\Lambda)P_{f}\bar{\Sigma}_{k} - \frac{1}{\eta}(RR)(\Lambda\Gamma^{kf}\Lambda)\bar{\Sigma}_{k}\bar{\Sigma}_{f} + \frac{4}{\eta^{2}}(\bar{\Lambda}\Gamma^{aj}R)(\Lambda\Gamma_{aj}\Lambda)(\bar{\Lambda}\Gamma^{cd}R)(\Lambda\Gamma_{ck}\Lambda)\bar{\Sigma}^{k}\bar{\Sigma}_{d} \notag \\
& - \frac{2}{\eta^{2}}(\bar{\Lambda}R)(\Lambda\Gamma^{kf}\Lambda)(\bar{\Lambda}\Gamma_{cd}R)(\Lambda\Gamma^{cd}\Lambda)\bar{\Sigma}_{k}\bar{\Sigma}_{f}
\end{align}
\begin{align}
M_{4} & = \frac{1}{\eta}(\bar{\Lambda}R)(\Lambda\Gamma^{jk}\Lambda)(\{Q , \bar{\Sigma}_{j}\})\bar{\Sigma}_{k} \notag \\
& = -  \frac{1}{\eta}(\bar{\Lambda}R)(\Lambda\Gamma^{jk}\Lambda)P_{j}\bar{\Sigma}_{k} + \frac{2}{\eta^{2}}(\bar{\Lambda}R)(\Lambda\Gamma^{jk}\Lambda)(\Lambda\Gamma^{fb}\Lambda)(\bar{\Lambda}\Gamma_{bj}\bar{\Lambda})P_{f}\bar{\Sigma}_{k} \notag \\
& - \frac{2}{\eta^{2}}(\bar{\Lambda}R)(\Lambda\Gamma^{jk}\Lambda)(\bar{\Lambda}\Gamma_{cj}R)(\Lambda\Gamma^{cf}\Lambda)\bar{\Sigma}_{f}\bar{\Sigma}_{k}\notag\\
& - \frac{2}{\eta^{3}}(\bar{\Lambda}R)(\Lambda\Gamma^{jk}\Lambda)(\bar{\Lambda}\Gamma_{cd}R)(\Lambda\Gamma^{cd}\Lambda)(\bar{\Lambda}\Gamma_{jg}\bar{\Lambda})(\Lambda\Gamma^{gf}\Lambda)\bar{\Sigma}_{f}\bar{\Sigma}_{k}\notag \\
& = - \frac{1}{\eta}(\bar{\Lambda}R)(\Lambda\Gamma^{jk}\Lambda)P_{j}\bar{\Sigma}_{k} + \frac{1}{\eta}(\bar{\Lambda}R)(\Lambda\Gamma^{kf}\Lambda)P_{f}\bar{\Sigma}_{k} \notag \\
& - \frac{1}{\eta^{2}}(\bar{\Lambda}R)(\Lambda\Gamma^{fk}\Lambda)(\bar{\Lambda}\Gamma^{cj}R)(\Lambda\Gamma_{cj}\Lambda)\bar{\Sigma}_{f}\bar{\Sigma}_{k} - \frac{1}{\eta^{2}}(\bar{\Lambda}R)(\bar{\Lambda}\Gamma^{cd}R)(\Lambda\Gamma_{cd}\Lambda)(\Lambda\Gamma^{kf}\Lambda)\bar{\Sigma}_{f}\bar{\Sigma}_{k} \notag \\
& = -\frac{2}{\eta}(\bar{\Lambda}R)(\Lambda\Gamma^{jk}\Lambda)P_{j}\bar{\Sigma}_{k}
\end{align}
\begin{align}
M_{5} & = -\frac{1}{\eta}(\bar{\Lambda}R)(\Lambda\Gamma^{jk}\Lambda)\bar{\Sigma}_{j}(\{Q , \bar{\Sigma}_{k}\}) \notag \\
& = \frac{1}{\eta}(\bar{\Lambda}R)(\Lambda\Gamma^{jk}\Lambda)\bar{\Sigma}_{j}P_{k} - \frac{2}{\eta^{2}}(\bar{\Lambda}R)(\Lambda\Gamma^{jk}\Lambda)\bar{\Sigma}_{j}(\Lambda\Gamma^{fb}\Lambda)(\bar{\Lambda}\Gamma_{bk}\bar{\Lambda})P_{f} \notag \\
& + \frac{2}{\eta^{2}}(\bar{\Lambda}R)(\Lambda\Gamma^{jk}\Lambda)\bar{\Sigma}_{j}(\bar{\Lambda}\Gamma_{ck}R)(\Lambda\Gamma^{cf}\Lambda)\bar{\Sigma}_{f}\notag\\
& + \frac{2}{\eta^{3}}(\bar{\Lambda}R)(\Lambda\Gamma^{jk}\Lambda)\bar{\Sigma}_{j}(\bar{\Lambda}\Gamma^{cd}R)(\Lambda\Gamma_{cd}\Lambda)(\bar{\Lambda}\Gamma_{kg}\bar{\Lambda})(\Lambda\Gamma^{gf}\Lambda)\bar{\Sigma}_{f}\notag \\
& = \frac{1}{\eta}(\bar{\Lambda}R)(\Lambda\Gamma^{kj}\Lambda)P_{j}\bar{\Sigma}_{k} - \frac{1}{\eta}(\bar{\Lambda}R)(\Lambda\Gamma^{fj}\Lambda)P_{f}\bar{\Sigma}_{j} \notag \\
& + \frac{1}{\eta^{2}}(\bar{\Lambda}R)(\Lambda\Gamma^{jf}\Lambda)\bar{\Sigma}_{j}(\bar{\Lambda}\Gamma^{ck}R)(\Lambda\Gamma_{ck}\Lambda)\bar{\Sigma}_{f} + \frac{1}{\eta^{2}}(\bar{\Lambda}R)(\Lambda\Gamma^{fj}\Lambda)\bar{\Sigma}_{j}(\bar{\Lambda}\Gamma^{cd}R)(\Lambda\Gamma_{cd}\Lambda)\bar{\Sigma}_{f}\notag \\
&= -\frac{2}{\eta}(\bar{\Lambda}R)(\Lambda\Gamma^{jk}\Lambda)P_{j}\bar{\Sigma}_{k}
\end{align}
Putting together all the terms in \eqref{section4eq27}:
\begin{align}
\{Q , b\} & = \sum_{i=1}^{5}M_{i} - \frac{4}{\eta^{2}}(\Lambda\Gamma^{fg}\Lambda)(\bar{\Lambda}\Gamma_{fg}R)(\bar{\Lambda}\Gamma^{aj}R)(\Lambda\Gamma_{ak}\Lambda)\bar{\Sigma}^{k}\bar{\Sigma}_{j} \notag \\
& -\frac{2}{\eta^{2}}(\Lambda\Gamma^{fg}\Lambda)(\bar{\Lambda}\Gamma_{fg}R)(\bar{\Lambda}R)(\Lambda\Gamma^{jk}\Lambda)\bar{\Sigma}_{j}\bar{\Sigma}_{k} + \frac{1}{\eta}(RR)(\Lambda\Gamma^{jk}\Lambda)\bar{\Sigma}_{j}\bar{\Sigma}_{k}\notag \\
& = -P^{2}
\end{align}
Recalling that $T = -P^{2}$ is the stress-energy tensor, we have checked that $\{Q , b\} = T$.

\subsection{$\{b,b\}$ = BRST-trivial}
In the $D=10$ case, the identity $\{\bar\Gamma^m, \bar \Gamma^n\}=0$ was crucial for showing that
$\{b,b\}=0$. However, in the $D=11$ case, it is shown in  Appendix \ref{apD3}  that 
$\{\bar\Sigma^j, \bar \Sigma^k\}$ is non-zero and is proportional to $R_\alpha$. This implies that 

\begin{equation}
\{b , b\} = R^{\alpha}G_{\alpha}(\Lambda, \bar{\Lambda}, R, W, D)
\end{equation}
for some $G_{\alpha}(\Lambda, \bar{\Lambda}, R, W, D)$.

\vspace{2mm}
Note that $[Q,\{b , b\}] = 0$ since $[b, T]=0$ where $T = -P_a P^a$. 
Since $Q = \Lambda^{\alpha}D_{\alpha} + R^{\alpha}\bar{W}_{\alpha}$, the quartet argument implies that the cohomology of $Q$ is independent of $R_\alpha$, which allows us to conclude that $\{b , b\} = $ BRST-trivial. It would be interesting to investigate if this BRST-triviality of $\{b,b\}$ is enough for the scattering amplitude prescription using the $b$-ghost to be consistent.

\section{An alternative $D=11$ ghost number zero vertex operator}\label{s4.3}
In the previous chapter we gave an argument against the existence of a ghost number zero vertex operator satisfying a standard descent equation \eqref{eq75} in the minimal formalism. We mentioned there that one possible way of fixing this problem is by extending the minimal formalism to the non-minimal one studied in this chapter. Indeed, we have shown the existence of a $D=11$ $b$-ghost in the non-minimal pure spinor framework. Using this object, one can then define a ghost number zero vertex operator through the stringy equation
\begin{eqnarray}\label{section4eq1000}
V^{(0)} &=& \{b, U^{(1)}\} 
\end{eqnarray}
where $U^{(1)}$ is given in \eqref{eq60} and $b$ is given in \eqref{section4eq16}, \eqref{section4eq28}. Since $\{Q , b\} = T$, it easily follows that $V^{(0)}$, defined by eqn. \eqref{section4eq1000}, satisfies the standard descent equation \eqref{eq75} as desired. However, this ghost number zero vertex operator will depend on non-minimal pure spinor variables in a complicated fashion as shown in the previous chapter.

\vspace{2mm}
We have succeeded in finding a considerably simpler form in \eqref{section4eq28} for the $D=11$ $b$-ghost than that of equation \eqref{section4eq16} which was presented in \cite{Karlsson:2014xva}. Although this simplified version is not strictly nilpotent, it satisfies the relation $\{b,b\} =$ BRST-trivial which may be good enough for consistency. 

\vspace{2mm}
It is natural to ask if the simplified $D=11$ b-ghost \eqref{section4eq28} is the same as the $b$-ghost presented in \eqref{section4eq16}. These two expressions are compared in Appendix \ref{apE} and we find that they coincide up to normal-ordering terms coming from the position of $N_{ab}$ in each expression. Note that the product of $N_{ab}$'s appears as an anticommutator in \eqref{section4eq16} whereas it appears as an simple ordinary product in \eqref{section4eq28}. However, because we have ignored normal-ordering questions in our analysis, we will not attempt to address this issue.

\chapter{Pure spinor master actions and equations of motion on ordinary superspace}\label{newchapter5}
Pure spinors $\lambda^\alpha$ in ten and eleven dimensions have been useful for constructing vertex operators and computing on-shell scattering amplitudes with manifest spacetime supersymmetry in super-Yang-Mills, supergravity and superstring theory \cite{HOWE1991141,Berkovits:2001rb,Berkovits:2002uc,Gomez:2013sla}.
After including non-minimal variables $(\bar\lambda_\alpha, r_\alpha)$, pure spinors have also been useful for constructing BRST-invariant off-shell actions for these maximally supersymmetric theories \cite{Cederwall:2010tn,Cederwall:2013vba,Cederwall:2011vy}.

\vspace{2mm}
These BRST-invariant actions have a very simple form and were constructed by Cederwall using superfields $\Psi(x^m, \theta^\alpha, \lambda^\alpha, \bar\lambda_\alpha, r_\alpha)$ which transform covariantly under spacetime supersymmetry and depend on both the usual superspace variables $(x^m, \theta^\alpha)$ and the non-minimal pure spinor variables $(\lambda^\alpha, \bar\lambda_\alpha, r_\alpha)$. Although the actions require a non-supersymmetric regulator to define integration over the non-minimal pure spinor variables, it is easy to show that the supersymmetry transformation of the regulator is BRST-trivial so the action is spacetime supersymmetric.

\vspace{2mm}
However, since the superfields $\Psi$ can depend in a non-trivial manner on the non-minimal variables, it is not obvious how to show that the solutions to the equations of motion correctly describe the usual on-shell $D=10$ and $D=11$ superfields which depend only on the $(x^m, \theta^\alpha)$ superspace variables.

\vspace{2mm}
In this chapter, an explicit procedure will be given for extracting the usual on-shell
$D=10$ and $D=11$ superfields from the equations of motion of the pure spinor actions for the cases of $D=10$ supersymmetric Born-Infeld and for $D=11$ supergravity. This procedure will be given explicitly to first order in the coupling constant in these two actions, but it is expected that the procedure generalizes to all orders in the coupling constant as well as to other types of actions constructed from pure spinor superfields\footnote{A similar procedure was used by Chang, Lin, Wang and Yin in \cite{Chang:2014nwa} to find the on-shell solution to abelian and non-abelian $D=10$ supersymmetric Born-Infeld. We thank Martin Cederwall for informing us of their work.}\label{foot1}.

\vspace{2mm}
The procedure consists in using BRST cohomology arguments to define a unique decomposition of the on-shell pure spinor superfield $\Psi(x^m, \theta^\alpha, \lambda^\alpha, \bar\lambda_\alpha, r_\alpha)$ into the sum of two terms as
$$\Psi(x^m, \theta^\alpha, \lambda^\alpha, \bar\lambda_\alpha, r_\alpha) =
\tilde\Psi(x^m, \theta^\alpha, \lambda^\alpha) +
\Lambda( x^m, \theta^\alpha, \lambda^\alpha, \bar\lambda_\alpha, r_\alpha)$$
where $\tilde\Psi(x^m, \theta^\alpha, \lambda^\alpha)$ is independent of the non-minimal variables and $\Lambda$ is constructed from the superfields in $\tilde\Psi$ and the non-minimal variables. Since $\tilde\Psi$ will have a fixed ghost number $g$ ($g$=1 for $D=10$ super-Born-Infeld and $g=3$ for $D=11$ supergravity), it can be expanded as $\tilde\Psi = \lambda^\alpha \tilde A_\alpha (x,\theta)$ or $\tilde\Psi = \lambda^\alpha\lambda^\beta\lambda^\gamma \tilde C_{\alpha\beta\gamma} (x,\theta)$, and it will be shown to first order in the coupling constant that $\tilde A_\alpha (x,\theta)$ and $\tilde C_{\alpha\beta\gamma}(x,\theta)$ correctly describe the on-shell spinor gauge superfield of $D=10$ super-Born-Infeld and the on-shell spinor 3-form superfield of $D=11$ supergravity.

\vspace{2mm}
We expect it should be possible to generalize this procedure to all orders in the coupling constant and to other types of pure spinor actions, but there is an important issue concerning these pure spinor actions which needs to be further investigated. If the superfields $\Psi$ in these actions are allowed to have poles of arbitrary order in the non-minimal pure spinor variables, the cohomology arguments used to define the on-shell superfields become invalid. This follows from the well-known property of non-minimal pure spinor variables that one can construct a state $\xi (\lambda, \bar\lambda, \theta, r)$ satisfying $Q\xi =1$ if $\xi$ is allowed to have poles of order $\lambda^{-11}$ in $D=10$ or poles of order 
$\lambda^{-23}$ in $D=11$. And if $\xi$ is allowed in the Hilbert space of states, all BRST cohomology becomes trivial since any state $V$ satisfying $QV=0$ can be expressed as $V = Q (\xi V)$.  

\vspace{2mm}
So in order for these actions to correctly describe the on-shell superfields, one needs to impose restrictions on the possible pole dependence of the superfields $\Psi$. But since the pole dependence of the product of superfields can be more singular than the pole dependence of individual superfields, it is not obvious how to restrict the pole dependence of the superfields in a manner which is consistent with the non-linear BRST transformations of the action.

\vspace{2mm}
In section \ref{s5.2} of this chapter, the $D=10$ pure spinor superparticle and the pure spinor actions for $D=10$ super-Maxwell and super-Yang-Mills will be reviewed. And in section \ref{s5.3}, these actions will be generalized to abelian $D=10$ supersymmetric Born-Infeld constructed in terms of a non-minimal pure spinor superfield $\Psi$. 
The super-Born-Infeld equations of motion take the simple form
\begin{eqnarray}
Q\Psi + k(\lambda\gamma^{m}\hat{\chi}\Psi)(\lambda\gamma^{n}\hat{\chi}\Psi)\hat{F}_{mn}\Psi &=& 0
\end{eqnarray}
where $k$ is the dimensionful coupling constant and $\hat\chi^\alpha$ and $\hat F_{mn}$ are operators depending in a complicated manner on the non-minimal variables.
After expanding $\Psi$ in powers of $k$ as $\Psi= \sum_{i=0}^{\infty}k^{i}\Psi_{i}$,
one finds that $\Psi_0$ satisfies the equation $Q\Psi_0=0$ with the super-Maxwell solution $\Psi_0 = \lambda^\alpha A_\alpha (x,\theta)$, and
$\Psi_1$ can be uniquely decomposed as 
$$\Psi_1(x, \theta, \lambda, \bar\lambda, r) =
\tilde\Psi_1(x, \theta, \lambda) +
\Lambda(A_{\alpha}, \lambda, \bar\lambda, r)$$
where $\tilde\Psi_1$ satisfies  \cite{Cederwall:2001td,Chang:2014nwa}
\begin{eqnarray}\label{eq01}
Q\tilde\Psi_1 + (\lambda\gamma^{m}\chi)(\lambda\gamma^{n}\chi){F}_{mn} = 0
\end{eqnarray}
and $\chi^\alpha$ and $F_{mn}$ are the linearized spinor and vector field-strengths constructed from the super-Maxwell superfield in $\Psi_0$. It is straightforward to show that \eqref{eq01} correctly describes the first-order correction of Born-Infeld to the super-Maxwell equations.

\vspace{2mm}
In section \ref{s5.4} of this chapter, the $D=11$ pure spinor superparticle and the pure spinor action for linearized $D=11$ supergravity will be reviewed. And in section \ref{s5.5}, this action will be generalized to the complete $D=11$ supergravity action constructed in terms of a non-minimal pure spinor superfield $\Psi$. 
The supergravity equations of motion take the form
\begin{eqnarray}
Q\Psi + \frac{\kappa}{2}(\lambda\Gamma_{ab}\lambda)R^{a}\Psi R^{b}\Psi + \frac{\kappa}{2}\Psi \{Q , T\}\Psi - \kappa^{2}(\lambda\Gamma_{ab}\lambda)T\Psi R^{a}\Psi R^{b}\Psi &=& 0
\end{eqnarray}
where $\kappa$ is the dimensionful coupling constant and $R^a$ and $T$ are operators depending in a complicated manner on the non-minimal variables.
After expanding $\Psi$ in powers of $\kappa$ as $\Psi= \sum_{i=0}^{\infty}\kappa^{i}\Psi_{i}$,
one finds that $\Psi_0$ satisfies the equation $Q\Psi_0=0$ with the linearized supergravity solution $\Psi_0 = \lambda^\alpha\lambda^\beta\lambda^\gamma C_{\alpha\beta\gamma}(x, \theta)$, and
$\Psi_1$ can be uniquely decomposed as 
$$\Psi_1(x, \theta, \lambda, \bar\lambda, r) =
\tilde\Psi_1(x, \theta, \lambda) +
\Lambda(C_{\alpha\beta\delta}, \lambda, \bar\lambda, r)$$
where $\tilde\Psi_1$ satisfies
\begin{eqnarray}\label{eq02}
Q\tilde\Psi_1 + \frac{1}{2}(\lambda\Gamma_{ab}\lambda)\Phi^{a}\Phi^{b} &=& 0
\end{eqnarray}
and $\Phi^a \equiv \lambda^\alpha E_\alpha^{(0)P}\hat{E}_P{}^a (x,\theta)$ is constructed from the linear deformation of the supergravity supervielbein and the background value of its respective inverse. It is straightforward to show that \eqref{eq02} correctly describes the first-order correction to the linearized supergravity equations.

\vspace{2mm}
Finally, Appendices \ref{apA} and \ref{apB} will contain some useful gamma matrix identities in $D=10$ and $D=11$, Appendix \ref{newapC} will briefly review the framework systematizing the construction of the pure spinor actions studied here, and Appendix \ref{apC} will explain the relation of the $D=11$ supergravity superfields $\Psi$ and $\Phi^a$.


\vspace{2mm}
\section{Ten-dimensional Pure Spinor Superparticle and Super Yang-Mills}\label{s5.2}
In this section we will review the pure spinor description for the ten-dimensional superparticle and its connection with ten-dimensional super-Maxwell. We will then discuss the generalization to the non-abelian case. 
\subsection{$D=10$ Pure spinor superparticle}
The ten-dimensional pure spinor superparticle action is given by \cite{Berkovits:2001rb,Berkovits:2002zk}
\begin{eqnarray}
S &=& \int d\tau \left[P_{m}\partial_{\tau}X^{m} + p_{\mu}\partial_{\tau}\theta^{\mu} + w_{\mu}\partial_{\tau}\lambda^{\mu}\right]
\end{eqnarray}
where $X^{m}$ is a ten-dimensional coordinate, $\theta^{\mu}$ is a ten-dimensional Majorana-Weyl spinor, $\lambda^{\mu}$ is a bosonic ten-dimensional Weyl spinor satisfying $\lambda\gamma^{m}\lambda = 0$; and $P_{m}$, $p_{\mu}$, $w_{\mu}$ are the conjugate momenta relative to $X^{m}$, $\theta^{\mu}$, $\lambda^{\mu}$ respectively. We are using Greek/Latin letters from the middle of the alphabet to denote ten-dimensional Majorana-Weyl spinor/vector indices. Furthermore, $(\gamma^{m})^{\mu\nu}$ and $(\gamma^{m})_{\mu\nu}$ are $16\times 16$ symmetric real matrices satisfying $(\gamma^{m})^{\mu\nu}(\gamma^{n})_{\nu\sigma} + (\gamma^{n})^{\mu\nu}(\gamma^{m})_{\nu\sigma} = 2\eta^{mn}\delta^{\mu}_{\sigma}$. The BRST operator is given by
\begin{eqnarray}
Q_{0} &=& \lambda^{\mu}d_{\mu}
\end{eqnarray}
where $d_{\mu} = p_\mu - (\gamma^m \theta)_\mu P_m$ are the fermionic constraints of the $D=10$ Brink-Schwarz superparticle \cite{BRINK1981310}. The physical spectrum is defined as the cohomology of the BRST operator $Q_{0}$. One can show that the ten-dimensional super-Maxwell physical fields are described by ghost number one states: $\Psi = \lambda^{\mu}A_{\mu}$. This can be easily seen since states in the cohomology satisfy the equation of motion and gauge invariance
\begin{eqnarray}
(\gamma^{mnpqr})^{\mu\nu}D_{\mu}A_{\nu} &=& 0\nonumber\\
\delta A_{\mu} &=& D_{\mu}\Lambda\label{eeeq104}
\end{eqnarray}
where $D_\mu = {\partial\over{\partial \theta^\mu}}- (\gamma^m\theta)_\mu \partial_m$.
These are indeed the superspace constraints describing ten-dimensional super-Maxwell. It can be shown that the remaining non-trivial cohomology is found at ghost number 0, 2 and 3 states; describing the super-Maxwell ghost, antifields and antighost, respectively, as dictated by BV quantization.

\subsection{$D=10$ Super-Maxwell}
In order to describe $D=10$ super-Maxwell \eqref{eeeq104} from a well-defined pure spinor action principle, one should introduce non-minimal pure spinor variables \cite{Berkovits:2005bt}. These non-minimal variables were studied in detail in \cite{Berkovits:2006vi} and consist of a pure spinor $\bar{\lambda}_{\mu}$ satisfying $\bar{\lambda}\gamma^{m}\bar{\lambda} = 0$, a fermionic spinor $r_{\mu}$ satisfying $\bar{\lambda}\gamma^{m}r = 0$ and their respective conjugate momenta $\bar{w}^{\mu}$, $s^{\mu}$. The non-minimal BRST operator is defined as $Q = Q_{0} + r_{\mu}\bar{w}^{\mu}$, so that these non-minimal variables will not affect the BRST cohomology. This means that one can always find a representative in the cohomology which is independent of non-minimal variables. 

Note that it will be assumed that the dependence on the non-minimal variables of the states is restricted to diverge slower than $(\lambda\bar\lambda)^{-11}$ when $\lambda^\mu \to 0$. Without this restriction, any BRST-closed operator is BRST-trivial since $Q(\xi V) = V$ where $\xi \equiv (\lambda\bar\lambda +r\theta)^{-1} (\bar\lambda\theta)$. Since the gauge transformation $\delta \Psi = Q\Lambda$  of super-Maxwell is linear, this restriction is easy to enforce by imposing a similar restriction on the gauge parameter $\Lambda$. However, for the non-linear gauge transformations discussed in the following sections for the super-Yang-Mills, supersymmetric Born-Infeld, and supergravity actions, it is unclear how to enforce this restriction. We shall ignore this subtlety here, but it is an important open problem to define the allowed set of states and gauge transformations for $\Psi$ and $\Lambda$ in these nonlinear actions.

\vspace{2mm}
Let $\mathcal{S}_{SM}$ be the following pure spinor action
\begin{eqnarray}\label{eeeq102}
\mathcal{S}_{SM} &=& \int [dZ] \,\Psi Q \Psi
\end{eqnarray}
where $[dZ] = [d^{10}x][d^{16}\theta][d\lambda][d\bar{\lambda}][d r] N$ is the integration measure, $\Psi$ is a pure spinor superfield (which can also depend on non-minimal variables) and $Q$ is the non-minimal BRST-operator. Let us explain what $[dZ]$ means. Firstly, $[d^{10}x][d^{16}\theta]$ is the usual measure on ordinary ten-dimensional superspace. The factors $[d\lambda][d\bar{\lambda}][d r]$ are given by
\begin{eqnarray}
\left[d\lambda\right]\lambda^{\mu}\lambda^{\nu}\lambda^{\rho} &=& (\epsilon T^{-1})^{\mu\nu\rho}_{\hspace{5mm}\sigma_{1}\ldots\sigma_{11}}d\lambda^{\sigma_{1}}\ldots d\lambda^{\sigma_{11}}\nonumber\\
\left[d\bar{\lambda}\right]\bar{\lambda}_{\mu}\bar{\lambda}_{\nu}\bar{\lambda}_{\rho} &=& (\epsilon T)_{\mu\nu\rho}^{\hspace{5mm}\sigma_{1}\ldots\sigma_{11}}d\bar{\lambda}_{\sigma_{1}}\ldots d\bar{\lambda}_{\sigma_{11}}\nonumber\\
\left[d r\right] &=& (\epsilon T^{-1})^{\mu\nu\rho}_{\hspace{5mm}\sigma_{1}\ldots\sigma_{11}}\bar{\lambda}_{\mu}\bar{\lambda}_{\nu}\bar{\lambda}_{\rho}(\frac{\partial}{\partial r_{\sigma_{1}}})\ldots (\frac{\partial}{\partial r_{\sigma_{11}}})
\end{eqnarray}
where the Lorentz-invariant tensors $(\epsilon T)_{\mu\nu\rho}{}^{\sigma_{1}\ldots\sigma_{11}}$ and $(\epsilon T^{-1})^{\mu\nu\rho}{}_{\sigma_{1}\ldots \sigma_{11}}$ were defined in \cite{Berkovits:2006vi}. They are symmetric and gamma-traceless in $(\mu,\nu,\rho)$ and are antisymmetric in $[\sigma_{1},\ldots,\sigma_{11}]$. $N = e^{-Q(\bar{\lambda} \theta)} = e^{(-\bar{\lambda}\lambda - r \theta)}$ is a regularization factor. Since the measure converges as $\lambda^{8}\bar{\lambda}^{11}$ when $\lambda \to 0$, the action is well-defined as long as the integrand diverges slower than $\lambda^{-8}\bar{\lambda}^{-11}$.

\vspace{2mm}
One can easily see that the equation of motion following from \eqref{eeeq102} is given by
\begin{eqnarray}
Q\Psi = 0
\end{eqnarray}
and since the measure factor $[dZ]$ picks out the top cohomology of the ten-dimensional pure spinor BRST operator, the transformation $\delta \Psi = Q\Lambda$ is a symmetry of the action \eqref{eeeq102}. Therefore, \eqref{eeeq102} describes $D=10$ super-Maxwell.

\subsection{$D=10$ Super Yang-Mills}
Let us define $\mathcal{S}_{SYM}$ to be
\begin{eqnarray}\label{eq222}
\mathcal{S}_{SYM} &=& \int [dZ] \,Tr(\frac{1}{2}\Psi Q\Psi + \frac{g}{3}\Psi \Psi \Psi)
\end{eqnarray}
where $[dZ]$ is the measure discussed above, $\Psi$ is a Lie-algebra valued generic pure spinor superfield, $Q$ is the non-minimal BRST operator and $g$ is the coupling constant. For $SU(n)$ gauge group, expand $\Psi$ in the form: $\Psi = \Psi^{a}T^{a}$, where $T^{a}$ are the Lie algebra generators and $a = 1,\ldots , n^{2}-1$. Using the conventions: $[T^{a}, T^{b}] = f^{abc}T^{c}$ with $f^{abc}$ totally antisymmetric, and $Tr(T^{a}T^{b})=\delta^{ab}$, one can rewrite \eqref{eq222} as follows
\begin{eqnarray}
\mathcal{S}_{SYM} &=& \int [dZ](\frac{1}{2}\Psi^{a}Q\Psi^{a} + \frac{g}{6}f^{abc}\Psi^{a}\Psi^{b}\Psi^{c})
\end{eqnarray}
The e.o.m following from this action is given by:
\begin{eqnarray}\label{eq225}
Q\Psi^{a} + \frac{g}{2}f^{abc}\Psi^{b}\Psi^{c} &=& 0
\end{eqnarray}
or in compact form
\begin{eqnarray}\label{eq223}
Q\Psi + g\Psi\Psi &=& 0
\end{eqnarray}
It turns out that \eqref{eq225} is invariant under the BRST symmetry
\begin{eqnarray}
\delta\Psi^{a} &=& Q\Lambda^{a} + f^{abc}\Psi^{b}\Lambda^{c}
\end{eqnarray}
or in compact form
\begin{eqnarray}\label{eq224}
\delta\Psi = Q\Lambda + [\Psi , \Lambda]
\end{eqnarray}
Since the equations \eqref{eq223}, \eqref{eq224} describe on-shell $D=10$ Super Yang-Mills on ordinary superspace \cite{Berkovits:2002zk}, one concludes that the action \eqref{eq222} describes $D=10$ Super Yang-Mills on a pure spinor superspace.

\section{Pure Spinor Description of Abelian Supersymmetric Born-Infeld}\label{s5.3}
In this section, we review the construction of the pure spinor action for supersymmetric abelian Born-Infeld and deduce the equations of motion on minimal pure spinor superspace to first order in the coupling.
\subsection{Physical operators}
In order to deform the quadratic super-Maxwell action to the supersymmetric Born-Infeld action, Cederwall introduced the ghost number -1 pure spinor operators \cite{Cederwall:2011vy} 
\begin{eqnarray}
\hat{A}_{\mu} &=& -\frac{1}{(\lambda\bar{\lambda})}[\frac{1}{8}(\gamma^{mn}\bar{\lambda})_{\mu}N_{mn} + \frac{1}{4}\bar{\lambda}_{\mu}N]\nonumber\\
\hat{A}_{m} &=& -\frac{1}{4(\lambda\bar{\lambda})}(\bar{\lambda}\gamma_{m}D) + \frac{1}{32(\lambda\bar{\lambda})^{2}}(\bar{\lambda}\gamma_{m}^{\hspace{2mm}np}r)N_{np}\nonumber\\
\hat{\chi}^{\mu} &=& \frac{1}{2(\lambda\bar{\lambda})}(\gamma^{m}\bar{\lambda})^{\mu}\Delta_{m}\nonumber\\
\hat{F}_{mn} &=& -\frac{1}{4(\lambda\bar{\lambda})}(r\gamma_{mn}\hat{\chi}) = \frac{1}{8(\lambda\bar{\lambda})}(\bar{\lambda}\gamma_{mn}^{\hspace{4mm}p}r)\Delta_{p}\label{eeq6}
\end{eqnarray}
where $\Delta_{m}$ is defined by
\begin{eqnarray}
\Delta_{m} &=& \partial_{m} + \frac{1}{4(\lambda\bar{\lambda})}(r\gamma_{m}D) - \frac{1}{32(\lambda\bar{\lambda})^{2}}(r\gamma_{mnp}r)N^{np}\label{eeq5}
\end{eqnarray}

These operators are constructed to satisfy 
\begin{eqnarray}
\left[ Q, \hat{A}_{\mu}\right] &=& -D_{\mu} - 2(\gamma^{m}\lambda)_{\mu}\hat{A}_{m}\nonumber\\
\{Q, \hat{A}_{m}\} &=& \partial_{m} - (\lambda\gamma_{m}\hat{\chi})\nonumber\\
\left[ Q, \hat{\chi}^{\mu}\right] &=& -\frac{1}{2}(\gamma^{mn}\lambda)^{\mu}\hat{F}_{mn}\nonumber\\
\{Q, \hat{F}_{mn}\} &=& 2(\lambda\gamma_{[m}\partial_{n]}\hat{\chi})
\end{eqnarray}
which mimic the superspace equations of motion of $D=10$ Super-Maxwell
\begin{eqnarray}
D_{\alpha}\Psi_0 + Q_{0} A_{\mu} + 2(\gamma^{m}\lambda)_{\mu}A_{m} &=& 0\nonumber\\
\partial_{m}\Psi_0 - Q_{0} A_{m} - (\lambda\gamma_{m}\chi) &=& 0\nonumber\\
Q_{0}\chi^{\mu} + \frac{1}{2}(\lambda\gamma_{mn})^{\mu}F^{mn} &=& 0\nonumber\\
Q_{0}F_{mn} - 2(\lambda\gamma_{[m}\partial_{n]}\chi) &=& 0 \label{eeq1}
\end{eqnarray}
with $\Psi_0 = \lambda^{\mu}A_{\mu}$.

\vspace{2mm}
If one acts with these operators on $\Psi_0$, they satisfy
 \begin{eqnarray}
\hat{A}_{\mu}\Psi_0 = A_{\mu}\hspace{2mm},\hspace{2mm}
\hat{A}_{m}\Psi_0 = A_{m}\hspace{2mm},\hspace{2mm}
\hat{\chi}^{\mu}\Psi_0 = \chi^{\mu}\hspace{2mm},\hspace{2mm}
\hat{F}_{mn}\Psi_0 = F_{mn}\label{eeq2}
 \end{eqnarray}
up to BRST-exact terms and certain ``shift-symmetry terms''  defined in \cite{Cederwall:2011vy,Cederwall:2013vba}. 
For example, the operator $\hat{A}_{\mu}$ acts as
\begin{eqnarray}\label{eeq3}
\hat{A}_{\mu}\Psi_0 &=& A_{\mu} - \frac{1}{2(\lambda\bar{\lambda})}(\lambda\gamma^{m})_{\mu}(\bar{\lambda}\gamma_{m}A)
\end{eqnarray}
where the shift symmetry is $\delta A_\mu = (\lambda\gamma_m)_\mu \phi^m$ for any $\phi^m$.
For $\hat{A}_{m}$ one finds that
\begin{eqnarray}
\hat{A}_{m}\Psi_0 &=& A_{m} - (\lambda\gamma_{m}\rho) + Q[\frac{1}{4(\lambda\bar{\lambda})}(\bar{\lambda}\gamma_{m}A)]
\end{eqnarray}
where the on-shell relation $D_{(\mu}A_{\nu)} = -(\gamma^{m})_{\mu\nu}A_{m}$ has been used, $\delta A_m = (\lambda \gamma_m)_\mu \rho^\mu$ is the shift symmetry, and 
\begin{equation}\label{eeq4}
\rho^{\mu} = \frac{1}{2(\lambda\bar{\lambda})}(\bar{\lambda}\gamma^{m})^{\mu}A_{m} + \frac{1}{8(\lambda\bar{\lambda})^{2}}(\bar{\lambda}\gamma^{m})^{\mu}(r\gamma_{m}A)
\end{equation}
Analogously, one can show a similar behavior for the other operators $\hat{\chi}^{\mu}$, $\hat{F}_{mn}$. 


\subsection{$D=10$ Abelian supersymmetric Born-Infeld}
The deformation to the linearized action \eqref{eeeq102} consistent with BRST symmetry is given by \cite{Cederwall:2011vy}
\begin{eqnarray}\label{eeq12}
\mathcal{S}_{SBI} &=& \int [dZ] \left[\frac{1}{2}\Psi Q\Psi + \frac{k}{4}\Psi(\lambda\gamma^{m}\hat{\chi}\Psi)(\lambda\gamma^{n}\hat{\chi}\Psi)\hat{F}_{mn}\Psi\right]
\end{eqnarray}
which is invariant under the BRST transformation
\begin{eqnarray}\label{newequationnew1}
\delta\Psi &=& Q\Lambda + k(\lambda\gamma^{m}\hat{\chi}\Psi)(\lambda\gamma^{n}\hat{\chi}\Psi)\hat{F}_{mn}\Lambda + 2k(\lambda\gamma^{m}\hat{\chi}\Psi)(\lambda\gamma^{n}\hat{\chi}\Lambda)\hat{F}_{mn}\Psi
\end{eqnarray}
for any ghost number 0 pure spinor superfield $\Lambda$.  Note that $k$ is a dimensionful parameter related to the string tension by $k = \alpha\ensuremath{'}^{2}$. The equation of motion coming from \eqref{eeq12} is
\begin{eqnarray}\label{eeq8}
Q\Psi + k(\lambda\gamma^{m}\hat{\chi}\Psi)(\lambda\gamma^{n}\hat{\chi}\Psi)\hat{F}_{mn}\Psi &=& 0
\end{eqnarray}
\vspace{2mm}
which can be written in terms of $\Delta_{m}$ as follows
\begin{eqnarray} \label{eeq9}
Q\Psi + \frac{k}{8(\lambda\bar{\lambda})^{2}}(\bar{\lambda}\gamma^{mnp}r)(\Delta_{m}\Psi)(\Delta_{n}\Psi)(\Delta_{p}\Psi) &=& 0\label{eeq9}
\end{eqnarray}

Since the equation of motion of \eqref{eeq9} for $\Psi$ depends explicitly on the non-minimal variables, it is not obvious how to extract from $\Psi$ the Born-Infeld superfield $\tilde A_\mu(x,\theta)$ which should be independent of the non-minimal variables. However, it will now be argued that there is a unique decomposition of the solution to  \eqref{eeq9} as
\begin{eqnarray}\label{eq1000}
\Psi(x,\theta,\lambda,\bar{\lambda},r) &=& \lambda^\mu \tilde A_\mu (x,\theta) + \Lambda (\tilde A_\mu, \lambda, \bar\lambda, r)
\end{eqnarray}
where $\tilde A_\mu(x,\theta)$ is the on-shell Born-Infeld superfield and $\Lambda$ depends on $\tilde A_\mu$ and on the non-minimal variables. This will be explicitly shown here to the leading Born-Infeld correction to super-Maxwell, and work is in progress on extending this to the complete Born-Infeld solution. As mentioned in footnote \ref{foot1}, a similar procedure was used in \cite{Chang:2014nwa} for the abelian and non-abelian Born-Infeld solutions.

\vspace{2mm}
To extract this leading-order correction to super-Maxwell from \eqref{eeq9}, we will first expand the pure spinor superfield $\Psi$ in positive powers of $k$:
\begin{eqnarray}\label{eq100}
\Psi(x,\theta,\lambda,\bar{\lambda},r) &=& \sum_{i=0}^{\infty}k^{i}\Psi_{i}
\end{eqnarray}
The replacement of \eqref{eq100} in \eqref{eeq9} gives us the following recursive relations
\begin{eqnarray}
Q\Psi_{0} &=& 0 \label{eq101}\\
Q\Psi_{1} &=& -\frac{1}{8(\lambda\bar{\lambda})^{2}}(\bar{\lambda}\gamma^{mnp}r)\Delta_{m}\Psi_{0}\Delta_{n}\Psi_{0}\Delta_{p}\Psi_{0}\label{eq102}\\
Q\Psi_{2} &=& -\frac{3}{8(\lambda\bar{\lambda})^{2}}(\bar{\lambda}\gamma^{mnp}r)\Delta_{m}\Psi_{1}\Delta_{n}\Psi_{0}\Delta_{p}\Psi_{0}\\
\vdots\nonumber
\end{eqnarray}

To determine $\Lambda$ in \eqref{eq1000}, first note that \eqref{eq101} has the solution $\Psi_0 = \lambda^\mu A_{0\,\mu}$ where $ A_{0\,\mu}$ is the super-Maxwell superfield which is independent of the non-minimal variables.
However, the solution $\Psi_1$ to \eqref{eq102} must depend on the non-minimal variables because the right-hand side of  \eqref{eq102} depends on these variables. To decompose the solution $\Psi_1$ to the form 
\begin{eqnarray}\label{eq1001}
\Psi_1(x,\theta,\lambda,\bar{\lambda},r) &=& \lambda^\mu A_{1\,\mu} (x,\theta) + \Lambda,
\end{eqnarray}
note that 
\eqref{eq102} implies
\begin{eqnarray}\label{eeeq99}
Q(\frac{1}{(\lambda\bar{\lambda})^{2}}(\bar{\lambda}\gamma^{mnp}r)\Delta_{m}\Psi_{0}\Delta_{n}\Psi_{0}\Delta_{p}\Psi_{0}) = 0
\end{eqnarray}
Since any BRST-closed expression can be expressed in terms of minimal variables up to a BRST-trivial term, there must exist a term $\Lambda$ such that 
\begin{eqnarray}\label{eqeqe34}
-\frac{1}{8(\lambda\bar{\lambda})^{2}}(\bar{\lambda}\gamma^{mnp}r)\Delta_{m}\Psi_{0}\Delta_{n}\Psi_{0}\Delta_{p}\Psi_{0} &=& Q\Lambda + F(\Psi_{0})
\end{eqnarray}
where $F({\Psi_{0}})$ is independent of non-minimal variables. This equation determines $\Lambda$ and $F(\Psi_0)$ up to the shift
\begin{eqnarray}\label{eqeqe33}
\delta \Lambda = H(\Psi_0) + Q\Omega, \quad \delta F(\Psi_0) = - Q H(\Psi_0)
\end{eqnarray}
 where $H(\Psi_0)$ only depends on the minimal variables.  But the BRST-trivial shift $F(\Psi_0) \to F(\Psi_0)  - QH(\Psi_0)$ can be cancelled by a redefinition of the field $\Psi_0 \to \Psi_0 - k H(\Psi_0)$.  So the ambiguity in defining $\Lambda$ in  \eqref{eqeqe34} does not affect the physical spectrum. 

\vspace{2mm}
In order to find $\Lambda$ and $F(\Psi_{0})$ in \eqref{eqeqe34}, first write $\Delta_{m}$ in the more convenient form
\begin{eqnarray}
\Delta_{m} &=& \partial_{m} - \{Q , \hat{A}_{m}\} + \bar{\lambda}\gamma_{m}\hat{\xi}
\end{eqnarray}
where $\hat{\xi}_{\mu}$ is an operator depending on $N_{mn}, D_{\mu}$, etc. Although it is not complicated to determine $\hat{\xi}_{\mu}$, this will not be relevant for our purposes as we will see later. Using the on-shell relation $\partial_{m}A_{\mu} - D_{\mu}A_{m} = (\gamma_{m}\chi)_{\mu}$, one finds that 
\begin{eqnarray}
\Delta_{m}\Psi_{0} &=& (\lambda\gamma_{m}\chi) + \lambda\gamma_{m}Q\rho + \bar{\lambda}\gamma_{m}\hat{\xi}\Psi_{0}
\end{eqnarray}
So
\begin{eqnarray}
-\frac{1}{8}(\bar{\lambda}\gamma^{mnp}r)\Delta_{m}\Psi_{0}\Delta_{n}\Psi_{0}\Delta_{p}\Psi_{0}&=& -\frac{1}{8(\lambda\bar{\lambda})^{2}}(\bar{\lambda}\gamma^{mnp}r)[(\lambda\gamma_{m}\chi)(\lambda\gamma_{n}\chi)(\lambda\gamma_{p}\chi)\nonumber\\
&& + 3(\lambda\gamma_{m}Q\rho)(\lambda\gamma_{n}\chi)(\lambda\gamma_{p}\chi)\nonumber\\
&& + 3(\lambda\gamma_{m}Q\rho)(\lambda\gamma_{n}Q\rho)(\lambda\gamma_{p}\chi)\nonumber\\
&&+ (\lambda\gamma_{m}Q\rho)(\lambda\gamma_{n}Q\rho)(\lambda\gamma_{p}Q\rho)]
\end{eqnarray}
The first term $H_{1} = -\frac{1}{8(\lambda\bar{\lambda})^{2}}(\bar{\lambda}\gamma_{mnp}r)(\lambda\gamma^{m}\chi)(\lambda\gamma^{n}\chi)(\lambda\gamma^{p}\chi)$ will provide us the term independent of non-minimal variables:
\begin{eqnarray}
H_{1} &=& -\frac{1}{8(\lambda\bar{\lambda})^{2}}(\bar{\lambda}\gamma_{mnp}r)(\lambda\gamma^{m}\chi)(\lambda\gamma^{n}\chi)(\lambda\gamma^{p}\chi)\nonumber\\
&=& \frac{1}{4(\lambda\bar{\lambda})}(r\gamma^{mn}\chi)(\lambda\gamma_{m}\chi)(\lambda\gamma_{n}\chi) - \frac{1}{8(\lambda\bar{\lambda})^{2}}(\lambda r)(\bar{\lambda}\gamma_{mn}\chi)(\lambda\gamma^{m}\chi)(\lambda\gamma^{n}\chi)\nonumber\\
&=& Q[\frac{1}{4(\lambda\bar{\lambda})}(\bar{\lambda}\gamma_{mn}\chi)(\lambda\gamma^{m}\chi)(\lambda\gamma^{n}\chi)] - F_{mn}(\lambda\gamma^{m}\chi)(\lambda\gamma^{n}\chi)
\end{eqnarray}
where the identity \eqref{ap1} was used. Analogous computations show us that the other terms are Q-exact:
\begin{eqnarray}
H_{2} &=& -\frac{3}{8(\lambda\bar{\lambda})^{2}}(\bar{\lambda}\gamma^{mnp}r)(\lambda\gamma_{m}Q\rho)(\lambda\gamma_{n}\chi)(\lambda\gamma_{p}\chi)\nonumber\\
&=& Q[\frac{6}{8(\lambda\bar{\lambda})}(\bar{\lambda}\gamma_{mn}Q\rho)(\lambda\gamma^{m}\chi)(\lambda\gamma^{n}\chi)]
\end{eqnarray}
\begin{eqnarray}
H_{3} &=& -\frac{3}{8(\lambda\bar{\lambda})^{2}}(\bar{\lambda}\gamma^{mnp}r)(\lambda\gamma_{m}Q\rho)(\lambda\gamma_{n}Q\rho)(\lambda\gamma_{p}\chi)\nonumber\\
&=& Q[\frac{6}{8(\lambda\bar{\lambda})}(\bar{\lambda}\gamma_{mn}Q\rho)(\lambda\gamma^{m}Q\rho)(\lambda\gamma^{n}\chi)]
\end{eqnarray}
\begin{eqnarray}
H_{4} &=& -\frac{1}{8(\lambda\bar{\lambda})^{2}}(\bar{\lambda}\gamma^{mnp}r)(\lambda\gamma_{m}Q\rho)(\lambda\gamma_{n}Q\rho)(\lambda\gamma_{p}Q\rho)\nonumber\\
&=& Q[\frac{2}{8(\lambda\bar{\lambda})}(\bar{\lambda}\gamma_{mn}Q\rho)(\lambda\gamma^{m}Q\rho)(\lambda\gamma^{n}Q\rho)]
\end{eqnarray}
Hence, one obtains
\begin{eqnarray}
-\frac{1}{8}(\bar{\lambda}\gamma^{mnp}r)\Delta_{m}\Psi_{0}\Delta_{n}\Psi_{0}\Delta_{p}\Psi_{0} &=& Q[\Lambda] - F_{mn}(\lambda\gamma^{m}\chi)(\lambda\gamma^{n}\chi)
\end{eqnarray}
where $\Lambda$ is defined by the expression
\begin{eqnarray}
\Lambda &=& \frac{1}{4(\lambda\bar{\lambda})}(\bar{\lambda}\gamma_{mn}\chi)(\lambda\gamma^{m}\chi)(\lambda\gamma^{n}\chi) + \frac{3}{4(\lambda\bar{\lambda})}(\bar{\lambda}\gamma_{mn}Q\rho)(\lambda\gamma^{m}\chi)(\lambda\gamma^{n}\chi)\nonumber\\
&& + \frac{3}{4(\lambda\bar{\lambda})}(\bar{\lambda}\gamma_{mn}Q\rho)(\lambda\gamma^{m}Q\rho)(\lambda\gamma^{n}\chi) + \frac{1}{4(\lambda\bar{\lambda})}(\bar{\lambda}\gamma_{mn}Q\rho)(\lambda\gamma^{m}Q\rho)(\lambda\gamma^{n}Q\rho)\nonumber\\ \label{eq1new}
\end{eqnarray}
Now, let us define the field $\tilde\Psi = \Psi_{0} + k(\Psi_{1} - \Lambda)$ 
which satisfies to first order in $k$ the equation of motion
\begin{eqnarray}
Q\tilde{\Psi} &=&Q (\Psi_0 + k(\Psi_1 - \Lambda)) =  - k F_{mn}(\lambda\gamma^{m}\chi)(\lambda\gamma^{n}\chi)\label{eeq11}
\end{eqnarray}
where $F_{mn}$, $\chi^{\mu}$ are the usual super-Maxwell superfields constructed from $A_{0\,\mu}$. Since the equation \eqref{eeq11} does not involve non-minimal variables, the solution is
\begin{eqnarray}
\tilde \Psi = \lambda^\mu \tilde {A}_{\mu} 
\end{eqnarray}
where $\tilde A_\mu\equiv A_{0\,\mu} + k A_{1\,\mu}$ satisfies
\begin{eqnarray}
\lambda^{\mu}\lambda^{\nu}\left [D_{\mu}\tilde{A}_{\nu} + k(\gamma^{m}\chi)_{\mu}(\gamma^{n}\chi)_{\nu}F_{mn}\right] &=& 0
\end{eqnarray}
This equation of motion coincides, at first order in $k$, with the abelian supersymmetric Born-Infeld equations of motion \cite{BERGSHOEFF1987371,JAMESGATES1987172,Berkovits:2002ag}. So it has been shown to first order in $k$ that 
\begin{eqnarray}
\Psi &=& \lambda^\mu \tilde A_\mu + k\Lambda
\end{eqnarray}
where $\tilde A_\mu(x, \theta)$ is the on-shell Born-Infeld superfield and $\Lambda$ depends on $A_{0\,\mu}$ and on the non-minimal variables. 

\section{Eleven-Dimensional Pure Spinor Superparticle and Supergravity}\label{s5.4}
In this section we review the eleven-dimensional pure spinor superparticle and its connection with linearized eleven-dimensional supergravity. 
\subsection{$D=11$ Pure spinor superparticle}
The eleven-dimensional pure spinor superparticle action is given by \cite{Berkovits:2002uc,Guillen:2017mte}
\begin{eqnarray}
S &=& \int d\tau \left[P_{m}\partial_{\tau}X^{m} + P_{\mu}\partial_{\tau}\theta^{\mu} + w_{\alpha}(\partial_{\tau}\lambda^{\alpha}+ \partial_\tau Z^M
\Omega_{M\beta}{}^\alpha \lambda^\beta)\right]
\end{eqnarray}
where $X^{m}$ is an eleven-dimensional coordinate, $\theta^{\mu}$ is an eleven-dimensional Majorana spinor, $Z^M=(X^m, \theta^\mu)$, $\lambda^{\alpha}$ is a bosonic eleven-dimensional Majorana spinor satisfying $\lambda\Gamma^{a}\lambda = 0$; $P_{m}$, $P_{\mu}$, $w_{\alpha}$ are the conjugate momenta relative to $X^{m}$, $\theta^{\mu}$, $\lambda^{\alpha}$ respectively, and $\Omega_{M\beta}{}^\alpha$ is the spin connection of the background. We are using Greek/Latin letters from the beginning of the alphabet to denote tangent-space eleven-dimensional spinor/vector indices, and
Greek/Latin letters from the middle of the alphabet to denote coordinate-space eleven-dimensional spinor/vector indices. Furthermore, capital letters from the
beginning of the alphabet will denote tangent-space indices (both spinor and vector) and capital letters from the
middle of the alphabet will denote coordinate-space indices (both spinor and vector). Finally, $(\Gamma^{a})^{\alpha\beta}$ and $(\Gamma^{a})_{\beta\delta}$ are $32\times 32$ symmetric matrices satisfying $(\Gamma^{a})^{\alpha\beta}(\Gamma^{b})_{\beta\delta}$ + $(\Gamma^{b})^{\alpha\beta}(\Gamma^{a})_{\beta\delta} = 2\eta^{ab}\delta^{\alpha}_{\delta}$. The BRST operator is given by
\begin{eqnarray}
Q_{0} &=& \lambda^{\alpha}d_{\alpha}
\end{eqnarray}
where
\begin{eqnarray}
d_\alpha &=& E_{\alpha}^{\hspace{2mm}M} (P_M +\Omega_{M\beta}{}^\gamma w_\gamma \lambda^\beta)
\end{eqnarray}
In a flat Minkowski background, $d_{\alpha} = P_{\alpha} - (\Gamma^{m}\theta)_{\alpha}P_{m}$ are the fermionic constraints of the $D=11$ Brink-Schwarz-like superparticle.

\vspace{2mm}
The physical spectrum is defined as the cohomology of the BRST operator $Q_{0}$. One can show that the eleven-dimensional linearized supergravity physical fields are described by ghost number three states: $\Psi = \lambda^{\alpha}\lambda^{\beta}\lambda^{\delta}C_{\alpha\beta\delta}$ \cite{Berkovits:2002uc}
where the physical state condition imposes the following equations of motion and gauge transformations for $C_{\alpha\beta\delta}$
\begin{eqnarray}\label{eqeqe1}
D_{(\alpha}C_{\beta\delta\epsilon)} &=& (\Gamma^{a})_{(\alpha\beta}C_{\vert a\vert \delta\epsilon)}\nonumber\\
\delta C_{\alpha\beta\delta} &=& D_{(\alpha}\Lambda_{\beta\delta)}
\end{eqnarray}
for some superfield $\Lambda_{\beta\delta}$. These are the superspace constraints describing eleven-dimensional linearized supergravity \cite{BRINK1980384}. It can be shown that the remaining non-trivial cohomology is found at ghost number 0, 1, 2, 4, 5, 6 and 7 states; describing the ghosts, antifields and antighosts as dictated by BV quantization of $D=11$ linearized supergravity.

\subsection{$D=11$ Linearized Supergravity}
In order to describe $D=11$ linearized supergravity \eqref{eqeqe1} from a pure spinor action principle, one should introduce eleven-dimensional non-minimal pure spinor variables \cite{Cederwall:2013vba}. These non-minimal variables were studied in detail in \cite{Cederwall:2012es, Cederwall:2009ez} and consist of a pure spinor $\bar{\lambda}_{\alpha}$ satisfying $\bar{\lambda}\Gamma^{a}\bar{\lambda} = 0$, a fermionic spinor $r_{\alpha}$ satisfying $\bar{\lambda}\Gamma^{a}r = 0$ and their respective conjugate momenta $\bar{w}^{\alpha}$, $s^{\alpha}$. The non-minimal BRST operator is defined as $Q = Q_{0} + r_{\alpha}\bar{w}^{\alpha}$, so that these non-minimal variables will not affect the BRST cohomology.

\vspace{2mm}
Let $\mathcal{S}_{LSG}$ be the following pure spinor action
\begin{eqnarray}\label{eeeeq102}
\mathcal{S}_{LSG} &=& \int [dZ] \,\Psi Q \Psi
\end{eqnarray}
where $[dZ] = [d^{11}x][d^{32}\theta][d\lambda][d\bar{\lambda}][d r] N$ is the integration measure, $\Psi$ is a pure spinor superfield (which, in general, can also depend on non-minimal variables) and $Q$ is the non-minimal BRST-operator. Let us explain what $[dZ]$ means. Firstly, $[d^{11}x][d^{32}\theta]$ is the usual measure on ordinary eleven-dimensional superspace. The factors $[d\lambda][d\bar{\lambda}][d r]$ are given by
\begin{eqnarray}
\left[d\lambda\right]\lambda^{\alpha_{1}}\ldots\lambda^{\alpha_{7}} &=& (\epsilon T^{-1})^{\alpha_{1}\ldots\alpha_{7}}_{\hspace{9mm}\beta_{1}\ldots\beta_{23}}d\lambda^{\beta_{1}}\ldots d\lambda^{\beta_{23}}\nonumber\\
\left[d\bar{\lambda}\right]\bar{\lambda}_{\alpha_{1}}\ldots\bar{\lambda}_{\alpha_{7}} &=& (\epsilon T)_{\alpha_{1}\ldots\alpha_{7}}^{\hspace{9mm}\beta_{1}\ldots\beta_{23}}d\bar{\lambda}_{\beta_{1}}\ldots d\bar{\lambda}_{\beta_{23}}\nonumber\\
\left[d r\right] &=& (\epsilon T^{-1})^{\alpha_{1}\ldots\alpha_{7}}_{\hspace{9mm}\beta_{1}\ldots\beta_{23}}\bar{\lambda}_{\alpha_{1}}\ldots\bar{\lambda}_{\alpha_{7}}(\frac{\partial}{\partial r_{\beta_{1}}})\ldots (\frac{\partial}{\partial r_{\beta_{23}}})
\end{eqnarray}
The Lorentz-invariant tensors $(\epsilon T)_{\alpha_{1}\ldots\alpha_{7}}^{\hspace{9mm}\beta_{1}\ldots\beta_{23}}$ and $(\epsilon T^{-1})^{\alpha_{1}\ldots \alpha_{7}}_{\hspace{9mm}\beta_{1}\ldots\beta_{23}}$ were defined in \cite{Cederwall:2009ez}. They are symmetric and gamma-traceless in $(\alpha_{1},\ldots,\alpha_{7})$ and are antisymmetric in $[\beta_{1},\ldots ,\beta_{23}]$. $N$ is a regularization factor which is given by $N = e^{-\lambda\bar{\lambda} - r\theta}$. Since the measure converges as $\lambda^{16}\bar{\lambda}^{23}$ when $\lambda\to 0$, the action is well-defined if the integrand diverges slower than  $\lambda^{-16}\bar{\lambda}^{-23}$.

\vspace{2mm}
One can easily see that the equation of motion following from \eqref{eeeeq102} is given by
\begin{eqnarray}
Q\Psi = 0
\end{eqnarray}
and since the measure factor $[dZ]$ picks out the top cohomology of the eleven-dimensional pure spinor BRST operator, the transformation $\delta \Psi = Q\Lambda$ is a symmetry of the action \eqref{eeeeq102}, that is a gauge symmetry of the theory. Therefore, \eqref{eeeeq102} describes $D=11$ linearized supergravity.


\section{Pure Spinor Description of Complete $D=11$ Supergravity}\label{s5.5}
As discussed in \cite{Cederwall:2013vba,Cederwall:2010tn} , the pure spinor BRST-invariant action for complete $D=11$ supergravity is given by 
\begin{eqnarray}\label{eq105}
\mathcal{S}_{SG} &=& {1\over{\kappa^2}}\int [dZ][\frac{1}{2}\Psi Q\Psi + \frac{1}{6}(\lambda\Gamma_{ab}\lambda)(1 - \frac{3}{2}T\Psi)\Psi R^{a}\Psi R^{b}\Psi]
\end{eqnarray}
which is invariant under the BRST symmetry 
\begin{eqnarray}
\delta\Psi &=& Q\Lambda + (\lambda\Gamma_{ab}\lambda)R^{a}\Psi R^{b}\Lambda + \frac{1}{2}\Psi\{Q,T\}\Lambda - \frac{1}{2}\Lambda\{Q,T\}\Psi - 2(\lambda\Gamma_{ab}\lambda)T\Psi R^{a}\Psi R^{b}\Lambda\nonumber\\
&& - (\lambda\Gamma_{ab}\lambda)(T\Lambda)R^{a}\Psi R^{b}\Psi
\end{eqnarray}
for any ghost number 2 pure spinor superfield $\Lambda$. Here $\kappa$ is the gravitational coupling constant, and $R^{a}$ and $T$ are ghost number -2 and -3 operators respectively, defined by the relations \cite{Cederwall:2009ez,Cederwall:2010tn}
\begin{eqnarray}
R^{a} &=& -8[\frac{1}{\eta}(\bar{\lambda}\Gamma^{ab}\bar{\lambda})\partial_{b} + \frac{1}{\eta^{2}}(\bar{\lambda}\Gamma^{ab}\bar{\lambda})(\bar{\lambda}\Gamma^{cd}r)(\lambda\Gamma_{bcd}D)\nonumber\\
&& - \frac{4}{\eta^{3}}(\bar{\lambda}\Gamma^{ab}\bar{\lambda})(\bar{\lambda}\Gamma^{cd}r)(\bar{\lambda}\Gamma^{ef}r)(\lambda\Gamma_{fb}\lambda)(\lambda\Gamma_{cde}w)\nonumber\\
&& + \frac{4}{\eta^{3}}(\bar{\lambda}\Gamma^{ac}\bar{\lambda})(\bar{\lambda}\Gamma^{de}r)(\bar{\lambda}\Gamma^{bf}r)(\lambda\Gamma_{fb}\lambda)(\lambda\Gamma_{cde}w)]\label{eeq13}\\
T &=& \frac{512}{\eta^{3}}(\bar{\lambda}\Gamma^{ab}\bar{\lambda})(\bar{\lambda}r)(rr)N_{ab}\label{eqq13},
\end{eqnarray}
and $\eta \equiv (\lambda \Gamma^{ab}\lambda)(\bar\lambda\Gamma_{ab}\bar\lambda)$.
Note that the action is invariant under the shift symmetry $\delta R^{a} = (\lambda\Gamma^{a}{\cal O})$ for any operator ${\cal O}$.

\vspace{2mm}
The equation of motion coming from the action \eqref{eq105} is
\begin{eqnarray}
Q\Psi + \frac{1}{2}\Psi\{Q,T\}\Psi + \frac{1}{2}(\lambda\Gamma_{ab}\lambda)(1 - 2T\Psi)R^{a}\Psi R^{b}\Psi &=& 0
\end{eqnarray}
To compare with the linearized equations, it is convenient to rescale $\Psi \to \kappa \Psi$ so that $\kappa$ drops out of the quadratic term in the action, and the e.o.m. takes the form
\begin{eqnarray}
Q\Psi + \frac{\kappa}{2}(\lambda\Gamma_{ab}\lambda)R^{a}\Psi R^{b}\Psi + \frac{\kappa}{2}\Psi \{Q , T\}\Psi - \kappa^{2}(\lambda\Gamma_{ab}\lambda)T\Psi R^{a}\Psi R^{b}\Psi &=& 0\label{eq106}
\end{eqnarray}  
In order to find the superspace equations of motion, we expand the pure spinor superfield $\Psi$ in positive powers of $\kappa$
\begin{eqnarray}
\Psi &=& \sum_{n=0}^{\infty} \kappa^{n}\Psi_{n}
\end{eqnarray}
where $\Psi_{0}$ is the linearized solution satisfying $Q\Psi_{0} = 0$, which describes linearized 11D supergravity. The recursive relations that one finds from eqn. \eqref{eq106} are:
\begin{eqnarray}
Q\Psi_{0} &=& 0\\
Q\Psi_{1} + \frac{1}{2}(\lambda\Gamma_{ab}\lambda)R^{a}\Psi_{0} R^{b}\Psi_{0} + \frac{1}{2}\Psi_{0} \{Q , T\}\Psi_{0} &=& 0\label{eq107}\\
\vdots\nonumber
\end{eqnarray}

The procedure will now be the same as that applied to the Born-Infeld case: We will first write the non-minimal contribution to \eqref{eq107} as a BRST-exact term $Q\Lambda$. We will then define a new superfield $\tilde\Psi = \Psi-\Lambda$, which will satisfy the equation $Q\tilde\Psi = G(\Psi_{0})$ where $G(\Psi_{0})$ is independent of non-minimal variables. We will finally identify $\tilde C_{\alpha\beta\gamma} = C_{0\,\alpha\beta\gamma} + \kappa C_{1\,\alpha\beta\gamma}$ in $\tilde \Psi = \lambda^\alpha\lambda^\beta \lambda^\gamma \tilde C_{\alpha\beta\gamma}$ as the first-order correction to the linearized D=11 superfield.

\vspace{2mm}
To find $\Lambda$ and $G(\Psi_0)$, the first step will be to write $R^{a}\Psi_{0}$ in terms of a superfield $\Phi^a(x, \theta, \lambda)$
depending only on minimal variables as 
\begin{eqnarray}\label{eeq15}
R^{a}\Psi_{0} &=& \Phi^{a}(x, \theta, \lambda) + Q(f^{a}) + \lambda\Gamma^a{\cal O}
\end{eqnarray} 
where $\lambda \Gamma^a {\cal O}$ is the shift symmetry of $R^a$. To linearized order in the supergravity deformation of the background, the superfield $\Phi^a$ can be expressed in terms of the super-vielbein $E_A{}^P$ and
its inverse $E_P{}^A$ as
\begin{eqnarray}\label{eeqq19}
\Phi^a &=& \lambda^\alpha E_\alpha^{(0)P} \hat E_P{}^a
\end{eqnarray}
where $E_A{}^P$ and $E_P{}^A$ have been expanded around their background values $\hat E_A{}^P$ and $\hat E_P{}^A$ as
\begin{eqnarray}
E_A{}^P &=& \hat E_A{}^P + \kappa E_A^{(0)P} + \kappa^2 E_A^{(1)P} + ...,\nonumber\\
E_P{}^{A} &=& \hat E_P{}^A + \kappa E_P^{(0)A} + \kappa^2 E_P^{(1)A} + ... .
\end{eqnarray}
For example, if one is expanding around the Minkowski space background, $\hat E_a{}^p = \delta_a^p$, $\hat E_\alpha{}^\mu = \delta_\alpha^\mu$ and $\hat E_\alpha{}^m = -
(\Gamma^m\theta)_\alpha$. Note that $E^{(0)P}_\alpha \hat E_P{}^a + \hat E_\alpha{}^P E^{(0) a}_P =0$, so one can also express $\Phi^a$ to linearized order in the deformation as 
\begin{eqnarray}\label{eeqq20}
\Phi^a &=& - \lambda^\alpha \hat E_\alpha{}^P E_P^{(0)a}
\end{eqnarray}
 
Since all of the supergravity fields are contained in $\Psi_0$, one should be able to describe $\Phi$ in terms of $\Psi_0$.  As discussed in \cite{Cederwall:2009ez}, this relation is given by \eqref{eeq15} and it will be explicitly shown in Appendix \ref{apC} that
\begin{eqnarray}
f^{a} &=& -\frac{24}{\eta^{2}}(\bar{\lambda}\Gamma^{ab}\bar{\lambda})(\bar{\lambda}\Gamma^{cd}r)(\lambda\Gamma_{bcd})^{\delta}C_{\delta\alpha\beta}\lambda^{\alpha}\lambda^{\beta} - \frac{24}{\eta}(\bar{\lambda}\Gamma^{ab}\bar{\lambda})C_{b\alpha\beta}\lambda^{\alpha}\lambda^{\beta}
\end{eqnarray}

Plugging eq. \eqref{eeq15} in \eqref{eq107} implies that
\begin{eqnarray}
Q\Psi_{1} + \frac{1}{2}(\lambda\Gamma_{ab}\lambda)[\Phi^{a} + Qf^{a}][\Phi^{b} + Qf^{b}] - Q\left[\frac{1}{2}\Psi_{0}T\Psi_{0}\right] &=& 0,
\end{eqnarray}
which implies that 
\begin{eqnarray}\label{eeq19}
Q(\Psi_1 - \Lambda) = - \frac{1}{2}(\lambda\Gamma_{ab}\lambda)\Phi^{a}\Phi^{b}
\end{eqnarray}
where
$$\Lambda =  \frac{1}{2}\Psi_{0}T\Psi_{0} + (\lambda\Gamma_{ab}\lambda)\Phi^{a}f^{b} - \frac{1}{2}(\lambda\Gamma_{ab}\lambda)f^{a}Qf^{b}.$$

Hence one can define the superfield $\tilde{\Psi}$:
\begin{eqnarray}
\tilde{\Psi} = \Psi_{0} + \kappa(\Psi_1 - \Lambda)
\end{eqnarray}
which will satisfy the following e.o.m at linear order in $\kappa$
\begin{eqnarray}
Q\tilde{\Psi} &=& - \frac{\kappa}{2}(\lambda\Gamma_{ab}\lambda)\Phi^{a}\Phi^{b}
\end{eqnarray}
which implies
\begin{eqnarray}\label{eq99}
\lambda^{\alpha}\lambda^{\beta}\lambda^{\delta}\lambda^{\epsilon}
[D_{\alpha}\tilde{C}_{\beta\delta\epsilon} +\frac{\kappa}{2}(\Gamma_{ab})_{\alpha\beta}E_{\delta}^{(0)P} \hat E_P{}^a E_{\epsilon}^{(0)Q}\hat E_Q{}^b] =0 
\end{eqnarray}
where $\tilde\Psi = \lambda^\alpha\lambda^\beta\lambda^\delta \tilde C_{\alpha\beta\delta}$.

\vspace{2mm}
This equation of motion \eqref{eq99} will now be shown to coincide with the $D=11$ supergravity equations of motion at first order in $\kappa$. The non-linear 
$D=11$ supergravity equations of motion can be expressed using pure spinors as 
\begin{eqnarray}\label{eqeqe181}
\lambda^\alpha \lambda^\beta \lambda^\gamma \lambda^\delta H_{\alpha\beta\gamma\delta} =0
\end{eqnarray}
where we
use the standard transformation rule from curved to tangent-space indices for the 4-form superfield strength:
\begin{eqnarray}\label{eqeqe301}
H_{\alpha\beta\delta\epsilon} &=& E_{\alpha}^{\hspace{2mm}M}E_{\beta}^{\hspace{2mm}N}E_{\delta}^{\hspace{2mm}P}E_{\epsilon}^{\hspace{2mm}Q}H_{MNPQ}
\end{eqnarray}
and $H_{MNPQ} = \nabla_{[M} C_{NPQ]}$. Furthermore, \eqref{eqeqe301} implies that one can choose conventional constraints (by appropriately defining $C_{\alpha\beta a}$ and $C_{\alpha a b}$) so that 
 $$H_{\alpha\beta\gamma\delta}= 
 H_{\alpha\beta\gamma a}= 0, \quad
 H_{\alpha\beta a b}= -{1\over{12}} (\Gamma_{ab})_{\alpha\beta}.$$
This is expected since
there are no physical supergravity fields
with the dimensions of $H_{\alpha\beta\gamma\delta}$, $H_{\alpha\beta\gamma a}$ and $H_{\alpha\beta a b}$. 
 

\vspace{2mm}
To perform an expansion in $\kappa$ and compare with \eqref{eq99}, define
\begin{eqnarray}
\hat{H}_{\alpha\beta\gamma\delta} &=& \hat{E}_{\alpha}^{\hspace{2mm}M}\hat{E}_{\beta}^{\hspace{2mm}N}\hat{E}_{\gamma}^{\hspace{2mm}P}\hat{E}_{\delta}^{\hspace{2mm}Q}H_{MNPQ}.
\end{eqnarray}\label{eqeq000} 
Eqn. \eqref{eqeqe181} implies that
$$
0 = \lambda^\alpha \lambda^\beta \lambda^\gamma \lambda^\delta (\hat H_{\alpha\beta\gamma\delta} +4\kappa \hat E_\alpha{}^M \hat E_\beta{}^N \hat E_\gamma{}^P E_\delta^{(0)Q}
H_{MNPQ} $$
\begin{eqnarray}
+6 
\kappa^2 \hat E_\alpha{}^M \hat E_\beta{}^N E_\gamma^{(0) P} E_\delta{}^{(0)Q} H_{MNPQ} + ...)
 \end{eqnarray}\label{eqq180}
 $$= \lambda^\alpha \lambda^\beta \lambda^\gamma \lambda^\delta (\hat H_{\alpha\beta\gamma\delta} +4\kappa \hat E_\alpha{}^M \hat E_\beta{}^N \hat E_\gamma{}^P E_\delta^{(0)Q}
E_M{}^A E_N{}^B E_P{}^C E_Q{}^D H_{ABCD}$$
\begin{eqnarray}
+6 
\kappa^2 \hat E_\alpha{}^M \hat E_\beta{}^N E_\gamma^{(0) P} E_\delta^{(0)Q} E_M{}^A E_N{}^B E_P{}^C E_Q{}^D H_{ABCD} + ...)
 \end{eqnarray}
 $$= \lambda^\alpha \lambda^\beta \lambda^\gamma \lambda^\delta (\hat H_{\alpha\beta\gamma\delta} +12\kappa^2  \hat E_\gamma{}^P E_P^{(0) a} E_\delta^{(0)Q}\hat E_Q{}^b
 H_{\alpha\beta a b}$$
 \begin{eqnarray}
+6 
\kappa^2  E_\gamma^{(0) P}\hat E_P{}^a E_\delta^{(0)Q} \hat E_Q{}^b H_{\alpha \beta a b} + ...)
 \end{eqnarray}
 \begin{eqnarray}\label{eqq22}
&=& 
 \lambda^\alpha \lambda^\beta \lambda^\gamma \lambda^\delta (\hat H_{\alpha\beta\gamma\delta}  + {1\over 2} 
\kappa^2  E_\gamma^{(0) P} \hat E_P{}^a E_\delta^{(0)Q} \hat E_Q{}^b (\Gamma_{ab})_{\alpha \beta} + ...)
\end{eqnarray}
 where $...$ denotes terms higher-order in $\kappa$.  Since 
 $$\lambda^\alpha \lambda^\beta \lambda^\gamma \lambda^\delta\hat H_{\alpha\beta\gamma\delta} = \kappa \lambda^\alpha \lambda^\beta \lambda^\gamma \lambda^\delta D_\alpha \tilde C_{\beta\gamma\delta},$$
 eqn. \eqref{eqq22} for the
 back-reaction to $\hat H_{\alpha\beta\gamma\delta}$ coincides with \eqref{eq99}.

\chapter{Final Remarks}
We have presented here the $D=11$ pure spinor superparticle through a series of similarity transformations among BRST cohomologies starting from the $D=11$ Brink-Schwarz-like superparticle in semi-light-cone gauge. Although this is an interesting way of relating both models, it is not the only one. Indeed, one can start with an eleven-dimensional particle subjected to a twistor-like constraint, which after some suitable gauge fixing, reduces to the $D=11$ pure spinor superparticle or the $D=11$ Brink-Schwarz superparticle \cite{Berkovits:2015yra}. In this novel approach, the pure spinor variables $\lambda^{\alpha}$ appear through the twistor-like constraints $(\lambda\gamma^{a})_{\alpha}P_{a} = 0$, and the coordinates $\theta^{\alpha}$ arise as the ghosts for the symmetries generated by these constraints.

\vspace{2mm}
We have also seen the light-cone gauge analysis of the ghost number three pure spinor BRST cohomology correctly reproduced the light-cone gauge linearized $D=11$ supergravity equations of motion found in \cite{Green:1999by}. This can also be interpreted as a proof that the ghost number three vertex operator $U^{(3)} = \lambda^{\alpha}\lambda^{\beta}\lambda^{\delta}C_{\alpha\beta\delta}$ indeed contains the physical fields of $D=11$ supergravity. 

\vspace{2mm}
In an attempt to calculate $D=11$ supergravity correlation functions, we successfully constructed a ghost number one vertex operator, which is BRST-closed once superfields are on-shell, but we failed when trying to construct a ghost number zero vertex operator satisfying a standard descent equation. We were able to fix this issue after introducing the non-minimal version of the $D=11$ pure spinor superparticle. In this, a composite $D=11$ $b$-ghost was presented and later simplified through the use an $SO(1,10)$ fermionic vector $\bar{\Sigma}^{a}$. After defining a ghost number zero vertex operator as
\begin{eqnarray}
V^{(0)} &=& \{b, U^{(1)}\}
\end{eqnarray}
and using that $\{Q, b\} = P^{2}$, $V^{(0)}$ is easily shown to satisfy a standard descent equation. However its explicit form has not been computed yet. It would be interesting to see what structure $V^{(0)}$ presents and if this can be split into a simple part depending on minimal and non-minimal variables plus a BRST-exact term. The non-BRST exact sector might be carefully analyzed and regularized using a similar procedure as in \cite{Berkovits:2006vi,Aisaka:2009yp,Cederwall:2012es}. This would give us a ghost number zero vertex operator which would be more tractable than the original one for N-point correlation functions computations. In addition, knowing an explicit formula for $V^{(0)}$ would be also useful for understanding how permutation invariance is realized in this framework.

\vspace{2mm}
Since the $D=11$ superparticle can be considered as the infinite tension limit of the $D=11$ supermembrane, understanding how to treat all of these issues might be useful to comprehend better the structure of $M$-theory. We hope to apply similar ideas to the ones presented in this thesis for constructing $D=11$ supermembrane vertex operators. We also pretend to investigate the connection between Matrix Theory and M-Theory in the pure spinor framework. Some progress on this was done in \cite{Oda:2003dz}, and we hope to benefit from the discoveries and obstacles found there in order to achieve our goal.

\vspace{2mm}
The pure spinor master actions for $D=10$ super-Born-Infeld and $D=11$ supergravity have been shown to correctly reproduce the equations of motion expected on ordinary superspace at first order in the coupling constant. The procedure to follow for higher order corrections is essentially the same as the one used in section 5. So, for instance, the second order correction for $\Psi$ in the $D=10$ super-Born-Infeld model obeys
\begin{eqnarray}
Q\Psi_{2} + 2(\lambda\gamma^{m}\hat{\chi}\Psi_{1})(\lambda\gamma^{n}\hat{\chi}\Psi_{0})\hat{F}_{mn}\Psi_{0} + (\lambda\gamma^{m}\hat{\chi}\Psi_{0})(\lambda\gamma^{n}\hat{\chi}\Psi_{0})\hat{F}_{mn}\Psi_{1} &=& 0
\end{eqnarray}
which can also be written in the form
\begin{eqnarray}\label{neweeq13}
Q\Psi_{2} + \frac{3}{8(\lambda\bar{\lambda})^{2}}(\bar{\lambda}\gamma^{ijk}r)\Delta_{i}\Psi_{1}\Delta_{j}\Psi_{0}\Delta_{k}\Psi_{0} &=& 0
\end{eqnarray}
As before, one can explicitly show that the second term in \eqref{neweeq13} is BRST-closed, which implies it should be possible to write it as: $Q \hat{\Lambda} + G(\Psi_0, \tilde{\Psi})$, where $\hat{\Lambda}$ is a function of non-minimal variables and G is a function depending only on minimal variables. Notice that the entire field $\Psi_{1}$ enters eqn. \eqref{neweeq13}, which means one should take into account the contribution coming from $\Lambda$ defined in eqn. \eqref{eq1new}, before expecting to find such a splitting. In addition, one should also use the equations of motion
\begin{eqnarray}
D_{\alpha}\tilde{A}_{\beta} + D_{\beta}\tilde{A}_{\alpha} + 2(\gamma^{i})_{\alpha\beta}\tilde{A}_{i} &=& - (\gamma^{i}\chi)_{\alpha}(\gamma^{j}\chi)_{\beta}F_{ij}\\
\partial_{j}\tilde{A}_{\delta} - D_{\delta}\tilde{A}_{j}  &=& (\gamma_{j}\tilde\chi)_{\delta}- (\gamma_{n}\chi)_{\delta}F_{mj}F^{mn} + \frac{1}{2}(\gamma^{n}\chi)_{\delta}(\chi\gamma_{j}\partial_{n}\chi)\nonumber\\
(\gamma_{jk})_{\delta}{}^{\alpha}D_{\alpha}\tilde{\chi}^{\delta} &=& -16\tilde{F}_{jk} - (\chi\gamma_{[j|mn|}\chi)\partial_{k]}F_{mn} + 16F_{kn}F_{mj}F^{mn}\nonumber\\
&& - 16F_{m[j}(\chi\gamma_{k]}\partial_{m}\chi) - 2(\chi\gamma_{jkn}\partial_{m}\chi)F^{mn}\nonumber\\ 
&& - 2(\chi\gamma_{m}\partial_{[j}\chi)F_{k]m} + (\chi\gamma_{ab[j}\partial_{k]}\chi)F_{ab}\\
D_{\alpha}\tilde{\chi}^{\alpha} &=& 2(\chi\gamma_{n}\partial_{m}\chi)F^{mn}
\end{eqnarray}
where $\tilde{A}_{\alpha}$, $\tilde{A}_{i}$, $\tilde{\chi}^{\alpha}$ are the first order corrections for the super-gauge field components and gluino field, and $\tilde{F}_{ij} = \partial_{i}\tilde{A}_{j} - \partial_{j}\tilde{A}_{i}$. The result thus obtained must match the $D=10$ super-Born-Infeld equation of motion at second order in the coupling constant \cite{Berkovits:2002ag}. This is work in progress, and we hope to apply the same line of reasoning for $D=11$ supergravity.

\vspace{2mm}
One can now use these models for computing scattering amplitudes using standard field-theory techniques. This was developed for $D=11$ supergravity in \cite{Cederwall:2012es}. In this framework, many interesting features and properties of $D=11$ supergravity amplitudes were shown to exist. However, some of the computations carried out there assumed that the $D=11$ $b$-ghost is nilpotent which we now know is not the case. It would be interesting to figure out how the non-nilpotency of the $b$-ghost affects their analysis. Likewise, it would be interesting to use the suggestions given in \cite{Karlsson:2014xva,Karlsson:2015qda} and perform a careful counting of zero modes in the pure spinor integration measures, which has a direct effect on the properties of $D=11$ supergravity amplitudes.

\begin{appendix}
\chapter{}\label{appendix1}
\section{$\Gamma$-matrices of $SO(10,1)$}\label{appA}

We will denote $SO(10,1)$ vector indices by $m, n, \ldots$ and $SO(9,1)$ vector indices by $\hat{m}, \hat{n}, \ldots$. In addition, we will denote $SO(10,1)$ spinor indices by $\alpha, \beta, \ldots$ and $SO(9,1)$ spinor indices by $\mu, \nu, \ldots$. As usual, we add a new matrix, $\Gamma^{10}$, to the set of $SO(9,1)$ gamma matrices $\{\Gamma^{\hat{m}}\}$, which is numerically equal to the chirality matrix $\Gamma^{(9,1)}$ in $D = (9,1)$:
\begin{equation}
\Gamma^{10} = \Gamma^{(9,1)} = \begin{pmatrix}
I_{16\times 16} & 0\\
0 & -I_{16\times 16}
\end{pmatrix}
\end{equation}
This matrix satisfies the properties $\{\Gamma^{m}, \Gamma^{10}\} = 0$, for $m = 0, \ldots , 9$, and $(\Gamma^{10})^{2} = 1$. The chirality matrix $\Gamma$ in $D = (10,1)$ is given by:
\begin{equation}
\Gamma = \Gamma^{0}\Gamma^{1}\ldots \Gamma^{9}\Gamma^{10} = \Gamma^{(9,1)}\Gamma^{10} = (\Gamma^{10})^{2} = 1
\end{equation}
which reflects the fact that we don't have Weyl (anti-Weyl) spinors in eleven dimensions. However, we can have Majorana spinors. Moreover, it is easy to see that $C = \Gamma^{0}$ satisfies the definition of the charge conjugation matrix\footnote{We know that for $D=(9,1)$, $C^{(9,1)} = \Gamma^{0}$ is the charge conjugation matrix, so we just need to show that $C = \Gamma^{0}$ obeys $C\Gamma^{10} = -(\Gamma^{10})^{T}C$, which is trivial since $\Gamma^{10}$ is symmetric and $\{\Gamma^{10}, \Gamma^{0}\} = 0$.} $C\Gamma^{m} = -(\Gamma^{m})^{T}C$.\\

\vspace{2mm}
For two Majorana spinors $\Theta$ and $\Psi$, we have $\bar{\Theta}\Gamma^{m}\Psi = \Theta^{T} C\Gamma^{m}\Psi$. This result can be viewed in terms of $SO(9,1)$ components:
\begin{equation}
\Theta^{T} C\Gamma^{m}\Psi =  \begin{pmatrix} \Theta^{\mu} & \Theta_{\mu}
\end{pmatrix} \begin{pmatrix}
\gamma^{\hat{m}}_{\mu\nu} & 0\\
0 & -(\gamma^{\hat{m}})^{\mu\nu}
\end{pmatrix} \begin{pmatrix}
\Psi^{\nu} & \Psi_{\nu}
\end{pmatrix},
\end{equation}
\begin{equation}
\Theta^{T} C\Gamma^{10}\Psi =  \begin{pmatrix} \Theta^{\mu} & \Theta_{\mu}
\end{pmatrix} \begin{pmatrix}
0 & -1\\
-1 & 0
\end{pmatrix} \begin{pmatrix}
\Psi^{\nu} & \Psi_{\nu}
\end{pmatrix},
\end{equation}
where $m = 0, \ldots, 9$ and $(\gamma^{\hat{m}})_{\mu\nu}$, and $(\gamma^{\hat{m}})^{\mu\nu}$ are the $SO(9,1)$ $\gamma$-matrices. 
It is useful to mention that the index structure of the charge conjugation matrix is $C_{\alpha\beta}$. So, the $\Gamma$-matrices have index structure $(\Gamma^{m})^{\alpha}_{\hspace{2mm}\beta}$ and when are multiplied by the charge conjugation matrix (or its inverse) we obtain the corresponding matrices $(\Gamma^{m})_{\alpha\beta}$ and $(\Gamma^{m})^{\alpha\beta}$.

\vspace{2mm}
Next we will show explicitly the form of the gamma matrices. For $D=(9,1)$, we have:

\begin{eqnarray}
\begin{aligned}
(\gamma^{0})^{\alpha\beta} &= \begin{pmatrix} 1_{8\times 8} & 0\\
0 & 1_{8\times 8}
\end{pmatrix} \hspace{8mm},\\
(\gamma^{9})^{\alpha\beta} &= \begin{pmatrix}1_{8\times 8} & 0\\
0 & -1_{8\times 8}
\end{pmatrix}\hspace{5.2mm},\\
(\gamma^{\hat{i}})^{\alpha\beta} &= \begin{pmatrix} 0 & \sigma^{\hat{i}}_{a\dot{a}}\\
\sigma^{\hat{i}}_{\dot{b}b} & 0\end{pmatrix}\hspace{12.6mm},\\
(\gamma^{+})^{\alpha\beta} &= \begin{pmatrix}\sqrt{2}_{8\times 8} & 0\\
0 & 0
\end{pmatrix}\hspace{10.5mm},\\
(\gamma^{-})^{\alpha\beta} &= \begin{pmatrix}0 & 0\\
0 & \sqrt{2}_{8\times 8}
\end{pmatrix}\hspace{10.5mm},\\
\end{aligned}\hspace{4mm}
\begin{aligned}
 (\gamma^{0})_{\alpha\beta} &= \begin{pmatrix} -1_{8\times 8} & 0\\
0 & -1_{8\times 8}
\end{pmatrix}\\
 (\gamma^{9})_{\alpha\beta} &= \begin{pmatrix} 1_{8\times 8} & 0\\
0 & -1_{8\times 8}
\end{pmatrix}\\
 (\gamma^{\hat{i}})_{\alpha\beta} &= \begin{pmatrix}0 & \sigma^{\hat{i}}_{a\dot{a}}\\
\sigma^{\hat{i}}_{\dot{b}b} & 0
\end{pmatrix}\\
 (\gamma^{+})_{\alpha\beta} &= \begin{pmatrix}0 & 0\\
0 & -\sqrt{2}_{8\times 8}
\end{pmatrix}\\
 (\gamma^{-})_{\alpha\beta} &= \begin{pmatrix}-\sqrt{2}_{8\times 8} & 0\\
0 & 0
\end{pmatrix},
\end{aligned}
\end{eqnarray}
where each entry is an $8\times 8$ matrix and $\hat{i}$ is a $SO(8)$ vector index. The matrices $\gamma^{\pm}$ are defined by
\begin{equation*}
\gamma^{\pm} = \frac{1}{\sqrt{2}}(\gamma^{0} \pm \gamma^{9})
\end{equation*}
The $\sigma^{\hat{i}}$ matrices are defined by

$$
\begin{array}{cl}
\sigma^{1}_{a\dot{a}} = \epsilon\otimes\epsilon\otimes\epsilon \hspace{10mm} & \sigma^{5}_{a\dot{a}} = \tau^{3}\otimes\epsilon\otimes 1\\
\sigma^{2}_{a\dot{a}} = 1\otimes\tau^{1}\otimes\epsilon \hspace{10mm} & \sigma^{6}_{a\dot{a}} = \epsilon\otimes 1\otimes\tau^{1}\\
\sigma^{3}_{a\dot{a}} = 1\otimes\tau^{3}\otimes\epsilon \hspace{10mm} & \sigma^{7}_{a\dot{a}} = \epsilon\otimes 1\otimes\tau^{3}\\
\sigma^{4}_{a\dot{a}} = \tau^{1}\otimes\epsilon\otimes 1 \hspace{10mm} & \sigma^{8}_{a\dot{a}} = 1\otimes 1\otimes 1
\end{array}
$$
where $\epsilon = i\tau^{2}$ and $\tau^{1}$, $\tau^{2}$, $\tau^{3}$ are the usual Pauli matrices. The $\sigma^{\hat{i}}_{a\dot{a}}$ are symmetric ($\sigma^{\hat{i}}_{a\dot{a}} = (\sigma^{\hat{i}}_{a\dot{a}})^{T}$) and satisfy the following relations:
\begin{eqnarray*}
\sigma^{\hat{i}}_{a\dot{a}}\sigma^{\hat{j}}_{b\dot{a}} + \sigma^{\hat{j}}_{a\dot{a}}\sigma^{\hat{i}}_{b\dot{a}} &=& 2\delta^{\hat{i}\hat{j}}\delta_{ab}\\
\sigma^{\hat{i}}_{a\dot{a}}\sigma^{\hat{j}}_{a\dot{b}} + \sigma^{\hat{j}}_{a\dot{a}}\sigma^{\hat{i}}_{a\dot{b}} &=& 2\delta^{\hat{i}\hat{j}}\delta_{\dot{a}\dot{b}}\\
\sigma^{\hat{i}}_{b\dot{a}}\sigma^{\hat{i}}_{a\dot{c}} + \sigma^{\hat{i}}_{a\dot{a}}\sigma^{\hat{i}}_{b\dot{c}} &=& 2\delta_{ab}\delta_{\dot{a}\dot{c}} 
\end{eqnarray*}

Similarly, for $D=(10,1)$, we have:
\begin{eqnarray}
\sbox0{$\begin{matrix}0&0\\0&-\sqrt{2}i\end{matrix}$}
\begin{aligned}
(\Gamma^{\hat{i}})^{\alpha\beta} &= \begin{pmatrix}
-i\gamma^{\hat{i}\,AB} & \makebox[\wd0]{\large $O$}\\
\makebox[\wd0]{\large $O$} & i\gamma^{\hat{i}}_{AB}
\end{pmatrix}\hspace{14mm},\\
(\Gamma^{11})^{\alpha\beta} &= \begin{pmatrix}
\makebox[\wd0]{\large $O$} & -i\\
-i & \makebox[\wd0]{\large $O$}
\end{pmatrix}\hspace{14mm},\\
(\Gamma^{+})^{\alpha\beta} &= \begin{pmatrix}
\begin{pmatrix}
-\sqrt{2}i & 0\\ 
0 & 0
\end{pmatrix} & \makebox[\wd0]{\large $O$}\\
\makebox[\wd0]{\large $O$} & \begin{pmatrix}
0 & 0\\
0 & -\sqrt{2}i
\end{pmatrix}
\end{pmatrix}\hspace{2mm},\\
(\Gamma^{-})^{\alpha\beta} &= \begin{pmatrix}
\begin{pmatrix}
0 & 0\\ 
0 & -\sqrt{2}i
\end{pmatrix} & \makebox[\wd0]{\large $O$}\\
\makebox[\wd0]{\large $O$} & \begin{pmatrix}
-\sqrt{2}i & 0\\
0 & 0
\end{pmatrix}
\end{pmatrix}\hspace{2mm},
\end{aligned}
\begin{aligned}
(\Gamma^{\hat{i}})_{\alpha\beta} &= \begin{pmatrix}
i\gamma^{\hat{i}}_{AB} & \makebox[\wd0]{\large $O$}\\
\makebox[\wd0]{\large $O$} & -i\gamma^{\hat{i}\,AB}
\end{pmatrix}\\
(\Gamma^{11})_{\alpha\beta} &= \begin{pmatrix}
\makebox[\wd0]{\large $O$} & i\\
i & \makebox[\wd0]{\large $O$}
\end{pmatrix}\\
(\Gamma^{+})_{\alpha\beta} &= \begin{pmatrix}
\begin{pmatrix}
0 & 0\\ 
0 & -\sqrt{2}i
\end{pmatrix} & \makebox[\wd0]{\large $O$}\\
\makebox[\wd0]{\large $O$} & \begin{pmatrix}
-\sqrt{2}i & 0\\
0 & 0
\end{pmatrix}
\end{pmatrix}\\
(\Gamma^{-})_{\alpha\beta} &= \begin{pmatrix}
\begin{pmatrix}
-\sqrt{2}i & 0\\ 
0 & 0
\end{pmatrix} & \makebox[\wd0]{\large $O$}\\
\makebox[\wd0]{\large $O$} & \begin{pmatrix}
0 & 0\\
0 & -\sqrt{2}i
\end{pmatrix}
\end{pmatrix}
\end{aligned}\nonumber\\
\end{eqnarray}
where $A$, $B$ are $SO(9)$ spinor indices. Notice that each $\Gamma$ matrix is $32\times 32$.

\vspace{2mm}
To construct the above representation of the $\Gamma$ matrices, we used a basis convenient for dealing with $SO(8)$ objects. Hence, an arbitrary $D=11$ spinor $\chi^{\alpha}$ is written in this basis as
\begin{equation}
\chi^{\alpha} = \begin{pmatrix}
\chi^{a}\\
\chi^{\dot{a}}\\
\bar{\chi}^{a}\\
\bar{\chi}^{\dot{a}}
\end{pmatrix}
\end{equation}
This was the convention used in \eqref{section2eq500}. This is useful when $SO(8)$ objects are our main concern, as in Section 3. However, when analyzing the light-cone gauge structure of the pure spinor cohomology and vertex operators, we need to deal with $SO(9)$ objects. So, we define the following change of basis matrix:
\begin{equation}
M_{cbm} = \left(\begin{matrix}
0 & 0 & 1 & 0\\
0 & 1 & 0 & 0\\
1 & 0 & 0 & 0\\
0 & 0 & 0 & 1
\end{matrix}\right)
\end{equation}
where each entry represents an $8\times 8$ matrix. Using this matrix we find the corresponding $\Gamma$ matrices in this new basis:
\begin{eqnarray}
\sbox0{$\begin{matrix}0&0\\0&-\sqrt{2}i\end{matrix}$}
\begin{aligned}
(\Gamma^{\hat{i}})^{\alpha\beta} &= i\begin{pmatrix}
\makebox[\wd0]{\large $O$} & (\gamma^{9}\gamma^{\hat{i}})^{AB}\\
-(\gamma^{9}\gamma^{\hat{i}})_{AB} & \makebox[\wd0]{\large $O$}
\end{pmatrix}\hspace{6.5mm},\\
(\Gamma^{11})^{\alpha\beta} &= -i\begin{pmatrix}
 \makebox[\wd0]{\large $O$} &  I^{AB}\\
I_{AB} &  \makebox[\wd0]{\large $O$}
\end{pmatrix}\hspace{6.5mm},\\
(\Gamma^{+})^{\alpha\beta} &= \left(\begin{array}{cc}
\makebox[\wd0]{$O$} & \makebox[\wd0]{$O$}\\
\makebox[\wd0]{$O$} & -\sqrt{2}i
\end{array}\right)\hspace{9mm},\\
(\Gamma^{-})^{\alpha\beta} &= \left(\begin{array}{cc}
-\sqrt{2}i & \makebox[\wd0]{$O$}\\
\makebox[\wd0]{$O$} & \makebox[\wd0]{$O$}
\end{array}\right)\hspace{9mm},\\
\end{aligned}
\begin{aligned}
(\Gamma^{\hat{i}})_{\alpha\beta} &= i\begin{pmatrix}
 \makebox[\wd0]{\large $O$} & -(\gamma^{9}\gamma^{\hat{i}})^{AB}\\
(\gamma^{9}\gamma^{\hat{i}})_{AB} &  \makebox[\wd0]{\large $O$}
\end{pmatrix}\\
(\Gamma^{11})_{\alpha\beta} &= i\begin{pmatrix}
\makebox[\wd0]{$O$} & I_{AB}\\
I^{AB} & \makebox[\wd0]{$O$}
\end{pmatrix}\\
(\Gamma^{+})_{\alpha\beta} &= 
\left(\begin{array}{cc}
-\sqrt{2}i & \makebox[\wd0]{$O$}\\
\makebox[\wd0]{$O$} & \makebox[\wd0]{$O$}
\end{array}\right)\\
(\Gamma^{-})_{\alpha\beta} &= \left(\begin{array}{cc}
\makebox[\wd0]{$O$} & \makebox[\wd0]{$O$}\\
\makebox[\wd0]{$O$} & -\sqrt{2}i 
\end{array}\right)
\end{aligned}
\end{eqnarray}
where $I_{AB}$ is the $SO(9)$ identity matrix, $A$, $B$ are $SO(9)$ spinor indices, and $\hat{i} = 1,\ldots,8$. Each entry in the above matrices is $16\times 16$.

\section{$SO(10,1) \rightarrow SO(3,1)\times SO(7)$}\label{appB}
Here we will explain the $\pm$ notation and construct explicitly a different representation for the $SO(10,1)$ gamma matrices. Let us define the \emph{raising} and \emph{lowering} $\Gamma$-matrices:
\begin{eqnarray}
\Gamma^{\pm 0+1} &=& \frac{1}{2}(\pm\Gamma^{0} + \Gamma^{1})\\
\Gamma^{2\pm 3i} &=& \frac{1}{2}(\Gamma^{2} \pm i\Gamma^{3})
\end{eqnarray}
These $\Gamma^{\pm}$ matrices act on an arbitrary spinor $\chi$ as follows: 
\begin{eqnarray}
\begin{aligned}
\Gamma^{0+1}|- + a> &= |+ + a>\hspace{1mm},\\
\Gamma^{0+1}|- - a> &= |+ - a>\hspace{1mm},\\
\Gamma^{-0+1}|+ + a> &= |- + a> \hspace{1mm},\\
\Gamma^{-0+1}|+ - a> &= |- - a>\hspace{1mm},\\
\Gamma^{j}|- - 0> &= |- - j>,
\end{aligned}\hspace{1mm}
&\begin{aligned}
\Gamma^{3+4i}|- - a> &= -|- + a>\hspace{3mm},\\
\Gamma^{3+4i}|+ - a> &= |+ + a>\hspace{6mm},\\
\Gamma^{3-4i}|- + a> &= -|- + a>\hspace{3mm},\\
\Gamma^{3-4i}|+ + a> &= |+ - a>\hspace{6mm},\\
\Gamma^{j}|- - j> &= |- - 0>\hspace{1.8mm},
\end{aligned}\hspace{1mm}\nonumber
&
\begin{aligned}
\Gamma^{j}|+ + 0> &= |+ + j> \\
\Gamma^{j}|+ + j> &= |+ + 0> \\
\Gamma^{j}|- + 0> &= -|- + j> \\
\Gamma^{j}|- + j> &= -|- + 0>\\
\Gamma^{j}|+ - 0> &= -|+ - j>
\end{aligned}\\
&
\begin{aligned}
\hspace{2mm}\Gamma^{j}|+ - j> &= -|+ - 0>\nonumber
\end{aligned}& \\
\end{eqnarray}
and any other relation vanishes. In these formulae we have made the identification $|\pm\pm a> = \chi^{\pm\pm a}$ with $a = 0, i$. It is clear that these relations are consistent with the $SO(10,1)$ Clifford algebra. With these rules, one can construct the respective representation:
\begin{equation}
\begin{aligned}
(\Gamma^{0+1})^{\alpha}_{\hspace{2mm}\beta} &= \begin{pmatrix}
0 & 0 & 0 & 1\\
0 & 0 & 0 & 0\\
0 & 1 & 0 & 0\\
0 & 0 & 0 & 0
\end{pmatrix}\hspace{2mm},
\end{aligned}\hspace{2mm}
\begin{aligned}
(\Gamma^{-0+1})^{\alpha}_{\hspace{2mm}\beta} &= \begin{pmatrix}
0 & 0 & 0 & 0\\
0 & 0 & 1 & 0\\
0 & 0 & 0 & 0\\
1 & 0 & 0 & 0
\end{pmatrix}
\end{aligned}
\end{equation}
Here and throughout this Appendix, each entry will represent an $8\times 8$ matrix unless otherwise stated. Now, it is easy to calculate the explicit form of the matrices $(\Gamma^{0})^{\alpha}_{\hspace{2mm}\beta}$, $(\Gamma^{1})^{\alpha}_{\hspace{2mm}\beta}$:
\begin{equation}
\begin{aligned}
(\Gamma^{0})^{\alpha}_{\hspace{2mm}\beta} &= \begin{pmatrix}
0 & 0 & 0 & 1\\
0 & 0 & -1 & 0\\
0 & 1 & 0 & 0\\
-1 & 0 & 0 & 0
\end{pmatrix}\hspace{2mm}, 
\end{aligned}\hspace{2mm}
\begin{aligned}
(\Gamma^{1})^{\alpha}_{\hspace{2mm}\beta} &= \begin{pmatrix}
0 & 0 & 0 & 1\\
0 & 0 & 1 & 0\\
0 & 1 & 0 & 0\\
1 & 0 & 0 & 0
\end{pmatrix}
\end{aligned}
\end{equation}
Similarly, we find
\begin{equation}
\begin{aligned}
(\Gamma^{2+3i})^{\alpha}_{\hspace{2mm}\beta} &= \begin{pmatrix}
0 & 0 & 1 & 0\\
0 & 0 & 0 & 0\\
0 & 0 & 0 & 0\\
0 & -1 & 0 & 0
\end{pmatrix}\hspace{5mm},\\
(\Gamma^{2})^{\alpha}_{\hspace{2mm}\beta} &= \begin{pmatrix}
0 & 0 & 1 & 0\\
0 & 0 & 0 & -1\\
1 & 0 & 0 & 0\\
0 & -1 & 0 & 0
\end{pmatrix}\hspace{2.8mm}, 
\end{aligned}\hspace{2mm}
\begin{aligned}
(\Gamma^{2-3i})^{\alpha}_{\hspace{2mm}\beta} &= \begin{pmatrix}
0 & 0 & 0 & 0\\
0 & 0 & 0 & -1\\
1 & 0 & 0 & 0\\
0 & 0 & 0 & 0
\end{pmatrix}\\
(\Gamma^{3})^{\alpha}_{\hspace{2mm}\beta} &= \begin{pmatrix}
0 & 0 & -i & 0\\
0 & 0 & 0 & -i\\
i & 0 & 0 & 0\\
0 & i & 0 & 0
\end{pmatrix}
\end{aligned}
\end{equation}
However, as already mentioned, there exists an antisymmetric metric tensor $C_{\alpha\beta}$ in $D=11$ dimensions which raises and lower indices. Let us define it as follows:
\begin{equation}
\begin{aligned}
C_{\alpha\beta} = \begin{pmatrix}
0 & -B & 0 & 0\\
B & 0 & 0 & 0\\
0 & 0 & 0 & -B\\
0 & 0 & B & 0
\end{pmatrix}\hspace{2mm},
\end{aligned}\hspace{2mm}
\begin{aligned}
(C^{-1})^{\alpha\beta} = \begin{pmatrix}
0 & B & 0 & 0\\
-B & 0 & 0 & 0\\
0 & 0 & 0 & B\\
0 & 0 & -B & 0
\end{pmatrix}
\end{aligned}
\end{equation}
where $B$ is a diagonal matrix with elements $B_{00}=1$, $B_{jj}=-1$. To preserve the original Clifford algebra we need to multiply the matrices $(\Gamma^{a})^{\alpha}_{\hspace{2mm}\beta}$ by $i$. Now we can find the matrices $(\Gamma^{m})_{\alpha\beta}$, $(\Gamma^{m})^{\alpha\beta}$:
\begin{equation}
\begin{aligned}
\Gamma^{-0+1}_{\alpha\beta} &=& \begin{pmatrix}
0 & 0 & -iB& 0 \\
0 & 0 & 0 & 0\\
-iB & 0 & 0 & 0\\
0 & 0 & 0 & 0
\end{pmatrix}\hspace{2mm},
\end{aligned}\hspace{2mm}
\begin{aligned}
\Gamma^{0+1}_{\alpha\beta} &=& \begin{pmatrix}
0 & 0 & 0 & 0 \\
0 & 0 & 0 & iB\\
0 & 0 & 0 & 0\\
0 & iB & 0 & 0
\end{pmatrix}
\end{aligned}
\end{equation}
and so
\begin{equation}
\begin{aligned}
\Gamma^{0}_{\alpha\beta} &=& \begin{pmatrix}
0 & 0 & iB & 0 \\
0 & 0 & 0 & iB\\
iB & 0 & 0 & 0\\
0 & iB & 0 & 0
\end{pmatrix}\hspace{2mm},
\end{aligned}\hspace{2mm}
\begin{aligned}
\Gamma^{1}_{\alpha\beta} &=& \begin{pmatrix}
0 & 0 & -iB & 0 \\
0 & 0 & 0 & iB\\
-iB & 0 & 0 & 0\\
0 & iB & 0 & 0
\end{pmatrix}
\end{aligned}
\end{equation}
By using $(\Gamma^{m})^{\alpha\beta} = C^{\alpha\delta}C^{\beta\lambda}(\Gamma^{m})_{\delta\lambda}$, we find the matrices $\Gamma^{0\,\alpha\beta}$, $\Gamma^{1\,\alpha\beta}$: 
\begin{equation}
\Gamma^{0\,\alpha\beta} = \begin{pmatrix}
0 & B & 0 & 0\\
-B & 0 & 0 & 0\\
0 & 0 & 0 & B\\
0 & 0 & -B & 0
\end{pmatrix}
\begin{pmatrix}
0 & 0 & iB & 0\\
0 & 0 & 0 & iB\\
iB & 0 & 0 & 0\\
0 & iB & 0 & 0
\end{pmatrix}
\begin{pmatrix}
0 & -B & 0 & 0\\
B & 0 & 0 & 0\\
0 & 0 & 0 & -B\\
0 & 0 & B & 0
\end{pmatrix} 
= \begin{pmatrix}
0 & 0 & iB & 0\\
0 & 0 & 0 & iB\\
iB & 0 & 0 & 0\\
0 & iB & 0 & 0
\end{pmatrix} 
\end{equation}
\begin{equation}
\Gamma^{1\,\alpha\beta} = \begin{pmatrix}
0 & 0 & iB & 0\\
0 & 0 & 0 & -iB\\
iB & 0 & 0 & 0\\
0 & -iB & 0 & 0
\end{pmatrix}
\end{equation}
Analogously we can find the remaining matrices,
\begin{equation}
\begin{aligned}
(\Gamma^{2+3i})_{\alpha\beta} &= \begin{pmatrix}
0 & 0 & 0 & 0\\
0 & 0 & iB & 0\\
0 & iB & 0 & 0\\
0 & 0 & 0 & 0
\end{pmatrix}\hspace{25mm},\\
(\Gamma^{2})_{\alpha\beta} &= \begin{pmatrix}
0 & 0 & 0 & iB\\
0 & 0 & iB & 0\\
0 & iB & 0 & 0\\
iB & 0 & 0 & 0
\end{pmatrix}\hspace{20.5mm},\\
(\Gamma^{2})^{\alpha\beta} &= \begin{pmatrix}
0 & 0 & 0 & -iB\\
0 & 0 & -iB & 0\\
0 & -iB & 0 & 0\\
-iB & 0 & 0 & 0
\end{pmatrix}\hspace{9.3mm},\\
(\Gamma^{i})_{\alpha\beta} &= \begin{pmatrix}
0 & -iBA & 0 & 0\\
iBA & 0 & 0 & 0\\
0 & 0 & 0 & iBA\\
0 & 0 & -iBA & 0
\end{pmatrix}\hspace{4.5mm}, 
\end{aligned}\hspace{2mm}
\begin{aligned}
(\Gamma^{2-3i})_{\alpha\beta} &= \begin{pmatrix}
0 & 0 & 0 & iB\\
0 & 0 & 0 & 0\\
0 & 0 & 0 & 0\\
iB & 0 & 0 & 0
\end{pmatrix}\\
(\Gamma^{3})_{\alpha\beta} &= \begin{pmatrix}
0 & 0 & 0 & -B\\
0 & 0 & B & 0\\
0 & B & 0 & 0\\
-B & 0 & 0 & 0
\end{pmatrix}\\
(\Gamma^{3})^{\alpha\beta} &= \begin{pmatrix}
0 & 0 & 0 & -B\\
0 & 0 & B & 0\\
0 & B & 0 & 0\\
-B & 0 & 0 & 0
\end{pmatrix}\\
(\Gamma^{i})^{\alpha\beta} &= \begin{pmatrix}
0 & iBA & 0 & 0\\
-iBA & 0 & 0 & 0\\
0 & 0 & 0 & -iBA\\
0 & 0 & iBA & 0
\end{pmatrix}
\end{aligned}
\end{equation}
where $A$ is an $8\times 8$ matrix with non-vanishing elements $A_{0j}=A_{j0}=1$. All these matrices are symmetric and satisfy the desired property: $\Gamma^{m}_{\alpha\beta}\Gamma^{n\,\beta\lambda} + \Gamma^{n}_{\alpha\beta}\Gamma^{m\,\beta\lambda} = 2\eta^{mn}\delta^{\lambda}_{\alpha} $. 

\vspace{2mm}
Finally, the product of two spinors $\chi^{\alpha}\rho_{\alpha}$ will be defined as follows:
\begin{eqnarray*}
\chi^{\alpha}C_{\alpha\beta}\rho^{\beta} &=& -\chi^{++0}\rho^{--0} + \chi^{++i}\rho^{--i} + \chi^{--0}\rho^{++0} - \chi^{--i}\rho^{++i} \\&&- \chi^{+-0}\rho^{-+0} + \chi^{+-i}\rho^{-+i} + \chi^{-+0}\rho^{+-0} - \chi^{-+i}\rho^{+-i}
\end{eqnarray*}
\section{Octonions and $SO(7)$ rotations}\label{appC}
In this Appendix we will show that any component of $C_{BCD}$ can be obtained from $C_{(+0)(+0)(+0)}$ by $SO(9)$ rotations. These rotations are defined by the operator 
\begin{equation}
R^{IJ} = \frac{1}{\sqrt{8\sqrt{2}P^{+}}}\hat{\bar{D}}\Gamma^{IJ}\hat{\bar{D}},
\end{equation}
which satisfy the algebra
\begin{equation}
[R^{IJ}, R^{KL}] = \eta^{IK}R^{JL} - \eta^{JK}R^{IL} - \eta^{IL}R^{JK} + \eta^{JL}R^{IK}
\end{equation}
Therefore, we can use this operator to rotate the \emph{ground state} $C_{(+0)(+0)(+0)}$. To do this let us first write the transformation rule for a general $C_{BCD}$ being acted on by $R^{IJ}$:
\begin{equation}
R^{IJ}C_{BCD} = \frac{1}{\sqrt{2}}(\Gamma^{IJ})_{B}^{\hspace{2mm}E}C_{ECD} + \frac{1}{\sqrt{2}}(\Gamma^{IJ})_{C}^{\hspace{2mm}E}C_{BED} + \frac{1}{\sqrt{2}}(\Gamma^{IJ})_{D}^{\hspace{2mm}E}C_{BCE}
\end{equation}
As explained above, only $\hat{\bar{D}}_{-i}$ and $\hat{\bar{D}}_{+0}$ will act non-trivially on $C_{(+0)(+0)(+0)}$. Thus, we have
\begin{equation}\label{eq600}
(\Gamma^{ij})_{(+k)(-0)}\hat{\bar{D}}_{-k}\hat{\bar{D}}_{+0}C_{(+0)(+0)(+0)} \propto (\Gamma^{ij})_{(+0)}^{\hspace{6mm}E}C_{E(+0)(+0)}
\end{equation}
To solve this equation we recall the notion of octonions \cite{Baez:2001dm}. 

\vspace{2mm}
The octonion mutiplication table can be written in the form
\begin{equation}
e_{i}e_{j} = -\delta_{ij} + \epsilon_{ijk}e_{k}
\end{equation}
which is equivalent to
\begin{equation}
e_{i}e_{j} = \delta_{ij} - i\epsilon_{ijk}e_{k}
\end{equation}
where $\epsilon_{ijk}$ is a totally antisymmetric tensor with value +1 when $(ijk) =$ $(123)$, $(145)$, $(176)$, $(246)$, $(257)$, $(347)$, $(365)$. Now we can identify these octonions as the gamma matrices of the $SO(7)$ Clifford algebra:
\begin{equation}
\Gamma^{i}\Gamma^{j} = \delta^{ij} - i\epsilon^{ijk}\Gamma^{k}
\end{equation}
This equation can be thought of as the 7-dimensional generalization of the 3-dimensional case
\begin{equation}
\tau^{i}\tau^{j} = \delta^{ij} + i e^{ijk}\tau^{k},
\end{equation}
where $\tau^{i}$ are the ordinary Pauli matrices.

\vspace{2mm}
Coming back to eqn. \eqref{eq600} and applying the octonion identity we obtain
\begin{eqnarray}
(\Gamma^{ij})_{(+k)(-0)}\hat{\bar{D}}_{-k}\hat{\bar{D}}_{+0}C_{(+0)(+0)(+0)} &\propto & (\Gamma^{ij})_{(+0)}^{\hspace{6mm}E}C_{E(+0)(+0)}\nonumber\\
\epsilon^{ijk}\hat{\bar{D}}_{-k}\hat{\bar{D}}_{+0}C_{(+0)(+0)(+0)} &\propto & \epsilon^{ijk}C_{(+k)(+0)(+0)}
\end{eqnarray}
Therefore, we have obtained the state $C_{(+i)(+0)(+0)}$. By acting with $R^{-k}$ on $C_{(+0)(+0)(+0)}$ we obtain the state $C_{(-i)(+0)(+0)}$:
\begin{eqnarray}
(\Gamma^{-k})_{(+i)(+j)}\hat{\bar{D}}_{-i}\hat{\bar{D}}_{-j}C_{(+0)(+0)(+0)} &\propto & (\Gamma^{-k})_{(+0)}^{\hspace{6mm}E}C_{E(+0)(+0)}\nonumber\\
\epsilon^{kij}\hat{\bar{D}}_{-i}\hat{\bar{D}}_{-j}C_{(+0)(+0)(+0)} &\propto & \delta^{kl}C_{(-l)(+0)(+0)}
\end{eqnarray}

In this way, one can obtain all states contained in $C_{ABC}$. The table below shows explicitly how this is done. For brevity, we include only one way to obtain each state. The dash ($-$) means that all states corresponding to an initial state have been already obtained from other initial states. Finally, since $C_{ABC}$ is completely symmetric, states related by symmetry to states on the table need not be included.

\begin{table}
\caption{States produced by the rotation operator $R^{IJ}$}
\begin{center}
    \begin{tabular}{  | l | l | p{8cm} |}
    \hline
    Initial state & States produced by $R^{ij}$ & States produced by $R^{-k}$  \\ \hline
    $C_{(+0)(+0)(+0)}$ & $C_{(+k)(+0)(+0)}$ & $C_{(-k)(+0)(+0)}$  
     \\ \hline
    $C_{(+k)(+0)(+0)}$ &  $C_{(+k)(+l)(+0)}$ & $C_{(-0)(+0)(+0)}$, $C_{(+l)(-j)(+0)}$  \\ \hline
    $C_{(+k)(+l)(+0)}$ &  
         $C_{(+k)(+l)(+r)}$ & 
          $C_{(+k)(-0)(+0)}$, $C_{(+k)(+l)(-r)}$ \\
    \hline
    $C_{(-0)(+0)(+0)}$ &  
    - & $C_{(-0)(-k)(+0)}$ \\
    \hline
    $C_{(+l)(-k)(+0)}$ &  $C_{(-j)(-r)(+0)}$ & $C_{(+l)(-r)(-k)}$  \\
    \hline
    $C_{(+l)(+k)(+r)}$ &  - & $C_{(-0)(+l)(+r)}$  \\
    \hline
    $C_{(+k)(-0)(+0)}$ &  - & $C_{(-0)(-0)(+0)}$,  $C_{(+k)(-0)(-r)}$ \\
    \hline
    $C_{(-0)(-k)(+0)}$ &  - & $C_{(-0)(-k)(-r)}$ \\
    \hline
    $C_{(-k)(-r)(+0)}$ &  - & $C_{(-k)(-r)(-t)}$ \\
    \hline
    $C_{(-0)(+l)(+r)}$ &  - & $C_{(-0)(-0)(+r)}$\\
    \hline 
    $C_{(-0)(-0)(+0)}$ &  - & $C_{(-0)(-0)(-r)}$\\
    \hline 
    $C_{(-0)(-0)(+r)}$ &  - & $C_{(-0)(-0)(-0)}$\\
    \hline 
    
    \end{tabular}
    
\end{center}
\end{table}

\chapter{}\label{appendix2}
\section{Review of the $D=10$ Brink-Schwarz superparticle}
The ten-dimensional Brink-Schwarz superparticle action is given by
\begin{equation}\label{newappendix1}
S = \int d\tau(P_{m}\Pi^{m} + e P^{m}P_{m})
\end{equation}
Throught all this Appendix we will use lowercase letters from the middle of the Greek/Latin alphabet to denote $SO(9,1)$ spinor/vector indices. In \eqref{newappendix1}, $\Pi^{m} = \partial_{\tau}X^{m} - i\partial_{\tau}\theta^{\mu}(\gamma^{m})_{\mu\nu}\theta^{\nu}$, $\theta^{\mu}$ is an $SO(9,1)$ Majorana-Weyl spinor, $P^{m}$ is the conjugate momentum associated to $X^{m}$ and $e$ is the Lagrange multiplier enforcing the massless constraint. The matrices $(\gamma^{m})_{\mu\nu}$, $(\gamma^{m})^{\mu\nu}$ are the $SO(9,1)$ Pauli matrices satisfying $(\gamma^{m})_{\mu\nu}(\gamma^{n})^{\nu\sigma} + (\gamma^{n})_{\mu\nu}(\gamma^{m})^{\nu\sigma} = 2\eta^{mn}\delta^{\sigma}_{\mu}$. In addition, these matrices satisfy the remarkable identity: $(\gamma^{m})_{\mu(\nu}(\gamma_{m})_{\sigma\rho)} = 0$. 

\vspace{2mm}
The action \eqref{newappendix1} is easily shown to be invariant under worldline diffeormorphisms. Furthermore, this action is also invariant under the following transformations \cite{SIEGEL1983397}
\begin{eqnarray*}
\mbox{SUSY transformations} &\rightarrow& \delta\theta^{\mu} = \epsilon^{\mu} \hspace{2mm} \mbox{,} \hspace{2mm} \delta X^{m} = i\theta^{\mu}(\gamma^{m})_{\mu\nu}\epsilon^{\nu} \hspace{2mm} \mbox{,} \hspace{2mm} \delta P_{m} = \delta e = 0\\
\mbox{$\kappa$ (local) transformations} &\rightarrow& \delta\theta^{\mu} = P^{m}(\gamma_{m})^{\mu\nu}\kappa_{\nu} \hspace{1mm},\hspace{1mm} \delta X^{m} = -i\theta^{\mu}(\gamma^{m})_{\mu\nu}\delta\theta^{\nu} \hspace{1mm},\hspace{1mm}\nonumber\\
&& \delta P_{m} = 0 \hspace{1mm},\hspace{1mm} \delta e = 2i(\partial_{\tau}\theta^{\mu})\kappa_{\mu}
\end{eqnarray*}

\vspace{2mm}
One can easily calculate the conjugate momentum associated to $\theta^{\mu}$:
\begin{equation}
p_{\mu} = -i(\gamma^{m})_{\mu\nu}\theta^{\nu}P_{m} 
\end{equation}
to realize that this model is a system with constraints, which are given by
\begin{equation}
d_{\mu} = p_{\mu} + i(\gamma^{m})_{\mu\nu}\theta^{\nu}P_{m}
\end{equation}
and satisfy the algebra
\begin{eqnarray}
\{d_{\mu}, d_{\nu}\} &=& -2(\gamma^{m})_{\mu\nu}P_{m} \label{eq32}
\end{eqnarray}
where $\{ , \}$ stands for a Poisson bracket. Since $P^{2} = 0$, there will be $8$ first-class constraints and $8$ second-class constraints. The $8$ first-class constraints are readily shown to be generated by $k^{\mu} = -iP^{m}(\gamma_{m})^{\mu\nu}d_{\nu}$. However, there is no simple way to covariantly write the $8$ second-class constraints.

\vspace{2mm}
As long as the physical spectrum is our main concern, one can quantize this model by choosing the so-called semi-light-cone gauge. This gauge is realized after imposing the relation $(\gamma^{+}\theta)_{\mu} = 0$, which can be always satisfied by using a suitable $\kappa$-transformation. Thus, the gauge-fixed Brink-Schwarz superparticle action takes the form
\begin{eqnarray}
S &=& \int d\tau\,(P_{m}\partial_{\tau}X^{m} + \frac{i}{2}S^{a}\partial_{\tau}S^{a} + e P^{m}P_{m})
\end{eqnarray}
where $a$ is an $SO(8)$ chiral spinor index, and $S^{a}$ is proportional to the $SO(8)$ chiral component of $\theta^{\mu}$.

\vspace{2mm}
One can now use the standard Dirac\'s procedure to quantize this system. The conjugate momentum associated to $S_{a}$ is found to be: $p_{a} = -\frac{i}{2}S_{a}$. Hence, the gauge-fixed theory also possesses constraints and they read
\begin{eqnarray}
\tilde{d}_{a} = p_{a} + \frac{i}{2}S_{a}
\end{eqnarray}
which satisfy the algebra
\begin{eqnarray}
\{\tilde{d}_{a}, \tilde{d}_{b}\} &=& \delta_{ab}
\end{eqnarray}
Using the constraint matrix $C_{ab} = \delta_{ab}$, one then learns that
\begin{eqnarray}
\{S_{a}, S_{b}\}_{D} &=& \{S_{a}, S_{b}\}_{P} - \sum_{e,f}\{S_{a}, \tilde{d}_{e}\}_{P}(C^{-1})^{ef}\{\tilde{d}_{f}, S_{b}\}_{P}\nonumber\\
\{S_{a}, S_{b}\}_{D} &=& \delta_{ab}\label{newappendix2}
\end{eqnarray}
where $\{\,\cdot\,\}_{D}$ means Dirac brackets. Therefore, the Hilbert space of the ten-dimensional Brink-Schwarz superparticle is described by an $SO(8)$ vector and an $SO(8)$ antichiral spinor, which realize the algebra \eqref{newappendix2} through the relations
\begin{eqnarray}
S^{a}\psi_{\hat{j}}(x) &=& \frac{1}{\sqrt{2}}\sigma^{a\dot{b}}_{\hat{j}}\psi_{\dot{b}}(x)\\
S_{a}\psi_{\dot{b}}(x) &=& \frac{1}{\sqrt{2}}\sigma^{\hat{j}}_{a\dot{b}}\psi_{\hat{j}}(x)
\end{eqnarray}
There are exactly the gluon and gluino physical degrees of freedom of $N=1$ $D=10$ super Yang-Mills.

\section{$D=10$ pure spinor superparticle}
This subsection is a brief review of \cite{Berkovits:2002zk,Berkovits:2004tw}. After introducing a new conjugate pair of variables $(\theta^{\mu}, p_{\nu})$ and a set of first-class constraints defined by
\begin{equation}
\hat{d}_{\mu} = d_{\mu} + \frac{1}{\sqrt{\sqrt{2}P^{+}}}(\gamma_{m}\gamma^{+}S)_{\mu}P^{m}
\end{equation}
where $d_{\mu} = p_{\mu} + i(\gamma^{m})_{\mu\nu}\theta^{\nu}P_{m}$ satisfy $\{d_{\mu}, d_{\nu}\} = -2(\gamma^{m})_{\mu\nu}P_{m}$,
the gauge-fixed Brink-Schwarz action reads
\begin{eqnarray}\label{appendixAAeq3}
S &=& \int d\tau (P_{m}\partial_{\tau}X^{m} + \frac{i}{2}S^{a}\partial_{\tau}S^{a} + e P^{m}P_{m} + p_{\mu}\partial_{\tau}\theta^{\mu} + f^{\mu}\hat{d}_{\mu})
\end{eqnarray}
Notice that the first-class constraints $\hat{d}_{\mu}$ generate a gauge symmetry which can be used to gauge away the new conjugate pair of variables. The algebra satisfied by $\hat{d}_{\mu}$ can be easily found to be
\begin{eqnarray}
\{\hat{d}_{\mu}, \hat{d}_{\nu}\} &=& -\frac{1}{P^{+}}P^{m}P_{m}(\gamma^{+})_{\mu\nu}
\end{eqnarray}
After fixing $e=-\frac{1}{2}$ and $f^{\mu} = 0$, the standard BRST method yields the action 
\begin{equation}
S = \int d\tau(P_{m}\partial_{\tau}X^{m} + \frac{i}{2}S^{a}\partial_{\tau}S^{a} + p_{\mu}\partial_{\tau}\theta^{\mu} + b\partial_{\tau}c + \hat{w}_{\mu}\partial_{\tau}\hat{\lambda}^{\mu} - \frac{1}{2}P^{m}P_{m})
\end{equation}
together with the BRST operator
\begin{eqnarray}
\hat{Q} = \hat{\lambda}^{\mu}\hat{d}_{\mu} + cP^{m}P_{m} + \frac{1}{2P^{+}}(\hat{\lambda}\gamma^{+}\hat{\lambda})b
\end{eqnarray}
One can then show the BRST cohomology of $\hat{Q}$ is equivalent to the BRST cohomology of $Q = \lambda^{\mu}d_{\mu}$ where $\lambda^{\mu}$ is a ten-dimensional pure spinor satisfying $\lambda\gamma^{m}\lambda = 0$. This proof can be made by using a two step argument, which can be illustrated as follows
\begin{eqnarray}\label{appendixAAeq1}
\mbox{$\hat{Q}$-cohomology} \xleftrightarrow[\text{equivalent to}]{} \mbox{$Q\ensuremath{'}$-cohomology} \xleftrightarrow[\text{equivalent to}]{} \mbox{$Q$-cohomology}
\end{eqnarray}
where $Q\ensuremath{'} = \lambda\ensuremath{'}^{\mu}\hat{d}_{\mu}$ and $\lambda\ensuremath{'}^{\mu}$ is a ten-dimensional bosonic spinor satisfying $\lambda\ensuremath{'}\gamma^{m}\lambda\ensuremath{'} = 0$.

\vspace{2mm}
The demonstration of \eqref{appendixAAeq1} can be easily understood from the following argument. If $V$ is a state annihilated by $Q\ensuremath{'}$, one then can always construct the state $\hat{V} =  V - 2P^{+}cW$, where $\hat{Q}V = (\hat{\lambda}\gamma^{+}\hat{\lambda})W + cP_m P^m V$, which is annihilated by $\hat{Q}$. Moreover, if a state $X$ is $Q\ensuremath{'}$-exact, one can always find the state $\hat{X} = X - 2P^{+}cW$, where $X = \hat{Q}\Omega - cP^2 \Omega + (\hat{\lambda}\gamma^{+}\hat{\lambda})Y$, for some $Y$, which is $\hat{Q}$-exact. One can readily reverse this argument to show that the first equivalence relation in \eqref{appendixAAe1} is indeed true. The last step is to prove that the BRST cohomology of the $Q\ensuremath{'}$ is equivalent to the BRST cohomology of $Q$. To show this one can use the pureness of the $SO(8)$ antichiral spinor $(\gamma^{+}\lambda\ensuremath{'})_{\dot{a}}$ to define an $U(4)$ subgroup which leaves $(\gamma^{+}\lambda\ensuremath{'})_{\dot{a}}$ invariant up to a scale factor. The fundamental representation of this $U(4)$ is defined by the relation $(\sigma^{\hat{j}})_{A \dot{a}}(\gamma^{+}\lambda\ensuremath{'})^{\dot{a}} = 0$ for all $\hat{j}=1,\ldots,8$, where $(\sigma^{\hat{j}})_{a\dot{a}}$ are the $SO(8)$ Pauli matrices. One can then write the $SO(8)$ chiral spinors $(\gamma^{-}\lambda\ensuremath{'})_{a}$, $(\gamma^{+}d)_{a}$, $S_{a}$ in terms of the fundamental and antifundamental representations of this $U(4)$ as folllows
\begin{eqnarray}
(\gamma^{-}\lambda\ensuremath{'})_{a} &\rightarrow & \left((\gamma^{-}\lambda\ensuremath{'})_{A}, (\gamma^{-}\lambda\ensuremath{'})_{\bar{A}}\right)\nonumber\\
(\gamma^{+}d)_{a} &\rightarrow & \left((\gamma^{+}d)_{A}, (\gamma^{+}d)_{\bar{A}}\right)\nonumber\\
S_{a} &\rightarrow & \left(S_{A}, S_{\bar{A}}\right)
\end{eqnarray}
where $A, \bar{A} = 1, \ldots , 4$. After making the shift
\begin{equation}
S_{A} \rightarrow S_{A} + i\frac{\sqrt{\sqrt{2}}}{2\sqrt{2P^{+}}}(\gamma^{+}d)_{A}
\end{equation}
and using the BCH formula, the BRST operator $Q\ensuremath{'}$ becomes
\begin{equation}\label{appendixAAeq2}
Q\ensuremath{'} \rightarrow \lambda\ensuremath{'}_{\dot{a}}d_{\dot{a}} + \lambda\ensuremath{'}_{A}d_{\bar{A}} + \sqrt{2\sqrt{2}P^{+}}\lambda\ensuremath{'}_{A}S_{\bar{A}}
\end{equation}
where $\lambda\ensuremath{'}^{\dot{a}}$ is an $SO(8)$ null spinor. If one defines a spinor $\lambda^{\mu} = [\lambda_{\dot{a}}, \lambda_{A}, \lambda_{\bar{A}}] = [\lambda\ensuremath{'}_{\dot{a}}, \lambda\ensuremath{'}_{A}, 0]$, eqn. \eqref{appendixAAeq2} becomes
\begin{equation}
Q\ensuremath{'} \rightarrow \lambda^{\mu}d_{\mu} + \sqrt{2\sqrt{2}P^{+}}\lambda\ensuremath{'}_{A}S_{\bar{A}}
\end{equation} 
The use of the quartet argument \cite{Kugo:1979gm} allow us to claim the Hilbert space will be independent of $\lambda\ensuremath{'}_{\bar{A}}$ and $S_{A}$, and its respective conjugate momenta $w\ensuremath{'}_{A}$ and $S_{\bar{A}}$. Therefore
\begin{equation}
Q\ensuremath{'} \rightarrow Q = \lambda^{\mu}d_{\mu}
\end{equation}
where $\lambda^{\mu}$ is a ten-dimensional pure spinor. In this way, the gauge-fixed Brink-Schwarz superparticle \eqref{appendixAAeq3} is physically equivalent to the pure spinor superparticle, whose action is given by
\begin{equation}
S = \int d\tau (P_{m}\partial_{\tau}X^{m} + p_{\mu}\dot\partial_{\tau}\theta^{\mu} + w_{\mu}\partial_{\tau}\lambda^{\mu} - \frac{1}{2}P^{m}P_{m})
\end{equation}
and the BRST operator reads $Q = \lambda^{\mu}d_{\mu}$, where $\lambda\gamma^{m}\lambda = 0$.

\section{Light-cone gauge equations of motion}
To see how the light-cone gauge description of $N=1$ $D=10$ super Yang-Mills emerges from the pure spinor framework, one should first perform similarity transformations on the pure spinor BRST operator. The first transformation is generated by the operator
\begin{eqnarray}
R &=& i\frac{P_{\hat{i}}\bar{N}^{\hat{i}}}{P^{+}}
\end{eqnarray}
where $\bar{N}^{\hat{i}} = -\frac{1}{\sqrt{2}}(\lambda^{a}(\sigma^{\hat{i}})_{a\dot{a}}\bar{w}^{\dot{a}})$. We are using $\hat{i}$ to denote $SO(8)$ vector indices, and $a$, $\dot{a}$ to denote $SO(8)$ Weyl spinor indices. As a result, the transformed BRST charge takes the form
\begin{eqnarray}
\hat{Q} &=& \lambda^{a}G_{a} + \lambda^{\dot{a}}\bar{d}_{\dot{a}}
\end{eqnarray}
where $G_{a} = \frac{P^{m}}{2P^{+}}(\gamma^{+}\gamma_{m})_{a}P_{m}$ satisfies
\begin{eqnarray}
\{G_{a}, \bar{d}_{\dot{a}}\} = 0\hspace{2mm},\hspace{2mm}\{G_{a}, G_{b}\} = -\frac{\eta_{ab}}{\sqrt{2}}\left(\frac{P^{2}}{P^{+}}\right)
\end{eqnarray}

\vspace{2mm}
The second similarity transformation is generated by the operator
\begin{eqnarray}
\hat{R} &=& \frac{i}{\sqrt{2}}(\theta^{a}(\sigma^{i})_{a\dot{a}}\bar{p}^{\dot{a}})\frac{P_{i}}{P^{+}}
\end{eqnarray}
which takes the BRST operator $\hat{Q}$ to the form
\begin{eqnarray}
\hat{\hat{Q}} &=& \lambda^{a}\left(p_{a} + \frac{i}{2\sqrt{2}}\frac{P^{2}}{P^{+}}\theta_{a}\right) + \bar{\lambda}^{\dot{a}}\left(\bar{p}_{\dot{a}} - \frac{i}{\sqrt{2}}P^{+}\bar{\theta}_{\dot{a}}\right)
\end{eqnarray}

\vspace{2mm}
One can now analyze the ghost number one sector of the BRST cohomology of $\hat{\hat{Q}}$. If one write the most general ghost number one vertex in the form
\begin{eqnarray}
\hat{U}^{(1)} &=& \lambda^{a}\hat{A}_{a} + \bar{\lambda}^{\dot{a}}\hat{\bar{A}}_{\dot{a}}
\end{eqnarray}
one can show that the $SO(8)$ component $\hat{\bar{A}}$ is pure gauge and thus one can set $\hat{\bar{A}}_{\dot{a}}=0$. In this gauge, one then finds that
\begin{eqnarray}
\hat{\bar{D}}_{\dot{a}}\hat{A}_{b} &=& (\sigma^{\hat{i}})_{b\dot{a}}\hat{A}_{i}
\end{eqnarray}
which is the constraint describing $N=1$ $D=10$ super Yang-Mills in light-cone gauge. The non-dependence on $\theta^{a}$ of $A_{b}$ and the massless constraint comes from the remaining component of the BRST cohomology condition on $\hat{U}^{(1)}$. 

\vspace{2mm}
Finally, one can obtain the ghost number one vertex operator for the BRST operator $Q = \lambda^{\mu}d_{\mu}$ by performing a similarity transformation on $\hat{U}^{(1)}$ generated by the operator $-(R+\hat{R})$.

\chapter{}\label{appendix3}
\section{$D=10$ ghost number one vertex operator}
The ghost number one vertex operator in the ten-dimensional pure spinor worldline framework can be constructed as a perturbation of the pure spinor BRST operator after coupling it to a super-Yang-Mills background. To see this, let us define the BRST operator as
\begin{eqnarray}\label{neqded1}
Q &=& \lambda^{\mu}\nabla_{\mu}
\end{eqnarray}
where $\nabla_{\mu} = D_{\mu} + A_{\mu}$, and $A_{\mu}$ is the fermionic component of the super-gauge-field of $D=10$ super-Yang-Mills. This operator is nilpotent as a consequence of the structure equations of $D=10$ super-Yang-Mills (see Appendix \ref{appendix4}). After expanding eqn. \eqref{neqded1} and converting it into a wordline vector with ghost number one, one learns that
\begin{eqnarray}
Q &=& Q_{0} + \lambda^{\mu}A_{\mu}
\end{eqnarray}
where $Q_{0} = \lambda^{\mu}d_{\mu}$ is the standard pure spinor BRST operator. Therefore, the ghost number one vertex operator will be defined to be
\begin{eqnarray}
U^{(1)} &=& \lambda^{\mu}A_{\mu}
\end{eqnarray}
This vertex is BRST-closed by construction. As a check, one can impose that $U^{(1)}$ belongs to the BRST cohomology of $Q_{0}$. This requirement gives rises to the conditions
\begin{eqnarray}
D_{\mu}A_{\nu} + D_{\nu}A_{\mu} = (\gamma^{m})_{\mu\nu}A_{m} \hspace{2mm}&,&\hspace{2mm} \delta A_{\mu} = D_{\mu}\Lambda
\end{eqnarray}
for some arbitrary superfield $\Lambda$. These are exactly the equations of motion of $D=10$ super-Maxwell on superspace.

\section{$D=10$ ghost number zero vertex operator}
In \cite{Bjornsson:2010wu,Bjornsson:2010wm}, a pure spinor worldline prescription was given for computing field theory correlation functions. This prescription makes use of ghost number one and zero vertex operators. In order for it to be well-defined, the ghost number zero vertex and $U^{(1)}$ must satisfy a standard descent equation. We saw in section 4 that it is not possible to write a ghost number zero vertex with this property in the eleven-dimensional minimal formalism. However, as a ramarkable fact, a ghost number zero vertex satisfying this requirement can be shown to exist in the ten-dimensional case.

\vspace{2mm}
Concretely, the ghost number zero vertex operator given by
\begin{eqnarray}
V^{(0)} &=& P^{m}A_{m}(x,\theta) + d_{\mu}\chi^{\mu}(x,\theta) + \frac{1}{2}N^{mn}F_{mn}(x,\theta)
\end{eqnarray}
where $N^{mn} = \frac{1}{2}(\lambda\gamma^{mn}w)$ and $A_{m}(x,\theta)$, $\chi^{\mu}(x,\theta)$, $F_{mn}(x,\theta)$ are the superfields describing $N=1$ $D=10$ super Yang-Mills on superspace, whose $\theta$-lowest components are exactly the gluon, gluino and gluon field strength polarizations, respectively \cite{Berkovits:2002zk,Berkovits:2001rb,Mafra:2009wq}; can be shown to obey the relation
\begin{eqnarray}\label{newappendixCeq1}
\{Q_{0}, V^{(0)}\} &=& [H, U^{(1)}]
\end{eqnarray}
where $H = P^{2}$. 

\chapter{}\label{appendix4}
\section{Review of superspace formulation of $D=10$ super-Yang-Mills}
In this appendix we will briefly review the superspace formulation of $D=10$ super-Yang-Mills. We refer to \cite{Siegel:1978yi,Witten:1985nt,HOWE1991141,Mafra:2009wq} for more details. We will use letters from middle of the Greek/Latin alphabet do denote $SO(1,9)$ spinor/vector indices. We will also let capital letters from the middle of the Latin alphabet stand for superspace indices. The covariant derivatives on superspace are then defined to be
\begin{eqnarray}
\nabla_{\mu} = D_{\mu} + A_{\mu} \hspace{2mm}&,&\hspace{2mm} \nabla_{m} = \partial_{m} + A_{m}
\end{eqnarray}
where $D_{\alpha} = \partial_{\alpha} + \frac{1}{2}(\theta\gamma^{m})_{\alpha}\partial_{m}$, and the super-gauge-connection is $A_{M} = (A_{\mu}, A_{m})$. The super-field-strength components are given by
\begin{eqnarray}
F_{\mu\nu} = \{\nabla_{\mu}, \nabla_{\nu}\} - (\gamma_{m})_{\mu\nu}\nabla_{m}\hspace{2mm},\hspace{2mm} F_{\mu m} = [\nabla_{\mu}, \nabla_{m}] \hspace{2mm},\hspace{2mm} F_{mn} = [\nabla_{m},\nabla_{n}]
\end{eqnarray}
Explicitly,
\begin{eqnarray}
F_{\mu\nu} &=& D_{\mu}A_{\nu} + D_{\nu}A_{\mu} + \{A_{\mu},A_{\nu}\} - (\gamma^{m})_{\mu\nu}A_{m}\\
F_{\mu m} &=& D_{\mu}A_{m} - \partial_{m}A_{\mu} + [A_{\mu}, A_{m}]\\
F_{mn} &=& \partial_{m}A_{n} - \partial_{n}A_{m} + [A_{m}, A_{n}]
\end{eqnarray}
These expressions are invariant under the gauge transformations
\begin{eqnarray}
\delta A_{\mu} = D_{\mu}\Lambda \hspace{2mm}&,&\hspace{2mm}\delta A_{m} = \partial_{m}\Lambda
\end{eqnarray}
for some arbitrary parameter $\Lambda$. In order to describe $D=10$ super-Yang-Mills one needs to impose constraints on the super-field-strength $F_{MN}$. It turns out that 
\begin{eqnarray}\label{ymeq1}
(\gamma^{mnpqr})^{\mu\nu}F_{\mu\nu} &=& 0
\end{eqnarray}
is enough for our purposes. Since it is always possible to redefine $A_{M}$ in such a way that $(\gamma^{m})^{\mu\nu}F_{\mu\nu} = 0$, eqn. \eqref{ymeq1} actually implies
\begin{eqnarray}\label{ymeq7}
F_{\mu\nu} &=& 0
\end{eqnarray}
Therefore, one finds that
\begin{eqnarray}
D_{\mu}A_{\nu} + D_{\nu}A_{\mu} + \{A_{\mu}, A_{\nu}\} &=& (\gamma^{m})_{\mu\nu}A_{m}
\end{eqnarray}
One can now use the super-Jacobi identity
\begin{eqnarray}
[\nabla_{[M},[\nabla_{N},\nabla_{P\}}\}\} &=& 0
\end{eqnarray}
to deduce the remaining equations of motion.
\begin{eqnarray}
(\mu\nu\rho) &:& (\gamma^{m})_{\mu\nu}F_{\rho m} +(\gamma^{m})_{\nu\rho}F_{\mu m} + (\gamma^{m})_{\rho\mu}F_{\nu m} = 0\nonumber\\
&&\rightarrow  F_{\mu m} = (\gamma_{m})_{\mu\nu}\chi^{\nu}\label{ymeq4}\\
(\mu\nu m) &:& (\gamma_{m})_{\mu\sigma}\nabla_{\nu}\chi^{\sigma} + (\gamma_{m})_{\nu\sigma}\nabla_{\mu}\chi^{\sigma} = (\gamma^{n})_{\mu\nu}F_{nm}\label{ymeq2}\\
(\mu mn) &:& (\gamma_{m})_{\mu\nu}\nabla_{n}\chi^{\nu} - (\gamma_{n})_{\mu\nu}\nabla_{m}\chi^{\nu} = \nabla_{\mu}F_{mn}\label{ymeq3}
\end{eqnarray}
From eqn. \eqref{ymeq4} one concludes that
\begin{eqnarray}
D_{\mu}A_{m} - \partial_{m}A_{\mu} &=& (\gamma_{m})_{\mu\nu}{}\chi^{\nu}
\end{eqnarray}
In addition, after multiplying by $(\gamma^{m})^{\mu\rho}(\gamma_{p})_{\tau\rho}(\gamma^{p})^{\nu\xi}$ on both sides of eqn. \eqref{ymeq2} and using that $\nabla_{\mu}\chi^{\mu} = 0$ one learns that
\begin{eqnarray}\label{ymeq5}
\nabla_{\mu}\chi^{\nu} &=& -\frac{1}{4}(\gamma^{mn})_{\mu}{}^{\nu}F_{mn}
\end{eqnarray}
The use of eqn. \eqref{ymeq5} allows us to write
\begin{eqnarray}
\{\nabla_{\mu},\nabla_{\nu}\}\chi^{\sigma} &=& -\frac{1}{4}(\gamma^{mn})_{\mu}{}^{\sigma}\nabla_{\nu}F_{mn} -\frac{1}{4}(\gamma^{mn})_{\nu}{}^{\sigma}\nabla_{\mu}F_{mn}
\end{eqnarray}
which after using eqns. \eqref{ymeq7}, \eqref{ymeq3} and multiplying by $\delta^{\mu}_{\sigma}$ yields
\begin{eqnarray}\label{ymeq8}
(\gamma^{m})_{\mu\nu}\partial_{m}\chi^{\nu} &=& 0
\end{eqnarray}
which is the equation of motion for the gluino superfield. Finally, one can deduce the equation of motion for the gluon super-field-strength by multiplying both sides of eqn. \eqref{ymeq8} by $(\gamma^{n})^{\mu\rho}\nabla_{\rho}$ and using eqns. \eqref{ymeq4}, \eqref{ymeq2} to obtain
\begin{eqnarray}
\nabla^{m}F_{mn} &=& -\frac{1}{2}(\gamma_{n})_{\mu\nu}\{\chi^{\mu},\chi^{\nu}\}
\end{eqnarray}

\chapter{}\label{appendixB}
\section{Review of superspace formulation of $D=11$ supergravity}
In this Appendix we will review the original superspace formulation of $D=11$ supergravity given in \cite{Brink:1980az}. This turns out to be useful to fix conventions and get consistently the equations of motion for the dynamical superfields, which in turn play a crucial role when constructing the ghost number one and zero vertex operators of sections \ref{sec3} and \ref{sec4}.  Let us start by fixing notation. Latin capital letters from the beginning/middle of the alphabet will be used to represent tangent/coordinate superspace indices. The vielbein and spin-connection will be defined to be 1-forms on superspace as follows
\begin{eqnarray}
E^{A} = d Z^{M}E_{M}{}^{A} \hspace{2mm}&,&\hspace{2mm} \Omega_{A}{}^{B} = dZ^{M}\Omega_{MA}{}^{B}
\end{eqnarray}
where $d Z^{M} = (dX^{m}, d\theta^{\mu})$. The existence of $\Omega_{A}{}^{B}$ allows one to introduce a super covariant derivative which will act on an arbitrary tensor $\mathcal{F}_{A_{1}\ldots A_{m}}{}^{B_{1}\ldots B_{n}}$ in the form
\begin{eqnarray}\label{section3eq12}
\mathcal{D}\mathcal{F}_{A_{1}\ldots A_{m}}{}^{B_{1}\ldots B_{n}} &=& d \mathcal{F}_{A_{1}\ldots A_{m}}{}^{B_{1}\ldots B_{n}} - \Omega_{A_{1}}{}^{C}\mathcal{F}_{C\ldots A_{m}}{}^{B_{1}\ldots B_{n}} - \ldots\nonumber\\
&& + \mathcal{F}_{A_{1}\ldots A_{m}}{}^{C\ldots B_{n}}\Omega_{C}{}^{B_{1}} + \ldots
\end{eqnarray}
where $d$ is the standard exterior derivative. Next one introduces the 2-form supertorsion as the covariant derivative of the 1-form supervielbein
\begin{eqnarray}\label{eq9}
T^{A} &=& \frac{1}{2}E^{A}E^{B}T_{BA}{}^{C} = \mathcal{D}E^{A}\nonumber\\
&=& d E^{A} + E^{B}\Omega_{B}{}^{A}
\end{eqnarray}
and the 2-form supercurvature as the covariant derivative of the 1-form super spin-connection
\begin{eqnarray}\label{eq11}
R_{A}{}^{B} &=& \frac{1}{2}E^{C}E^{D}R_{DC,A}{}^{B} = \mathcal{D}\Omega_{A}{}^{B}\nonumber\\
&=& d\Omega_{A}{}^{B} + \Omega_{A}{}^{C}\Omega_{C}{}^{B}
\end{eqnarray}
As usual, we will constrain the super spin-connection components to satisfy
\begin{eqnarray}
\Omega_{\alpha\beta} &=& \frac{1}{4}(\Gamma^{mn})_{\alpha\beta}\Omega_{mn}
\end{eqnarray}
and all the other components to vanish. This choice automatically implies that
\begin{eqnarray}\label{eq21}
R_{DC,\alpha\beta} &=& \frac{1}{4}(\Gamma^{mn})_{\alpha\beta}R_{DC,mn}
\end{eqnarray}
Using \eqref{section3eq12}, \eqref{eq9}, \eqref{eq11} one easily finds the so-called Bianchi identities
\begin{eqnarray}
\mathcal{D}T^{A} = E^{B} R_{B}{}^{A} \hspace{2mm}&,&\hspace{2mm} \mathcal{D}R_{A}{}^{B} = 0
\end{eqnarray}
which in component notation read
\begin{eqnarray}
R_{[BD,C\}}{}^{A} - \nabla_{[B}T_{DC\}}{}^{A} - T_{[BD}{}^{F}T_{|F|C\}}{}^{A} &=& 0\label{eq15}\\
\nabla_{[F}R_{DC\},A}{}^{B} + T_{[FD}{}^{E}R_{|E|C\},A}{}^{B} &=& 0\label{eq16}
\end{eqnarray}
where $[\cdot,\cdot\}$ is a graded antisymmetrization. 

\vspace{2mm}
Furthermore, a 4-form superfield can also be introduced
\begin{eqnarray}
H &=& \frac{1}{4!}E^{D}E^{C}E^{B}E^{A}H_{ABCD}
\end{eqnarray}
which will be required to satisfy $dH = 0$. This condition gives rise to a new identity, which in component notation takes the form
\begin{eqnarray}\label{eq17}
\nabla_{[F}H_{ABCD\}} + 2T_{[FA}{}^{E}H_{|E|BCD\}} &=& 0
\end{eqnarray}

\vspace{2mm}
In order to put the theory on-shell we will impose the standard conventional and dynamical constraints, namely
\begin{eqnarray}\label{eq18}
H_{\alpha abc} = H_{\alpha\beta\delta a} = H_{\alpha\beta\delta\epsilon} = T_{ab}{}^{c} = T_{\alpha\beta}{}^{\delta} = T_{a\alpha}{}^{c} = 0\nonumber\\
T_{\alpha\beta}{}^{a} = (\Gamma^{a})_{\alpha\beta} \hspace{6mm},\hspace{6mm} H_{\alpha\beta ab} = (\Gamma_{ab})_{\alpha\beta}\hspace{10mm}
\end{eqnarray}
In this way, the only dynamical superfields of $D=11$ supergravity are $H_{abcd}$, $T_{a\alpha}{}^{\beta}$, $T_{ab}{}^{\alpha}$. To see how this works one should solve the identities \eqref{eq15}, \eqref{eq16}, \eqref{eq17} by plugging \eqref{eq18} into them. For instance, from eqn. \eqref{eq17} one gets
\begin{eqnarray}
(\alpha\beta\delta\gamma a)&:& \hspace{4mm} 3T_{(\alpha\beta}{}^{A}H_{|A|\delta\gamma)a} = 0\nonumber\\
&&\rightarrow 3 (\Gamma^{a})_{(\alpha\beta}(\Gamma_{ab})_{\delta\gamma)} = 0\\
(\alpha\beta cde)&:& \hspace{4mm}2[T_{\alpha\beta}{}^{A}H_{Acde} - 6T_{(\alpha [c}{}^{A}H_{|A|\beta) de]} + 3T_{[cd}{}^{A}H_{|A\alpha\beta| e]}] = 0\nonumber\\
&&\rightarrow (\Gamma^{a})_{\alpha\beta}H_{acde} - 6T_{(\alpha[c}{}^{\delta}(\Gamma_{de]})_{\beta)\delta} = 0\label{eq19}\\
(\alpha bcde) &:& \hspace{4mm}\nabla_{\alpha}H_{bcde} + 2(3 T_{[bc}{}^{\beta}H_{|\beta\alpha| de]}) = 0\nonumber\\
&& \rightarrow  \nabla_{\alpha}H_{bcde} + 6(\Gamma_{[de})_{\alpha\beta}T_{bc]}{}^{\beta} = 0\label{eq20}\\
(abcde)&:& \hspace{4mm}\nabla_{[a}H_{bcde]} = 0
\end{eqnarray}
The first equation is just a consistency check. The second equation \eqref{eq19} tells us that
\begin{eqnarray}
H_{abcd} &=& \frac{3}{16}(\Gamma^{a}\Gamma_{[de})^{\alpha}{}_{\delta}T_{|\alpha| c]}{}^{\delta}\nonumber\\
&=& \frac{3}{8}\eta_{a[d}(\Gamma_{e})^{\alpha}{}_{\delta}T_{|\alpha| c]}{}^{\delta} + \frac{3}{16}(\Gamma_{a[de})^{\alpha}{}_{\delta}T_{|\alpha|c]}{}^{\delta}
\end{eqnarray} 
which implies that $(\Gamma_{a})^{\alpha}{}_{\delta}T_{\alpha b}^{\delta} = 0$ and
\begin{eqnarray}
H_{abcd} &=& \frac{3}{16}(\Gamma_{a[de})^{\alpha}{}_{\delta}T_{|\alpha|c]}{}^{\delta}
\end{eqnarray}
This implies that $T_{\alpha a}{}^{\beta}$ can be written in terms of $H_{abcd}$. Using symmetry arguments one finds that
\begin{eqnarray}
T_{\alpha a}{}^{\delta} &=& c_{3}(\Gamma^{bcd})_{\alpha}{}^{\delta}H_{abcd} + c_{5}(\Gamma_{abcde})_{\alpha}{}^{\delta}H^{bcde}
\end{eqnarray}
The use of eqn. \eqref{eq19} tells us that
\begin{eqnarray}
\frac{1}{6}(\Gamma^{a})_{\alpha\beta}H_{acde} &=& c_{3}(\Gamma^{bfg}\Gamma_{[de})_{(\alpha\beta)}H_{c]bfg} + c_{5}(\Gamma_{[c|abfg|}\Gamma_{de]})_{(\alpha\beta)}H^{abfg}\nonumber\\
&=& -6c_{3}(\Gamma^{g})_{\alpha\beta}H_{cdeg} + c_{3}(\Gamma^{[de|bfg|})_{\alpha\beta}H_{c]bfg} - 8c_{5}(\Gamma^{[de|bfg|})_{\alpha\beta}H_{c]bfg}\nonumber\\
\end{eqnarray}
which leads us to conclude that $c_{3} = \frac{1}{36}$ and $c_{5} = \frac{c_{3}}{8} = \frac{1}{288}$. Thus one can write
\begin{eqnarray}\label{eq24}
T_{\alpha a}{}^{\delta} &=& \frac{1}{36}[(\Gamma^{bcd})_{\alpha}{}^{\delta}H_{abcd} + \frac{1}{8}(\Gamma_{abcde})_{\alpha}{}^{\delta}H^{bcde}]
\end{eqnarray}
Moreover, $H_{abcd}$ and $T_{ab}{}^{\alpha}$ are related to each other via eqn. \eqref{eq20}
\begin{eqnarray}
\nabla_{\alpha}H_{bcde} &=& -6(\Gamma_{[de})_{\alpha\beta}T_{bc]}{}^{\beta}
\end{eqnarray}
Next, one can use the Bianchi identity \eqref{eq15} together with eqn. \eqref{eq21} to find
\begin{eqnarray}
(\alpha\beta\delta)(\gamma)&:& \hspace{4mm} \frac{1}{4}(\Gamma^{ab})_{(\delta}{}^{\gamma}R_{\alpha\beta),ab} + (\Gamma^{a})_{(\alpha\beta}T_{\delta)a}{}^{\gamma} = 0\label{eq22}\\
(a\alpha\beta)(\gamma)&:&\hspace{4mm}(\Gamma^{bc})_{(\beta}{}^{\gamma}R_{|a|\alpha),bc} - 4\nabla_{(\alpha}T_{\beta)a}{}^{\gamma} - 2(\Gamma^{b})_{\alpha\beta}T_{ba}{}^{\gamma} = 0\label{eq26}\\
(\alpha\beta b)(c) &:& \hspace{4mm} R_{(\alpha\beta),b}{}^{c} + 2(\Gamma^{c})_{\gamma(\beta}T_{\alpha)b}{}^{\gamma} = 0\label{eq23}\\
(ab\alpha)(\beta)&:& \hspace{4mm}\frac{1}{4}(\Gamma^{cd})_{\alpha}{}^{\beta}R_{ab,cd} + 2\nabla_{[a}T_{|\alpha|b]}{}^{\beta} - \nabla_{\alpha}T_{ab}{}^{\beta} - 2T_{\alpha [a}{}^{\delta}T_{|\delta| b]}{}^{\beta} = 0\label{eq40}\\
(\alpha ab)(c)&:& \hspace{4mm} R_{\alpha [a,b]}{}^{c} - \frac{1}{2}(\Gamma^{c})_{\gamma\alpha}T_{ab}{}^{\gamma} = 0\label{eq27}\\
(abc)(\alpha)&:&\hspace{4mm} \nabla_{[a}T_{bc]}{}^{\alpha} + T_{[ab}{}^{\gamma}T_{|\gamma|c]}{}^{\alpha} = 0\\
(abc)(d)&:& \hspace{4mm} R_{[ab,c]}{}^{d} = 0
\end{eqnarray}
The eqns. \eqref{eq22}, \eqref{eq23} imply that 
\begin{eqnarray}\label{eq25}
\frac{1}{2}(\Gamma^{ab})_{(\delta}{}^{\gamma}(\Gamma_{a})_{|\epsilon|\beta}T_{\alpha)b}{}^{\epsilon} + (\Gamma^{a})_{(\alpha\beta}T_{\delta)a}{}^{\gamma} &=& 0
\end{eqnarray}
After replacing \eqref{eq24} in \eqref{eq25}, one gets
\begin{eqnarray}\label{eq255}
H_{cdef}[\frac{3}{2}(\Gamma^{cd})_{(\delta}{}^{\gamma}(\Gamma^{ef})_{\alpha\beta)} + \frac{1}{16}(\Gamma^{ab})_{(\delta}{}^{\gamma}(\Gamma_{abcdef})_{\alpha\beta)} - (\Gamma^{c})_{(\alpha\beta}(\Gamma^{def})_{\delta)}{}^{\gamma} - \frac{1}{8}(\Gamma^{a})_{(\alpha\beta}(\Gamma_{acdef})_{\delta)}{}^{\gamma}] = 0\nonumber\\
\end{eqnarray}
which is an identity as can be shown by multiplying on both sides of \eqref{eq255} by $(\Gamma^{a})^{\alpha}$, $(\Gamma^{ab})^{\alpha\beta}$, $(\Gamma^{abcde})^{\alpha\beta}$\footnote{The GAMMA package \cite{Gran:2001yh} turns out to be useful for this type of computations.}.

\vspace{2mm}
Using eqns. \eqref{eq26}, \eqref{eq27} one gets a set of constraints on $T_{ab}{}^{\alpha}$. Let us see how this works. The use of eqn. \eqref{eq27} allows us to write
\begin{eqnarray}
(\Gamma^{bc})_{\beta}{}^{\gamma}R_{a\alpha, bc} &=& - (\Gamma^{bc})_{\beta}{}^{\gamma}(\Gamma_{c})_{\alpha\delta}T_{ab}{}^{\delta} + \frac{1}{2}(\Gamma^{bc})_{\beta}{}^{\gamma}(\Gamma_{a})_{\alpha\delta}T_{bc}{}^{\delta}
\end{eqnarray}
Plugging this expression into eqn. \eqref{eq26} ones arrives at the relation
\begin{eqnarray}\label{eq256}
- (\Gamma^{bc})_{(\beta}{}^{\gamma}(\Gamma_{c})_{\alpha)\delta}T_{ab}{}^{\delta} + \frac{1}{2}(\Gamma^{bc})_{(\beta}{}^{\gamma}(\Gamma_{a})_{\alpha)\delta}T_{bc}{}^{\delta} - 4\nabla_{(\alpha}T_{\beta)a}{}^{\gamma} - 2(\Gamma^{b})_{\alpha\beta}T_{ba}{}^{\gamma} &=& 0\nonumber\\
\end{eqnarray}
Moreover, from eqns. \eqref{eq24}, \eqref{eq20} one finds
\begin{eqnarray}
\nabla_{(\alpha}T_{\beta)a}{}^{\gamma} &=& -\frac{1}{6}[(\Gamma^{bcd})_{(\beta}{}^{\gamma}(\Gamma_{[cd})_{\alpha)\delta}T_{ab]}{}^{\delta} + \frac{1}{8}(\Gamma_{abcde})_{(\beta}{}^{\gamma}(\Gamma_{de})_{\alpha)\delta}T_{bc}{}^{\delta}]
\end{eqnarray}
Thus eqn. \eqref{eq256} becomes
\begin{eqnarray}
- (\Gamma^{bc})_{(\beta}{}^{\gamma}(\Gamma_{c})_{\alpha)\delta}T_{ab}{}^{\delta} + \frac{1}{2}(\Gamma^{bc})_{(\beta}{}^{\gamma}(\Gamma_{a})_{\alpha)\delta}T_{bc}{}^{\delta} - 2(\Gamma^{b})_{\alpha\beta}T_{ba}{}^{\gamma}\nonumber\\
+ \frac{2}{3}[(\Gamma^{bcd})_{(\beta}{}^{\gamma}(\Gamma_{[cd})_{\alpha)\delta}T_{ab]}{}^{\delta} + \frac{1}{8}(\Gamma_{abcde})_{(\beta}{}^{\gamma}(\Gamma_{de})_{\alpha)\delta}T_{bc}{}^{\delta}] = 0
\end{eqnarray}
After multiplying on both sides by $(\Gamma^{a})^{\alpha\beta}$, $(\Gamma^{ab})^{\alpha\beta}$, $(\Gamma^{abcde})^{\alpha\beta}$ one arrives at
\begin{eqnarray}\label{eq90}
(\Gamma^{abc})_{\alpha\beta}T_{bc}{}^{\beta} = (\Gamma^{ab})_{\alpha\beta}T_{ab}{}^{\beta} = (\Gamma^{b})_{\alpha\beta}T_{ab}{}^{\beta} = 0
\end{eqnarray}
Using this result and eqn. \eqref{eq20} one learns that
\begin{eqnarray}\label{eq258}
(\Gamma^{cd}\Gamma_{[ab})_{\alpha\beta}T_{cd]}{}^{\beta} = -7 T_{ab\,\alpha} \hspace{2mm}&,& T_{ab}{}^{\alpha} = \frac{1}{42}(\Gamma^{cd})^{\alpha\beta}\nabla_{\beta}H_{abcd}
\end{eqnarray}
Plugging this back into eqn. \eqref{eq20}, one finds that
\begin{eqnarray}
\nabla_{\alpha}H_{abcd} &=& \frac{1}{7}(\Gamma_{[cd}\Gamma^{ef})_{\alpha}{}^{\delta}\nabla_{\delta}H_{ab]ef}
\end{eqnarray}

\vspace{2mm}
On the other hand, after multiplying by
$(\Gamma^{cd})_{\beta}{}^{\alpha}$ and $\eta^{bd}$ on both sides of eqn. \eqref{eq40} one obtains
\begin{eqnarray}\label{eq42}
R_{ac} &=& \frac{1}{8}\nabla_{\alpha}T_{ac}{}^{\alpha} - \frac{1}{8}(\Gamma^{cb})_{\beta}{}^{\alpha}T_{\alpha[a}{}^{\delta}T_{|\delta|b]}{}^{\beta}
\end{eqnarray}
The first term vanishes as a consequence of eqns. \eqref{eq23}, \eqref{eq24}. The second term in \eqref{eq42} takes the simple form
\begin{eqnarray}
(\Gamma^{cb})_{\beta}{}^{\alpha}T_{\alpha[a}{}^{\delta}T_{|\delta|b]}{}^{\beta} &=& \frac{2}{3}H_{adef}H_{c}{}^{def} - \frac{1}{18}\eta_{ac}H_{defg}H^{defg}
\end{eqnarray}
Thus, the graviton e.o.m is given by
\begin{eqnarray}
R_{ac} &=& -\frac{1}{12}H_{adef}H_{c}{}^{def} + \frac{1}{144}\eta_{ac}H_{defg}H^{defg}
\end{eqnarray}
Finally, one can obtain the e.o.m for the 4-form field strength by multiplying on both sides of eqn. \eqref{eq40} by $(\Gamma^{c})_{\beta}{}^{\alpha}$ and using \eqref{eq258} to get
\begin{eqnarray}
\frac{1}{42}(\Gamma^{c})_{\beta}{}^{\alpha}(\Gamma^{de})^{\beta\delta}\nabla_{\alpha}\nabla_{\delta}H_{abde} - \frac{1}{1296}\epsilon_{abcdefghijk}H^{defg}H^{hijk} &=& 0
\end{eqnarray}
So after antisymmetrizing in $(a,b,c)$ one concludes that
\begin{eqnarray}
\nabla^{d}H_{dabc} + \frac{1}{1192}\epsilon_{abcdefghijk}H^{defg}H^{hijk} &=& 0
\end{eqnarray}
where the identity $(\Gamma^{b})^{\beta\alpha}\nabla_{\alpha}H_{bcde} = - \frac{3}{7}(\Gamma_{[c})^{\beta}{}_{\delta}(\Gamma^{fg})^{\delta\gamma}\nabla_{\gamma}H_{de]fg}$ was used.

\chapter{}\label{appendix5}
\section{$D=11$ pure spinor identities}\label{apapB}
We list some pure spinor identities in eleven dimensions:
\begin{eqnarray}
(\bar{\Lambda}\Gamma^{ab}\bar{\Lambda})(\Gamma_{b}\bar{\Lambda})_{\alpha} &=& 0 \label{newapp1}\\
(\bar{\Lambda}\Gamma^{[ab}\bar{\Lambda})(\bar{\Lambda}\Gamma^{c]d}\bar{\Lambda}) &=& 0 \label{newapp2}\\
(\bar{\Lambda}\Gamma^{[ab}\bar{\Lambda})(\bar{\Lambda}\Gamma^{cd]}\bar{\Lambda}) &=& 0 \label{newapp3}\\ 
(\bar{\Lambda}\Gamma^{[ab}\bar{\Lambda})(\bar{\Lambda}\Gamma^{cd]}R) &=& 0 \label{newapp4}\\
(\bar{\Lambda}\Gamma_{ij}R)(\bar{\Lambda}\Gamma_{k}^{\hspace{2mm}j}R) &=& (\bar{\Lambda}\Gamma_{ik}R)(\bar{\Lambda}R) + \frac{1}{2}(\bar{\Lambda}\Gamma_{ik}\bar{\Lambda})(RR) \label{newapp12}\\
(\bar{\Lambda}\Gamma_{ab}R)(\bar{\Lambda}\Gamma_{cd}R)f^{ac}g^{bd} &=& 0 \label{newapp15}\\
(\Lambda\Gamma_{sk}\Lambda)(\Lambda\Gamma^{abcdk}\Lambda) &=& 0 \label{newapp11}\\
(\Gamma_{i}\Lambda)_{\alpha}(\Lambda\Gamma^{abcdi}\Lambda) &=& 6(\Gamma^{[ab}\Lambda)_{\alpha}(\Lambda\Gamma^{cd]}\Lambda) \label{newapp23}\\
(\Gamma_{ij}\Lambda)_{\alpha}(\Lambda\Gamma^{abcij}\Lambda) &=& -18(\Gamma^{[a}\Lambda)_{\alpha}(\Lambda\Gamma^{bc]}\Lambda)\label{newapp24}
\end{eqnarray}
where $f^{ac}$, $g^{bd}$ are antisymmetric in $(a,c)$, $(b,d)$ respectively. In addition, using \eqref{app4} it can be shown that
\begin{eqnarray}
L^{(1)}_{ab,cd}f^{abc} &=& (\bar{\Lambda}\Gamma_{ab}\bar{\Lambda})(\bar{\Lambda}\Gamma_{cd}R)f^{abc} \label{newapp5}\\
L^{(1)}_{ab,cd}f^{abc} &=& -(\bar{\Lambda}\Gamma_{cd}\bar{\Lambda})(\bar{\Lambda}\Gamma_{ab}R)f^{abc} \label{newapp6}\\
L^{(2)}_{ab,cd,ef}f^{abce} &=& (\bar{\Lambda}\Gamma_{ab}\bar{\Lambda})(\bar{\Lambda}\Gamma_{cd}R)(\bar{\Lambda}\Gamma_{ef}R)f^{abce} \label{newapp7}
\end{eqnarray}
where $f^{abc}, f^{abce}$ are antisymmetric in all of their indices.\\
Other useful identities:
\begin{eqnarray}
L^{(1)\hspace{2mm}d}_{ad,c} &=& (\bar{\Lambda}\Gamma_{ac}\bar{\Lambda})(\bar{\Lambda}R) \label{newapp10}\\
L^{(2)\hspace{5mm}a}_{ab,cd,e} &=& \frac{1}{3}[2(\bar{\Lambda}\Gamma_{eb}\bar{\Lambda})(\bar{\Lambda}\Gamma_{cd}R)(\bar{\Lambda}R) - (\bar{\Lambda}\Gamma_{cd}\bar{\Lambda})(\bar{\Lambda}\Gamma_{ab}R)(\bar{\Lambda}\Gamma_{e}^{\hspace{2mm}a}R)] \label{newapp8}\\
L^{(2)\hspace{5mm}c}_{ab,cd,e} &=& \frac{1}{3}[(\bar{\Lambda}\Gamma_{ab}\bar{\Lambda})(\bar{\Lambda}\Gamma_{cd}R)(\bar{\Lambda}\Gamma_{e}^{\hspace{2mm}c}R) - 2(\bar{\Lambda}\Gamma_{ed}\bar{\Lambda})(\bar{\Lambda}\Gamma_{ab}R)(\bar{\Lambda}R)] \label{newapp9}\\
(\bar{\Lambda}\Gamma_{ab}\bar{\Lambda})\bar{\Sigma}^{b} &=& 0 \label{newapp16}
\end{eqnarray}
Some useful commutation relations\footnote{It is worth mentioning that these identities were obtained by considering a canonical commutation relation between $\Lambda^{\alpha}$ and $W_{\beta}$. A fair computation must take into account the correct commutation relation between the pure spinor variables. However, this will not be a problem for us, since all of them will be used in expressions which are gauge invariant under the pure spinor constraint, and so all the non-canonical contributions will exactly cancel out.} 
\begin{align}
[\bar{\Sigma}^{i}, \eta] & = 0 \label{newapp17}\\
[\bar{\Sigma}^{j}, (\Lambda\Gamma^{mn}\Lambda)] & = \frac{2}{\eta^{2}}(\bar{\Lambda}\Gamma_{ef}\bar{\Lambda})(\bar{\Lambda}\Gamma_{gh}R)[(\Lambda\Gamma^{efgmj}\Lambda)(\Lambda\Gamma^{hn}\Lambda) - (\Lambda\Gamma^{efgnj}\Lambda)(\Lambda\Gamma^{hm}\Lambda)] \label{newapp18}\\
\{\bar{\Sigma}^{j}, (\bar{\Lambda}\Gamma_{mn}R)(\Lambda\Gamma^{mn}\Lambda)\} & = 0 \label{newapp25}\\
[Q, \eta] & = -2(\Lambda\Gamma^{mn}\Lambda)(\bar{\Lambda}\Gamma_{mn}R) \label{newapp19}\\
[Q, (\bar{\Lambda}\Gamma^{ab}\bar{\Lambda})] & = -2(\bar{\Lambda}\Gamma^{ab}R) \label{newapp20}\\
[Q, N^{hi}] & = (\Lambda\Gamma^{hi}D) \label{newapp21}\\
[Q, D_{\beta}] & = -2(\Gamma^{m}\Lambda)_{\beta}P_{m} \label{newapp22}\\
[N^{hi}, \eta] & = -2(\bar{\Lambda}\Gamma_{ab}\bar{\Lambda})[-2\eta^{ai}(\Lambda\Gamma^{bh}\Lambda) + 2\eta^{ah}(\Lambda\Gamma^{bi}\Lambda)] \notag\\
& = 4(\bar{\Lambda}\Gamma^{i}_{\hspace{2mm}b}\bar{\Lambda})(\Lambda\Gamma^{bh}\Lambda) - 4(\bar{\Lambda}\Gamma^{h}_{\hspace{2mm}b}\bar{\Lambda})(\Lambda\Gamma^{bi}\Lambda) \label{newapp13}\\
[N^{hi}, (\Lambda\Gamma^{lmnpq}\Lambda)] & = -2\eta^{iq}(\Lambda\Gamma^{chlmn}\Lambda) + 2\eta^{in}(\Lambda\Gamma^{chlmq}\Lambda) - 2\eta^{im}(\Lambda\Gamma^{chlnq}\Lambda)  \notag \\
& + 2\eta^{il}(\Lambda\Gamma^{chmnq}\Lambda) + 2\eta^{hq}(\Lambda\Gamma^{cilmn}\Lambda) - 2\eta^{hn}(\Lambda\Gamma^{cilmq}\Lambda) \notag \\
& + 2\eta^{hm}(\Lambda\Gamma^{cilnq}\Lambda) - 2\eta^{hl}(\Lambda\Gamma^{cimnq}\Lambda) + 2\eta^{ci}(\Lambda\Gamma^{hlmnq}\Lambda)\notag \\
& - 2\eta^{ch}(\Lambda\Gamma^{ilmnq}\Lambda) \label{newapp14}\\
[\Lambda\Gamma^{a}W, \Lambda\Gamma^{b}W] & = -2N^{ab} \label{newapp42}\\
[\Lambda\Gamma^{a}W, \Lambda\Gamma^{mns}W] & = -2\Lambda\Gamma^{amns}W\label{newapp43}\\
[\Lambda\Gamma^{abc}W, \Lambda\Gamma^{mnp}W] & = 4\delta^{bc}_{np}N^{am} - 4\delta^{bc}_{mp}N^{an} + 4\delta^{bc}_{mn}N^{ap} - 4\delta^{ac}_{np}N^{bm} + 4\delta^{ac}_{mp}N^{bn}\notag\\
& - 4\delta^{ac}_{mn}N^{bp} + 4\delta^{ab}_{np}N^{cm} - 4\delta^{ab}_{mp}N^{cn} + 4\delta^{ab}_{mn}N^{cp} - 2\Lambda\Gamma^{abcmnp}W\label{newapp41}
\end{align}

\section{$U(5)$-covariant analysis of the nilpotency of the $D=10$ $b$-ghost}\label{apI}
In this Appendix, we give a simple argument proving the nilpotecy of the $D=10$ $b$-ghost through the use of an $U(5)$-decomposition of pure spinor variables \cite{Berkovits:2002zk}. Let us start by showing that $\{\bar{\Gamma}^{m}, \bar{\Gamma}^{n}\} = 0$. This can bee seen from eqn. \eqref{neweqnonew} and by choosing the only non-zero component of $\bar{\lambda}_{\mu}$ to be $\bar{\lambda}_{-----} \neq 0$. Then $r_{-++++}$, $r_{+-+++}$, $r_{++-++}$, $r_{+++-+}$, $r_{++++-}$ vanish as a consequence of $\bar{\lambda}\gamma^{m}r = 0$. This implies that the only components of $d_{\mu}$ and $N_{mn}$ appearing in \eqref{neweqnonew} are $d_{\nu}$ with $\nu = \{(-++++), (+-+++), (++-++), (+++-+), (++++-)\}$ and $N^{pq}$ with $p,q = \{(1-2i), (3-4i), (5-6i), (7-8i), (9-10i)\}$. Since all the components of $\bar{\lambda}_{\mu}$ with two plus signs are zero, the commutator $[N_{mn}, \lambda^{\mu}\bar{\lambda}_{\mu}]$ vanishes. Likewise the commutator $[N_{mn},N_{pq}]$ vanishes for $(p,q,m,n)$ in $\{(1-2i), (3-4i), (5-6i), (7-8i), (9-10i)\}$ because the metric components are zero for any combination of these values. Thus we see that $\{\bar{\Gamma}^{m},\bar{\Gamma}^{n}\} = 0$. Moreover, since the only contribution of $\lambda$ in $(\lambda\bar{\lambda})$ is $\lambda^{+++++}$, the commutator $[\bar{\Gamma}^{m}, (\lambda\bar{\lambda})] = 0$. One can now use the constraint $(\gamma_{m}\bar{\lambda})^{\mu}\bar{\Gamma}^{m} = 0$ to conclude that the only non-zero components of $\bar{\Gamma}^{m}$ are $\bar{\Gamma}^{n}$ with $n=\{(1+2i),(3+4i),(5+6i),(7+8i),(9+10i)\}$. This in turn implies that the term $(\lambda\gamma^{mn}r)$ in \eqref{neweeq6new} is non-zero only for the cases $m,n = \{(1-2i), (3-4i), (5-6i), (7-8i), (9-10i)\}$. Then, since $w_{\mu}$ can only appear with two or four plus signs in $N^{pq}$ and $\lambda^{\mu}$ appears in $(\lambda\gamma^{mn}r)$ at least with two plus signs, the only relevant situation is when $w_{\mu}$ has two plus signs and $\lambda^{\mu}$ has three plus signs, however when this occurs the only components of $r_{\mu}$ which contribute are those with one minus sign making the whole expression vanishes. Therefore, the commutator $[\bar{\Gamma}^{m}, (\lambda\gamma^{pq}r)]\bar{\Gamma}_{p}\bar{\Gamma}_{q} = 0$. This immediately implies that $\{b,b\} = 0$.

\section{$U(1)\times U(1)\times SO(7)$-covariant analysis of the nilpotency of the $D=11$ $b$-ghost}\label{apII}
We could think of applying a non-Lorentz covariant analysis similar to the one carried out for the $D=10$ case in Appendix \ref{apI} in order to check that $\{b,b\} = 0$. However, as we will see below, this analysis is not conclusive for the $D=11$ case since it does not give us the non-zero value of $\{\Sigma^{i}, \bar{\Sigma}^{j}\}$. Let us start by choosing the only non-zero component of $\bar{\Lambda}^{\alpha}$ to be $\bar{\Lambda}^{++0}$. This implies that the only non-vanishing components of $\bar{\Sigma}^{i}_{0}$ are $\bar{\Sigma}^{j}_{0}$ with $j = \{(1-2i),(3-4i),l\}$ and $l = 1 ,\ldots, 7$. These expressions are written explicitly in \eqref{eq4}, \eqref{eq5}, \eqref{eq6}. From these formulae, one can see that the only relevant contributions will come from $\{\bar{\Sigma}^{(1-2i)}_{0},\bar{\Sigma}^{l}_{0}\}$, $\{\bar{\Sigma}^{(3-4i)}_{0},\bar{\Sigma}^{l}_{0}\}$. However, these terms vanish as a result of the constraint algebra $\{D_{\alpha},D_{\beta}\} = -2(\Gamma^{a})_{\alpha\beta}P_{a}$. Therefore $\{\bar{\Sigma}_{0}^{i}, \bar{\Sigma}_{0}^{j}\} = 0$. Now let us focus on the anticommutator $\{\bar{\Sigma}_{0}^{i}, \bar{\Sigma}^{j}_{1}\}$ where $\bar{\Sigma}^{j}_{1}$ is the $D_{\alpha}$-independent part of $\bar{\Sigma}^{j}$. This object can be read from eqn. \eqref{section4eeq22}
\begin{eqnarray}
\bar{\Sigma}^{i}_{1} &=& \frac{2}{\eta^{2}}(\bar{\Lambda}\Gamma_{ab}\bar{\Lambda})(\bar{\Lambda}\Gamma_{cd}R)(\Lambda\Gamma^{abcki}\Lambda)N^{d}_{\hspace{2mm}k} + \frac{2}{3\eta^{2}}(\bar{\Lambda}\Gamma_{ab}\bar{\Lambda})(\bar{\Lambda}\Gamma_{c}^{\hspace{2mm}i}R)(\Lambda\Gamma^{abcqj}\Lambda)N_{qj}\nonumber\\
&& - \frac{2}{3\eta^{2}}(\bar{\Lambda}\Gamma_{ac}\bar{\Lambda})(\bar{\Lambda}R)(\Lambda\Gamma^{aicqj}\Lambda)N_{qj}
\end{eqnarray}
The only components of $\Lambda^{\alpha}$ contributing to $\bar{\Sigma}^{i}_{0}$ are $\Lambda^{--0}$, $\Lambda^{--l}$ with $l=1,\ldots,7$. Hence the only way to obtain a non-zero result is if the terms $W^{++0}$, $W^{++l}$ are present in $\bar{\Sigma}_{1}^{j}$. With our choice of the direction of $\bar{\Lambda}^{\alpha}$ we can show that the only $N_{ab}$'s contributing to $\bar{\Sigma}_{1}^{j}$ are:
\begin{eqnarray}
\mbox{From the first term of $\bar{\Sigma}_{1}^{j}$} &\rightarrow & N_{(1+2i)(3+4i)}, N_{(1+2i)(l)}, N_{(3+4i),l}\\
\mbox{From the second term of $\bar{\Sigma}_{1}^{j}$} &\rightarrow &N_{(1+2i)(3+4i)}, N_{(1+2i)(l)}, N_{(3+4i),l}, N_{lk}\\
\mbox{From the third term of $\bar{\Sigma}_{1}^{j}$} &\rightarrow &N_{(1+2i)(3+4i)}, N_{(1+2i)(l)}, N_{(3+4i),l}, N_{lk}
\end{eqnarray} 
with $l,k=1,\ldots 7$. It is easy to see that the first term of $\bar{\Sigma}_{1}^{j}$ will not contribute. However, the other two terms containing the ghost currents $N_{lk}$ will do, since they contain terms $W^{++k}$. Thus, this shows that $\{\bar{\Sigma}^{i}, \bar{\Sigma}^{j}\}$ might not vanish. Due to this, one needs to take into account numerical factors and signs in order to use them in the computation of $\{b, b\}$. We do this in Appendix \ref{apD3} in a Lorentz-covariant way.

\section{The $b$-ghost and $\bar{\Sigma}^{a}$ have the same $D_{\alpha}$'s}\label{apapC}
We should figure out which are the $D_{\alpha}$'s appearing in the expressions for $\bar{\Sigma}^{i}$ and the $b$-ghost. For this, we will decompose the eleven dimensional Lorentz group in the following way: $SO(10,1)\rightarrow SO(3,1)\times SO(7)$. In addition, we will conveniently choose the special direction for $\bar{\Lambda}_{\alpha}$ to be $\bar{\Lambda}^{++0}\neq 0$. Thus,
\begin{eqnarray}
\bar{\Lambda}\Gamma^{a}R = 0 &\rightarrow & R^{+-0} = R^{-+0} = R^{--j} = 0\hspace{3mm},\hspace{3mm} \mbox{where $j=1,\ldots,7$}
\end{eqnarray}
On the other hand, from the pure spinor constraint $\Lambda\Gamma^{a}\Lambda = 0$ we have:
\begin{eqnarray}
\Lambda^{-+0} &=& -\frac{\Lambda^{-+j}\Lambda^{--j}}{\Lambda^{--0}} \label{eq7}\\
\Lambda^{+-0} &=& -\frac{\Lambda^{+-j}\Lambda^{--j}}{\Lambda^{--0}}\\
\Lambda^{++j} &=& \frac{1}{\Lambda^{--0}}[\Lambda^{--j}\Lambda^{++0} - \Lambda^{-+j}\Lambda^{+-0} + \Lambda^{+-j}\Lambda^{-+0}]
\end{eqnarray}
where $j = 1, \ldots, 7$ and we have assumed that $\Lambda^{--0}\neq 0$. This allows us to expand the quadratic term in $D_{\alpha}$ in the b-ghost in terms of these components:
\begin{eqnarray}
b_{1} &\propto & \frac{(\bar{\Lambda}^{++0}\bar{\Lambda}^{++0})(\bar{\Lambda}^{++0}R^{--0})}{(\bar{\Lambda}^{++0}\bar{\Lambda}^{++0})^{2}(\Lambda^{--0}\Lambda^{--0} + \Lambda^{--k}\Lambda^{-k})^{2}}\{[\Lambda^{--0}D^{+-0} + \Lambda^{--k}D^{+-k} + \Lambda^{+-0}D^{--0} \nonumber \\
&& + \Lambda^{+-k}D^{--k}]\times [\Lambda^{--0}D^{-+0} + \Lambda^{--k}D^{-+k} - \Lambda^{-+0}D^{--0} - \Lambda^{-+k}D^{--k}]\} \nonumber \\
&& + \frac{(\bar{\Lambda}^{++0}\bar{\Lambda}^{++0})(\bar{\Lambda}^{++0}R^{--0})}{(\bar{\Lambda}^{++0}\bar{\Lambda}^{++0})^{2}(\Lambda^{--0}\Lambda^{--0} + \Lambda^{--k}\Lambda^{--k})^{2}}\{[\Lambda^{--0}D^{-+0} + \Lambda^{--k}D^{-+k} + \Lambda^{-+0}D^{--0} \nonumber \\
&& + \Lambda^{-+k}D^{--k}]\times[\Lambda^{--0}D^{+-0} + \Lambda^{--k}D^{+-k} - \Lambda^{+-0}D^{--0} - \Lambda^{+-k}D^{--k}]\}\nonumber \\
&& + \frac{(\bar{\Lambda}^{++0}\bar{\Lambda}^{++0})(\bar{\Lambda}^{++0}R^{+-j})}{(\bar{\Lambda}^{++0}\bar{\Lambda}^{++0})^{2}(\Lambda^{--0}\Lambda^{--0} + \Lambda^{--k}\Lambda^{-k})^{2}}\{[\Lambda^{--0}D^{-+0} + \Lambda^{--k}D^{-+k} \nonumber \\ 
&& + \Lambda^{-+0}D^{--0} + \Lambda^{-+k}D^{--k}]\times[\Lambda^{--0}D^{--j} - \Lambda^{--j}D^{--0}]\}\nonumber \\
&& - \frac{(\bar{\Lambda}^{++0}\bar{\Lambda}^{++0})(\bar{\Lambda}^{++0}R^{-+j})}{(\bar{\Lambda}^{++0}\bar{\Lambda}^{++0})^{2}(\Lambda^{--0}\Lambda^{--0} + \Lambda^{--k}\Lambda^{--k})^{2}}\{[\Lambda^{--0}D^{+-0} + \Lambda^{--k}D^{+-k} \nonumber \\
&& + \Lambda^{+-0}D^{--0} + \Lambda^{+-k}D^{--k}]\times[\Lambda^{--0}D^{--j} - \Lambda^{--j}D^{--0}]\}
\end{eqnarray}
Now, we write $\bar{\Sigma}^{a}_{0}$ in the convenient form:
\begin{eqnarray}
\bar{\Sigma}^{a}_{0} &=& \frac{1}{2\eta}[2(\bar{\Lambda}\Gamma^{ba}\bar{\Lambda})(\Lambda\Gamma_{b}D) + (\bar{\Lambda}\Gamma_{bc}\bar{\Lambda})(\Lambda\Gamma^{bca}D)]
\end{eqnarray}
After using the particular direction chosen above, the $SO(3,1)\times SO(7)$ components of $\bar{\Sigma}^{a}_{0}$ read
\begin{eqnarray}
\bar{\Sigma}_{0}^{1+2i} &=& 0\\
\bar{\Sigma}_{0}^{3+4i} &=& 0\\
\bar{\Sigma}_{0}^{1-2i} &\propto & \frac{\bar{\Lambda}^{++0}\bar{\Lambda}^{++0}}{(\bar{\Lambda}^{++0}\bar{\Lambda}^{++0})(\Lambda^{--0}\Lambda^{--0} + \Lambda^{--k}\Lambda^{--k})}(\Lambda^{--0}D^{+-0} + \Lambda^{--k}D^{+-k}) \nonumber\\
\label{eq4}\\
\bar{\Sigma}_{0}^{3-4i} &\propto & \frac{\bar{\Lambda}^{++0}\bar{\Lambda}^{++0}}{(\bar{\Lambda}^{++0}\bar{\Lambda}^{++0})(\Lambda^{--0}\Lambda^{--0} + \Lambda^{--k}\Lambda^{--k})}(\Lambda^{--0}D^{-+0} + \Lambda^{--k}D^{-+k})\nonumber\\
\label{eq5}\\
\bar{\Sigma}_{0}^{j} &\propto & \frac{\bar{\Lambda}^{++0}\bar{\Lambda}^{++0}}{(\bar{\Lambda}^{++0}\bar{\Lambda}^{++0})(\Lambda^{--0}\Lambda^{--0} + \Lambda^{--k}\Lambda^{--k})}(\Lambda^{--0}D^{--j} - \Lambda^{--j}D^{--0})\nonumber\\
\label{eq6}
\end{eqnarray}
where $k,j = 1,\ldots, 7$. 
Therefore, after using the pure spinor constraint, we see that the expression for $b_{1}$ contains the same combinations of $D_{\alpha}$'s as those contained in the expression for $\bar{\Sigma}_{0}^{a}$.
\section{$D_{\alpha}$ in terms of $\bar{\Sigma}_{0}^{j}$}\label{apD}
Let us define the quantity:
\begin{equation}
H_{\alpha} = (\Lambda\Gamma_{i})_{\alpha}\bar{\Sigma}_{0}^{i} =\frac{1}{2\eta}(\Gamma_{i}\Lambda)_{\alpha}(\bar{\Lambda}\Gamma_{ab}\bar{\Lambda})(\Lambda\Gamma^{ab}\Gamma^{i}D)
\end{equation} 
Now we will assume that there exist a matrix $(M^{-1})_{\alpha}^{\hspace{2mm}\beta}$ such that:
\begin{equation}
D_{\alpha} = (M^{-1})_{\alpha}^{\hspace{2mm}\beta}H_{\beta}
\end{equation}
and let us check that the following ansatz for $(M^{-1})_{\alpha}^{\hspace{2mm}\beta}$:
\begin{equation}
(M^{-1})_{\alpha}^{\hspace{2mm}\beta} = 2\delta_{\alpha}^{\hspace{1mm}\beta} + \frac{2}{\eta}(\Lambda\Gamma_{m})_{\alpha}(\bar{\Lambda}\Gamma^{mn}\bar{\Lambda})(\Lambda\Gamma_{n})^{\beta}
\end{equation}
is indeed correct. This can be seen easily as follows
\begin{eqnarray*}
H_{\alpha} &=& \frac{1}{2\eta}(\Gamma_{i}\Lambda)_{\alpha}(\bar{\Lambda}\Gamma_{ab}\bar{\Lambda})(\Lambda\Gamma^{ab}\Gamma^{i}M^{-1}H)\\
&=& \frac{1}{\eta}(\Gamma_{i}\Lambda)_{\alpha}(\bar{\Lambda}\Gamma_{ab}\bar{\Lambda})[(\Lambda\Gamma^{ab}\Gamma^{i}\Gamma^{j}\Lambda)\bar{\Sigma}_{0\,j} + \frac{1}{\eta}(\Lambda\Gamma^{ab}\Gamma^{i}\Gamma^{m}\Lambda)(\bar{\Lambda}\Gamma_{mn}\bar{\Lambda})(\Lambda\Gamma^{nj}\Lambda)\bar{\Sigma}_{0\,j}]\\
&=& (\Gamma^{j}\Lambda)_{\alpha}\bar{\Sigma}_{0\,j} - \frac{2}{\eta}(\Gamma_{i}\Lambda)_{\alpha}(\bar{\Lambda}\Gamma^{i}_{\hspace{1mm}b}\bar{\Lambda})(\Lambda\Gamma^{bj}\Lambda)\bar{\Sigma}_{0\,j} + \frac{1}{\eta^{2}}[\eta(\Gamma_{i}\Lambda)_{\alpha}(\bar{\Lambda}\Gamma^{i}_{\hspace{1mm}n}\bar{\Lambda})(\Lambda\Gamma^{nj}\Lambda)\bar{\Sigma}_{0\,j}\\
& & -2(\Gamma_{i}\Lambda)_{\alpha}(\bar{\Lambda}\Gamma^{i}_{\hspace{1mm}b}\bar{\Lambda})(\Lambda\Gamma^{b}_{\hspace{1mm}m}\Lambda)(\bar{\Lambda}\Gamma^{m}_{\hspace{2mm}n}\bar{\Lambda})(\Lambda\Gamma^{nj}\Lambda)\bar{\Sigma}_{0\,j}]\\
&=& (\Gamma^{j}\Lambda)_{\alpha}\bar{\Sigma}_{0\,j}
\end{eqnarray*}
where the identity \eqref{newapp3} was used. 
Therefore we have the relation:
\begin{equation}\label{newapp50}
D_{\alpha} = 2(\Lambda\Gamma^{c})_{\alpha}\bar{\Sigma}_{0\,c} + \frac{2}{\eta}(\Lambda\Gamma^{m})_{\alpha}(\bar{\Lambda}\Gamma_{mn}\bar{\Lambda})(\Lambda\Gamma^{nj}\Lambda)\bar{\Sigma}_{0\,j}
\end{equation}
Furthermore, from the constraint $(\bar{\Lambda}\Gamma_{ab}\bar{\Lambda})\bar{\Sigma}_{0}^{b} = 0$, one immediately concludes that
\begin{equation}
\bar{\Sigma}_{0}^{k} = \frac{1}{\eta}(\bar{\Lambda}\Gamma_{ij}\bar{\Lambda})(\Lambda\Gamma^{ijk}H) 
\end{equation}
which is the inverse relation between $\bar{\Sigma}^{k}$ and $H_{\alpha}$.
\section{The $D_{\alpha}$'s in $\{Q, \bar{\Sigma}^{i}\}$ are gauge invariant}\label{apD1}
We will show that the $D_{\alpha}$'s appearing in \eqref{neweq26} are invariant under the gauge transformations \eqref{eq400}. This will allows us to conclude that these $D_{\alpha}$'s are the same as those contained in the definition of $\bar{\Sigma}^{i}$. In this Appendix and the next ones we have used the GAMMA package \cite{Gran:2001yh} due to the heavy manipulation of gamma matrix identities which our computations demanded. Let us call $I^{i}$ to the terms containing $D_{\alpha}$'s explicitly in \eqref{neweq26}. The identities \eqref{newapp23}, \eqref{newapp24} allow us simplify this object:
\begin{align}
I^{i} & = -\frac{1}{\eta}(\bar{\Lambda}\Gamma_{ab}R)(\Lambda\Gamma^{ab}\Gamma^{i}D) - \frac{2}{\eta^{2}}(\bar{\Lambda}\Gamma_{ab}\bar{\Lambda})(\bar{\Lambda}\Gamma_{cd}R)(\Lambda\Gamma^{abcki}\Lambda)(\Lambda\Gamma^{d}_{\hspace{1mm}k}D) \notag\\
& - \frac{2}{3\eta^{2}}(\bar{\Lambda}\Gamma_{ab}\bar{\Lambda})(\bar{\Lambda}\Gamma_{c}^{\hspace{2mm}i}R)(\Lambda\Gamma^{abcdk}\Lambda)(\Lambda\Gamma_{dk}D)- \frac{2}{3\eta^{2}}(\bar{\Lambda}\Gamma_{ab}\bar{\Lambda})(\bar{\Lambda}R)(\Lambda\Gamma^{iabdk}\Lambda)(\Lambda\Gamma_{dk}D)\notag\\
& = -\frac{1}{\eta}(\bar{\Lambda}\Gamma_{ab}R)(\Lambda\Gamma^{abi}D) - \frac{2}{\eta}(\bar{\Lambda}\Gamma^{ai}R)(\Lambda\Gamma_{a}D) - \frac{2}{\eta^{2}}(\bar{\Lambda}\Gamma_{ab}\bar{\Lambda})(\bar{\Lambda}\Gamma_{cd}R)(\Lambda\Gamma^{abcik}\Lambda)(\Lambda\Gamma_{k}^{\hspace{1mm}d}D) \notag \\
& + \frac{8}{\eta^{2}}(\bar{\Lambda}\Gamma_{ab}\bar{\Lambda})(\bar{\Lambda}\Gamma_{c}^{\hspace{2mm}i}R)(\Lambda\Gamma^{bc}\Lambda)(\Lambda\Gamma^{a}D) + \frac{4}{\eta}(\bar{\Lambda}\Gamma^{ci}R)(\Lambda\Gamma_{c}D) \notag \\
& + \frac{8}{\eta^{2}}(\bar{\Lambda}\Gamma_{ab}\bar{\Lambda})(\bar{\Lambda}R)(\Lambda\Gamma^{bi}\Lambda)(\Lambda\Gamma^{a}D) + \frac{4}{\eta}(\bar{\Lambda}R)(\Lambda\Gamma^{i}D) \label{newneweq23}
\end{align}
The third term of this expression requires more careful manipulations:
\begin{align}
I^{*\,i} & = -\frac{2}{\eta^{2}}(\bar{\Lambda}\Gamma_{ab}\bar{\Lambda})(\bar{\Lambda}\Gamma_{cd}R)(\Lambda\Gamma^{abcik}\Lambda)(\Lambda\Gamma_{k}^{\hspace{2mm}d}D) \notag \\
& = -\frac{4}{\eta^{2}}(\bar{\Lambda}\Gamma_{ac}\bar{\Lambda})(\bar{\Lambda}R)(\Lambda\Gamma^{ci}\Lambda)(\Lambda\Gamma^{a}D) - \frac{2}{\eta^{2}}(\bar{\Lambda}\Gamma_{ab}\bar{\Lambda})(\bar{\Lambda}\Gamma_{cd}R)(\Lambda\Gamma^{ci}\Lambda)(\Lambda\Gamma^{abd}D) \notag \\
& + \frac{4}{\eta^{2}}(\bar{\Lambda}\Gamma_{bc}\bar{\Lambda})(\bar{\Lambda}R)(\Lambda\Gamma^{bi}\Lambda)(\Lambda\Gamma^{c}D) + \frac{4}{\eta^{2}}(\bar{\Lambda}\Gamma_{ab}\bar{\Lambda})(\bar{\Lambda}\Gamma_{cd}R)(\Lambda\Gamma^{bi}\Lambda)(\Lambda\Gamma^{acd}D) \notag \\
& - \frac{4}{\eta^{2}}(\bar{\Lambda}\Gamma_{ab}\bar{\Lambda})(\bar{\Lambda}\Gamma_{c}^{\hspace{2mm}i}R)(\Lambda\Gamma^{bc}\Lambda)(\Lambda\Gamma^{a}D) - \frac{4}{\eta}(\bar{\Lambda}R)(\Lambda\Gamma^{i}D)\notag\\
& - \frac{4}{\eta^{2}}(\bar{\Lambda}\Gamma_{ab}\bar{\Lambda})(\bar{\Lambda}\Gamma_{cd}R)(\Lambda\Gamma^{bc}\Lambda)(\Lambda\Gamma^{aid}D) - \frac{2}{\eta}(\bar{\Lambda}\Gamma^{ci}R)(\Lambda\Gamma_{c}D) - \frac{2}{\eta}(\bar{\Lambda}\Gamma_{cd}R)(\Lambda\Gamma^{cid}D) 
\end{align}
Furthermore, the identity \eqref{newapp4} allows us to cast this result as
\begin{align}
I^{*\,i} & = -\frac{8}{\eta^{2}}(\bar{\Lambda}\Gamma_{ac}\bar{\Lambda})(\bar{\Lambda}R)(\Lambda\Gamma^{ci}\Lambda)(\Lambda\Gamma^{a}D) - \frac{2}{\eta^{2}}(\bar{\Lambda}\Gamma_{ab}\bar{\Lambda})(\bar{\Lambda}\Gamma_{cd}R)(\Lambda\Gamma^{ci}\Lambda)(\Lambda\Gamma^{abd}D) \notag \\
& + \frac{4}{\eta^{2}}(\bar{\Lambda}\Gamma_{ab}\bar{\Lambda})(\bar{\Lambda}\Gamma_{cd}R)(\Lambda\Gamma^{bi}\Lambda)(\Lambda\Gamma^{acd}D) - \frac{4}{\eta^{2}}(\bar{\Lambda}\Gamma_{ab}\bar{\Lambda})(\bar{\Lambda}\Gamma_{c}^{\hspace{2mm}i}R)(\Lambda\Gamma^{bc}\Lambda)(\Lambda\Gamma^{a}D) \notag \\
& - \frac{4}{\eta}(\bar{\Lambda}R)(\Lambda\Gamma^{i}D) + \frac{1}{\eta}(\bar{\Lambda}\Gamma_{ad}R)(\Lambda\Gamma^{aid}D) + \frac{1}{\eta^{2}}(\bar{\Lambda}\Gamma_{ad}\bar{\Lambda})(\bar{\Lambda}\Gamma_{bc}R)(\Lambda\Gamma^{bc}\Lambda)(\Lambda\Gamma^{aid}\Lambda) \notag \\
& - \frac{2}{\eta}(\bar{\Lambda}\Gamma^{ci}R)(\Lambda\Gamma_{c}D) - \frac{2}{\eta}(\bar{\Lambda}\Gamma_{cd}R)(\Lambda\Gamma^{cid}D) 
\end{align}
Plugging this result into \eqref{newneweq23}, we find
\begin{align}
I^{i} & = \frac{4}{\eta^{2}}(\bar{\Lambda}\Gamma_{ab}\bar{\Lambda})(\bar{\Lambda}\Gamma_{c}^{\hspace{2mm}i}R)(\Lambda\Gamma^{bc}\Lambda)(\Lambda\Gamma^{a}D) + \frac{2}{\eta^{2}}(\bar{\Lambda}\Gamma_{ab}\bar{\Lambda})(\bar{\Lambda}\Gamma_{cd}R)(\Lambda\Gamma^{ci}\Lambda)(\Lambda\Gamma^{abd}D) \notag \\
& + \frac{1}{\eta^{2}}(\bar{\Lambda}\Gamma_{ad}\bar{\Lambda})(\bar{\Lambda}\Gamma_{bc}R)(\Lambda\Gamma^{bc}\Lambda)(\Lambda\Gamma^{aid}D) \label{newneweq25}
\end{align}
After applying the transformation \eqref{eq400} and using the identities \eqref{newapp2}, \eqref{newapp3}, \eqref{newapp4} one can show that this expression is invariant under \eqref{eq400} as mentioned above.

\vspace{2mm}
Therefore we can replace the inverse relation \eqref{newapp50} in \eqref{neweq26}. After doing this for each term in \eqref{newneweq25}, we get:
\begin{align}
I_{1}^{i} 
& = -\frac{2}{\eta}(\bar{\Lambda}\Gamma^{ci}R)(\Lambda\Gamma_{ck}\Lambda)\bar{\Sigma}_{0}^{k}
\end{align}

\begin{align}
I_{2}^{i} 
& = \frac{4}{\eta}(\bar{\Lambda}\Gamma_{cd}R)(\Lambda\Gamma^{ci}\Lambda)\bar{\Sigma}_{0}^{d} + \frac{2}{\eta}(\bar{\Lambda}R)(\Lambda\Gamma^{ik}\Lambda)\bar{\Sigma}_{0\,k}
\end{align}

\begin{align}
I_{3}^{i} 
&= -\frac{2}{\eta}(\bar{\Lambda}\Gamma^{cd}R)(\Lambda\Gamma_{cd}\Lambda)\bar{\Sigma}_{0}^{i} - \frac{2}{\eta^{2}}(\bar{\Lambda}\Gamma_{cd}R)(\Lambda\Gamma^{cd}\Lambda)(\bar{\Lambda}\Gamma^{in}\bar{\Lambda})(\Lambda\Gamma^{nk}\Lambda)\bar{\Sigma}_{0\,k}
\end{align}
Replacing these expressions in \eqref{newneweq25} and putting all together in \eqref{neweq26} we obtain
\begin{align}
\{Q, \bar{\Sigma}^{i}\} & = -P^{i} - \frac{2}{\eta}[(\bar{\Lambda}\Gamma^{mb}\bar{\Lambda})(\Lambda\Gamma_{b}^{\hspace{1mm}i}\Lambda) - (\bar{\Lambda}\Gamma^{ib}\bar{\Lambda})(\Lambda\Gamma_{b}^{\hspace{1mm}m}\Lambda)]P_{m} + \frac{4}{\eta}(\bar{\Lambda}\Gamma_{mn}R)(\Lambda\Gamma^{mn}\Lambda)\bar{\Sigma}^{i} \notag \\
&- \frac{2}{\eta}(\bar{\Lambda}\Gamma_{mn}R)(\Lambda\Gamma^{mn}\Lambda)\bar{\Sigma}^{i}_{0} - \frac{2}{\eta}(\bar{\Lambda}\Gamma^{ci}R)(\Lambda\Gamma_{ck}\Lambda)\bar{\Sigma}_{0}^{k} + \frac{4}{\eta}(\bar{\Lambda}\Gamma_{cd}R)(\Lambda\Gamma^{ci}\Lambda)\bar{\Sigma}_{0}^{d} \notag \\
& + \frac{2}{\eta}(\bar{\Lambda}R)(\Lambda\Gamma^{ik}\Lambda)\bar{\Sigma}_{0\,k} -\frac{2}{\eta}(\bar{\Lambda}\Gamma^{cd}R)(\Lambda\Gamma_{cd}\Lambda)\bar{\Sigma}_{0}^{i} - \frac{2}{\eta^{2}}(\bar{\Lambda}\Gamma_{cd}R)(\Lambda\Gamma^{cd}\Lambda)(\bar{\Lambda}\Gamma^{in}\bar{\Lambda})(\Lambda\Gamma^{nk}\Lambda)\bar{\Sigma}_{0\,k} \notag \\
& - \frac{4}{\eta^{2}}(\bar{\Lambda}\Gamma_{ab}R)(\bar{\Lambda}\Gamma_{cd}R)(\Lambda\Gamma^{abcki}\Lambda)N^{d}_{\hspace{1mm}k} - \frac{4}{3\eta^{2}}(\bar{\Lambda}\Gamma_{ab}R)(\bar{\Lambda}\Gamma_{c}^{\hspace{2mm}i}R)(\Lambda\Gamma^{abcdk}\Lambda)N_{dk} \notag \\
& - \frac{4}{3\eta^{2}}(\bar{\Lambda}\Gamma_{ab}R)(\bar{\Lambda}R)(\Lambda\Gamma^{iabdk}\Lambda)N_{dk} - \frac{2}{3\eta^{2}}(\bar{\Lambda}\Gamma_{ab}\bar{\Lambda})(RR)(\Lambda\Gamma^{iabdk}\Lambda)N_{dk}
\end{align}

\section{Cancellation of all of the $N_{ab}$ contributions in eqn. \eqref{section4eq401}}\label{apD2}

We will show this cancellation in two steps. First we will simplify the expression depending explicitly on $\bar{\Sigma}^{i}_{0}$ and then simplify the expression depending explicitly on $N_{ab}$. Finally we will see that these two expressions identically cancel out. We start with the following equation
\begin{eqnarray}
J^{i} &=& \frac{4}{\eta}(\bar{\Lambda}\Gamma^{ab}R)(\Lambda\Gamma_{ab}\Lambda)(\bar{\Sigma}^{i}-\bar{\Sigma}_{0}^{i}) - \frac{2}{\eta}(\bar{\Lambda}\Gamma^{ci}R)(\Lambda\Gamma_{ck}\Lambda)\bar{\Sigma}_{0}^{k} + \frac{4}{\eta}(\bar{\Lambda}\Gamma_{cd}R)(\Lambda\Gamma^{ci}\Lambda)\bar{\Sigma}_{0}^{d} \nonumber \\
&& + \frac{2}{\eta}(\bar{\Lambda}R)(\Lambda\Gamma^{ik}\Lambda)\bar{\Sigma}_{0\,k} - \frac{2}{\eta^{2}}(\bar{\Lambda}\Gamma^{cd}R)(\Lambda\Gamma_{cd}\Lambda)(\bar{\Lambda}\Gamma^{if}\bar{\Lambda})(\Lambda\Gamma_{fk}\Lambda)\bar{\Sigma}_{0}^{k} \label{eeq35}
\end{eqnarray}
One can show that the term proportional to $\bar{\Lambda}R$ can be cast as
\begin{eqnarray}
J_{1}^{i} &=& \frac{8}{\eta^{3}}(\Lambda\Gamma^{bi}\Lambda)(\bar{\Lambda}R)(\bar{\Lambda}\Gamma_{ab}\bar{\Lambda})(\bar{\Lambda}\Gamma_{ck}R)(\Lambda\Gamma^{ck}\Lambda)(\Lambda\Gamma^{a}W) \nonumber \\
&& - \frac{24}{\eta^{2}}(\Lambda\Gamma^{ik}\Lambda)(\bar{\Lambda}R)(\bar{\Lambda}\Gamma_{ck}R)(\Lambda\Gamma^{c}W) \nonumber \\
&& - \frac{16}{\eta^{2}}(\bar{\Lambda}\Gamma^{ab}R)(\Lambda\Gamma_{ab}\Lambda)(\bar{\Lambda}R)(\Lambda\Gamma^{i}W)
\end{eqnarray}
The use of the identity \eqref{newapp24} allows us to write the term proportional to $(\bar{\Lambda}\Gamma^{ci}R)$ in the form
\begin{eqnarray}
J_{2}^{i} &=& -\frac{16}{\eta^{3}}(\bar{\Lambda}\Gamma^{ef}R)(\Lambda\Gamma_{ef}\Lambda)(\bar{\Lambda}\Gamma_{ab}\bar{\Lambda})(\bar{\Lambda}\Gamma_{c}{}^{i}R)(\Lambda\Gamma^{bc}\Lambda)(\Lambda\Gamma^{a}W)\nonumber\\
&& - \frac{12}{\eta^{2}}(\bar{\Lambda}\Gamma^{ab}R)(\Lambda\Gamma_{ab}\Lambda)(\bar{\Lambda}\Gamma_{c}{}^{i}R)(\Lambda\Gamma^{c}W)\nonumber \\
&& - \frac{8}{\eta^{2}}(\bar{\Lambda}\Gamma^{ci}R)(\Lambda\Gamma_{ck}\Lambda)(\bar{\Lambda}\Gamma_{f}{}^{k}R)(\Lambda\Gamma^{f}W) 
\end{eqnarray}
Finally, with a little of algebra and the use of the identities \eqref{newapp4}, \eqref{newapp23} one gets the following result
\begin{eqnarray}
J^{i} &=& -\frac{4}{\eta^{2}}(\bar{\Lambda}\Gamma^{ef}R)(\Lambda\Gamma_{ef}\Lambda)(\bar{\Lambda}\Gamma_{bd}R)(\Lambda\Gamma^{bdi}W) + \frac{4}{\eta^{2}}(\bar{\Lambda}\Gamma_{cd}R)(\Lambda\Gamma^{ci}\Lambda)(\bar{\Lambda}\Gamma_{ef}R)(\Lambda\Gamma^{efd}W)\nonumber\\
&& -\frac{4}{\eta^{2}}(\bar{\Lambda}\Gamma^{ef}R)(\Lambda\Gamma_{ef}\Lambda)(\bar{\Lambda}\Gamma_{c}{}^{i}R)(\Lambda\Gamma^{c}W)- \frac{8}{\eta^{2}}(\bar{\Lambda}\Gamma_{c}{}^{i}R)(\Lambda\Gamma^{ck}\Lambda)(\bar{\Lambda}\Gamma_{fk}R)(\Lambda\Gamma^{f}W)\nonumber \\
\end{eqnarray}

Now we will simplify the expressions containing $N_{ab}$ explicitly:
\begin{eqnarray}\label{eqap1}
S^{i} &=& -\frac{4}{\eta^{2}}(\bar{\Lambda}\Gamma_{ab}R)(\bar{\Lambda}\Gamma_{cd}R)(\Lambda\Gamma^{abcki}\Lambda)N^{d}{}_{k} - \frac{4}{3\eta^{2}}(\bar{\Lambda}\Gamma_{ab}R)(\bar{\Lambda}\Gamma_{c}{}^{i}R)(\Lambda\Gamma^{abcdk}\Lambda)N_{dk}\nonumber \\
&& - \frac{4}{3\eta^{2}}(\bar{\Lambda}\Gamma_{ab}R)(\bar{\Lambda}R)(\Lambda\Gamma^{iabdk}\Lambda)N_{dk} - \frac{2}{3\eta^{2}}(\bar{\Lambda}\Gamma_{ab}\bar{\Lambda})(RR)(\Lambda\Gamma^{iabdk}\Lambda)N_{dk}
\end{eqnarray}
The first term in \eqref{eqap1} can be written as follows
\begin{eqnarray}
S_{1}^{i} &=& -\frac{8}{\eta^{2}}(\bar{\Lambda}\Gamma_{ac}R)(\bar{\Lambda}R)(\Lambda\Gamma^{ci}\Lambda)(\Lambda\Gamma^{a}W) - \frac{4}{\eta^{2}}(\bar{\Lambda}\Gamma_{ac}\bar{\Lambda})(RR)(\Lambda\Gamma^{ci}\Lambda)(\Lambda\Gamma^{a}W)\nonumber \\
&& -\frac{4}{\eta^{2}}(\bar{\Lambda}\Gamma_{ab}R)(\bar{\Lambda}\Gamma_{cd}R)(\Lambda\Gamma^{ci}\Lambda)(\Lambda\Gamma^{abd}W) - \frac{8}{\eta^{2}}(\bar{\Lambda}\Gamma_{ab}\bar{\Lambda})(\bar{\Lambda}\Gamma_{c}{}^{i}R)(\Lambda\Gamma^{ca}\Lambda)(\Lambda\Gamma^{b}W)\nonumber\\
&& +\frac{8}{\eta^{2}}(\bar{\Lambda}\Gamma_{ca}R)(\bar{\Lambda}R)(\Lambda\Gamma^{ca}\Lambda)(\Lambda\Gamma^{i}W) - \frac{4}{\eta}(RR)(\Lambda\Gamma^{i}W)\nonumber\\
&& -\frac{4}{\eta^{2}}(\bar{\Lambda}\Gamma^{ab}R)(\Lambda\Gamma_{ab}\Lambda)(\bar{\Lambda}\Gamma_{c}{}^{i}R)(\Lambda\Gamma^{c}W) + \frac{4}{\eta^{2}}(\bar{\Lambda}\Gamma^{ab}R)(\Lambda\Gamma_{ab}\Lambda)(\bar{\Lambda}\Gamma_{cd}R)(\Lambda\Gamma^{cdi}W)\nonumber\\
&& - \frac{8}{\eta^{2}}(\bar{\Lambda}\Gamma_{ac}R)(\bar{\Lambda}R)(\Lambda\Gamma^{ia}\Lambda)(\Lambda\Gamma^{c}W) - \frac{4}{\eta^{2}}(\bar{\Lambda}\Gamma_{ac}\bar{\Lambda})(RR)(\Lambda\Gamma^{ia}\Lambda)(\Lambda\Gamma^{c}W)\nonumber\\
&& + \frac{8}{\eta^{2}}(\bar{\Lambda}\Gamma_{ab}R)(\bar{\Lambda}\Gamma_{cd}R)(\Lambda\Gamma^{ia}\Lambda)(\Lambda\Gamma^{bcd}W)
\end{eqnarray}
The last three terms in \eqref{eqap1} can be put into the form:
\begin{eqnarray}
S_{2}^{i} &=& \frac{16}{\eta^{2}}(\bar{\Lambda}\Gamma_{ab}R)(\bar{\Lambda}\Gamma_{c}{}^{i}R)(\Lambda\Gamma^{bc}\Lambda)(\Lambda\Gamma^{a}W) + \frac{8}{\eta^{2}}(\bar{\Lambda}\Gamma_{ab}R)(\Lambda\Gamma^{ab}\Lambda)(\bar{\Lambda}\Gamma_{c}{}^{i}R)(\Lambda\Gamma^{c}W)\nonumber\\
&& +\frac{8}{\eta^{2}}(\bar{\Lambda}\Gamma_{ab}R)(\Lambda\Gamma^{ab}\Lambda)(\bar{\Lambda}R)(\Lambda\Gamma^{i}W) + \frac{16}{\eta^{2}}(\bar{\Lambda}\Gamma_{ab}R)(\bar{\Lambda}R)(\Lambda\Gamma^{bi}\Lambda)(\Lambda\Gamma^{a}W)\nonumber\\
&& + \frac{4}{\eta}(RR)(\Lambda\Gamma^{i}W) + \frac{8}{\eta^{2}}(\bar{\Lambda}\Gamma_{ab}\bar{\Lambda})(RR)(\Lambda\Gamma^{bi}\Lambda)(\Lambda\Gamma^{a}W)
\end{eqnarray}
After summing $S_{1}^{i} + S_{2}^{i}$ we obtain
\begin{eqnarray}
S^{i} &=& \frac{4}{\eta^{2}}(\bar{\Lambda}\Gamma^{ef}R)(\Lambda\Gamma_{ef}\Lambda)(\bar{\Lambda}\Gamma_{bd}R)(\Lambda\Gamma^{bdi}W) - \frac{4}{\eta^{2}}(\bar{\Lambda}\Gamma_{cd}R)(\Lambda\Gamma^{ci}\Lambda)(\bar{\Lambda}\Gamma_{ef}R)(\Lambda\Gamma^{efd}W)\nonumber\\
&& + \frac{4}{\eta^{2}}(\bar{\Lambda}\Gamma^{ef}R)(\Lambda\Gamma_{ef}\Lambda)(\bar{\Lambda}\Gamma_{c}{}^{i}R)(\Lambda\Gamma^{c}W) + \frac{8}{\eta^{2}}(\bar{\Lambda}\Gamma_{c}{}^{i}R)(\Lambda\Gamma^{ck}\Lambda)(\bar{\Lambda}\Gamma_{fk}R)(\Lambda\Gamma^{f}W)\nonumber \\
\end{eqnarray}
Thus we have a full cancellation $J^{i} + S^{i} = 0$.

\section{Calculation of $\{\bar{\Sigma}^{i}, {\bar{\Sigma}}^{j}\}$}\label{apD3}
The object $\bar{\Sigma}^{i}$ has a part depending on $D_{\alpha}$ and other part depending on $N_{ab}$, as can be seen in \eqref{section4eeq22}. The part depending on $N_{ab}$ will be called $\bar{\Sigma}^{i}_{1}$ and, as before, we use $\bar{\Sigma}^{i}_{0}$ to denote the part depending on $D_{\alpha}$. Therefore
\begin{equation}
\bar{\Sigma}^{i} = \bar{\Sigma}^{i}_{0} + \bar{\Sigma}^{i}_{1}
\end{equation}
It is easy to see that $\{\bar{\Sigma}^{i}_{0}, \bar{\Sigma}^{j}_{0}\} = 0$\footnote{A simple non-Lorentz covariant argument for this can be found in Appendix \ref{apII}.}:
\begin{eqnarray}
\{\bar{\Sigma}^{i}_{0}, \bar{\Sigma}^{j}_{0}\} &=& \{\frac{1}{2\eta}(\bar{\Lambda}\Gamma_{ab}\bar{\Lambda})(\Lambda\Gamma^{ab}\Gamma^{i}D), \frac{1}{2\eta}(\bar{\Lambda}\Gamma_{cd}\bar{\Lambda})(\Lambda\Gamma^{cd}\Gamma^{j}D)\}\nonumber\\
&=& \frac{1}{4\eta^{2}}(\bar{\Lambda}\Gamma_{ab}\bar{\Lambda})(\bar{\Lambda}\Gamma_{cd}\bar{\Lambda})(\Lambda\Gamma^{ab}\Gamma^{i})^{\alpha}(\Lambda\Gamma^{cd}\Gamma^{j})^{\beta}\{D_{\alpha}, D_{\beta}\}\nonumber\\
&=& -\frac{1}{2\eta^{2}}(\bar{\Lambda}\Gamma_{ab}\bar{\Lambda})(\bar{\Lambda}\Gamma_{cd}\bar{\Lambda})(\Lambda\Gamma^{ab}\Gamma^{i}\Gamma^{f}\Gamma^{j}\Gamma^{cd}\Lambda)P_{f}\nonumber\\
&=& 0
\end{eqnarray}
where eqn. \eqref{newapp3} was used. 

\vspace{2mm}
To compute the anticommutator $\{\bar{\Sigma}_{0}^{i}, \bar{\Sigma}_{1}^{j}\}$, we write $\bar{\Sigma}_{1}^{j}$ in the more convenient way:
\begin{eqnarray}
\bar{\Sigma}_{1}^{j} &=& \frac{2}{\eta^{2}}(\bar{\Lambda}\Gamma_{ab}\bar{\Lambda})(\bar{\Lambda}\Gamma_{cd}R)(\Lambda\Gamma^{abckj}\Lambda)N^{d}{}_{k} + \frac{2}{3\eta^{2}}(\bar{\Lambda}\Gamma_{ab}\bar{\Lambda})(\bar{\Lambda}\Gamma_{c}{}^{j}R)(\Lambda\Gamma^{abcdk}\Lambda)N_{dk} \nonumber\\
&& + \frac{2}{3\eta^{2}}(\bar{\Lambda}\Gamma_{ab}\bar{\Lambda})(\bar{\Lambda}R)(\Lambda\Gamma^{jabdk}\Lambda)N_{dk}
\end{eqnarray}
and denote each term by $\bar{\Sigma}^{j(1)}_{1}$, $\bar{\Sigma}^{j(2)}_{1}$, $\bar{\Sigma}^{j(3)}_{1}$, respectively. It can be shown that $\{\bar{\Sigma}^{i}_{0}, \bar{\Sigma}_{1}^{j(1)}\} = 0$. Now, we rewrite $\bar{\Sigma}_{1}^{j(2)}$, $\bar{\Sigma}_{1}^{j(3)}$ in the more convenient way:
\begin{eqnarray}
\bar{\Sigma}^{j(2)}_{1} &=& -\frac{4}{\eta}(\bar{\Lambda}\Gamma^{cj}R)(\Lambda\Gamma_{c}W) - \frac{8}{\eta^{2}}(\bar{\Lambda}\Gamma_{ab}\bar{\Lambda})(\bar{\Lambda}\Gamma_{c}{}^{j}R)(\Lambda\Gamma_{ca}\Lambda)(\Lambda\Gamma^{b}W)\\
\bar{\Sigma}^{j(3)}_{1} &=& -\frac{4}{\eta}(\bar{\Lambda}R)(\Lambda\Gamma^{j}W) - \frac{8}{\eta^{2}}(\bar{\Lambda}\Gamma_{ab}\bar{\Lambda})(\bar{\Lambda}R)(\Lambda\Gamma^{ja}\Lambda)(\Lambda\Gamma^{b}W)
\end{eqnarray}
where we used the identity \eqref{newapp24}. Therefore,
\begin{eqnarray}
\{\bar{\Sigma}^{i}_{0}, \bar{\Sigma}^{j(2)}_{1}\} &=& \frac{2}{\eta^{2}}(\bar{\Lambda}\Gamma_{ab}\bar{\Lambda})(\bar{\Lambda}\Gamma_{c}{}^{j}R)(\Lambda\Gamma^{c}\Gamma^{ab}\Gamma^{i}D)\nonumber\\
&& + \frac{4}{\eta^{3}}(\bar{\Lambda}\Gamma_{ef}\bar{\Lambda})(\bar{\Lambda}\Gamma_{ab}\bar{\Lambda})(\bar{\Lambda}\Gamma_{c}{}^{j}R)(\Lambda\Gamma^{ca}\Lambda)(\Lambda\Gamma^{b}\Gamma^{ef}\Gamma^{i}D)\nonumber\\
&=& -\frac{8}{\eta^{3}}(\bar{\Lambda}\Gamma_{ab}\bar{\Lambda})(\bar{\Lambda}\Gamma_{c}{}^{j}R)(\Lambda\Gamma^{ca}\Lambda)(\bar{\Lambda}\Gamma^{if}\bar{\Lambda})(\Lambda\Gamma^{b}{}_{f}D)\nonumber\\
&& + \frac{4}{\eta^{3}}(\bar{\Lambda}\Gamma_{a}{}^{i}\bar{\Lambda})(\bar{\Lambda}\Gamma_{c}{}^{j}R)(\Lambda\Gamma^{ca}\Lambda)(\bar{\Lambda}\Gamma_{ef}\bar{\Lambda})(\Lambda\Gamma^{ef}D)\nonumber\\
&& + \frac{4}{\eta^{2}}(\bar{\Lambda}\Gamma_{c}{}^{j}R)(\bar{\Lambda}\Gamma^{ci}\bar{\Lambda})(\Lambda D) - \frac{4}{\eta^{2}}(\bar{\Lambda}\Gamma^{cj}R)(\bar{\Lambda}\Gamma^{if}\bar{\Lambda})(\Lambda\Gamma_{cf}D) \nonumber \\
&& - \frac{4}{\eta^{2}}(\bar{\Lambda}\Gamma^{cj}R)(\bar{\Lambda}\Gamma_{cf}\bar{\Lambda})(\Lambda\Gamma^{if}D) + \frac{2}{\eta^{2}}(\bar{\Lambda}\Gamma^{ij}R)(\bar{\Lambda}\Gamma^{ab}\bar{\Lambda})(\Lambda\Gamma_{ab}D)\nonumber \\
&& + \frac{2}{\eta^{2}}(\bar{\Lambda}\Gamma_{c}{}^{j}R)(\bar{\Lambda}\Gamma_{ab}\bar{\Lambda})(\Lambda\Gamma^{ciab}D)\nonumber\\
&=& \frac{4}{\eta^{2}}(\bar{\Lambda}R)(\bar{\Lambda}\Gamma^{ij}\bar{\Lambda})(\Lambda D)- \frac{4}{\eta^{2}}(\bar{\Lambda}\Gamma^{cj}R)(\bar{\Lambda}\Gamma^{if}\bar{\Lambda})(\Lambda\Gamma_{cf}D)\nonumber\\
&& + \frac{4}{\eta^{2}}(\bar{\Lambda}R)(\bar{\Lambda}\Gamma^{j}{}_{f}\bar{\Lambda})(\Lambda\Gamma^{if}D) + \frac{2}{\eta^{2}}(\bar{\Lambda}\Gamma^{ij}R)(\bar{\Lambda}\Gamma_{ab}\bar{\Lambda})(\Lambda\Gamma^{ab}D)\nonumber\\
&& + \frac{2}{\eta^{2}}(\bar{\Lambda}\Gamma_{c}{}^{j}R)(\bar{\Lambda}\Gamma_{ab}\bar{\Lambda})(\Lambda\Gamma^{ciab}D)
\end{eqnarray}
Analogously, one finds for $\bar{\Sigma}^{j(3)}_{1}$ that
\begin{eqnarray}
\{\bar{\Sigma}_{0}^{i}, \bar{\Sigma}^{j(3)}_{1}\} &=& \frac{2}{\eta^{2}}(\bar{\Lambda}R)(\bar{\Lambda}\Gamma_{ab}\bar{\Lambda})(\Lambda\Gamma^{j}\Gamma^{ab}\Gamma^{i}D)\nonumber\\
&& + \frac{4}{\eta^{3}}(\bar{\Lambda}\Gamma_{ab}\bar{\Lambda})(\bar{\Lambda}R)(\Lambda\Gamma^{ja}\Lambda)(\bar{\Lambda}\Gamma_{ef}\bar{\Lambda})(\Lambda\Gamma^{b}\Gamma^{ef}\Gamma^{i}D)\nonumber\\
&=& - \frac{8}{\eta^{3}}(\bar{\Lambda}\Gamma_{ab}\bar{\Lambda})(\bar{\Lambda}R)(\Lambda\Gamma^{ja}\Lambda)(\bar{\Lambda}\Gamma^{i}{}_{f}\bar{\Lambda})(\Lambda\Gamma^{bf}D)\nonumber\\
&& + \frac{4}{\eta^{3}}(\bar{\Lambda}\Gamma_{a}{}^{i}\bar{\Lambda})(\bar{\Lambda}R)(\Lambda\Gamma^{ja}\Lambda)(\bar{\Lambda}\Gamma_{ef}\bar{\Lambda})(\Lambda\Gamma^{ef}D)\nonumber\\
&& -\frac{4}{\eta^{2}}(\bar{\Lambda}R)(\bar{\Lambda}\Gamma^{ij}\bar{\Lambda})(\Lambda D) - \frac{4}{\eta^{2}}(\bar{\Lambda}R)(\bar{\Lambda}\Gamma^{j}{}_{f}\bar{\Lambda})(\Lambda\Gamma^{if}D)\nonumber\\
&& - \frac{4}{\eta^{2}}(\bar{\Lambda}R)(\bar{\Lambda}\Gamma^{i}{}_{f}\bar{\Lambda})(\Lambda\Gamma^{jf}D) + \frac{4}{\eta^{2}}\eta^{ij}(\bar{\Lambda}R)(\bar{\Lambda}\Gamma_{ab}\bar{\Lambda})(\Lambda\Gamma^{ab}D)\nonumber\\
&& -\frac{4}{\eta^{2}}(\bar{\Lambda}R)(\bar{\Lambda}\Gamma_{ab}\bar{\Lambda})(\Lambda\Gamma^{ijab}D)
\end{eqnarray}
In this manner, one learns that
\begin{eqnarray}
\{\bar{\Sigma}^{i}_{0}, \bar{\Sigma}^{j}_{1}\} &=& -\frac{4}{\eta^{2}}(\bar{\Lambda}\Gamma_{cj}R)(\bar{\Lambda}\Gamma^{if}\bar{\Lambda})(\Lambda\Gamma^{cf}D) + \frac{2}{\eta^{2}}(\bar{\Lambda}\Gamma^{ij}R)(\bar{\Lambda}\Gamma_{ef}\bar{\Lambda})(\Lambda\Gamma^{ef}D)\nonumber\\
&& + \frac{2}{\eta^{2}}(\bar{\Lambda}\Gamma_{c}{}^{j}R)(\bar{\Lambda}\Gamma_{ef}\bar{\Lambda})(\Lambda\Gamma^{cief}D) - \frac{4}{\eta^{2}}(\bar{\Lambda}R)(\bar{\Lambda}\Gamma^{i}{}_{f}\bar{\Lambda})(\Lambda\Gamma^{jf}D)\nonumber\\
&& + \frac{2}{\eta^{2}}\eta^{ij}(\bar{\Lambda}R)(\bar{\Lambda}\Gamma_{ef}\bar{\Lambda})(\Lambda\Gamma^{ef}D) - \frac{2}{\eta}(\bar{\Lambda}R)(\bar{\Lambda}\Gamma_{ef}\bar{\Lambda})(\Lambda\Gamma^{ijef}D)\label{endeq500a}
\end{eqnarray}
In a similar way, one obtains
\begin{eqnarray}
\{\bar{\Sigma}^{i}_{1}, \bar{\Sigma}^{j}_{0}\} &=& -\frac{4}{\eta^{2}}(\bar{\Lambda}\Gamma_{ci}R)(\bar{\Lambda}\Gamma^{jf}\bar{\Lambda})(\Lambda\Gamma^{cf}D) + \frac{2}{\eta^{2}}(\bar{\Lambda}\Gamma^{ji}R)(\bar{\Lambda}\Gamma_{ab}\bar{\Lambda})(\Lambda\Gamma^{ab}D)\nonumber\\
&& + \frac{2}{\eta^{2}}(\bar{\Lambda}\Gamma_{c}{}^{i}R)(\bar{\Lambda}\Gamma_{ab}\bar{\Lambda})(\Lambda\Gamma^{cjab}D) - \frac{4}{\eta^{2}}(\bar{\Lambda}R)(\bar{\Lambda}\Gamma^{j}{}_{f}\bar{\Lambda})(\Lambda\Gamma^{if}D)\nonumber\\
&& + \frac{2}{\eta^{2}}\eta^{ji}(\bar{\Lambda}R)(\bar{\Lambda}\Gamma_{ab}\bar{\Lambda})(\Lambda\Gamma^{ab}D) - \frac{2}{\eta}(\bar{\Lambda}R)(\bar{\Lambda}\Gamma_{ab}\bar{\Lambda})(\Lambda\Gamma^{jiab}D) \nonumber\\\label{endeq501}
\end{eqnarray}
Therefore, the sum of \eqref{endeq500a} and \eqref{endeq501} gives
\begin{eqnarray}
\{\bar{\Sigma}^{i}_{0}, \bar{\Sigma}^{j}_{1}\} + \{\bar{\Sigma}^{i}_{1}, \bar{\Sigma}^{j}_{0}\} &=& -\frac{4}{\eta^{2}}(\bar{\Lambda}\Gamma^{if}\bar{\Lambda})(\bar{\Lambda}\Gamma^{cj}R)(\Lambda\Gamma_{cf}D) -\frac{4}{\eta^{2}}(\bar{\Lambda}\Gamma^{jf}\bar{\Lambda})(\bar{\Lambda}\Gamma^{ci}R)(\Lambda\Gamma_{cf}D) \nonumber \\
&& + \frac{2}{\eta^{2}}(\bar{\Lambda}\Gamma_{ab}\bar{\Lambda})(\bar{\Lambda}\Gamma_{c}{}^{j}R)(\Lambda\Gamma^{ciab}D) + \frac{2}{\eta^{2}}(\bar{\Lambda}\Gamma_{ab}\bar{\Lambda})(\bar{\Lambda}\Gamma_{c}{}^{i}R)(\Lambda\Gamma^{cjab}D)\nonumber \\
&&-\frac{4}{\eta^{2}}(\bar{\Lambda}\Gamma^{i}{}_{f}\bar{\Lambda})(\bar{\Lambda}R)(\Lambda\Gamma^{jf}D) -\frac{4}{\eta^{2}}(\bar{\Lambda}\Gamma^{j}{}_{f}\bar{\Lambda})(\bar{\Lambda}R)(\Lambda\Gamma^{if}D)\nonumber\\
&& +\frac{4}{\eta^{2}}\eta^{ij}(\bar{\Lambda}\Gamma^{ab}\bar{\Lambda})(\bar{\Lambda}R)(\Lambda\Gamma_{ab}D)
\end{eqnarray}
Another convenient expression for $\bar{\Sigma}_{1}^{j}$, which will turn out to be useful for us, can be written as follows
\begin{eqnarray}
\bar{\Sigma}^{j}_{1} &=& \frac{2}{\eta^{2}}(\bar{\Lambda}\Gamma_{ab}\bar{\Lambda})(\bar{\Lambda}\Gamma_{cd}R)(\Lambda\Gamma^{abckj}\Lambda)N^{d}{}_{k} - \frac{8}{\eta^{2}}(\bar{\Lambda}\Gamma_{ab}\bar{\Lambda})(\bar{\Lambda}\Gamma_{c}{}^{j}R)(\Lambda\Gamma^{bc}\Lambda)(\Lambda\Gamma^{a}W) \nonumber\\
&&-\frac{4}{\eta}(\bar{\Lambda}\Gamma^{cj}R)(\Lambda\Gamma_{c}W) - \frac{4}{\eta}(\bar{\Lambda}R)(\Lambda\Gamma^{j}W) - \frac{8}{\eta^{2}}(\bar{\Lambda}\Gamma_{ab}\bar{\Lambda})(\bar{\Lambda}R)(\Lambda\Gamma^{ja}\Lambda)(\Lambda\Gamma^{b}W)\nonumber \\\label{endeq502}
\end{eqnarray}
Lets us call $Y_{1}^{j}$ to the first term of this expression and expand it as follows
\begin{eqnarray}
Y_{1}^{j} &=& (\frac{6}{24})(\bar{\Lambda}\Gamma_{ab}\bar{\Lambda})(\bar{\Lambda}\Gamma_{cd}R)[4(\Lambda\Gamma^{cj}\Lambda)(\Lambda\Gamma^{ab}\Gamma^{d}W) - 4(\Lambda\Gamma^{ca}\Lambda)(\Lambda\Gamma^{jb}\Gamma^{d}W)\nonumber\\
&& + 4(\Lambda\Gamma^{cb}\Lambda)(\Lambda\Gamma^{ja}\Gamma^{d}W) + 4(\Lambda\Gamma^{ab}\Lambda)(\Lambda\Gamma^{cj}\Gamma^{d}W)\nonumber\\
&& - 4(\Lambda\Gamma^{jb}\Lambda)(\Lambda\Gamma^{ca}\Gamma^{d}W) + 4(\Lambda\Gamma^{ja}\Lambda)(\Lambda\Gamma^{cb}\Gamma^{d}W)] \nonumber \\
&=& \frac{2}{\eta^{2}}(\bar{\Lambda}\Gamma_{ab}\bar{\Lambda})(\bar{\Lambda}\Gamma_{cd}R)[(\Lambda\Gamma^{cj}\Lambda)(\Lambda\Gamma^{ab}\Gamma^{d}W) - 2(\Lambda\Gamma^{ca}\Lambda)(\Lambda\Gamma^{jb}\Gamma^{d}W)\nonumber\\
&& + (\Lambda\Gamma^{ab}\Lambda)(\Lambda\Gamma^{cj}\Gamma^{d}W) - 2(\Lambda\Gamma^{jb}\Lambda)(\Lambda\Gamma^{ca}\Gamma^{d}W)]\nonumber\\
&=& \frac{2}{\eta^{2}}(\bar{\Lambda}\Gamma_{ab}\bar{\Lambda})(\bar{\Lambda}\Gamma_{cd}R)[2\eta^{bd}(\Lambda\Gamma^{cj}\Lambda)(\Lambda\Gamma^{a}W) + (\Lambda\Gamma^{cj}\Lambda)(\Lambda\Gamma^{abd}W)\nonumber\\
&& + 2\eta^{dj}(\Lambda\Gamma^{ca}\Lambda)(\Lambda\Gamma^{b}W) -2\eta^{bd}(\Lambda\Gamma^{ca}\Lambda)(\Lambda\Gamma^{j}W) - 2(\Lambda\Gamma^{ca}\Lambda)(\Lambda\Gamma^{bdj}W)\nonumber\\
&& + \eta^{dj}(\Lambda\Gamma^{ab}\Lambda)(\Lambda\Gamma^{c}W) -\eta^{cd}(\Lambda\Gamma^{ab}\Lambda)(\Lambda\Gamma^{j}W) - (\Lambda\Gamma^{ab}\Lambda)(\Lambda\Gamma^{cdj}W)\nonumber\\
&& + 2\eta^{cd}(\Lambda\Gamma^{jb}\Lambda)(\Lambda\Gamma^{a}W) -2\eta^{ad}(\Lambda\Gamma^{jb}\Lambda)(\Lambda\Gamma^{c}W) + 2(\Lambda\Gamma^{jb}\Lambda)(\Lambda\Gamma^{acd}W)]\nonumber\\
&=&\frac{4}{\eta^{2}}(\bar{\Lambda}\Gamma_{ac}\bar{\Lambda})(\bar{\Lambda}R)(\Lambda\Gamma^{cj}\Lambda)(\Lambda\Gamma^{a}W) + \frac{2}{\eta^{2}}(\bar{\Lambda}\Gamma_{ab}\bar{\Lambda})(\bar{\Lambda}\Gamma_{cd}R)(\Lambda\Gamma^{cj}\Lambda)(\Lambda\Gamma^{abd}W) \nonumber\\
&& + \frac{4}{\eta^{2}}(\bar{\Lambda}\Gamma^{ab}\bar{\Lambda})(\bar{\Lambda}\Gamma^{cj}R)(\Lambda\Gamma_{ca}\Lambda)(\Lambda\Gamma_{b}W) + \frac{4}{\eta}(\bar{\Lambda}R)(\Lambda\Gamma^{j}W)\nonumber\\
&& - \frac{4}{\eta^{2}}(\bar{\Lambda}\Gamma_{ab}\bar{\Lambda})(\bar{\Lambda}\Gamma_{cd}R)(\Lambda\Gamma^{ca}\Lambda)(\Lambda\Gamma^{bdj}W) + \frac{2}{\eta}(\bar{\Lambda}\Gamma^{cj}R)(\Lambda\Gamma_{c}W)\nonumber\\
&& - \frac{2}{\eta}(\bar{\Lambda}\Gamma_{cd}R)(\Lambda\Gamma^{cdj}W) - \frac{4}{\eta^{2}}(\bar{\Lambda}\Gamma_{ab}\bar{\Lambda})(\bar{\Lambda}R)(\Lambda\Gamma^{jb}\Lambda)(\Lambda\Gamma^{c}W)\nonumber\\
&& + \frac{4}{\eta^{2}}(\bar{\Lambda}\Gamma_{ab}\bar{\Lambda})(\bar{\Lambda}\Gamma_{cd}R)(\Lambda\Gamma^{jb}\Lambda)(\Lambda\Gamma^{acd}W)\nonumber\\
&=&\frac{4}{\eta^{2}}(\bar{\Lambda}\Gamma_{ac}\bar{\Lambda})(\bar{\Lambda}R)(\Lambda\Gamma^{cj}\Lambda)(\Lambda\Gamma^{a}W) + \frac{2}{\eta^{2}}(\bar{\Lambda}\Gamma_{ab}\bar{\Lambda})(\bar{\Lambda}\Gamma_{cd}R)(\Lambda\Gamma^{cj}\Lambda)(\Lambda\Gamma^{abd}W) \nonumber\\
&& + \frac{2}{\eta^{2}}(\bar{\Lambda}\Gamma^{ac}\bar{\Lambda})(\bar{\Lambda}\Gamma^{bj}R)(\Lambda\Gamma_{ca}\Lambda)(\Lambda\Gamma_{b}W) +  \frac{2}{\eta^{2}}(\bar{\Lambda}\Gamma^{bj}\bar{\Lambda})(\bar{\Lambda}\Gamma^{ac}R)(\Lambda\Gamma_{ca}\Lambda)(\Lambda\Gamma_{b}W) \nonumber\\
&& + \frac{4}{\eta^{2}}(\bar{\Lambda}\Gamma^{cj}\bar{\Lambda})(\bar{\Lambda}\Gamma^{ba}R)(\Lambda\Gamma_{ca}\Lambda)(\Lambda\Gamma_{b}W) + \frac{4}{\eta}(\bar{\Lambda}R)(\Lambda\Gamma^{j}W)\nonumber\\
&& - \frac{1}{\eta^{2}}(\bar{\Lambda}\Gamma_{ac}\bar{\Lambda})(\bar{\Lambda}\Gamma_{bd}R)(\Lambda\Gamma^{ca}\Lambda)(\Lambda\Gamma^{bdj}W) - \frac{1}{\eta^{2}}(\bar{\Lambda}\Gamma_{bd}\bar{\Lambda})(\bar{\Lambda}\Gamma_{ac}R)(\Lambda\Gamma^{ca}\Lambda)(\Lambda\Gamma^{bdj}W)\nonumber\\
&& + \frac{2}{\eta}(\bar{\Lambda}\Gamma^{cj}R)(\Lambda\Gamma_{c}W) - \frac{2}{\eta}(\bar{\Lambda}\Gamma_{cd}R)(\Lambda\Gamma^{cdj}W)\nonumber\\
&& -\frac{4}{\eta^{2}}(\bar{\Lambda}\Gamma_{cb}\bar{\Lambda})(\bar{\Lambda}R)(\Lambda\Gamma^{jb}\Lambda)(\Lambda\Gamma^{c}W) - \frac{4}{\eta^{2}}(\bar{\Lambda}\Gamma_{ab}\bar{\Lambda})(\bar{\Lambda}\Gamma_{cd}R)(\Lambda\Gamma^{cj}\Lambda)(\Lambda\Gamma^{abd}W)\nonumber\\
&=& \frac{8}{\eta^{2}}(\bar{\Lambda}\Gamma_{ac}\bar{\Lambda})(\bar{\Lambda}R)(\Lambda\Gamma^{cj}\Lambda)(\Lambda\Gamma^{a}W) - \frac{2}{\eta^{2}}(\bar{\Lambda}\Gamma_{ab}\bar{\Lambda})(\bar{\Lambda}\Gamma_{cd}R)(\Lambda\Gamma^{cj}\Lambda)(\Lambda\Gamma^{abd}W)\nonumber\\
&& - \frac{1}{\eta}(\bar{\Lambda}\Gamma_{cd}R)(\Lambda\Gamma^{cdj}W) + \frac{2}{\eta^{2}}(\bar{\Lambda}\Gamma^{bj}\bar{\Lambda})(\bar{\Lambda}\Gamma^{ac}R)(\Lambda\Gamma_{ca}\Lambda)(\Lambda\Gamma_{b}W)\nonumber\\
&& + \frac{4}{\eta^{2}}(\bar{\Lambda}\Gamma^{cj}\bar{\Lambda})(\bar{\Lambda}\Gamma^{ba}R)(\Lambda\Gamma_{ca}\Lambda)(\Lambda\Gamma_{b}W) + \frac{4}{\eta}(\bar{\Lambda}R)(\Lambda\Gamma^{j}W)\nonumber\\
&& - \frac{1}{\eta^{2}}(\bar{\Lambda}\Gamma_{bd}\bar{\Lambda})(\bar{\Lambda}\Gamma_{ac}R)(\Lambda\Gamma^{ca}\Lambda)(\Lambda\Gamma^{bdj}W)
\end{eqnarray}
Plugging this result into eqn. \eqref{endeq502} and simplifying, one obtains
\begin{eqnarray}
\bar{\Sigma}_{1}^{j} &=& - \frac{2}{\eta^{2}}(\bar{\Lambda}\Gamma_{ab}\bar{\Lambda})(\bar{\Lambda}\Gamma_{cd}R)(\Lambda\Gamma^{cj}\Lambda)(\Lambda\Gamma^{abd}W) - \frac{1}{\eta}(\bar{\Lambda}\Gamma_{cd}R)(\Lambda\Gamma^{cdj}W)\nonumber\\
&& + \frac{2}{\eta^{2}}(\bar{\Lambda}\Gamma^{bj}\bar{\Lambda})(\bar{\Lambda}\Gamma^{ac}R)(\Lambda\Gamma_{ca}\Lambda)(\Lambda\Gamma_{b}W) + \frac{4}{\eta^{2}}(\bar{\Lambda}\Gamma^{cj}\bar{\Lambda})(\bar{\Lambda}\Gamma^{ba}R)(\Lambda\Gamma_{ca}\Lambda)(\Lambda\Gamma_{b}W)\nonumber\\
&& - \frac{1}{\eta^{2}}(\bar{\Lambda}\Gamma_{bd}\bar{\Lambda})(\bar{\Lambda}\Gamma_{ac}R)(\Lambda\Gamma^{ca}\Lambda)(\Lambda\Gamma^{bdj}W)  - \frac{8}{\eta^{2}}(\bar{\Lambda}\Gamma_{ab}\bar{\Lambda})(\bar{\Lambda}\Gamma_{c}^{\hspace{2mm}j}R)(\Lambda\Gamma^{bc}\Lambda)(\Lambda\Gamma^{a}W)\nonumber\\
&& -\frac{4}{\eta}(\bar{\Lambda}\Gamma^{cj}R)(\Lambda\Gamma_{c}W)\nonumber\\
&=& - \frac{2}{\eta^{2}}(\bar{\Lambda}\Gamma_{ab}\bar{\Lambda})(\bar{\Lambda}\Gamma_{cd}R)(\Lambda\Gamma^{cj}\Lambda)(\Lambda\Gamma^{abd}W) - \frac{1}{\eta}(\bar{\Lambda}\Gamma_{cd}R)(\Lambda\Gamma^{cdj}W)\nonumber\\
&& + \frac{2}{\eta^{2}}(\bar{\Lambda}\Gamma^{bj}\bar{\Lambda})(\bar{\Lambda}\Gamma^{ac}R)(\Lambda\Gamma_{ca}\Lambda)(\Lambda\Gamma_{b}W) + \frac{4}{\eta^{2}}(\bar{\Lambda}\Gamma^{cj}\bar{\Lambda})(\bar{\Lambda}\Gamma^{ba}R)(\Lambda\Gamma_{ca}\Lambda)(\Lambda\Gamma_{b}W)\nonumber\\
&& - \frac{1}{\eta^{2}}(\bar{\Lambda}\Gamma_{bd}\bar{\Lambda})(\bar{\Lambda}\Gamma_{ac}R)(\Lambda\Gamma^{ca}\Lambda)(\Lambda\Gamma^{bdj}W) + \frac{4}{\eta^{2}}(\bar{\Lambda}\Gamma^{ab}\bar{\Lambda})(\bar{\Lambda}\Gamma^{fj}R)(\Lambda\Gamma_{ab}\Lambda)(\Lambda\Gamma^{f}W) \nonumber\\
&& + \frac{4}{\eta^{2}}(\bar{\Lambda}\Gamma^{aj}\bar{\Lambda})(\bar{\Lambda}\Gamma^{bc}R)(\Lambda\Gamma_{bc}\Lambda)(\Lambda\Gamma_{a}W)  + \frac{8}{\eta^{2}}(\bar{\Lambda}\Gamma^{cj}\bar{\Lambda})(\bar{\Lambda}\Gamma^{ab}R)(\Lambda\Gamma_{bc}\Lambda)(\Lambda\Gamma_{a}W)\nonumber\\
&& - \frac{4}{\eta}(\bar{\Lambda}\Gamma^{cj}R)(\Lambda\Gamma_{c}W)\nonumber\\
&=& - \frac{2}{\eta^{2}}(\bar{\Lambda}\Gamma_{ab}\bar{\Lambda})(\bar{\Lambda}\Gamma_{cd}R)(\Lambda\Gamma^{cj}\Lambda)(\Lambda\Gamma^{abd}W) - \frac{1}{\eta}(\bar{\Lambda}\Gamma_{cd}R)(\Lambda\Gamma^{cdj}W)\nonumber\\
&& + \frac{2}{\eta^{2}}(\bar{\Lambda}\Gamma^{bj}\bar{\Lambda})(\bar{\Lambda}\Gamma^{ca}R)(\Lambda\Gamma_{ca}\Lambda)(\Lambda\Gamma_{b}W) 
- \frac{4}{\eta^{2}}(\bar{\Lambda}\Gamma^{cj}\bar{\Lambda})(\bar{\Lambda}\Gamma^{ba}R)(\Lambda\Gamma_{ca}\Lambda)(\Lambda\Gamma_{b}W)\nonumber\\
&& + \frac{1}{\eta^{2}}(\bar{\Lambda}\Gamma_{bd}\bar{\Lambda})(\bar{\Lambda}\Gamma_{ac}R)(\Lambda\Gamma^{ac}\Lambda)(\Lambda\Gamma^{bdj}W)
\end{eqnarray}
This expression is invariant under the gauge symmetry generated by the pure spinor constraint, as it should be. Now, let us make the following definitions:
\begin{eqnarray}
W_{1}^{j} &=& -\frac{2}{\eta^{2}}(\bar{\Lambda}\Gamma_{ab}\bar{\Lambda})(\bar{\Lambda}\Gamma_{cd}R)(\Lambda\Gamma^{cj}\Lambda)(\Lambda\Gamma^{abd}W)\\
W_{2}^{j} &=& -\frac{1}{\eta}(\bar{\Lambda}\Gamma_{cd}R)(\Lambda\Gamma^{cdj}W)\\
W_{3}^{j} &=& \frac{2}{\eta^{2}}(\bar{\Lambda}\Gamma^{bj}\bar{\Lambda})(\bar{\Lambda}\Gamma^{ca}R)(\Lambda\Gamma_{ca}\Lambda)(\Lambda\Gamma_{b}W)\\
W_{4}^{j} &=& -\frac{4}{\eta^{2}}(\bar{\Lambda}\Gamma^{cj}\bar{\Lambda})(\bar{\Lambda}\Gamma^{ba}R)(\Lambda\Gamma_{ca}\Lambda)(\Lambda\Gamma_{b}W)\\
W_{5}^{j} &=& \frac{1}{\eta^{2}}(\bar{\Lambda}\Gamma_{bd}\bar{\Lambda})(\bar{\Lambda}\Gamma_{ac}R)(\Lambda\Gamma^{ac}\Lambda)(\Lambda\Gamma^{bdj}W)
\end{eqnarray}
Hence, we should calculate the anticommutator between each pair of these variables $W_{\{1,2,3,4,5\}}^{j}$. Explicitly, one has that
\begin{eqnarray}
\{W_{1}^{i}, W_{1}^{j}\} &=& \{-\frac{2}{\eta^{2}}(\bar{\Lambda}\Gamma_{ef}\bar{\Lambda})(\bar{\Lambda}\Gamma_{gh}R)(\Lambda\Gamma^{gi}\Lambda)(\Lambda\Gamma^{efh}W),-\frac{2}{\eta^{2}}(\bar{\Lambda}\Gamma_{ab}\bar{\Lambda})(\bar{\Lambda}\Gamma_{cd}R)(\Lambda\Gamma^{cj}\Lambda)(\Lambda\Gamma^{abd}W)\} \nonumber \\
&=& \frac{4}{\eta^{4}}(\bar{\Lambda}\Gamma_{ef}\bar{\Lambda})(\bar{\Lambda}\Gamma_{gh}R)(\bar{\Lambda}\Gamma_{ab}\bar{\Lambda})(\bar{\Lambda}\Gamma_{cd}R)[(\Lambda\Gamma^{gi}\Lambda)(\Lambda\Gamma^{efh}W), (\Lambda\Gamma^{cj}\Lambda)(\Lambda\Gamma^{abd}W)] \nonumber \\
&=& \frac{4}{\eta^{4}}(\bar{\Lambda}\Gamma_{ef}\bar{\Lambda})(\bar{\Lambda}\Gamma_{gh}R)(\bar{\Lambda}\Gamma_{ab}\bar{\Lambda})(\bar{\Lambda}\Gamma_{cd}R)\{(\Lambda\Gamma^{gi}\Lambda)(\Lambda\Gamma^{cj}\Lambda)[(\Lambda\Gamma^{efh}W), (\Lambda\Gamma^{abd}W)] \nonumber\\
&& + (\Lambda\Gamma^{gi}\Lambda)[(\Lambda\Gamma^{efh}W),(\Lambda\Gamma^{cj}\Lambda)](\Lambda\Gamma^{abd}W) + (\Lambda\Gamma^{cj}\Lambda)[(\Lambda\Gamma^{gi}\Lambda),(\Lambda\Gamma^{abd}W)](\Lambda\Gamma^{efh}W)\}\nonumber\\
&=& \frac{64}{\eta^{4}}(\bar{\Lambda}\Gamma_{ef}\bar{\Lambda})(\bar{\Lambda}\Gamma_{gh}R)(\bar{\Lambda}\Gamma_{ab}\bar{\Lambda})(\bar{\Lambda}\Gamma_{cd}R)(\Lambda\Gamma^{gi}\Lambda)(\Lambda\Gamma^{cj}\Lambda)[\delta^{eh}_{ad}N^{fd}]\nonumber\\
&=& -\frac{32}{\eta^{4}}(\bar{\Lambda}\Gamma_{df}\bar{\Lambda})(\bar{\Lambda}\Gamma_{ga}R)(\bar{\Lambda}\Gamma_{ab}\bar{\Lambda})(\bar{\Lambda}\Gamma_{cd}R)(\Lambda\Gamma^{gi}\Lambda)(\Lambda\Gamma^{cj}\Lambda)N^{fd}\nonumber\\
&=& 0
\end{eqnarray}
because of the identities \eqref{newapp41} and $(\bar{\Lambda}R)(\bar{\Lambda}R)=0$.
\begin{eqnarray}
\{W_{2}^{i}, W_{2}^{j}\} &=& \{-\frac{1}{\eta}(\bar{\Lambda}\Gamma_{ab}R)(\Lambda\Gamma^{abi}W), -\frac{1}{\eta}(\bar{\Lambda}\Gamma_{cd}R)(\Lambda\Gamma^{cdj}W)\}\nonumber\\
&=& \frac{1}{\eta^{2}}(\bar{\Lambda}\Gamma_{ab}R)(\bar{\Lambda}\Gamma_{cd}R)[(\Lambda\Gamma^{abi}W), (\Lambda\Gamma^{cdj}W)]\nonumber\\
&=& \frac{1}{\eta^{2}}(\bar{\Lambda}\Gamma_{ab}R)(\bar{\Lambda}\Gamma_{cd}R)(4\delta^{ab}_{cd}N^{ij} - 8\delta^{ab}_{cj}N^{id} - 8\delta^{ai}_{cd}N^{bj} - 16\delta^{bi}_{cj}N^{ad})\nonumber\\
&=& -\frac{8}{\eta^{2}}(\bar{\Lambda}\Gamma^{cj}R)(\bar{\Lambda}\Gamma_{cd}R)N^{id} - \frac{8}{\eta^{2}}(\bar{\Lambda}\Gamma_{ab}R)(\bar{\Lambda}\Gamma^{ai}R)N^{bj}\nonumber\\
&& -\frac{8}{\eta^{2}}(\bar{\Lambda}\Gamma_{a}{}^{c}R)(\bar{\Lambda}\Gamma_{cd}R)\eta^{ij}N^{ad} + \frac{8}{\eta^{2}}(\bar{\Lambda}\Gamma^{aj}R)(\bar{\Lambda}\Gamma^{id}R)N_{ad}
\end{eqnarray}
The use of the identity \eqref{newapp12} allows us to write
\begin{eqnarray}
\{W_{2}^{i}, W_{2}^{j}\} &=& -\frac{8}{\eta^{2}}[(\bar{\Lambda}\Gamma^{jd}R)(\bar{\Lambda}R) + \frac{1}{2}(\bar{\Lambda}\Gamma^{jd}\bar{\Lambda})(RR)]N^{i}{}_{d}\nonumber \\
&& +\frac{8}{\eta^{2}}[(\bar{\Lambda}\Gamma^{ai}R)(\bar{\Lambda}R) + \frac{1}{2}(\bar{\Lambda}\Gamma^{ai}\bar{\Lambda})(RR)]N^{j}{}_{a}\nonumber \\
&& +\frac{8}{\eta^{2}}\eta^{ij}[(\bar{\Lambda}\Gamma^{ad}R)(\bar{\Lambda}R) + \frac{1}{2}(\bar{\Lambda}\Gamma^{ad}\bar{\Lambda})(RR)]N_{ad}\nonumber \\
&& + \frac{8}{\eta^{2}}(\bar{\Lambda}\Gamma^{aj}R)(\bar{\Lambda}\Gamma^{id}R)N_{ad}\nonumber\\
&=& -\frac{8}{\eta^{2}}(\bar{\Lambda}\Gamma^{jd}R)(\bar{\Lambda}R)N^{i}{}_{d} - \frac{4}{\eta^{2}}(\bar{\Lambda}\Gamma^{jd}\bar{\Lambda})(RR)N^{i}{}_{d}\nonumber\\
&& +\frac{8}{\eta^{2}}(\bar{\Lambda}\Gamma^{di}R)(\bar{\Lambda}R)N^{j}{}_{d} + \frac{4}{\eta^{2}}(\bar{\Lambda}\Gamma^{di}\bar{\Lambda})(RR)N^{j}{}_{d}\nonumber\\
&& + \frac{8}{\eta^{2}}\eta^{ij}(\bar{\Lambda}\Gamma^{ad}R)(\bar{\Lambda}R)N_{ad} + \frac{4}{\eta^{2}}\eta^{ij}(\bar{\Lambda}\Gamma^{ad}\bar{\Lambda})(RR)N_{ad}\nonumber\\
&& + \frac{8}{\eta^{2}}(\bar{\Lambda}\Gamma^{aj}R)(\bar{\Lambda}\Gamma^{id}R)N_{ad}
\end{eqnarray}
\begin{eqnarray}
\{W_{3}^{i}, W_{3}^{j}\} &=& \{\frac{2}{\eta^{2}}(\bar{\Lambda}\Gamma^{ei}\bar{\Lambda})(\bar{\Lambda}\Gamma^{gf}R)(\Lambda\Gamma_{gf}\Lambda)(\Lambda\Gamma_{e}W), \frac{2}{\eta^{2}}(\bar{\Lambda}\Gamma^{bj}\bar{\Lambda})(\bar{\Lambda}\Gamma^{ca}R)(\Lambda\Gamma_{ca}\Lambda)(\Lambda\Gamma_{b}W)\}\nonumber\\
&=& -\frac{8}{\eta^{2}}(\bar{\Lambda}\Gamma^{ei}\bar{\Lambda})(\bar{\Lambda}\Gamma^{fg}R)(\bar{\Lambda}\Gamma^{bj})\bar{\Lambda}(\bar{\Lambda}\Gamma^{ca}R)(\Lambda\Gamma_{fg}\Lambda)(\Lambda\Gamma_{ca}\Lambda)N_{eb}\nonumber\\
&=& 0
\end{eqnarray}
because of the identities \eqref{newapp42} and $(\Lambda\Gamma_{mn}\Lambda)(\bar{\Lambda}\Gamma^{mn}R)(\Lambda\Gamma_{ab}\Lambda)(\bar{\Lambda}\Gamma^{ab}R) = 0$.
\begin{eqnarray}
\{W_{4}^{i}, W_{4}^{j}\} &=& \{-\frac{4}{\eta^{2}}(\bar{\Lambda}\Gamma^{ei}\bar{\Lambda})(\bar{\Lambda}\Gamma^{gf}R)(\Lambda\Gamma_{ef}\Lambda)(\Lambda\Gamma_{g}W), -\frac{4}{\eta^{2}}(\bar{\Lambda}\Gamma^{cj}\bar{\Lambda})(\bar{\Lambda}\Gamma^{ba}R)(\Lambda\Gamma_{ca}\Lambda)(\Lambda\Gamma_{b}W)\}\nonumber\\
&=& -\frac{32}{\eta^{4}}(\bar{\Lambda}\Gamma^{ei}\bar{\Lambda})(\bar{\Lambda}\Gamma^{gf}R)(\bar{\Lambda}\Gamma^{cj}\bar{\Lambda})(\bar{\Lambda}\Gamma^{ba}R)(\Lambda\Gamma_{ef}\Lambda)(\Lambda\Gamma_{ca}\Lambda)N_{gb}
\end{eqnarray}
which follows directly from eqn. \eqref{newapp42}
\begin{eqnarray}
\{W_{5}^{i}, W_{5}^{j}\} &=& \{\frac{1}{\eta^{2}}(\bar{\Lambda}\Gamma_{ef}\bar{\Lambda})(\bar{\Lambda}\Gamma^{gh}R)(\Lambda\Gamma_{gh}\Lambda)(\Lambda\Gamma^{efi}W), \frac{1}{\eta^{2}}(\bar{\Lambda}\Gamma_{bd}\bar{\Lambda})(\bar{\Lambda}\Gamma^{ac}R)(\Lambda\Gamma_{ac}\Lambda)(\Lambda\Gamma^{bdj}W)\}\nonumber\\
&=& \frac{1}{\eta^{4}}(\bar{\Lambda}\Gamma_{ef}\bar{\Lambda})(\bar{\Lambda}\Gamma_{bd}\bar{\Lambda})(\bar{\Lambda}\Gamma^{gh}R)(\bar{\Lambda}\Gamma^{ac}R)(\Lambda\Gamma_{gh}\Lambda)(\Lambda\Gamma_{ac}\Lambda)[(\Lambda\Gamma^{efi}W), (\Lambda\Gamma^{bdj}W)]\nonumber\\ 
&=& \frac{8}{\eta^{4}}(\bar{\Lambda}\Gamma^{ej}\bar{\Lambda})(\bar{\Lambda}\Gamma^{id}\bar{\Lambda})(\bar{\Lambda}\Gamma^{gh}R)(\bar{\Lambda}\Gamma^{ac}R)(\Lambda\Gamma_{gh}\Lambda)(\Lambda\Gamma_{ac}\Lambda)N_{ed}\nonumber\\
&=& 0
\end{eqnarray}
because of the identity $(\Lambda\Gamma_{mn}\Lambda)(\bar{\Lambda}\Gamma^{mn}R)(\Lambda\Gamma_{ab}\Lambda)(\bar{\Lambda}\Gamma^{ab}R) = 0$.

\begin{eqnarray}
\{W_{1}^{i}, W_{2}^{j}\} &=& \{-\frac{2}{\eta^{2}}(\bar{\Lambda}\Gamma_{ab}\bar{\Lambda})(\bar{\Lambda}\Gamma_{ef}R)(\Lambda\Gamma^{ei}\Lambda)(\Lambda\Gamma^{abf}W),-\frac{1}{\eta}(\bar{\Lambda}\Gamma_{cd}R)(\Lambda\Gamma^{cdj}W)\}\nonumber\\
&=& \frac{2}{\eta^{3}}(\bar{\Lambda}\Gamma_{ab}\bar{\Lambda})(\bar{\Lambda}\Gamma_{ef}R)(\bar{\Lambda}\Gamma_{cd}R)[(\Lambda\Gamma^{ei}\Lambda)(\Lambda\Gamma^{abf}W), (\Lambda\Gamma^{cdj}W)]\nonumber\\
&=& \frac{2}{\eta^{3}}(\bar{\Lambda}\Gamma_{ab}\bar{\Lambda})(\bar{\Lambda}\Gamma_{ef}R)(\bar{\Lambda}\Gamma_{cd}R)\{(\Lambda\Gamma^{ei}\Lambda)[(\Lambda\Gamma^{abf}W), (\Lambda\Gamma^{cdj}W)] \nonumber\\
&& + [(\Lambda\Gamma^{ei}\Lambda), (\Lambda\Gamma^{cdj}W)](\Lambda\Gamma^{abf}W)\}\nonumber\\
&=& \frac{2}{\eta^{3}}(\bar{\Lambda}\Gamma_{ab}\bar{\Lambda})(\bar{\Lambda}\Gamma_{ef}R)(\bar{\Lambda}\Gamma_{cd}R)(\Lambda\Gamma^{ei}\Lambda)[-8\delta^{ab}_{cj}N^{fd} + 8\delta^{bf}_{cd}N^{aj} - 16\delta^{bf}_{cj}N^{ad}]\nonumber\\
&& +\frac{4}{\eta^{3}}(\bar{\Lambda}\Gamma_{ab}\bar{\Lambda})(\bar{\Lambda}\Gamma_{ef}R)(\bar{\Lambda}\Gamma_{cd}R)(\Lambda\Gamma^{cdjei}\Lambda)(\Lambda\Gamma^{abf}W)\nonumber\\
&=& -\frac{16}{\eta^{3}}(\bar{\Lambda}\Gamma^{cj}\bar{\Lambda})(\bar{\Lambda}\Gamma_{ef}R)(\bar{\Lambda}\Gamma_{cd}R)(\Lambda\Gamma^{ei}\Lambda)N^{fd}\nonumber\\
&& +\frac{16}{\eta^{3}}(\bar{\Lambda}\Gamma_{ab}\bar{\Lambda})(\bar{\Lambda}\Gamma_{ef}R)(\bar{\Lambda}\Gamma^{bf}R)(\Lambda\Gamma^{ei}\Lambda)N^{aj}\nonumber\\
&& - \frac{16}{\eta^{3}}[(\bar{\Lambda}\Gamma_{ac}\bar{\Lambda})(\bar{\Lambda}\Gamma_{e}{}^{j}R)(\bar{\Lambda}\Gamma^{cd}R)(\Lambda\Gamma^{ei}\Lambda) - (\bar{\Lambda}\Gamma_{a}{}^{j}\bar{\Lambda})(\bar{\Lambda}\Gamma_{ec}R)(\bar{\Lambda}\Gamma^{cd}R)(\Lambda\Gamma^{ei}\Lambda)]N^{a}{}_{d}\nonumber \\
&& +\frac{4}{\eta^{3}}(\bar{\Lambda}\Gamma_{ab}\bar{\Lambda})(\bar{\Lambda}\Gamma_{ef}R)(\bar{\Lambda}\Gamma_{cd}R)(\Lambda\Gamma^{cdjei}\Lambda)(\Lambda\Gamma^{abf}W)\nonumber\\
&=& \frac{32}{\eta^{3}}(\bar{\Lambda}\Gamma_{d}{}^{j}\bar{\Lambda})(\bar{\Lambda}\Gamma_{ef}R)(\bar{\Lambda}R)(\Lambda\Gamma^{ei}\Lambda)N^{fd} + \frac{16}{\eta^{3}}(\bar{\Lambda}\Gamma_{fd}\bar{\Lambda})(\bar{\Lambda}\Gamma^{ej}R)(\bar{\Lambda}R)(\Lambda\Gamma_{e}{}^{i}\Lambda)N^{fd}\nonumber\\
&& -\frac{16}{\eta^{3}}(\bar{\Lambda}\Gamma_{d}{}^{j}\bar{\Lambda})(\bar{\Lambda}\Gamma_{ef}\bar{\Lambda})(RR)(\Lambda\Gamma^{ei}\Lambda)N^{df}\nonumber \\
&& +\frac{4}{\eta^{3}}(\bar{\Lambda}\Gamma_{ab}\bar{\Lambda})(\bar{\Lambda}\Gamma_{ef}R)(\bar{\Lambda}\Gamma_{cd}R)(\Lambda\Gamma^{cdjei}\Lambda)(\Lambda\Gamma^{abf}W)
\end{eqnarray}

\begin{eqnarray}
\{W_{1}^{i}, W_{3}^{j}\} &=& \{-\frac{2}{\eta^{2}}(\bar{\Lambda}\Gamma_{ef}\bar{\Lambda})(\bar{\Lambda}\Gamma_{gh}R)(\Lambda\Gamma^{gi}\Lambda)(\Lambda\Gamma^{efh}W), \frac{2}{\eta^{2}}(\bar{\Lambda}\Gamma^{bj}\bar{\Lambda})(\bar{\Lambda}\Gamma^{ca}R)(\Lambda\Gamma_{ca}\Lambda)(\Lambda\Gamma_{b}W)\}\nonumber\\
&=& \frac{8}{\eta^{4}}(\bar{\Lambda}\Gamma_{ef}\bar{\Lambda})(\bar{\Lambda}\Gamma_{gh}R)(\Lambda\Gamma^{gi}\Lambda)(\bar{\Lambda}\Gamma^{bj}\bar{\Lambda})(\bar{\Lambda}\Gamma^{ca}R)(\Lambda\Gamma_{ca}\Lambda)(\Lambda\Gamma^{efh}{}_{b}W)\nonumber\\
&=& 0
\end{eqnarray}
where we have used eqn. \eqref{newapp2}.

\begin{eqnarray}
\{W_{1}^{i}, W_{4}^{j}\} &=& \{-\frac{2}{\eta^{2}}(\bar{\Lambda}\Gamma_{ef}\bar{\Lambda})(\bar{\Lambda}\Gamma_{gh}R)(\Lambda\Gamma^{gi}\Lambda)(\Lambda\Gamma^{efh}W), -\frac{4}{\eta^{2}}(\bar{\Lambda}\Gamma^{cj}\bar{\Lambda})(\bar{\Lambda}\Gamma^{ba}R)(\Lambda\Gamma_{ca}\Lambda)(\Lambda\Gamma_{b}W)\}\nonumber\\
&=& \frac{8}{\eta^{4}}(\bar{\Lambda}\Gamma_{ef}\bar{\Lambda})(\bar{\Lambda}\Gamma_{gh}R)(\bar{\Lambda}\Gamma^{cj}\bar{\Lambda})(\bar{\Lambda}\Gamma^{ba}R)(\Lambda\Gamma^{gi}\Lambda)\{[(\Lambda\Gamma^{efh}W),(\Lambda\Gamma_{ca}\Lambda)](\Lambda\Gamma_{b}W)\nonumber\\
&& + (\Lambda\Gamma_{ca}\Lambda)[(\Lambda\Gamma^{efh}W),(\Lambda\Gamma_{b}W)]\}\nonumber\\
&=& -\frac{16}{\eta^{4}}(\bar{\Lambda}\Gamma_{ef}\bar{\Lambda})(\bar{\Lambda}\Gamma_{gh}R)(\bar{\Lambda}\Gamma^{cj}\bar{\Lambda})(\bar{\Lambda}\Gamma^{ba}R)(\Lambda\Gamma^{gi}\Lambda)(\Lambda\Gamma^{efh}{}_{ca}\Lambda)(\Lambda\Gamma_{b}W)\nonumber\\ 
&&-\frac{16}{\eta^{3}}(\bar{\Lambda}\Gamma_{ef}\bar{\Lambda})(\bar{\Lambda}\Gamma_{gh}R)(\bar{\Lambda}\Gamma^{bj}R)(\Lambda\Gamma^{gi}\Lambda)(\Lambda\Gamma^{efh}{}_{b}W)\nonumber\\
&=&-\frac{16}{\eta^{3}}(\bar{\Lambda}\Gamma_{ef}\bar{\Lambda})(\bar{\Lambda}\Gamma_{gh}R)(\bar{\Lambda}\Gamma^{bj}R)(\Lambda\Gamma^{gi}\Lambda)(\Lambda\Gamma^{efh}{}_{b}W)
\end{eqnarray}
because of the identity \eqref{newapp2}.

\begin{eqnarray}
\{W_{1}^{i}, W_{5}^{j}\} &=& \{-\frac{2}{\eta}(\bar{\Lambda}\Gamma_{ef}\bar{\Lambda})(\bar{\Lambda}\Gamma_{gh}R)(\Lambda\Gamma^{gi}\Lambda)(\Lambda\Gamma^{efh}W), \frac{1}{\eta^{2}}(\bar{\Lambda}\Gamma_{bd}\bar{\Lambda})(\bar{\Lambda}\Gamma_{ac}R)(\Lambda\Gamma^{ac}\Lambda)(\Lambda\Gamma^{bdj}W)\}\nonumber\\
&=& -\frac{2}{\eta^{4}}(\bar{\Lambda}\Gamma_{ef}\bar{\Lambda})(\bar{\Lambda}\Gamma_{gh}R)(\bar{\Lambda}\Gamma_{bd}\bar{\Lambda})(\bar{\Lambda}\Gamma_{ac}R)[(\Lambda\Gamma^{gi}\Lambda)(\Lambda\Gamma^{efh}W),(\Lambda\Gamma^{ac}\Lambda)(\Lambda\Gamma^{bdj}W)]\nonumber\\
&=& -\frac{2}{\eta^{4}}(\bar{\Lambda}\Gamma_{ef}\bar{\Lambda})(\bar{\Lambda}\Gamma_{gh}R)(\bar{\Lambda}\Gamma_{bd}\bar{\Lambda})(\bar{\Lambda}\Gamma_{ac}R)\{(\Lambda\Gamma^{gi}\Lambda)(\Lambda\Gamma^{ac}\Lambda)[(\Lambda\Gamma^{efh}W),(\Lambda\Gamma^{bdj}W)] \nonumber\\
&& + (\Lambda\Gamma^{ac}\Lambda)[(\Lambda\Gamma^{gi}\Lambda),(\Lambda\Gamma^{bdj}W)](\Lambda\Gamma^{efh}W)\}\nonumber\\
&=& -\frac{4}{\eta^{4}}(\bar{\Lambda}\Gamma_{ef}\bar{\Lambda})(\bar{\Lambda}\Gamma_{gh}R)(\bar{\Lambda}\Gamma_{bd}\bar{\Lambda})(\bar{\Lambda}\Gamma_{ac}R)(\Lambda\Gamma^{ac}\Lambda)(\Lambda\Gamma^{bdjgi}\Lambda)(\Lambda\Gamma^{efh}W)\nonumber\\
&& - \frac{2}{\eta^{4}}(\bar{\Lambda}\Gamma_{ef}\bar{\Lambda})(\bar{\Lambda}\Gamma_{gh}R)(\bar{\Lambda}\Gamma_{bd}\bar{\Lambda})(\bar{\Lambda}\Gamma_{ac}R)(\Lambda\Gamma^{gi}\Lambda)(\Lambda\Gamma^{ac}\Lambda)[16\delta^{eh}_{bj}N^{fd}]\nonumber\\
&=& \frac{16}{\eta^{4}}(\bar{\Lambda}\Gamma^{jf}\bar{\Lambda})(\bar{\Lambda}\Gamma_{gb}R)(\bar{\Lambda}\Gamma^{bd}\bar{\Lambda})(\bar{\Lambda}\Gamma^{ac}R)(\Lambda\Gamma_{ac}\Lambda)(\Lambda\Gamma^{gi}\Lambda)N_{fd}\nonumber\\
&=& \frac{8}{\eta^{4}}(\bar{\Lambda}\Gamma^{j}{}_{g}\bar{\Lambda})(\bar{\Lambda}\Gamma^{fd}\bar{\Lambda})(\bar{\Lambda}R)(\bar{\Lambda}\Gamma^{ac}R)(\Lambda\Gamma_{ac}\Lambda)(\Lambda\Gamma^{gi}\Lambda)N_{fd}
\end{eqnarray}

\begin{eqnarray}
\{W_{2}^{i}, W_{3}^{j}\} &=& \{-\frac{1}{\eta}(\bar{\Lambda}\Gamma_{ef}R)(\Lambda\Gamma^{efi}W) , \frac{2}{\eta^{2}}(\bar{\Lambda}\Gamma_{b}{}^{j}\bar{\Lambda})(\bar{\Lambda}\Gamma^{ca}R)(\Lambda\Gamma_{ca}\Lambda)(\Lambda\Gamma^{b}W)\}\nonumber\\
&=& \frac{4}{\eta^{3}}(\bar{\Lambda}\Gamma_{ef}R)(\bar{\Lambda}\Gamma_{b}{}^{j}\bar{\Lambda})(\bar{\Lambda}\Gamma^{ca}R)(\Lambda\Gamma_{ca}\Lambda)(\Lambda\Gamma^{efib}W)
\end{eqnarray}

\begin{eqnarray}
\{W_{2}^{i}, W_{4}^{j}\} &=& \{-\frac{1}{\eta}(\bar{\Lambda}\Gamma_{ef}R)(\Lambda\Gamma^{efi}W), -\frac{4}{\eta^{2}}(\bar{\Lambda}\Gamma^{cj}\bar{\Lambda})(\bar{\Lambda}\Gamma^{ba}R)(\Lambda\Gamma_{ca}\Lambda)(\Lambda\Gamma_{b}W)\}\nonumber\\
&=&\frac{4}{\eta^{3}}(\bar{\Lambda}\Gamma_{ef}R)(\bar{\Lambda}\Gamma_{c}{}^{j}\bar{\Lambda})(\bar{\Lambda}\Gamma_{ba}R)[-2(\Lambda\Gamma^{ca}\Lambda)(\Lambda\Gamma^{efib}W) - 2(\Lambda\Gamma^{efica}\Lambda)(\Lambda\Gamma^{b}W)]\nonumber\\
&=& -\frac{8}{\eta^{3}}(\bar{\Lambda}\Gamma_{ef}R)(\bar{\Lambda}\Gamma_{c}{}^{j}\bar{\Lambda})(\bar{\Lambda}\Gamma_{ba}R)(\Lambda\Gamma^{ca}\Lambda)(\Lambda\Gamma^{efib}W)\nonumber\\
&&  -\frac{8}{\eta^{3}}(\bar{\Lambda}\Gamma_{ef}R)(\bar{\Lambda}\Gamma_{c}{}^{j}\bar{\Lambda})(\bar{\Lambda}\Gamma_{ba}R)(\Lambda\Gamma^{efica}\Lambda)(\Lambda\Gamma^{b}W)
\end{eqnarray}

\begin{eqnarray}
\{W_{2}^{i}, W_{5}^{j}\} &=& \{-\frac{1}{\eta}(\bar{\Lambda}\Gamma_{ef}R)(\Lambda\Gamma^{efi}W), \frac{1}{\eta^{2}}(\bar{\Lambda}\Gamma_{bd}\bar{\Lambda})(\bar{\Lambda}\Gamma_{ac}R)(\Lambda\Gamma^{ac}\Lambda)(\Lambda\Gamma^{bdj}W)\}\nonumber\\
&=& - \frac{1}{\eta^{3}}(\bar{\Lambda}\Gamma_{ef}R)(\bar{\Lambda}\Gamma_{bd}\bar{\Lambda})(\bar{\Lambda}\Gamma_{ac}R)(\Lambda\Gamma^{ac}\Lambda)[(\Lambda\Gamma^{efi}W), (\Lambda\Gamma^{bdj}W)]\nonumber\\
&=& - \frac{1}{\eta^{3}}(\bar{\Lambda}\Gamma_{ef}R)(\bar{\Lambda}\Gamma_{bd}\bar{\Lambda})(\bar{\Lambda}\Gamma_{ac}R)(\Lambda\Gamma^{ac}\Lambda)[-8\delta^{ef}_{bj}N^{id} - 8\delta^{ei}_{bd}N^{fj} + 16\delta^{ei}_{bj}N^{fd}]\nonumber\\
&=& \frac{8}{\eta^{3}}(\bar{\Lambda}\Gamma^{bj}R)(\bar{\Lambda}\Gamma_{bd}\bar{\Lambda})(\bar{\Lambda}\Gamma_{ac}R)(\Lambda\Gamma^{ac}\Lambda)N^{id}\nonumber\\
&& + \frac{8}{\eta^{3}}(\bar{\Lambda}\Gamma_{ef}R)(\bar{\Lambda}\Gamma^{ei}\bar{\Lambda})(\bar{\Lambda}\Gamma_{ac}R)(\Lambda\Gamma^{ac}\Lambda)N^{fj}\nonumber\\
&& -\frac{16}{\eta^{3}}(\frac{1}{2})[\eta^{ij}(\bar{\Lambda}\Gamma^{bf}R)(\bar{\Lambda}\Gamma_{b}{}^{d}\bar{\Lambda}) - (\bar{\Lambda}\Gamma^{jf}R)(\bar{\Lambda}\Gamma^{id}\bar{\Lambda})](\bar{\Lambda}\Gamma_{ac}R)(\Lambda\Gamma^{ac}\Lambda)N_{fd}\nonumber\\
&=& -\frac{8}{\eta^{3}}(\bar{\Lambda}\Gamma^{jd}\bar{\Lambda})(\bar{\Lambda}R)(\bar{\Lambda}\Gamma_{ac}R)(\Lambda\Gamma^{ac}\Lambda)N^{i}{}_{d}\nonumber\\
&& -\frac{8}{\eta^{3}}(\bar{\Lambda}\Gamma^{id}\bar{\Lambda})(\bar{\Lambda}R)(\bar{\Lambda}\Gamma_{ac}R)(\Lambda\Gamma^{ac}\Lambda)N^{j}{}_{d}\nonumber\\
&& + \frac{8}{\eta^{3}}(\bar{\Lambda}\Gamma^{fd}\bar{\Lambda})(\bar{\Lambda}R)(\bar{\Lambda}\Gamma_{ac}R)(\Lambda\Gamma^{ac}\Lambda)\eta^{ij}N_{fd}\nonumber\\
&&+\frac{8}{\eta{3}}(\bar{\Lambda}\Gamma^{jf}R)(\bar{\Lambda}\Gamma^{id}\bar{\Lambda})(\bar{\Lambda}\Gamma_{ac}R)(\Lambda\Gamma^{ac}\Lambda)N_{fd}
\end{eqnarray}

\begin{eqnarray}
\{W_{3}^{i}, W_{4}^{j}\} &=& \{\frac{2}{\eta^{2}}(\bar{\Lambda}\Gamma^{fi}\bar{\Lambda})(\bar{\Lambda}\Gamma^{ge}R)(\Lambda\Gamma_{ge}\Lambda)(\Lambda\Gamma_{f}W),-\frac{4}{\eta^{2}}(\bar{\Lambda}\Gamma^{cj}\bar{\Lambda})(\bar{\Lambda}\Gamma^{ba}R)(\Lambda\Gamma_{ca}\Lambda)(\Lambda\Gamma_{b}W)\}\nonumber\\
&=& \frac{16}{\eta^{4}}(\bar{\Lambda}\Gamma^{fi}\bar{\Lambda})(\bar{\Lambda}\Gamma^{ge}R)(\Lambda\Gamma_{ge}\Lambda)(\bar{\Lambda}\Gamma^{cj}\bar{\Lambda})(\bar{\Lambda}\Gamma^{ba}R)(\Lambda\Gamma_{ca}\Lambda)N_{fb}
\end{eqnarray}

\begin{eqnarray}
\{W_{3}^{i}, W_{5}^{j}\} &=& \{\frac{2}{\eta^{2}}(\bar{\Lambda}\Gamma^{fi}\bar{\Lambda})(\bar{\Lambda}\Gamma^{ge}R)(\Lambda\Gamma_{ge}\Lambda)(\Lambda\Gamma_{f}W), \frac{1}{\eta^{2}}(\bar{\Lambda}\Gamma_{bd}\bar{\Lambda})(\bar{\Lambda}\Gamma_{ac}R)(\Lambda\Gamma^{ac}\Lambda)(\Lambda\Gamma^{bdj}W)\}\nonumber\\
&=& 0
\end{eqnarray}

\begin{eqnarray}
\{W_{4}^{i}, W_{5}^{j}\} &=& \{-\frac{4}{\eta^{2}}(\bar{\Lambda}\Gamma^{gi}\bar{\Lambda})(\bar{\Lambda}\Gamma^{fe}R)(\Lambda\Gamma_{ge}\Lambda)(\Lambda\Gamma_{f}W), \frac{1}{\eta^{2}}(\bar{\Lambda}\Gamma_{bd}\bar{\Lambda})(\bar{\Lambda}\Gamma_{ac}R)(\Lambda\Gamma^{ac}\Lambda)(\Lambda\Gamma^{bdj}W)\}\nonumber\\
&=& -\frac{8}{\eta^{4}}(\bar{\Lambda}\Gamma^{gi}\bar{\Lambda})(\bar{\Lambda}\Gamma_{f}{}^{e}R)(\bar{\Lambda}\Gamma_{bd}\bar{\Lambda})(\bar{\Lambda}\Gamma_{ac}R)(\Lambda\Gamma_{ge}\Lambda)(\Lambda\Gamma^{ac}\Lambda)(\Lambda\Gamma^{bdjf}W)\nonumber\\
&=& -\frac{4}{\eta^{3}}(\bar{\Lambda}\Gamma_{f}{}^{i}R)(\bar{\Lambda}\Gamma_{bd}\bar{\Lambda})(\bar{\Lambda}\Gamma_{ac}R)(\Lambda\Gamma^{ac}\Lambda)(\Lambda\Gamma^{bdjf}W)
\end{eqnarray}
All in all, we arrive at the result:
\begin{eqnarray}
\{\bar{\Sigma}^{i}, \bar{\Sigma}^{j}\} &=& -\frac{4}{\eta^{2}}(\bar{\Lambda}\Gamma^{if}\bar{\Lambda})(\bar{\Lambda}\Gamma^{cj}R)(\Lambda\Gamma_{cf}D) -\frac{4}{\eta^{2}}(\bar{\Lambda}\Gamma^{jf}\bar{\Lambda})(\bar{\Lambda}\Gamma^{ci}R)(\Lambda\Gamma_{cf}D)\nonumber\\
&& + \frac{2}{\eta^{2}}(\bar{\Lambda}\Gamma_{ef}\bar{\Lambda})(\bar{\Lambda}\Gamma_{c}{}^{j}R)(\Lambda\Gamma^{cief}D) + \frac{2}{\eta^{2}}(\bar{\Lambda}\Gamma_{ef}\bar{\Lambda})(\bar{\Lambda}\Gamma_{c}{}^{i}R)(\Lambda\Gamma^{cjef}D)\nonumber \\
&&-\frac{4}{\eta^{2}}(\bar{\Lambda}\Gamma^{i}{}_{f}\bar{\Lambda})(\bar{\Lambda}R)(\Lambda\Gamma^{jf}D) -\frac{4}{\eta^{2}}(\bar{\Lambda}\Gamma^{j}{}_{f}\bar{\Lambda})(\bar{\Lambda}R)(\Lambda\Gamma^{if}D) +\frac{4}{\eta^{2}}\eta^{ij}(\bar{\Lambda}\Gamma^{ef}\bar{\Lambda})(\bar{\Lambda}R)(\Lambda\Gamma_{ef}D)\nonumber\\
&& -\frac{8}{\eta^{2}}(\bar{\Lambda}\Gamma^{jd}R)(\bar{\Lambda}R)N^{i}{}_{d} - \frac{4}{\eta^{2}}(\bar{\Lambda}\Gamma^{jd}\bar{\Lambda})(RR)N^{i}{}_{d}\nonumber\\
&& +\frac{8}{\eta^{2}}(\bar{\Lambda}\Gamma^{di}R)(\bar{\Lambda}R)N^{j}{}_{d} + \frac{4}{\eta^{2}}(\bar{\Lambda}\Gamma^{di}\bar{\Lambda})(RR)N^{j}{}_{d}\nonumber\\
&& + \frac{8}{\eta^{2}}\eta^{ij}(\bar{\Lambda}\Gamma^{ed}R)(\bar{\Lambda}R)N_{ed} + \frac{4}{\eta^{2}}\eta^{ij}(\bar{\Lambda}\Gamma^{ed}\bar{\Lambda})(RR)N_{ed} + \frac{8}{\eta^{2}}(\bar{\Lambda}\Gamma^{ej}R)(\bar{\Lambda}\Gamma^{id}R)N_{ed}\nonumber \\
&& + \frac{32}{\eta^{3}}(\bar{\Lambda}\Gamma_{d}{}^{j}\bar{\Lambda})(\bar{\Lambda}\Gamma_{gh}R)(\bar{\Lambda}R)(\Lambda\Gamma^{gi}\Lambda)N^{hd} + \frac{16}{\eta^{3}}(\bar{\Lambda}\Gamma_{hd}\bar{\Lambda})(\bar{\Lambda}\Gamma^{gj}R)(\bar{\Lambda}R)(\Lambda\Gamma_{g}{}^{i}\Lambda)N^{hd}\nonumber\\
&& -\frac{16}{\eta^{3}}(\bar{\Lambda}\Gamma_{d}{}^{j}\bar{\Lambda})(\bar{\Lambda}\Gamma_{gh}\bar{\Lambda})(RR)(\Lambda\Gamma^{gi}\Lambda)N^{dh} + \frac{32}{\eta^{3}}(\bar{\Lambda}\Gamma_{d}{}^{i}\bar{\Lambda})(\bar{\Lambda}\Gamma_{gh}R)(\bar{\Lambda}R)(\Lambda\Gamma^{gj}\Lambda)N^{hd}\nonumber\\
&& + \frac{16}{\eta^{3}}(\bar{\Lambda}\Gamma_{hd}\bar{\Lambda})(\bar{\Lambda}\Gamma^{gi}R)(\bar{\Lambda}R)(\Lambda\Gamma_{g}{}^{j}\Lambda)N^{hd} -\frac{16}{\eta^{3}}(\bar{\Lambda}\Gamma_{d}{}^{i}\bar{\Lambda})(\bar{\Lambda}\Gamma_{gh}\bar{\Lambda})(RR)(\Lambda\Gamma^{gj}\Lambda)N^{dh}\nonumber\\
&& -\frac{16}{\eta^{3}}(\bar{\Lambda}\Gamma_{ef}\bar{\Lambda})(\bar{\Lambda}\Gamma_{gh}R)(\bar{\Lambda}\Gamma^{bj}R)(\Lambda\Gamma^{gi}\Lambda)(\Lambda\Gamma^{efh}{}_{b}W)\nonumber\\
&& -\frac{16}{\eta^{3}}(\bar{\Lambda}\Gamma_{ef}\bar{\Lambda})(\bar{\Lambda}\Gamma_{gh}R)(\bar{\Lambda}\Gamma^{bi}R)(\Lambda\Gamma^{gj}\Lambda)(\Lambda\Gamma^{efh}{}_{b}W)\nonumber\\
&& + \frac{8}{\eta^{4}}(\bar{\Lambda}\Gamma^{j}{}_{g}\bar{\Lambda})(\bar{\Lambda}\Gamma^{fd}\bar{\Lambda})(\bar{\Lambda}R)(\bar{\Lambda}\Gamma^{ac}R)(\Lambda\Gamma_{ac}\Lambda)(\Lambda\Gamma^{gi}\Lambda)N_{fd}\nonumber\\
&& + \frac{8}{\eta^{4}}(\bar{\Lambda}\Gamma^{i}{}_{g}\bar{\Lambda})(\bar{\Lambda}\Gamma^{fd}\bar{\Lambda})(\bar{\Lambda}R)(\bar{\Lambda}\Gamma^{ac}R)(\Lambda\Gamma_{ac}\Lambda)(\Lambda\Gamma^{gj}\Lambda)N_{fd}\nonumber\\
&& + \frac{4}{\eta^{3}}(\bar{\Lambda}\Gamma_{ef}R)(\bar{\Lambda}\Gamma_{b}{}^{j}\bar{\Lambda})(\bar{\Lambda}\Gamma^{ca}R)(\Lambda\Gamma_{ca}\Lambda)(\Lambda\Gamma^{efib}W)\nonumber\\
&& + \frac{4}{\eta^{3}}(\bar{\Lambda}\Gamma_{ef}R)(\bar{\Lambda}\Gamma_{b}{}^{i}\bar{\Lambda})(\bar{\Lambda}\Gamma^{ca}R)(\Lambda\Gamma_{ca}\Lambda)(\Lambda\Gamma^{efjb}W)\nonumber\\
&&  -\frac{8}{\eta^{3}}(\bar{\Lambda}\Gamma_{ef}R)(\bar{\Lambda}\Gamma_{c}{}^{j}\bar{\Lambda})(\bar{\Lambda}\Gamma_{ba}R)(\Lambda\Gamma^{ca}\Lambda)(\Lambda\Gamma^{efib}W)\nonumber\\
&&  -\frac{8}{\eta^{3}}(\bar{\Lambda}\Gamma_{ef}R)(\bar{\Lambda}\Gamma_{c}{}^{j}\bar{\Lambda})(\bar{\Lambda}\Gamma_{ba}R)(\Lambda\Gamma^{efica}\Lambda)(\Lambda\Gamma^{b}W)\nonumber\\
&&  -\frac{8}{\eta^{3}}(\bar{\Lambda}\Gamma_{ef}R)(\bar{\Lambda}\Gamma_{c}{}^{i}\bar{\Lambda})(\bar{\Lambda}\Gamma_{ba}R)(\Lambda\Gamma^{ca}\Lambda)(\Lambda\Gamma^{efjb}W)\nonumber\\
&&  -\frac{8}{\eta^{3}}(\bar{\Lambda}\Gamma_{ef}R)(\bar{\Lambda}\Gamma_{c}{}^{i}\bar{\Lambda})(\bar{\Lambda}\Gamma_{ba}R)(\Lambda\Gamma^{efjca}\Lambda)(\Lambda\Gamma^{b}W)\nonumber\\
&& -\frac{16}{\eta^{3}}(\bar{\Lambda}\Gamma^{jd}\bar{\Lambda})(\bar{\Lambda}R)(\bar{\Lambda}\Gamma_{ac}R)(\Lambda\Gamma^{ac}\Lambda)N^{i}{}_{d}\nonumber\\
&& -\frac{16}{\eta^{3}}(\bar{\Lambda}\Gamma^{id}\bar{\Lambda})(\bar{\Lambda}R)(\bar{\Lambda}\Gamma_{ac}R)(\Lambda\Gamma^{ac}\Lambda)N^{j}{}_{d}\nonumber\\
&& + \frac{16}{\eta^{3}}(\bar{\Lambda}\Gamma^{fd}\bar{\Lambda})(\bar{\Lambda}R)(\bar{\Lambda}\Gamma_{ac}R)(\Lambda\Gamma^{ac}\Lambda)\eta^{ij}N_{fd}\nonumber\\
&& +\frac{16}{\eta{3}}(\bar{\Lambda}\Gamma^{jf}R)(\bar{\Lambda}\Gamma^{id}\bar{\Lambda})(\bar{\Lambda}\Gamma_{ac}R)(\Lambda\Gamma^{ac}\Lambda)N_{fd}\nonumber\\ 
&& +\frac{16}{\eta^{4}}(\bar{\Lambda}\Gamma^{fi}\bar{\Lambda})(\bar{\Lambda}\Gamma^{ge}R)(\Lambda\Gamma_{ge}\Lambda)(\bar{\Lambda}\Gamma^{cj}\bar{\Lambda})(\bar{\Lambda}\Gamma^{ba}R)(\Lambda\Gamma_{ca}\Lambda)N_{fb}\nonumber\\
&& + \frac{16}{\eta^{4}}(\bar{\Lambda}\Gamma^{fj}\bar{\Lambda})(\bar{\Lambda}\Gamma^{ge}R)(\Lambda\Gamma_{ge}\Lambda)(\bar{\Lambda}\Gamma^{ci}\bar{\Lambda})(\bar{\Lambda}\Gamma^{ba}R)(\Lambda\Gamma_{ca}\Lambda)N_{fb}\nonumber\\
&& -\frac{4}{\eta^{3}}(\bar{\Lambda}\Gamma_{f}{}^{i}R)(\bar{\Lambda}\Gamma_{bd}\bar{\Lambda})(\bar{\Lambda}\Gamma_{ac}R)(\Lambda\Gamma^{ac}\Lambda)(\Lambda\Gamma^{bdjf}W)\nonumber\\
&& -\frac{4}{\eta^{3}}(\bar{\Lambda}\Gamma_{f}{}^{j}R)(\bar{\Lambda}\Gamma_{bd}\bar{\Lambda})(\bar{\Lambda}\Gamma_{ac}R)(\Lambda\Gamma^{ac}\Lambda)(\Lambda\Gamma^{bdif}W)
\end{eqnarray}

Therefore we conclude that $\{\bar{\Sigma}^{i}, \bar{\Sigma}^{j}\}$ depends linearly and quadratically on $R_{\alpha}$. This allows us to find the $R_{\alpha}$-dependence of $\{b , b\}$ which turns out to be of the form:
\begin{equation}\label{eq503}
\{b,b\} = R^{\alpha}f^{(1)}_{\alpha} + \ldots + R^{\alpha}R^{\beta}R^{\delta}R^{\sigma}R^{\rho}R^{\lambda}f^{(6)}_{\alpha\beta\delta\sigma\rho\lambda}
\end{equation}
where $f^{(i)}_{\alpha_{1}\ldots\alpha_{i}}$ for $i=1,\ldots ,6$ are functions of pure spinor variables $\Lambda^{\alpha}, \bar{\Lambda}_{\alpha}, W_{\alpha}$ and the fermionic constraints $D_{\alpha}$.

\vspace{2mm}
This can be used to check that $\{b,b\} = Q\Omega$ where $\Omega$ is an arbitrary function of pure spinor variables and the constraints $D_{\alpha}$. To see this let us expand $\Omega$ in terms of $R^{\alpha}$:
\begin{equation}
\Omega = \Omega^{(0)} + R^{\alpha}\Omega_{\alpha}^{(1)} + R^{\alpha\beta}\Omega^{(2)}_{\alpha\beta} + \ldots + R^{\alpha_{1}\ldots\alpha_{23}}\Omega^{(23)}_{\alpha_{1}\ldots\alpha_{23}} 
\end{equation}
Thus the action of the BRST operator $Q = Q_{0} + R^{\alpha}\bar{W}_{\alpha}$ on $\Omega$ gives us
\begin{eqnarray}
Q\Omega = Q_{0}\Omega^{(0)} + R^{\alpha}(\frac{\partial}{\partial\bar{\Lambda}^{\alpha}}\Omega^{(0)} + Q_{0}\Omega^{(1)}_{\alpha}) + R^{\alpha}R^{\beta}(\frac{\partial}{\partial\bar{\Lambda}^{\alpha}}\Omega_{\beta}^{(1)} + Q_{0}\Omega_{\alpha\beta}^{(2)}) + \ldots
\end{eqnarray} 
The comparison of this result with the equation \eqref{eq503} determines the functions $\Omega^{(k)}$ for $k=1,\ldots,23$:
\begin{eqnarray}
0 &=& Q_{0}\Omega^{(0)}\\
f^{(1)}_{\alpha} &=& \frac{\partial}{\partial \bar{\Lambda}^{\alpha}}\Omega^{(0)} + Q_{0}\Omega^{(1)}_{\alpha}\\
f^{(2)}_{\alpha\beta} &=& \frac{\partial}{\partial \bar{\Lambda}^{\alpha}}\Omega^{(1)}_{\beta} + Q_{0}\Omega^{(2)}_{\alpha\beta}\\
&\vdots & \nonumber
\end{eqnarray}
Therefore if we make the following definitions:
\begin{eqnarray}
\Omega^{(0)} &=& \bar{\Lambda}^{\alpha}f_{\alpha}^{(1)}\\
\Omega^{(1)}_{\beta} &=& \bar{\Lambda}^{\alpha}f_{\alpha\beta}^{(2)}\\
\Omega^{(2)}_{\beta\delta} &=& \bar{\Lambda}^{\alpha}f_{\alpha\beta\delta}^{(3)}\\
&\vdots & \nonumber\\
\Omega^{(5)}_{\beta\delta\sigma\rho\lambda} &=& \bar{\Lambda}^{\alpha}f_{\alpha\beta\delta\sigma\rho\lambda}^{(6)}\\
\Omega^{(6)}_{\beta\delta\sigma\rho\lambda\gamma} &=& 0\\
&\vdots & \nonumber\\
\Omega^{(23)} &=& 0
\end{eqnarray}
the equations above are automatically satisfied.
\section{Expanding the simplified $D=11$ $b$-ghost}\label{apE}
In this Appendix we will show explicitly the terms contained in $O(\bar{\Sigma}^{2})$ in the expression for the simplified $D=11$ $b$-ghost. 
We will work with the expression
\begin{equation}\label{endeq14}
b_{simpl} = P^{i}\bar{\Sigma}_{i} -\frac{4}{\eta^{2}}(\bar{\Lambda}\Gamma_{ab}\bar{\Lambda})(\bar{\Lambda}\Gamma_{cd}R)(\Lambda\Gamma^{aj}\Lambda)\bar{\Sigma}_{j}[(\Lambda\Gamma^{bd}\Lambda)\bar{\Sigma}^{c} + \frac{1}{\eta}(\Lambda\Gamma^{bd}\Lambda)(\bar{\Lambda}\Gamma^{cf}\bar{\Lambda})(\Lambda\Gamma_{fk}\Lambda)\bar{\Sigma}^{k}]
\end{equation}
It is useful to write $\bar{\Sigma}^{i}_{0}$ in the convenient way:
\begin{equation}
\bar{\Sigma}^{i}_{0} = \frac{1}{2\eta}[(\bar{\Lambda}\Gamma_{ab}\bar{\Lambda})(\Lambda\Gamma^{i}\Gamma^{ab}D) + 4(\bar{\Lambda}\Gamma^{ai}\bar{\Lambda})(\Lambda\Gamma_{a}D)]
\end{equation}
Therefore, we have
\begin{equation}\label{endeq13}
(\Lambda\Gamma_{ij}\Lambda)\bar{\Sigma}^{j}_{0} = \frac{2}{\eta}(\Lambda\Gamma_{ij}\Lambda)(\bar{\Lambda}\Gamma^{aj}\bar{\Lambda})(\Lambda\Gamma_{a}D)
\end{equation}
which is a direct consequence of \eqref{newapp1}. Using eqns. \eqref{section4eeq22}, \eqref{endeq13}, one can write \eqref{endeq14} as
\begin{align}\label{endeq19}
b_{simpl} & = P^{i}\bar{\Sigma}_{i} - \frac{4}{\eta^{2}}(\bar{\Lambda}\Gamma_{ab}\bar{\Lambda})(\bar{\Lambda}\Gamma_{cd}R)(\Lambda\Gamma^{bd}\Lambda)(\Lambda\Gamma^{a}{}_{j}\Lambda)[\bar{\Sigma}_{0}^{j} +  \frac{2}{\eta^{2}}(\bar{\Lambda}\Gamma_{ef}\bar{\Lambda})(\bar{\Lambda}\Gamma_{gh}R)(\Lambda\Gamma^{efgij}\Lambda)N^{h}{}_{i} \notag \\
& + \frac{2}{3\eta^{2}}(\bar{\Lambda}\Gamma_{ef}\bar{\Lambda})(\bar{\Lambda}\Gamma_{g}{}^{j}R)(\Lambda\Gamma^{efghi}\Lambda)N_{hi} \notag - \frac{2}{3\eta^{2}}(\bar{\Lambda}\Gamma_{eg}\bar{\Lambda})(\bar{\Lambda}R)(\Lambda\Gamma^{ejghi}\Lambda)N_{hi}]\times\\
& \{\bar{\Sigma}_{0}^{c} + \frac{2}{\eta^{2}}(\bar{\Lambda}\Gamma_{lf}\bar{\Lambda})(\bar{\Lambda}\Gamma_{gh}R)(\Lambda\Gamma^{lfgec}\Lambda)N^{h}{}_{e} + \frac{2}{3\eta^{2}}(\bar{\Lambda}\Gamma_{le}\bar{\Lambda})(\bar{\Lambda}\Gamma_{f}{}^{c}R)(\Lambda\Gamma^{lefgh}\Lambda)N_{gh} \notag \\
&  - \frac{2}{3\eta^{2}}(\bar{\Lambda}\Gamma_{lf}\bar{\Lambda})(\bar{\Lambda}R)(\Lambda\Gamma^{lcfgh}\Lambda)N_{gh} + \frac{1}{\eta}(\bar{\Lambda}\Gamma^{c}{}_{h}\bar{\Lambda})(\Lambda\Gamma^{h}{}_{k}\Lambda)[\bar{\Sigma}_{0}^{k} + \frac{2}{\eta^{2}}(\bar{\Lambda}\Gamma_{gh}\bar{\Lambda})(\bar{\Lambda}\Gamma_{ed}R)(\Lambda\Gamma^{ghefk}\Lambda)N^{d}{}_{f} \notag \\
&  + \frac{2}{3\eta^{2}}(\bar{\Lambda}\Gamma_{ge}\bar{\Lambda})(\bar{\Lambda}\Gamma_{f}{}^{k}R)(\Lambda\Gamma^{gefcd}\Lambda)N_{cd} \notag - \frac{2}{3\eta^{2}}(\bar{\Lambda}\Gamma_{hf}\bar{\Lambda})(\bar{\Lambda}R)(\Lambda\Gamma^{hkfgh}\Lambda)N_{gh}]\}\\
\end{align}
The contributions proportional to $D^{2}$ are:
\begin{eqnarray*}
b^{(2)}_{simp} &=& -\frac{4}{\eta^{2}}(\bar{\Lambda}\Gamma_{ab}\bar{\Lambda})(\bar{\Lambda}\Gamma_{cd}R)(\Lambda\Gamma^{aj}\Lambda)\bar{\Sigma}_{0\,j}[(\Lambda\Gamma^{bd}\Lambda)\bar{\Sigma}_{0}^{c} + \frac{1}{\eta}(\Lambda\Gamma^{bd}\Lambda)(\bar{\Lambda}\Gamma^{cf}\bar{\Lambda})(\Lambda\Gamma_{fk}\Lambda)\bar{\Sigma}_{0}^{k}]\\
&=& -\frac{4}{\eta^{2}}(\bar{\Lambda}\Gamma_{ab}\bar{\Lambda})(\bar{\Lambda}\Gamma_{cd}R)(\Lambda\Gamma^{aj}\Lambda)(\Lambda\Gamma^{bd}\Lambda)\bar{\Sigma}_{0\,j}\bar{\Sigma}_{0}^{c} \\
&& -\frac{4}{\eta^{3}}(\bar{\Lambda}\Gamma_{ab}\bar{\Lambda})(\bar{\Lambda}\Gamma_{cd}R)(\Lambda\Gamma^{aj}\Lambda)(\Lambda\Gamma^{bd}\Lambda)(\bar{\Lambda}\Gamma^{cf}\bar{\Lambda})(\Lambda\Gamma_{fk}\Lambda)\bar{\Sigma}_{0\,j}\bar{\Sigma}_{0}^{k}\\
&=& -\frac{8}{\eta^{3}}(\bar{\Lambda}\Gamma_{ab}\bar{\Lambda})(\bar{\Lambda}\Gamma_{cd}\bar{\Lambda})(\Lambda\Gamma^{bd}\Lambda)(\Lambda\Gamma^{aj}\Lambda)(\bar{\Lambda}\Gamma_{fj}\bar{\Lambda})(\Lambda\Gamma^{f}D)\bar{\Sigma}_{0}^{c}\\
&& - \frac{2}{\eta^{2}}(\bar{\Lambda}\Gamma_{ab}\bar{\Lambda})(\bar{\Lambda}R)(\Lambda\Gamma^{bk}\Lambda)(\Lambda\Gamma^{aj}\Lambda)\bar{\Sigma}_{0\,j}\bar{\Sigma}_{0\,k}\\
&=& -\frac{2}{\eta^{3}}(\bar{\Lambda}\Gamma_{ac}R)(\Lambda\Gamma^{aj}\Lambda)(\bar{\Lambda}\Gamma_{fj}\bar{\Lambda})(\Lambda\Gamma^{f}D)[(\bar{\Lambda}\Gamma_{gh}\bar{\Lambda})(\Lambda\Gamma^{ghc}D) + 2(\bar{\Lambda}\Gamma_{f}{}^{c}\bar{\Lambda})(\Lambda\Gamma^{f}D)]\\
&& - \frac{1}{\eta}(\bar{\Lambda}R)(\Lambda\Gamma^{jk}\Lambda)\bar{\Sigma}_{0\,j}\bar{\Sigma}_{0\,k}\\
&=& -\frac{2}{\eta^{3}}(\bar{\Lambda}\Gamma_{ac}R)(\Lambda\Gamma^{aj}\Lambda)(\bar{\Lambda}\Gamma_{fj}\bar{\Lambda})(\bar{\Lambda}\Gamma_{gh}\bar{\Lambda})(\Lambda\Gamma^{f}D)(\Lambda\Gamma^{ghc}D)\\
&& - \frac{4}{\eta^{3}}(\bar{\Lambda}R)(\Lambda\Gamma^{aj}\Lambda)(\bar{\Lambda}\Gamma_{fj}\bar{\Lambda})(\bar{\Lambda}\Gamma^{ga}\bar{\Lambda})(\Lambda\Gamma^{f}D)(\Lambda\Gamma^{g}D)\\
&& -\frac{4}{\eta^{3}}(\bar{\Lambda}R)(\Lambda\Gamma^{jk}\Lambda)(\bar{\Lambda}\Gamma_{fj}\bar{\Lambda})(\bar{\Lambda}\Gamma_{ek}\bar{\Lambda})(\Lambda\Gamma^{f}D)(\Lambda\Gamma^{e}D)\\
&=& -\frac{1}{\eta^{2}}(\bar{\Lambda}\Gamma_{gh}\bar{\Lambda})(\bar{\Lambda}\Gamma_{fc}R)(\Lambda\Gamma^{f}D)(\Lambda\Gamma^{ghc}D) + \frac{2}{\eta^{2}}(\bar{\Lambda}R)(\bar{\Lambda}\Gamma^{fr}\bar{\Lambda})(\Lambda\Gamma^{f}D)(\Lambda\Gamma^{r}D)\\
&& - \frac{2}{\eta^{2}}(\bar{\Lambda}R)(\bar{\Lambda}\Gamma^{fg}\bar{\Lambda})(\Lambda\Gamma_{f}D)(\Lambda\Gamma_{g}D)\\
&=& -\frac{1}{\eta^{2}}(\bar{\Lambda}\Gamma_{gh}\bar{\Lambda})(\bar{\Lambda}\Gamma_{fc}R)(\Lambda\Gamma^{f}D)(\Lambda\Gamma^{ghc}D)\\
&=& \frac{1}{\eta^{2}}L^{(1)}_{ac,bd}(\Lambda\Gamma^{a}D)(\Lambda\Gamma^{cbd}D)
\end{eqnarray*}
where the identities \eqref{newapp3}, \eqref{newapp4} were used. 

\vspace{2mm}
Now let us move to the terms proportional to $\eta^{-3}$. We will calculate it in two steps. First we focus on the part proportional to $(\Lambda\Gamma^{a}D)$, which will be called $K_{1}$, and then we will simplify the part proportional to $(\Lambda\Gamma^{bcd}D)$, which will be called $K_{2}$. Thus,
\begin{align}
K_{1} &= - \frac{4}{\eta^{2}}(\bar{\Lambda}\Gamma_{ab}\bar{\Lambda})(\bar{\Lambda}\Gamma_{cd}R)(\Lambda\Gamma^{bd}\Lambda)(\Lambda\Gamma^{a}{}_{j}\Lambda)\bar{\Sigma}_{0}^{j}\times\{\frac{2}{\eta^{2}}(\bar{\Lambda}\Gamma_{le}\bar{\Lambda})(\bar{\Lambda}\Gamma_{fg}R)(\Lambda\Gamma^{lefhc}\Lambda)N^{g}{}_{h} \notag \\
& + \frac{2}{3\eta^{2}}(\bar{\Lambda}\Gamma_{le}\bar{\Lambda})(\bar{\Lambda}\Gamma_{f}{}^{c}R)(\Lambda\Gamma^{lefgh}\Lambda)N_{gh} - \frac{2}{3\eta^{2}}(\bar{\Lambda}\Gamma_{lf}\bar{\Lambda})(\bar{\Lambda}R)(\Lambda\Gamma^{lcfgh}\Lambda)N_{gh}\notag \\
& + \frac{1}{\eta}(\bar{\Lambda}\Gamma^{c}{}_{h}\bar{\Lambda})(\Lambda\Gamma^{h}{}_{k}\Lambda)[\frac{2}{\eta^{2}}(\bar{\Lambda}\Gamma_{ge}\bar{\Lambda})(\bar{\Lambda}\Gamma_{fd}R)(\Lambda\Gamma^{gefbk}\Lambda)N^{d}{}_{b} \notag \\
&  + \frac{2}{3\eta^{2}}(\bar{\Lambda}\Gamma_{ef}\bar{\Lambda})(\bar{\Lambda}\Gamma_{d}{}^{k}R)(\Lambda\Gamma^{efdgh}\Lambda)N_{gh} \notag - \frac{2}{3\eta^{2}}(\bar{\Lambda}\Gamma_{ef}\bar{\Lambda})(\bar{\Lambda}R)(\Lambda\Gamma^{ekfgh}\Lambda)N_{gh}]\}\\ 
\end{align}
One can now use eqn. \eqref{endeq13} to find that
\begin{align}
K_{1} &=
- \frac{4}{\eta^{2}}(\bar{\Lambda}\Gamma_{ab}\bar{\Lambda})(\bar{\Lambda}\Gamma_{cd}R)(\Lambda\Gamma^{bd}\Lambda)(\Lambda\Gamma^{a}_{\hspace{2mm}j}\Lambda)\bar{\Sigma}_{0}^{j}\times\{\frac{2}{\eta^{2}}(\bar{\Lambda}\Gamma_{le}\bar{\Lambda})(\bar{\Lambda}\Gamma_{fg}R)(\Lambda\Gamma^{lefhc}\Lambda)N^{g}{}_{h} \notag \\
& + \frac{2}{3\eta^{2}}(\bar{\Lambda}\Gamma_{le}\bar{\Lambda})(\bar{\Lambda}\Gamma_{f}{}^{c}R)(\Lambda\Gamma^{lefgh}\Lambda)N_{gh} - \frac{2}{3\eta^{2}}(\bar{\Lambda}\Gamma_{le}\bar{\Lambda})(\bar{\Lambda}R)(\Lambda\Gamma^{lcefg}\Lambda)N_{fg}\notag \\
& + \frac{1}{\eta}(\bar{\Lambda}\Gamma^{c}{}_{f}\bar{\Lambda})(\Lambda\Gamma^{f}{}_{k}\Lambda)[\frac{2}{3\eta^{2}}(\bar{\Lambda}\Gamma_{gh}\bar{\Lambda})(\bar{\Lambda}\Gamma_{i}{}^{k}R)(\Lambda\Gamma^{ghide}\Lambda)N_{de}]\} \notag \\
&- \frac{8}{\eta^{3}}(\bar{\Lambda}\Gamma_{ab}\bar{\Lambda})(\bar{\Lambda}\Gamma_{cd}R)(\Lambda\Gamma^{bd}\Lambda)(\Lambda\Gamma^{aj}\Lambda)(\bar{\Lambda}\Gamma_{ej}\bar{\Lambda})(\Lambda\Gamma^{e}D)\times\{\frac{2}{\eta^{2}}(\bar{\Lambda}\Gamma_{lf}\bar{\Lambda})(\bar{\Lambda}\Gamma_{gh}R)(\Lambda\Gamma^{lfgic}\Lambda)N^{h}{}_{i} \notag \\
& + \frac{2}{3\eta^{2}}(\bar{\Lambda}\Gamma_{le}\bar{\Lambda})(\bar{\Lambda}\Gamma_{f}{}^{c}R)(\Lambda\Gamma^{lefgh}\Lambda)N_{gh} - \frac{2}{3\eta^{2}}(\bar{\Lambda}\Gamma_{lf}\bar{\Lambda})(\bar{\Lambda}R)(\Lambda\Gamma^{lcfgh}\Lambda)N_{gh}\notag \\
& + \frac{2}{3\eta^{3}}(\bar{\Lambda}\Gamma^{c}{}_{f}\bar{\Lambda})(\Lambda\Gamma^{f}{}_{k}\Lambda)(\bar{\Lambda}\Gamma_{gh}\bar{\Lambda})(\bar{\Lambda}\Gamma_{e}{}^{k}R)(\Lambda\Gamma^{gheij}\Lambda)N_{ij}\} \notag \\
& -\frac{4}{\eta^{2}}(\bar{\Lambda}\Gamma_{eb}\bar{\Lambda})(\bar{\Lambda}\Gamma_{cd}R)(\Lambda\Gamma^{e}D)(\Lambda\Gamma^{bd}\Lambda)\times\{\frac{2}{\eta^{2}}(\bar{\Lambda}\Gamma_{lf}\bar{\Lambda})(\bar{\Lambda}\Gamma_{gi}R)(\Lambda\Gamma^{lfgjc}\Lambda)N^{i}{}_{j} \notag \\
& + \frac{2}{3\eta^{2}}(\bar{\Lambda}\Gamma_{le}\bar{\Lambda})(\bar{\Lambda}\Gamma_{f}{}^{c}R)(\Lambda\Gamma^{lefgh}\Lambda)N_{gh} - \frac{2}{3\eta^{2}}(\bar{\Lambda}\Gamma_{lf}\bar{\Lambda})(\bar{\Lambda}R)(\Lambda\Gamma^{lcfgh}\Lambda)N_{gh}\notag \\
& + \frac{2}{3\eta^{3}}(\bar{\Lambda}\Gamma^{c}{}_{e}\bar{\Lambda})(\Lambda\Gamma^{e}{}_{k}\Lambda)(\bar{\Lambda}\Gamma_{fg}\bar{\Lambda})(\bar{\Lambda}\Gamma_{h}{}^{k}R)(\Lambda\Gamma^{fghij}\Lambda)N_{ij}\}\label{newend1}
\end{align}
The use of eqns. \eqref{newapp2}, \eqref{newapp4} allows us to write the following identity
\begin{align}
(\bar{\Lambda}\Gamma_{eb}\bar{\Lambda})(\bar{\Lambda}\Gamma_{cd}R)(\Lambda\Gamma^{bd}\Lambda) & = [\frac{1}{2}(\bar{\Lambda}\Gamma_{ec}\bar{\Lambda})(\bar{\Lambda}\Gamma_{bd}R) + \frac{1}{2}(\bar{\Lambda}\Gamma_{bd}\bar{\Lambda})(\bar{\Lambda}\Gamma_{ec}R) + (\bar{\Lambda}\Gamma_{cd}\bar{\Lambda})(\bar{\Lambda}\Gamma_{be}R)](\Lambda\Gamma^{bd}\Lambda)
\end{align}
which can be used in \eqref{newend1} to have
\begin{align}
K_{1} & =
 -\frac{4}{\eta^{3}}(\bar{\Lambda}\Gamma_{ec}R)(\Lambda\Gamma^{e}D)(\bar{\Lambda}\Gamma_{lf}\bar{\Lambda})(\bar{\Lambda}\Gamma_{gi}R)(\Lambda\Gamma^{lfgjc}\Lambda)N^{i}{}_{j} \notag  \\
& -\frac{4}{3\eta^{3}}(\bar{\Lambda}\Gamma_{ec}R)(\Lambda\Gamma^{e}D)(\bar{\Lambda}\Gamma_{lf}\bar{\Lambda})(\bar{\Lambda}\Gamma_{g}{}^{c}R)(\Lambda\Gamma^{lfgij}\Lambda)N_{ij} \notag \\
& + \frac{4}{3\eta^{3}}(\bar{\Lambda}\Gamma_{ec}R)(\Lambda\Gamma^{e}D)(\bar{\Lambda}\Gamma^{lf}\bar{\Lambda})(\bar{\Lambda}R)(\Lambda\Gamma^{lcfij}\Lambda)N_{ij} \notag \\
& +\frac{2}{3\eta^{3}}(\bar{\Lambda}\Gamma^{ec}R)(\Lambda\Gamma_{e}D)(\bar{\Lambda}\Gamma_{cf}\bar{\Lambda})(\bar{\Lambda}\Gamma_{gh}R)(\Lambda\Gamma^{ghfij}\Lambda)N_{ij} \notag \\
& = -\frac{4}{\eta^{3}}(\bar{\Lambda}\Gamma_{le}\bar{\Lambda})(\bar{\Lambda}\Gamma_{fi}R)(\bar{\Lambda}\Gamma^{dc}R)(\Lambda\Gamma^{lefjc}\Lambda)(\Lambda\Gamma^{d}D)N_{ij} \notag  \\
& + \frac{4}{3\eta^{3}}(\bar{\Lambda}\Gamma_{le}\bar{\Lambda})(\bar{\Lambda}\Gamma_{gc}R)(\bar{\Lambda}\Gamma_{f}{}^{c}R)(\Lambda\Gamma^{lefij}\Lambda)(\Lambda\Gamma^{g}D))N_{ij} \notag \\
& - \frac{4}{3\eta^{3}}(\bar{\Lambda}\Gamma^{le}\bar{\Lambda})(\bar{\Lambda}\Gamma_{fc}R)(\bar{\Lambda}R)(\Lambda\Gamma^{lceij}\Lambda)(\Lambda\Gamma^{f}D)N_{ij} \notag \\
& - \frac{2}{3\eta^{3}}(\bar{\Lambda}\Gamma_{cf}\bar{\Lambda})(\bar{\Lambda}\Gamma^{ec}R)(\bar{\Lambda}\Gamma_{gh}R)(\Lambda\Gamma^{ghfij}\Lambda)(\Lambda\Gamma_{e}D)N_{ij} \notag \\
&= \frac{4}{\eta^{3}}(\bar{\Lambda}\Gamma_{le}\bar{\Lambda})(\bar{\Lambda}\Gamma_{ch}R)(\bar{\Lambda}\Gamma_{f}{}^{i}R)(\Lambda\Gamma^{lecfj}\Lambda)(\Lambda\Gamma^{h}D)N_{ji} \notag\\
& -\frac{2}{\eta^{3}}(\bar{\Lambda}\Gamma_{lf}\bar{\Lambda})(\bar{\Lambda}\Gamma_{ge}R)(\bar{\Lambda}R)(\Lambda\Gamma^{lefij}\Lambda)(\Lambda\Gamma^{g}D)N_{ij} \notag \\
&= \frac{4}{\eta^{3}}(\bar{\Lambda}\Gamma_{le}\bar{\Lambda})(\bar{\Lambda}\Gamma_{cg}R)(\bar{\Lambda}\Gamma_{f}{}^{i}R)(\Lambda\Gamma^{lecfj}\Lambda)(\Lambda\Gamma^{g}D)N_{ji} \notag\\
& -\frac{2}{\eta^{3}}(\bar{\Lambda}\Gamma_{le}\bar{\Lambda})(\bar{\Lambda}\Gamma_{fg}R)(\bar{\Lambda}R)(\Lambda\Gamma^{lefij}\Lambda)(\Lambda\Gamma^{g}D)N_{ij}\label{eq403}
\end{align}

\vspace{2mm}
Now let us move on to $K_{2}$. One has that
\begin{align}
K_{2} & = -\frac{4}{\eta^{2}}(\bar{\Lambda}\Gamma_{ab}\bar{\Lambda})(\bar{\Lambda}\Gamma_{cd}R)(\Lambda\Gamma^{bd}\Lambda)(\Lambda\Gamma^{a}{}_{j}\Lambda)[\frac{2}{\eta^{2}}(\bar{\Lambda}\Gamma_{ef}\bar{\Lambda})(\bar{\Lambda}\Gamma_{gh}R)(\Lambda\Gamma^{efgij}\Lambda)N^{h}{}_{i} \notag \\
& + \frac{2}{3\eta^{2}}(\bar{\Lambda}\Gamma_{ef}\bar{\Lambda})(\bar{\Lambda}\Gamma_{g}{}^{j}R)(\Lambda\Gamma^{efghi}\Lambda)N_{hi} - \frac{2}{3\eta^{2}}(\bar{\Lambda}\Gamma_{eg}\bar{\Lambda})(\bar{\Lambda}R)(\Lambda\Gamma^{ejghi}\Lambda)N_{hi}]\times \notag \\
& [\bar{\Sigma}_{0}^{c} + \frac{1}{\eta}(\bar{\Lambda}\Gamma^{cl}\bar{\Lambda})(\Lambda\Gamma_{l}{}^{k}\Lambda)\bar{\Sigma}_{0\,k}] \notag \\
&= -\frac{2}{\eta}(\bar{\Lambda}\Gamma_{ac}R)(\Lambda\Gamma^{a}{}_{j}\Lambda)[\frac{2}{3\eta^{2}}(\bar{\Lambda}\Gamma_{ef}\bar{\Lambda})(\bar{\Lambda}\Gamma_{g}{}^{j}R)(\Lambda\Gamma^{efghi}\Lambda)N_{hi}]\times \notag \\
& [\bar{\Sigma}_{0}^{c} + \frac{1}{\eta}(\bar{\Lambda}\Gamma^{cl}\bar{\Lambda})(\Lambda\Gamma_{l}{}^{k}\Lambda)\bar{\Sigma}_{0\,k}]
\end{align}
where we used eqns. \eqref{newapp4}, \eqref{newapp11}. Therefore,

\begin{align}
K_{2} & = 
-\frac{4}{3\eta^{3}}(\bar{\Lambda}\Gamma_{ac}R)(\Lambda\Gamma^{a}{}_{j}\Lambda)[(\bar{\Lambda}\Gamma_{ef}\bar{\Lambda})(\bar{\Lambda}\Gamma_{g}{}^{j}R)(\Lambda\Gamma^{efghi}\Lambda)N_{hi}]\times \notag \\
& [\frac{1}{2\eta}(\bar{\Lambda}\Gamma_{kl}\bar{\Lambda})(\Lambda\Gamma^{klc}D)] \notag\\
&= -\frac{2}{3\eta^{4}}(\bar{\Lambda}\Gamma_{ac}R)(\Lambda\Gamma^{a}{}_{j}\Lambda)[(\bar{\Lambda}\Gamma_{ef}\bar{\Lambda})(\bar{\Lambda}\Gamma_{g}{}^{j}R)(\Lambda\Gamma^{efghi}\Lambda)N_{hi}]\times \notag \\
& (\bar{\Lambda}\Gamma_{kl}\bar{\Lambda})(\Lambda\Gamma^{klc}D) \notag\\
&= \frac{2}{3\eta^{4}}(\bar{\Lambda}\Gamma_{ac}R)(\Lambda\Gamma^{a}{}_{j}\Lambda)[(\bar{\Lambda}\Gamma_{g}{}^{j}\bar{\Lambda})(\bar{\Lambda}\Gamma_{ef}R)(\Lambda\Gamma^{efghi}\Lambda)N_{hi}]\times \notag \\
& (\bar{\Lambda}\Gamma_{kl}\bar{\Lambda})(\Lambda\Gamma^{klc}D) \notag \\
& = \frac{1}{3\eta^{3}}(\bar{\Lambda}\Gamma_{cg}\bar{\Lambda})(\bar{\Lambda}\Gamma_{kl}R)(\bar{\Lambda}\Gamma_{ef}R)(\Lambda\Gamma^{efghi}\Lambda)N_{hi}(\Lambda\Gamma^{klc}D) \notag \\
& = -\frac{1}{3\eta^{3}}(\bar{\Lambda}\Gamma_{cg}\bar{\Lambda})(\bar{\Lambda}\Gamma_{ef}R)(\bar{\Lambda}\Gamma_{kl}R)(\Lambda\Gamma^{efghi}\Lambda)N_{hi}(\Lambda\Gamma^{ckl}D) \notag \\
& = \frac{1}{3\eta^{3}}(\bar{\Lambda}\Gamma_{ef}\bar{\Lambda})(\bar{\Lambda}\Gamma_{cg}R)(\bar{\Lambda}\Gamma_{kl}R)(\Lambda\Gamma^{efghi}\Lambda)N_{hi}(\Lambda\Gamma^{ckl}D) \notag \\
& = -\frac{1}{3\eta^{3}}(\bar{\Lambda}\Gamma_{ef}\bar{\Lambda})(\bar{\Lambda}\Gamma_{gc}R)(\bar{\Lambda}\Gamma_{kl}R)(\Lambda\Gamma^{efghi}\Lambda)N_{hi}(\Lambda\Gamma^{ckl}D) \label{neweq21}
\end{align}
One then concludes that
\begin{eqnarray}
b_{simp}^{(3)} &=& \frac{4}{\eta^{3}}(\bar{\Lambda}\Gamma_{ab}\bar{\Lambda})(\bar{\Lambda}\Gamma_{cd}R)(\bar{\Lambda}\Gamma_{e}{}^{g}R)(\Lambda\Gamma^{abcef}\Lambda)(\Lambda\Gamma^{d}D)N_{fg} \nonumber\\
&& -\frac{2}{\eta^{3}}(\bar{\Lambda}\Gamma_{ab}\bar{\Lambda})(\bar{\Lambda}\Gamma_{cd}R)(\bar{\Lambda}R)(\Lambda\Gamma^{abcef}\Lambda)(\Lambda\Gamma^{d}D)N_{ef} \nonumber \\
&& -\frac{1}{3\eta^{3}}(\bar{\Lambda}\Gamma_{ef}\bar{\Lambda})(\bar{\Lambda}\Gamma_{gc}R)(\bar{\Lambda}\Gamma_{ab}R)(\Lambda\Gamma^{efghi}\Lambda)N_{hi}(\Lambda\Gamma^{cab}D)\label{eq404}
\end{eqnarray}

\vspace{2mm}
Now, let us simplify the terms proportional to $\eta^{-4}$:
\begin{align}
b_{simp}^{(4)} &=
 -\frac{4}{\eta^{2}}(\bar{\Lambda}\Gamma_{ab}\bar{\Lambda})(\bar{\Lambda}\Gamma_{cd}R)(\Lambda\Gamma^{bd}\Lambda)(\Lambda\Gamma^{aj}\Lambda)(\frac{2}{3\eta^{2}})(\bar{\Lambda}\Gamma_{ef}\bar{\Lambda})(\bar{\Lambda}\Gamma_{gj}R)(\Lambda\Gamma^{efghi}\Lambda)N_{hi}[\frac{2}{\eta^{2}}(\bar{\Lambda}\Gamma_{lm}\bar{\Lambda})\times \notag\\
&(\bar{\Lambda}\Gamma_{np}R)(\Lambda\Gamma^{lmnqc}\Lambda)N_{pq} + \frac{2}{3\eta^{2}}(\bar{\Lambda}\Gamma_{lm}\bar{\Lambda})(\bar{\Lambda}\Gamma_{n}^{\hspace{2mm}c}R)(\Lambda\Gamma^{lmnpq}\Lambda)N_{pq}\notag\\
& - \frac{2}{3\eta^{2}}(\bar{\Lambda}\Gamma_{ln}\bar{\Lambda})(\bar{\Lambda}R)(\Lambda\Gamma^{lcnpq}\Lambda)N_{pq} + \frac{2}{3\eta^{3}}(\bar{\Lambda}\Gamma_{cs}\bar{\Lambda})(\Lambda\Gamma^{s}_{\hspace{2mm}k}\Lambda)(\bar{\Lambda}\Gamma_{rt}\bar{\Lambda})(\bar{\Lambda}\Gamma_{uk}R)(\Lambda\Gamma^{rtupq}\Lambda)N_{pq}] \notag\\
& = -\frac{4}{3\eta^{3}}(\bar{\Lambda}\Gamma_{ac}R)(\Lambda\Gamma^{aj}\Lambda)(\bar{\Lambda}\Gamma_{ef}\bar{\Lambda})(\bar{\Lambda}\Gamma_{gj}R)(\Lambda\Gamma^{efghi}\Lambda)N_{hi}[\frac{2}{\eta^{2}}(\bar{\Lambda}\Gamma_{lm}\bar{\Lambda})(\bar{\Lambda}\Gamma_{np}R)(\Lambda\Gamma^{lmnqc}\Lambda)N_{pq} \notag \\
& + \frac{2}{3\eta^{2}}(\bar{\Lambda}\Gamma_{lm}\bar{\Lambda})(\bar{\Lambda}\Gamma_{n}^{\hspace{2mm}c}R)(\Lambda\Gamma^{lmnpq}\Lambda)N_{pq} - \frac{2}{3\eta^{2}}(\bar{\Lambda}\Gamma_{ln}\bar{\Lambda})(\bar{\Lambda}R)(\Lambda\Gamma^{lcnpq}\Lambda)N_{pq} \notag \\ 
& + \frac{1}{3\eta{2}}(\bar{\Lambda}\Gamma_{rt}\bar{\Lambda})(\bar{\Lambda}\Gamma^{c}_{\hspace{2mm}u}R)(\Lambda\Gamma^{rtupq}\Lambda)N_{pq}] \notag\\
&= \frac{2}{3\eta^{2}}(\bar{\Lambda}\Gamma_{gc}R)(\bar{\Lambda}\Gamma_{ef}R)(\Lambda\Gamma^{efghi}\Lambda)N_{hi}[\frac{2}{\eta^{2}}(\bar{\Lambda}\Gamma_{lm}\bar{\Lambda})(\bar{\Lambda}\Gamma_{np}R)(\Lambda\Gamma^{lmnqc}\Lambda)N_{pq} \notag \\
& + \frac{1}{3\eta^{2}}(\bar{\Lambda}\Gamma_{lm}\bar{\Lambda})(\bar{\Lambda}\Gamma_{n}^{\hspace{2mm}c}R)(\Lambda\Gamma^{lmnpq}\Lambda)N_{pq} - \frac{2}{3\eta^{2}}(\bar{\Lambda}\Gamma_{ln}\bar{\Lambda})(\bar{\Lambda}R)(\Lambda\Gamma^{lcnpq}\Lambda)N_{pq}] \notag \\
&= \frac{2}{3\eta^{2}}(\bar{\Lambda}\Gamma_{gc}R)(\bar{\Lambda}\Gamma_{ef}R)(\Lambda\Gamma^{efghi}\Lambda)N_{hi}[\frac{2}{\eta^{2}}(\bar{\Lambda}\Gamma_{lm}\bar{\Lambda})(\bar{\Lambda}\Gamma_{np}R)(\Lambda\Gamma^{lmnqc}\Lambda)N_{pq} \notag \\
& - \frac{1}{3\eta^{2}}(\bar{\Lambda}\Gamma_{n}^{\hspace{2mm}c}\bar{\Lambda})(\bar{\Lambda}\Gamma_{lm}R)(\Lambda\Gamma^{lmnpq}\Lambda)N_{pq} - \frac{2}{3\eta^{2}}(\bar{\Lambda}\Gamma_{ln}\bar{\Lambda})(\bar{\Lambda}R)(\Lambda\Gamma^{lcnpq}\Lambda)N_{pq}] \notag \\
& = \frac{4}{3\eta^{4}}(\bar{\Lambda}\Gamma_{lm}\bar{\Lambda})(\bar{\Lambda}\Gamma_{np}R)(\bar{\Lambda}\Gamma_{gc}R)(\bar{\Lambda}\Gamma_{ef}R)(\Lambda\Gamma^{efghi}\Lambda)N_{hi}(\Lambda\Gamma^{lmnqc}\Lambda)N_{pq} \notag \\
& -\frac{2}{9\eta^{4}}(\bar{\Lambda}R)(\bar{\Lambda}\Gamma_{ef}R)(\bar{\Lambda}\Gamma_{ng}\bar{\Lambda})(\bar{\Lambda}\Gamma_{lm}R)(\Lambda\Gamma^{efghi}\Lambda)N_{hi}(\Lambda\Gamma^{lmnpq}\Lambda)N_{pq} \notag\\
& - \frac{4}{9\eta^{4}}(\bar{\Lambda}\Gamma_{gm}R)(\bar{\Lambda}\Gamma_{ef}R)(\bar{\Lambda}\Gamma_{ln}\bar{\Lambda})(\bar{\Lambda}R)(\Lambda\Gamma^{efghi}\Lambda)N_{hi}(\Lambda\Gamma^{lmnpq}\Lambda)N_{pq} \notag \\
& = \frac{4}{3\eta^{4}}(\bar{\Lambda}\Gamma_{lm}\bar{\Lambda})(\bar{\Lambda}\Gamma_{np}R)(\bar{\Lambda}\Gamma_{gc}R)(\bar{\Lambda}\Gamma_{ef}R)(\Lambda\Gamma^{efghi}\Lambda)N_{hi}(\Lambda\Gamma^{lmnqc}\Lambda)N_{pq} \notag \\
& +\frac{2}{9\eta^{4}}(\bar{\Lambda}\Gamma_{ef}R)(\bar{\Lambda}\Gamma_{lm}\bar{\Lambda})(\bar{\Lambda}\Gamma_{ng}R)(\bar{\Lambda}R)(\Lambda\Gamma^{efghi}\Lambda)N_{hi}(\Lambda\Gamma^{lmnpq}\Lambda)N_{pq} \notag\\
& - \frac{4}{9\eta^{4}}(\bar{\Lambda}\Gamma_{gm}R)(\bar{\Lambda}\Gamma_{ef}R)(\bar{\Lambda}\Gamma_{ln}\bar{\Lambda})(\bar{\Lambda}R)(\Lambda\Gamma^{efghi}\Lambda)N_{hi}(\Lambda\Gamma^{lmnpq}\Lambda)N_{pq} \notag \\
& = \frac{4}{3\eta^{4}}(\bar{\Lambda}\Gamma_{lm}\bar{\Lambda})(\bar{\Lambda}\Gamma_{np}R)(\bar{\Lambda}\Gamma_{gc}R)(\bar{\Lambda}\Gamma_{ef}R)(\Lambda\Gamma^{efghi}\Lambda)N_{hi}(\Lambda\Gamma^{lmnqc}\Lambda)N_{pq} \notag \\
& - \frac{2}{3\eta^{4}}(\bar{\Lambda}\Gamma_{gm}R)(\bar{\Lambda}\Gamma_{ef}R)(\bar{\Lambda}\Gamma_{ln}\bar{\Lambda})(\bar{\Lambda}R)(\Lambda\Gamma^{efghi}\Lambda)N_{hi}(\Lambda\Gamma^{lmnpq}\Lambda)N_{pq} \notag \\
& = \frac{4}{3\eta^{4}}(\bar{\Lambda}\Gamma_{lm}\bar{\Lambda})(\bar{\Lambda}\Gamma_{np}R)(\bar{\Lambda}\Gamma_{gc}R)(\bar{\Lambda}\Gamma_{ef}R)(\Lambda\Gamma^{efghi}\Lambda)N_{hi}(\Lambda\Gamma^{lmnqc}\Lambda)N_{pq} \notag \\
& - \frac{2}{3\eta^{4}}(\bar{\Lambda}\Gamma_{ln}\bar{\Lambda})(\bar{\Lambda}\Gamma_{mg}R)(\bar{\Lambda}\Gamma_{ef}R)(\bar{\Lambda}R)(\Lambda\Gamma^{efghi}\Lambda)N_{hi}(\Lambda\Gamma^{lnmpq}\Lambda)N_{pq} \notag \\
& = \frac{4}{3\eta^{4}}(\bar{\Lambda}\Gamma_{ef}\bar{\Lambda})(\bar{\Lambda}\Gamma_{gc}R)(\bar{\Lambda}\Gamma_{lm}R)(\bar{\Lambda}\Gamma_{np}R)(\Lambda\Gamma^{efghi}\Lambda)N_{hi}(\Lambda\Gamma^{clmnq}\Lambda)N_{qp} \notag \\
& - \frac{2}{3\eta^{4}}(\bar{\Lambda}\Gamma_{ef}\bar{\Lambda})(\bar{\Lambda}\Gamma_{gm}R)(\bar{\Lambda}\Gamma_{ln}R)(\bar{\Lambda}R)(\Lambda\Gamma^{efghi}\Lambda)N_{hi}(\Lambda\Gamma^{lnmpq}\Lambda)N_{pq}
\end{align} 
Or equivalently,
\begin{eqnarray}
b_{simp}^{(4)} &=& \frac{4}{3\eta^{4}}(\bar{\Lambda}\Gamma_{ef}\bar{\Lambda})(\bar{\Lambda}\Gamma_{gc}R)(\bar{\Lambda}\Gamma_{lj}R)(\bar{\Lambda}\Gamma_{ka}R)(\Lambda\Gamma^{efghi}\Lambda)N_{hi}(\Lambda\Gamma^{cljkb}\Lambda)N_{ba} \nonumber \\
&& - \frac{2}{3\eta^{4}}(\bar{\Lambda}\Gamma_{ef}\bar{\Lambda})(\bar{\Lambda}\Gamma_{ga}R)(\bar{\Lambda}\Gamma_{ld}R)(\bar{\Lambda}R)(\Lambda\Gamma^{efghi}\Lambda)N_{hi}(\Lambda\Gamma^{ldabc}\Lambda)N_{bc} \nonumber\\
\label{neweq20new}
\end{eqnarray}
In order to compare the two expressions for the $b$-ghost, we should move all of the $N_{ab}$'s at the end of the expressions showed above. 

\vspace{2mm}
Let us start with the term proportional to $\eta^{-3}$. One should put the ghost current $N^{hi}$ to the right hand side of $(\Lambda\Gamma^{cab}D)$. For this purpose, let us compute the commutator between $N^{hi}$ and $(\Lambda\Gamma^{cab}D)$ with the symmetry properties written in \eqref{neweq21}:
\begin{eqnarray}
[N^{hi}, (\Lambda\Gamma^{cab}D)] &=& -2\eta^{hi}_{ab}(\Lambda\Gamma^{c}D) - 4\eta^{ca}_{hi}(\Lambda\Gamma^{b}D) - 4\eta^{ha}(\Lambda\Gamma^{cib}D) - 2\eta^{ch}(\Lambda\Gamma^{iab}D)\nonumber \\
&& + (\Lambda\Gamma^{chiab}D)
\end{eqnarray}
The use of the identities \eqref{newapp4}, \eqref{newapp15} allows us to claim all terms will vanish except for the last one. Therefore,
\begin{align}
K_{2} & = -\frac{1}{3\eta^{3}}(\bar{\Lambda}\Gamma_{ef}\bar{\Lambda})(\bar{\Lambda}\Gamma_{gc}R)(\bar{\Lambda}\Gamma_{ab}R)(\Lambda\Gamma^{efghi}\Lambda)(\Lambda\Gamma^{cab}D)N_{hi} \notag\\
\end{align}

\vspace{2mm}
We now move on to analyzing the terms proportional to $\eta^{-4}$. One should move $N_{hi}$ to the 
right hand side of $(\Lambda\Gamma^{clabd}\Lambda)$:
\begin{align}
[N^{hi}, (\Lambda\Gamma^{clabd}\Lambda)] & = 4\eta^{hd}(\Lambda\Gamma^{cilab}\Lambda) - 4\eta^{hb}(\Lambda\Gamma^{cilad}\Lambda) - 8\eta^{hl}(\Lambda\Gamma^{ciabd}\Lambda) - 4\eta^{ch}(\Lambda\Gamma^{ilabd}\Lambda)
\end{align}
These are the only relevant terms in \eqref{neweq20new}, as a direct consequence of eqn. \eqref{newapp14} and the symmetry properties of the expression where this term appears in \eqref{neweq20new}. The last two terms do not contribute because of eqns. \eqref{newapp4}, \eqref{newapp15}. So we are left with:
\begin{align}
Z_{1} & = \frac{4}{3\eta^{4}}(\bar{\Lambda}\Gamma_{ef}\bar{\Lambda})(\bar{\Lambda}\Gamma_{gc}R)(\bar{\Lambda}\Gamma_{la}R)(\bar{\Lambda}\Gamma_{bj}R)(\Lambda\Gamma^{efghi}\Lambda)(\Lambda\Gamma^{clabk}\Lambda)N_{hi}N_{kj} \notag \\
& + \frac{16}{3\eta^{4}}(\bar{\Lambda}\Gamma_{ef}\bar{\Lambda})(\bar{\Lambda}\Gamma_{gc}R)(\bar{\Lambda}\Gamma_{la}R)(\bar{\Lambda}\Gamma_{bj}R)[(\Lambda\Gamma^{efgki}\Lambda)(\Lambda\Gamma^{cilab}\Lambda) - (\Lambda\Gamma^{efgbi}\Lambda)(\Lambda\Gamma^{cilak}\Lambda)]N_{kj} \notag \\
& = \frac{4}{3\eta^{4}}(\bar{\Lambda}\Gamma_{ef}\bar{\Lambda})(\bar{\Lambda}\Gamma_{gc}R)(\bar{\Lambda}\Gamma_{la}R)(\bar{\Lambda}\Gamma_{bj}R)(\Lambda\Gamma^{efghi}\Lambda)(\Lambda\Gamma^{clabk}\Lambda)N_{hi}N_{kj} \notag \\
& + \frac{16}{3\eta^{4}}[(\bar{\Lambda}\Gamma_{ef}\bar{\Lambda})(\bar{\Lambda}\Gamma_{gc}R)(\bar{\Lambda}\Gamma_{la}R)(\bar{\Lambda}\Gamma_{bj}R)(\Lambda\Gamma^{efgki}\Lambda)(\Lambda\Gamma^{cilab}\Lambda) \notag \\
& - \frac{16}{3\eta^{4}}(\bar{\Lambda}\Gamma_{la}\bar{\Lambda})(\bar{\Lambda}\Gamma_{cg}R)(\bar{\Lambda}\Gamma_{ef}R)(\bar{\Lambda}\Gamma_{bj}R)(\Lambda\Gamma^{lacbi}\Lambda)(\Lambda\Gamma^{giefk}\Lambda)]N_{kj} \notag \\
& = \frac{4}{3\eta^{4}}(\bar{\Lambda}\Gamma_{ef}\bar{\Lambda})(\bar{\Lambda}\Gamma_{gc}R)(\bar{\Lambda}\Gamma_{la}R)(\bar{\Lambda}\Gamma_{bj}R)(\Lambda\Gamma^{efghi}\Lambda)(\Lambda\Gamma^{clabk}\Lambda)N_{hi}N_{kj} \notag \\
& + \frac{16}{3\eta^{4}}[(\bar{\Lambda}\Gamma_{ef}\bar{\Lambda})(\bar{\Lambda}\Gamma_{gc}R)(\bar{\Lambda}\Gamma_{la}R)(\bar{\Lambda}\Gamma_{bj}R)(\Lambda\Gamma^{efgki}\Lambda)(\Lambda\Gamma^{cilab}\Lambda) \notag \\
& - \frac{16}{3\eta^{4}}(\bar{\Lambda}\Gamma_{gc}\bar{\Lambda})(\bar{\Lambda}\Gamma_{la}R)(\bar{\Lambda}\Gamma_{ef}R)(\bar{\Lambda}\Gamma_{bj}R)(\Lambda\Gamma^{clabi}\Lambda)(\Lambda\Gamma^{efgik}\Lambda)]N_{kj} \notag \\
& = \frac{4}{3\eta^{4}}(\bar{\Lambda}\Gamma_{ef}\bar{\Lambda})(\bar{\Lambda}\Gamma_{gc}R)(\bar{\Lambda}\Gamma_{la}R)(\bar{\Lambda}\Gamma_{bj}R)(\Lambda\Gamma^{efghi}\Lambda)(\Lambda\Gamma^{clabk}\Lambda)N_{hi}N_{kj} \notag \\
& + \frac{16}{3\eta^{4}}[(\bar{\Lambda}\Gamma_{ef}\bar{\Lambda})(\bar{\Lambda}\Gamma_{gc}R)(\bar{\Lambda}\Gamma_{la}R)(\bar{\Lambda}\Gamma_{bj}R)(\Lambda\Gamma^{efgki}\Lambda)(\Lambda\Gamma^{cilab}\Lambda) \notag \\
& - \frac{16}{3\eta^{4}}(\bar{\Lambda}\Gamma_{ef}\bar{\Lambda})(\bar{\Lambda}\Gamma_{gc}R)(\bar{\Lambda}\Gamma_{la}R)(\bar{\Lambda}\Gamma_{bj}R)(\Lambda\Gamma^{cliab}\Lambda)(\Lambda\Gamma^{efgki}\Lambda)]N_{kj} \notag \\
& = \frac{4}{3\eta^{4}}(\bar{\Lambda}\Gamma_{ef}\bar{\Lambda})(\bar{\Lambda}\Gamma_{gc}R)(\bar{\Lambda}\Gamma_{la}R)(\bar{\Lambda}\Gamma_{bj}R)(\Lambda\Gamma^{efghi}\Lambda)(\Lambda\Gamma^{clabk}\Lambda)N_{hi}N_{kj}
\end{align}
Now let us make the same procedure with the last term in \eqref{neweq20new}. The relevant commutation relation is:
\begin{align}
[N^{hi}, (\Lambda\Gamma^{labjk}\Lambda)] & = -8\eta^{hj}(\Lambda\Gamma^{ilabk}\Lambda) - 4\eta^{ha}(\Lambda\Gamma^{ilbjk}\Lambda) + 8\eta^{hl}(\Lambda\Gamma^{iabjk}\Lambda)
\end{align}
Once again, we have obtained this result by using the identity \eqref{newapp14} and the symmetry properties of the expression where this term appears in \eqref{neweq20new}. After applying \eqref{newapp4}, \eqref{newapp15}, the last two terms vanish and we obtain
\begin{align}
Z_{2} & = - \frac{2}{3\eta^{4}}(\bar{\Lambda}\Gamma_{ef}\bar{\Lambda})(\bar{\Lambda}\Gamma_{ga}R)(\bar{\Lambda}\Gamma_{lb}R)(\bar{\Lambda}R)(\Lambda\Gamma^{efghi}\Lambda)(\Lambda\Gamma^{lbajk}\Lambda)N_{hi}N_{jk} \notag \\
& + \frac{16}{3\eta^{4}}(\bar{\Lambda}\Gamma_{ef}\bar{\Lambda})(\bar{\Lambda}\Gamma_{ga}R)(\bar{\Lambda}\Gamma_{lb}R)(\bar{\Lambda}R)(\Lambda\Gamma^{efgji}\Lambda)(\Lambda\Gamma^{ilabk}\Lambda)N_{jk}\label{endend1}
\end{align}
The last term in \eqref{endend1} vanishes as can be seen from the following computation:
\begin{align}
M & = \frac{16}{3\eta^{4}}(\bar{\Lambda}\Gamma_{ef}\bar{\Lambda})(\bar{\Lambda}\Gamma_{ga}R)(\bar{\Lambda}\Gamma_{lb}R)(\bar{\Lambda}R)(\Lambda\Gamma^{efgji}\Lambda)(\Lambda\Gamma^{ilabk}\Lambda)N_{jk} \notag \\
& =  -\frac{16}{3\eta^{4}}(\bar{\Lambda}\Gamma_{lb}\bar{\Lambda})(\bar{\Lambda}\Gamma_{ag}R)(\bar{\Lambda}\Gamma_{ef}R)(\bar{\Lambda}R)(\Lambda\Gamma^{lbaki}\Lambda)(\Lambda\Gamma^{iegfj}\Lambda)N_{jk} \notag \\
& = \frac{16}{3\eta^{4}}(\bar{\Lambda}\Gamma_{ef}\bar{\Lambda})(\bar{\Lambda}\Gamma_{ag}R)(\bar{\Lambda}\Gamma_{lb}R)(\bar{\Lambda}R)(\Lambda\Gamma^{labki}\Lambda)(\Lambda\Gamma^{efgji}\Lambda)N_{jk} \notag \\
& = -\frac{16}{3\eta^{4}}(\bar{\Lambda}\Gamma_{ef}\bar{\Lambda})(\bar{\Lambda}\Gamma_{ga}R)(\bar{\Lambda}\Gamma_{lb}R)(\bar{\Lambda}R)(\Lambda\Gamma^{labki}\Lambda)(\Lambda\Gamma^{efgji}\Lambda)N_{jk} \notag \\
& = -M \notag \\
\rightarrow M & = 0
\end{align}
Therefore,
\begin{equation}
Z_{2} = - \frac{2}{3\eta^{4}}(\bar{\Lambda}\Gamma_{ef}\bar{\Lambda})(\bar{\Lambda}\Gamma_{ga}R)(\bar{\Lambda}\Gamma_{lb}R)(\bar{\Lambda}R)(\Lambda\Gamma^{efghi}\Lambda)(\Lambda\Gamma^{lbajk}\Lambda)N_{hi}N_{jk}
\end{equation}
This means that $b^{(4)}_{simpl}$ does not change after moving $N_{mn}$ to the right end of each term in \eqref{neweq20new}.

\vspace{2mm}
Therefore, one concludes that $b_{simpl}^{(3)}$ changes by the factor $$- \frac{1}{3\eta^{3}}
(\bar{\Lambda}\Gamma_{ef}\bar{\Lambda})(\bar{\Lambda}\Gamma_{gc}R)(\bar{\Lambda}\Gamma_{ab}R)(\Lambda\Gamma^{efghi}\Lambda)(\Lambda\Gamma^{chiab}D)$$ and $b_{simpl}^{(4)}$ does not receive any contribution. In this manner, the simplified $b$-ghost takes the following form:
\begin{eqnarray}
b_{simpl} &=& P^{i}[\frac{1}{2}\eta^{-1}(\bar{\Lambda}\Gamma_{ab}\bar{\Lambda})(\Lambda\Gamma^{ab}\Gamma_{i}D) + \eta^{-2}L^{(1)}_{ab,cd}[2(\Lambda\Gamma^{abc}{}_{ki}\Lambda)N^{dk}\nonumber \\
&& 
+ \frac{2}{3}(\eta^{b}{}_{p}\eta^{d}{}_{i} - \eta^{bd}\eta_{pi})(\Lambda\Gamma^{apcfj}\Lambda)N_{fj}]] + \frac{1}{\eta^{2}}L^{(1)}_{ac,ef}(\Lambda\Gamma^{a}D)(\Lambda\Gamma^{cef}D) + \nonumber\\
&& + \frac{4}{\eta^{3}}(\bar{\Lambda}\Gamma_{ab}\bar{\Lambda})(\bar{\Lambda}\Gamma_{cg}R)(\bar{\Lambda}\Gamma_{d}{}^{f}R)(\Lambda\Gamma^{abcde}\Lambda)(\Lambda\Gamma^{g}D)N_{ef} \nonumber\\
&& -\frac{2}{\eta^{3}}(\bar{\Lambda}\Gamma_{ab}\bar{\Lambda})(\bar{\Lambda}\Gamma_{cd}R)(\bar{\Lambda}R)(\Lambda\Gamma^{abcef}\Lambda)(\Lambda\Gamma^{d}D)N_{ef} \nonumber\\
&& -\frac{1}{3\eta^{3}}(\bar{\Lambda}\Gamma_{ef}\bar{\Lambda})(\bar{\Lambda}\Gamma_{gc}R)(\bar{\Lambda}\Gamma_{ab}R)(\Lambda\Gamma^{efghi}\Lambda)(\Lambda\Gamma^{cab}D)N_{hi} \nonumber \\
&& + \frac{2}{3\eta^{4}}(\bar{\Lambda}\Gamma_{ef}\bar{\Lambda})(\bar{\Lambda}\Gamma_{gc}R)(\bar{\Lambda}\Gamma_{la}R)(\bar{\Lambda}\Gamma_{bj}R)(\Lambda\Gamma^{efghi}\Lambda)(\Lambda\Gamma^{clabd}\Lambda)\{N_{hi}, N_{dj}\} \nonumber \\
&& - \frac{1}{3\eta^{4}}(\bar{\Lambda}\Gamma_{ef}\bar{\Lambda})(\bar{\Lambda}\Gamma_{ga}R)(\bar{\Lambda}\Gamma_{lb}R)(\bar{\Lambda}R)(\Lambda\Gamma^{efghi}\Lambda)(\Lambda\Gamma^{lbacd}\Lambda)\{N_{hi}, N_{cd}\}\nonumber \\
&& - \frac{1}{3\eta^{3}}(\bar{\Lambda}\Gamma_{ef}\bar{\Lambda})(\bar{\Lambda}\Gamma_{gc}R)(\bar{\Lambda}\Gamma_{ab}R)(\Lambda\Gamma^{efghi}\Lambda)(\Lambda\Gamma^{chiab}D) \label{newapp30}
\end{eqnarray}
where we have written the anticommutator instead of the ordinary product of $N_{ab}$'s after making use of the relation $2N_{ab}N_{cd} = [N_{ab}, N_{cd}] + \{N_{ab}, N_{cd}\}$, and the fact that the contribution coming from the commutator vanishes because of the identity \eqref{newapp11}. 
%

\vspace{2mm}
This result should be compared with the expansion of the $b$-ghost in \eqref{section4eq16}:
\begin{align}\label{newapp2000}
b & = \frac{1}{2}\eta^{-1}(\bar{\Lambda}\Gamma_{ab}\bar{\Lambda})(\Lambda\Gamma^{ab}\Gamma^{i}D)P_{i} + \eta^{-2}L^{(1)}_{ab,cd}[(\Lambda\Gamma^{a}D)(\Lambda\Gamma^{bcd}D) + 2(\Lambda\Gamma^{abc}{}_{ij}\Lambda)N^{di}P^{j} \notag \\
& \frac{2}{3}(\eta^{b}{}_{p}\eta^{d}{}_{q} - \eta^{bd}\eta_{ef})(\Lambda\Gamma^{aecij}\Lambda)N_{ij}P^{f}] - \frac{1}{3}\eta^{-3}L^{(2)}_{ab,cd,ef}\{(\Lambda\Gamma^{abcij}\Lambda)(\Lambda\Gamma^{def}D)N_{ij} \notag \\
 & - 12[ (\Lambda\Gamma^{abcei}\Lambda)\eta^{fj} - \frac{2}{3}\eta^{f[a}(\Lambda\Gamma^{bce]ij}\Lambda](\Lambda\Gamma^{d}D)N_{ij}\} \notag \\
& + \frac{4}{3}\eta^{-4}L^{(3)}_{ab,cd,ef,gh}(\Lambda\Gamma^{abcij}\Lambda)[(\Lambda\Gamma^{defgk}\Lambda)\eta^{hl} - \frac{2}{3}\eta^{h[d}(\Lambda\Gamma^{efg]kl}\Lambda)]\{N_{ij},N_{kl}\}
\end{align}
Using eqn. \eqref{newapp5}, the quadratic term in $D_{\alpha}$ takes the form
\begin{equation}
b^{(2)} = \frac{1}{\eta^{2}}(\bar{\Lambda}\Gamma_{ab}\bar{\Lambda})(\bar{\Lambda}\Gamma_{cd}R)(\Lambda\Gamma^{a}D)(\Lambda\Gamma^{bcd}D)
\end{equation} 

\vspace{2mm}
We will now calculate the term proportional to $\eta^{-3}$ in two steps. First, let us focus on the term proportional to $(\Lambda\Gamma^{a}D)$, which will be called $K\ensuremath{'}_{1}$, and then on the term proportional to $(\Lambda\Gamma^{abc}D)$, which will be called $K\ensuremath{'}_{2}$. Thus,
\begin{eqnarray*}
K\ensuremath{'}_{1} &=& \frac{4}{\eta^{3}}L^{(2)}_{ab,cd,ef}[(\Lambda\Gamma^{abcei}\Lambda)(\Lambda\Gamma^{d}D)\eta^{fj}N_{ij} - \frac{2}{3}\eta^{f[a}(\Lambda\Gamma^{bce]ij}\Lambda)(\Lambda\Gamma^{d}D)N_{ij}]\\
&=& \frac{4}{\eta^{3}}(\bar{\Lambda}\Gamma_{ab}\bar{\Lambda})(\bar{\Lambda}\Gamma_{cd}R)(\bar{\Lambda}\Gamma_{e}^{\hspace{2mm}j}R)(\Lambda\Gamma^{abcei}\Lambda)(\Lambda\Gamma^{d}D)N_{ij}\\
&& - \frac{8}{3\eta^{3}}L^{(2)}_{ab,cd,ef}(\frac{1}{4})[\eta^{fa}(\Lambda\Gamma^{bceij}\Lambda) - \eta^{fb}(\Lambda\Gamma^{aceij}\Lambda) + \eta^{fc}(\Lambda\Gamma^{abeij}\Lambda)](\Lambda\Gamma^{d}D)N_{ij}\\
&=& \frac{4}{\eta^{3}}(\bar{\Lambda}\Gamma_{ab}\bar{\Lambda})(\bar{\Lambda}\Gamma_{cd}R)(\bar{\Lambda}\Gamma_{e}^{\hspace{2mm}j}R)(\Lambda\Gamma^{abcei}\Lambda)(\Lambda\Gamma^{d}D)N_{ij}\\
&& - \frac{2}{3\eta^{3}}L^{(2)}_{ab,cd,ef}[2\eta^{fa}(\Lambda\Gamma^{bceij}\Lambda) + \eta^{fc}(\Lambda\Gamma^{abeij}\Lambda)](\Lambda\Gamma^{d}D)N_{ij}\\
&=& \frac{4}{\eta^{3}}(\bar{\Lambda}\Gamma_{ab}\bar{\Lambda})(\bar{\Lambda}\Gamma_{cd}R)(\bar{\Lambda}\Gamma_{e}^{\hspace{2mm}j}R)(\Lambda\Gamma^{abcei}\Lambda)(\Lambda\Gamma^{d}D)N_{ij}\\
&& - \frac{2}{3\eta^{3}}[2L^{(2)\hspace{5mm}a}_{ab,cd,e}(\Lambda\Gamma^{bceij}\Lambda) + L^{(2)\hspace{5mm}c}_{ab,cd,e}(\Lambda\Gamma^{abeij}\Lambda)](\Lambda\Gamma^{d}D)N_{ij}\\
&=& \frac{4}{\eta^{3}}(\bar{\Lambda}\Gamma_{ab}\bar{\Lambda})(\bar{\Lambda}\Gamma_{cd}R)(\bar{\Lambda}\Gamma_{e}^{\hspace{2mm}j}R)(\Lambda\Gamma^{abcei}\Lambda)(\Lambda\Gamma^{d}D)N_{ij}\\
&& - \frac{2}{9\eta^{3}}\{[4(\bar{\Lambda}\Gamma_{eb}\bar{\Lambda})(\bar{\Lambda}\Gamma_{cd}R)(\bar{\Lambda}R) - 2(\bar{\Lambda}\Gamma_{cd}\bar{\Lambda})(\bar{\Lambda}\Gamma_{ab}R)(\bar{\Lambda}\Gamma_{e}^{\hspace{2mm}a}R)](\Lambda\Gamma^{bceij}\Lambda)\\
&& + [(\bar{\Lambda}\Gamma_{ab}\bar{\Lambda})(\bar{\Lambda}\Gamma_{cd}R)(\bar{\Lambda}\Gamma_{e}^{\hspace{2mm}c}R) - 2(\bar{\Lambda}\Gamma_{ed}\bar{\Lambda})(\bar{\Lambda}\Gamma_{ab}R)(\bar{\Lambda}R)](\Lambda\Gamma^{abeij}\Lambda)\}(\Lambda\Gamma^{d}D)N_{ij}\\
&=&\frac{4}{\eta^{3}}(\bar{\Lambda}\Gamma_{ab}\bar{\Lambda})(\bar{\Lambda}\Gamma_{cd}R)(\bar{\Lambda}\Gamma_{e}^{\hspace{2mm}j}R)(\Lambda\Gamma^{abcei}\Lambda)(\Lambda\Gamma^{d}D)N_{ij}\\
&& - \frac{2}{9\eta^{3}}\{[6(\bar{\Lambda}\Gamma_{eb}\bar{\Lambda})(\bar{\Lambda}\Gamma_{cd}R)(\bar{\Lambda}R) - 2(\bar{\Lambda}\Gamma_{cd}\bar{\Lambda})(\bar{\Lambda}\Gamma_{ab}R)(\bar{\Lambda}\Gamma_{e}^{\hspace{2mm}a}R)](\Lambda\Gamma^{bceij}\Lambda)\\
&& + (\bar{\Lambda}\Gamma_{ab}\bar{\Lambda})(\bar{\Lambda}\Gamma_{cd}R)(\bar{\Lambda}\Gamma_{e}^{\hspace{2mm}c}R)(\Lambda\Gamma^{abeij}\Lambda)\}(\Lambda\Gamma^{d}D)N_{ij}\\
&=& \frac{4}{\eta^{3}}(\bar{\Lambda}\Gamma_{ab}\bar{\Lambda})(\bar{\Lambda}\Gamma_{cd}R)(\bar{\Lambda}\Gamma_{e}^{\hspace{2mm}j}R)(\Lambda\Gamma^{abcei}\Lambda)(\Lambda\Gamma^{d}D)N_{ij}\\
&& -\frac{2}{9\eta^{3}}\{[6(\bar{\Lambda}\Gamma_{eb}\bar{\Lambda})(\bar{\Lambda}\Gamma_{cd}R)(\bar{\Lambda}R) - 2(\bar{\Lambda}\Gamma_{ce}\bar{\Lambda})(\bar{\Lambda}\Gamma_{db}R)(\bar{\Lambda}R) - (\bar{\Lambda}\Gamma_{cb}\bar{\Lambda})(\bar{\Lambda}\Gamma_{ad}R)(\bar{\Lambda}\Gamma_{e}^{\hspace{2mm}a}R)\\
&& - (\bar{\Lambda}\Gamma_{ed}\bar{\Lambda})(\bar{\Lambda}\Gamma_{cb}R)(\bar{\Lambda}R)](\Lambda\Gamma^{bceij}\Lambda) - (\bar{\Lambda}\Gamma_{cb}\bar{\Lambda})(\bar{\Lambda}\Gamma_{ad}R)(\bar{\Lambda}\Gamma_{e}^{\hspace{2mm}a}R)(\Lambda\Gamma^{bceij}\Lambda)\}(\Lambda\Gamma^{d}D)N_{ij}\\
&=& \frac{4}{\eta^{3}}(\bar{\Lambda}\Gamma_{ab}\bar{\Lambda})(\bar{\Lambda}\Gamma_{cd}R)(\bar{\Lambda}\Gamma_{e}^{\hspace{2mm}j}R)(\Lambda\Gamma^{abcei}\Lambda)(\Lambda\Gamma^{d}D)N_{ij}\\
&& -\frac{2}{9\eta^{3}}\{[6(\bar{\Lambda}\Gamma_{eb}\bar{\Lambda})(\bar{\Lambda}\Gamma_{cd}R)(\bar{\Lambda}R) - 2(\bar{\Lambda}\Gamma_{ce}\bar{\Lambda})(\bar{\Lambda}\Gamma_{db}R)(\bar{\Lambda}R) - 2(\bar{\Lambda}\Gamma_{cb}\bar{\Lambda})(\bar{\Lambda}\Gamma_{ad}R)(\bar{\Lambda}\Gamma_{e}^{\hspace{2mm}a}R)\\
&& - (\bar{\Lambda}\Gamma_{ed}\bar{\Lambda})(\bar{\Lambda}\Gamma_{cb}R)(\bar{\Lambda}R)](\Lambda\Gamma^{bceij}\Lambda)\}(\Lambda\Gamma^{d}D)N_{ij}\\
&=& \frac{4}{\eta^{3}}(\bar{\Lambda}\Gamma_{ab}\bar{\Lambda})(\bar{\Lambda}\Gamma_{cd}R)(\bar{\Lambda}\Gamma_{e}^{\hspace{2mm}j}R)(\Lambda\Gamma^{abcei}\Lambda)(\Lambda\Gamma^{d}D)N_{ij}\\
&& -\frac{2}{9\eta^{3}}\{[6(\bar{\Lambda}\Gamma_{eb}\bar{\Lambda})(\bar{\Lambda}\Gamma_{cd}R)(\bar{\Lambda}R) - 2(\bar{\Lambda}\Gamma_{ce}\bar{\Lambda})(\bar{\Lambda}\Gamma_{db}R)(\bar{\Lambda}R) - 2(\bar{\Lambda}\Gamma_{cb}\bar{\Lambda})(\bar{\Lambda}\Gamma_{ed}R)(\bar{\Lambda}R)\\
&& + (\bar{\Lambda}\Gamma_{cb}\bar{\Lambda})(\bar{\Lambda}\Gamma_{ed}R)(\bar{\Lambda}R)](\Lambda\Gamma^{bceij}\Lambda)\}(\Lambda\Gamma^{d}D)N_{ij}
\end{eqnarray*}
where eqns. \eqref{newapp4}, \eqref{newapp7}, \eqref{newapp8}, \eqref{newapp9} were used. By using the antisymmetry in $(b,c,e)$, one then shows that:
\begin{eqnarray}\label{endeq15}
K\ensuremath{'}_{1} &=& \frac{4}{\eta^{3}}(\bar{\Lambda}\Gamma_{ab}\bar{\Lambda})(\bar{\Lambda}\Gamma_{cd}R)(\bar{\Lambda}\Gamma_{e}^{\hspace{2mm}j}R)(\Lambda\Gamma^{abcei}\Lambda)(\Lambda\Gamma^{d}D)N_{ij}\nonumber \\
&& - \frac{2}{\eta^{3}}(\bar{\Lambda}\Gamma_{bc}\bar{\Lambda})(\bar{\Lambda}\Gamma_{ed}R)(\bar{\Lambda}R)(\Lambda\Gamma^{bceij}\Lambda)(\Lambda\Gamma^{d}D)N_{ij}
\end{eqnarray}

\vspace{2mm}
On the other hand, $K\ensuremath{'}_{2}$ is easily recognized to be
\begin{equation}\label{newneweq17}
K\ensuremath{'}_{2} = -\frac{1}{3\eta^{2}}(\bar{\Lambda}\Gamma_{ab}\bar{\Lambda})(\bar{\Lambda}\Gamma_{cd}R)(\bar{\Lambda}\Gamma_{ef}R)(\Lambda\Gamma^{abcpq}\Lambda)(\Lambda\Gamma^{def}D)N_{pq}
\end{equation}
Therefore, one learns that
\begin{eqnarray}
b^{(3)} &=& \frac{4}{\eta^{3}}(\bar{\Lambda}\Gamma_{ab}\bar{\Lambda})(\bar{\Lambda}\Gamma_{cd}R)(\bar{\Lambda}\Gamma_{e}{}^{j}R)(\Lambda\Gamma^{abcei}\Lambda)(\Lambda\Gamma^{d}D)N_{ij}\nonumber \\
&& - \frac{2}{\eta^{3}}(\bar{\Lambda}\Gamma_{bc}\bar{\Lambda})(\bar{\Lambda}\Gamma_{ed}R)(\bar{\Lambda}R)(\Lambda\Gamma^{bceij}\Lambda)(\Lambda\Gamma^{d}D)N_{ij}\nonumber \\
&& -\frac{1}{3\eta^{2}}(\bar{\Lambda}\Gamma_{ab}\bar{\Lambda})(\bar{\Lambda}\Gamma_{cd}R)(\bar{\Lambda}\Gamma_{ef}R)(\Lambda\Gamma^{abcgh}\Lambda)(\Lambda\Gamma^{def}D)N_{gh}
\end{eqnarray}
which should be compared with the analog expression in $b_{simpl}$, eqn. \eqref{eq404}.

\vspace{2mm}
The last term to be simplified is the one proportional to $\eta^{-4}$ in \eqref{newapp2000}:
\begin{equation}
b^{(4)} = \frac{4}{3}\eta^{-4}L^{(3)}_{ab,cd,ef,gh}(\Lambda\Gamma^{abcij}\Lambda)[(\Lambda\Gamma^{defgk}\Lambda)\eta^{hl} - \frac{2}{3}\eta^{h[d}(\Lambda\Gamma^{efg]kl}\Lambda)]\{N_{ij}, N_{kl}\}
\end{equation}
It is more convenient to do this in two steps. First, we will focus on the first term, which will be called $P_{1}$, and then on the second term, which will be called $P_{2}$:
\begin{align}\label{endeq18}
P_{1} &= \frac{4}{3}\eta^{-4}L^{(3)}_{ab,cd,ef,gh}(\Lambda\Gamma^{abcij}\Lambda)(\Lambda\Gamma^{defgk}\Lambda)\eta^{hl}\{N_{ij}, N_{kl}\} \notag \\
& = \frac{4}{3\eta^{4}}(\frac{1}{4})[(\bar{\Lambda}\Gamma_{ab}\bar\Lambda)(\bar{\Lambda}\Gamma_{cd}R)(\bar{\Lambda}\Gamma_{ef}R)(\bar{\Lambda}\Gamma_{gh}R) - (\bar{\Lambda}\Gamma_{cd}\bar\Lambda)(\bar{\Lambda}\Gamma_{ab}R)(\bar{\Lambda}\Gamma_{ef}R)(\bar{\Lambda}\Gamma_{gh}R) + \notag \\
& + (\bar{\Lambda}\Gamma_{ef}\bar\Lambda)(\bar{\Lambda}\Gamma_{ab}R)(\bar{\Lambda}\Gamma_{cd}R)(\bar{\Lambda}\Gamma_{gh}R) - (\bar{\Lambda}\Gamma_{gh}\bar\Lambda)(\bar{\Lambda}\Gamma_{ab}R)(\bar{\Lambda}\Gamma_{cd}R)(\bar{\Lambda}\Gamma_{ef}R)](\Lambda\Gamma^{abcij}\Lambda)\times \notag \\
& (\Lambda\Gamma^{defgk}\Lambda)\eta^{hl}\{N_{ij}, N_{kl}\} \notag \\
& = \frac{2}{3\eta^{4}}[(\bar{\Lambda}\Gamma_{ab}\bar\Lambda)(\bar{\Lambda}\Gamma_{cd}R)(\bar{\Lambda}\Gamma_{ef}R)(\bar{\Lambda}\Gamma_{gh}R) + (\bar{\Lambda}\Gamma_{ef}\bar\Lambda)(\bar{\Lambda}\Gamma_{ab}R)(\bar{\Lambda}\Gamma_{cd}R)(\bar{\Lambda}\Gamma_{gh}R)](\Lambda\Gamma^{abcij}\Lambda)\times \notag \\
& (\Lambda\Gamma^{defgk}\Lambda)\eta^{hl}\{N_{ij}, N_{kl}\} \notag \\
& = \frac{2}{3\eta^{4}}[(\bar{\Lambda}\Gamma_{ab}\bar\Lambda)(\bar{\Lambda}\Gamma_{cd}R)(\bar{\Lambda}\Gamma_{ef}R)(\bar{\Lambda}\Gamma_{gh}R) - (\bar{\Lambda}\Gamma_{cd}\bar\Lambda)(\bar{\Lambda}\Gamma_{ab}R)(\bar{\Lambda}\Gamma_{ef}R)(\bar{\Lambda}\Gamma_{gh}R)](\Lambda\Gamma^{abcij}\Lambda)\times \notag \\
& (\Lambda\Gamma^{defgk}\Lambda)\eta^{hl}\{N_{ij}, N_{kl}\} \notag \\
& = \frac{4}{3\eta^{4}}(\bar{\Lambda}\Gamma_{ab}\bar\Lambda)(\bar{\Lambda}\Gamma_{cd}R)(\bar{\Lambda}\Gamma_{ef}R)(\bar{\Lambda}\Gamma_{gh}R)(\Lambda\Gamma^{abcij}\Lambda)(\Lambda\Gamma^{defgk}\Lambda)\eta^{hl}\{N_{ij}, N_{kl}\}
\end{align}
where the identities \eqref{newapp5} and \eqref{newapp6} were used repeatedly. Let us now focus on $P_{2}$:
\begin{align}
P_{2} & = -\frac{8}{9\eta^{4}}L^{(3)}_{ab,cd,ef,gh}(\Lambda\Gamma^{abcij}\Lambda)\eta^{h[d}(\Lambda\Gamma^{efg]kl}\Lambda)\{N_{ij}, N_{kl}\} \notag \\
& = -\frac{2}{9\eta^{4}}L^{(3)}_{ab,cd,ef,gh}(\Lambda\Gamma^{abcij}\Lambda)[\eta^{hd}(\Lambda\Gamma^{efgkl}\Lambda) - 2\eta^{he}(\Lambda\Gamma^{dfgkl}\Lambda)]\{N_{ij}, N_{kl}\} \notag \\
& = -\frac{2}{9\eta^{4}}[L^{(3)}_{ab,cd,ef,gh}\eta^{hd}(\Lambda\Gamma^{abcij}\Lambda)(\Lambda\Gamma^{efgkl}\Lambda) - 2L^{(3)}_{ab,cd,ef,gh}\eta^{he}(\Lambda\Gamma^{abcij}\Lambda)(\Lambda\Gamma^{dfgkl}\Lambda)]\{N_{ij}, N_{kl}\}
\end{align}
One can simplify each term separately as follows:
\begin{align}
(P\ensuremath{'}_{2})^{ijkl} & = L^{(3)}_{ab,cd,ef,gh}\eta^{hd}(\Lambda\Gamma^{abcij}\Lambda)(\Lambda\Gamma^{efgkl}\Lambda) \notag \\
& = \frac{1}{4}[(\bar{\Lambda}\Gamma_{ab}\bar{\Lambda})(\bar{\Lambda}\Gamma_{cd}R)(\bar{\Lambda}\Gamma_{ef}R)(\bar{\Lambda}\Gamma_{g}^{\hspace{2mm}d}R) - (\bar{\Lambda}\Gamma_{cd}\bar{\Lambda})(\bar{\Lambda}\Gamma_{ab}R)(\bar{\Lambda}\Gamma_{ef}R)(\bar{\Lambda}\Gamma_{g}^{\hspace{2mm}d}R) \notag \\
& + (\bar{\Lambda}\Gamma_{ef}\bar{\Lambda})(\bar{\Lambda}\Gamma_{ab}R)(\bar{\Lambda}\Gamma_{cd}R)(\bar{\Lambda}\Gamma_{g}^{\hspace{2mm}d}R) - (\bar{\Lambda}\Gamma_{g}{}^{d}\bar{\Lambda})(\bar{\Lambda}\Gamma_{ab}R)(\bar{\Lambda}\Gamma_{cd}R)(\bar{\Lambda}\Gamma_{ef}R)](\Lambda\Gamma^{abcij}\Lambda)(\Lambda\Gamma^{efgkl}\Lambda) \notag \\
& = \frac{1}{4}[ - 2(\bar{\Lambda}\Gamma_{cg}\bar{\Lambda})(\bar{\Lambda}\Gamma_{ab}R)(\bar{\Lambda}\Gamma_{ef}R)(\bar{\Lambda}R) - 2(\bar{\Lambda}\Gamma_{g}{}^{d}\bar{\Lambda})(\bar{\Lambda}\Gamma_{ab}R)(\bar{\Lambda}\Gamma_{cd}R)(\bar{\Lambda}\Gamma_{ef}R)](\Lambda\Gamma^{abcij}\Lambda)(\Lambda\Gamma^{efgkl}\Lambda) \notag \\
& = \frac{1}{4}[ - 2(\bar{\Lambda}\Gamma_{cg}\bar{\Lambda})(\bar{\Lambda}\Gamma_{ab}R)(\bar{\Lambda}\Gamma_{ef}R)(\bar{\Lambda}R) - 2(\bar{\Lambda}\Gamma_{gc}\bar{\Lambda})(\bar{\Lambda}\Gamma_{ab}R)(\bar{\Lambda}R)(\bar{\Lambda}\Gamma_{ef}R)](\Lambda\Gamma^{abcij}\Lambda)(\Lambda\Gamma^{efgkl}\Lambda) \notag \\
& = (\bar{\Lambda}\Gamma_{ab}\bar{\Lambda})(\bar{\Lambda}\Gamma_{cg}R)(\bar{\Lambda}\Gamma_{ef}R)(\bar{\Lambda}R)(\Lambda\Gamma^{abcij}\Lambda)(\Lambda\Gamma^{efgkl}\Lambda)
\end{align} 
\begin{align}
(P\ensuremath{''}_{2})^{ijkl} & = -2L^{(3)}_{ab,cd,ef,gh}(\Lambda\Gamma^{dfgkl}\Lambda)(\Lambda\Gamma^{abcij}\Lambda) \notag \\
& = -\frac{1}{2}[(\bar{\Lambda}\Gamma_{ab}\bar{\Lambda})(\bar{\Lambda}\Gamma_{cd}R)(\bar{\Lambda}\Gamma_{ef}R)(\bar{\Lambda}\Gamma_{g}^{\hspace{2mm}e}R) - (\bar{\Lambda}\Gamma_{cd}\bar{\Lambda})(\bar{\Lambda}\Gamma_{ab}R)(\bar{\Lambda}\Gamma_{ef}R)(\bar{\Lambda}\Gamma_{g}^{\hspace{2mm}e}R) \notag \\
& + (\bar{\Lambda}\Gamma_{ef}\bar{\Lambda})(\bar{\Lambda}\Gamma_{ab}R)(\bar{\Lambda}\Gamma_{cd}R)(\bar{\Lambda}\Gamma_{g}^{\hspace{2mm}e}R) - (\bar{\Lambda}\Gamma_{g}^{\hspace{2mm}e}\bar{\Lambda})(\bar{\Lambda}\Gamma_{ab}R)(\bar{\Lambda}\Gamma_{cd}R)(\bar{\Lambda}\Gamma_{ef}R)](\Lambda\Gamma^{dfgkl}\Lambda)(\Lambda\Gamma^{abcij}\Lambda) \notag \\
&=[(\bar{\Lambda}\Gamma_{cd}\bar{\Lambda})(\bar{\Lambda}\Gamma_{ab}R)(\bar{\Lambda}\Gamma_{ef}R)(\bar{\Lambda}\Gamma_{g}^{\hspace{2mm}e}R) + (\bar{\Lambda}\Gamma_{fg}\bar{\Lambda})(\bar{\Lambda}\Gamma_{ab}R)(\bar{\Lambda}\Gamma_{cd}R)(\bar{\Lambda}R)](\Lambda\Gamma^{dfgkl}\Lambda)(\Lambda\Gamma^{abcij}\Lambda) \notag \\
&=[-(\bar{\Lambda}\Gamma_{cd}\bar{\Lambda})(\bar{\Lambda}\Gamma_{ab}R)(\bar{\Lambda}\Gamma_{fg}R)(\bar{\Lambda}R) + (\bar{\Lambda}\Gamma_{fg}\bar{\Lambda})(\bar{\Lambda}\Gamma_{ab}R)(\bar{\Lambda}\Gamma_{cd}R)(\bar{\Lambda}R)](\Lambda\Gamma^{dfgkl}\Lambda)(\Lambda\Gamma^{abcij}\Lambda) \notag \\
& = 2(\bar{\Lambda}\Gamma_{fg}\bar{\Lambda})(\bar{\Lambda}\Gamma_{ab}R)(\bar{\Lambda}\Gamma_{cd}R)(\bar{\Lambda}R)(\Lambda\Gamma^{dfgkl}\Lambda)(\Lambda\Gamma^{abcij}\Lambda) \notag \\
& = 2(\bar{\Lambda}\Gamma_{fg}\bar{\Lambda})(\bar{\Lambda}\Gamma_{dc}R)(\bar{\Lambda}\Gamma_{ab}R)(\bar{\Lambda}R)(\Lambda\Gamma^{fgdkl}\Lambda)(\Lambda\Gamma^{cabij}\Lambda)
\end{align}
Therefore
\begin{align}
P_{2} & = -\frac{2}{3\eta^{4}}(\bar{\Lambda}\Gamma_{ab}\bar{\Lambda})(\bar{\Lambda}\Gamma_{cg}R)(\bar{\Lambda}\Gamma_{ef}R)(\bar{\Lambda}R)(\Lambda\Gamma^{abcij}\Lambda)(\Lambda\Gamma^{efgkl}\Lambda)\{N_{ij}, N_{kl}\}
\end{align}
Hence, one concludes that
\begin{eqnarray}
b^{(4)} & =& \frac{4}{3\eta^{4}}(\bar{\Lambda}\Gamma_{ab}\bar\Lambda)(\bar{\Lambda}\Gamma_{cd}R)(\bar{\Lambda}\Gamma_{ef}R)(\bar{\Lambda}\Gamma_{gh}R)(\Lambda\Gamma^{abcij}\Lambda)(\Lambda\Gamma^{defgk}\Lambda)\eta^{hl}\{N_{ij}, N_{kl}\} \nonumber \\
&& -\frac{2}{3\eta^{4}}(\bar{\Lambda}\Gamma_{ab}\bar{\Lambda})(\bar{\Lambda}\Gamma_{cg}R)(\bar{\Lambda}\Gamma_{ef}R)(\bar{\Lambda}R)(\Lambda\Gamma^{abcij}\Lambda)(\Lambda\Gamma^{efgkl}\Lambda)\{N_{ij}, N_{kl}\}\nonumber\\
\end{eqnarray}
which should be compared with the analog expression in $b_{simpl}$, eqn. \eqref{neweq20new}.

From our previous computations, we see that the difference between the $b$-ghost proposed in \cite{Cederwall:2012es}, eqn. \eqref{newapp2000}, and its simplified form presented here, eqn. \eqref{newapp30}, is the non-zero extra term proportional to $(\Lambda\Gamma^{chiab}D)$. This might be related to normal-ordering ambiguities.

\chapter{}\label{appendix6}
\section{$D=10$ gamma matrix identities}\label{apA}
In $D=10$ dimensions, one has chiral and antichiral spinors which have been denoted here by $\chi^{\mu}$ and $\chi_{\mu}$ respectively. The product of two spinors can be decomposed into two forms depending on the chiralities of the spinors used:
\begin{eqnarray}
\xi_{\mu}\chi^{\nu} &=& \frac{1}{16}\delta^{\nu}_{\mu}(\xi\chi) - \frac{1}{2!16}(\gamma^{mn})^{\nu}_{\hspace{2mm}\mu}(\xi\gamma_{mn}\chi) + \frac{1}{4!16}(\gamma^{mnpq})^{\nu}_{\hspace{2mm}\mu}(\xi\gamma_{mnpq}\chi)\\
\xi^{\mu}\chi^{\nu} &=& \frac{1}{16}\gamma_{m}^{\mu\nu}(\xi\gamma^{m}\chi) + \frac{1}{3!16}(\gamma_{mnp})^{\mu\nu}(\xi\gamma^{mnp}\chi) + \frac{1}{5!32}\gamma^{\mu\nu}_{mnpqr}(\xi\gamma^{mnpqr}\chi) \label{ap4}
\end{eqnarray}
The 1-form and 5-form are symmetric, and the 3-form is antisymmetric. Furthermore, it is true that $(\gamma^{mn})^{\mu}_{\hspace{2mm}\nu} = -(\gamma^{mn})_{\nu}^{\hspace{2mm}\mu}$, $(\gamma^{mnpq})^{\mu}_{\hspace{2mm}\nu} = (\gamma^{mnpq})_{\nu}^{\hspace{2mm}\mu}$.

\vspace{2mm}
Two particularly useful identities are:
\begin{eqnarray}
(\gamma^{m})_{(\mu\nu}(\gamma_{m})_{\rho)\sigma}  &=& 0\\
(\gamma^{mn})^{\mu}{}_{\nu}(\gamma_{mn})^{\rho}{}_{\sigma} &=& 4(\gamma^{m})^{\mu\rho}(\gamma_{m})_{\nu\sigma} - 2\delta^{\mu}_{\nu}\delta^{\rho}_{\sigma} - 8\delta^{\mu}_{\sigma}\delta^{\rho}_{\nu} \label{ap1}
\end{eqnarray}
From \ref{ap1} we can deduce the following:
\begin{eqnarray}
(\gamma^{mn})^{\mu}{}_{\nu}(\gamma_{mnp})^{\rho\sigma} &=& 2(\gamma^{m})^{\mu\rho}(\gamma_{pm})^{\sigma}_{\hspace{2mm}\nu} + 6(\gamma_{p})^{\mu\rho}\delta^{\sigma}_{\nu} - (\rho \leftrightarrow \sigma)\label{ap2} \\
(\gamma^{mn})^{\mu}{}_{\nu}(\gamma_{mnp})_{\rho\sigma} &=& -2(\gamma^{m})_{\nu\sigma}(\gamma_{pm})^{\mu}{}_{\rho} + 6(\gamma_{p})_{\nu\sigma}\delta^{\mu}_{\rho} - (\rho \leftrightarrow \sigma)\\
(\gamma_{mnp})^{\mu\nu}(\gamma^{mnp})^{\rho\sigma} &=& 12[(\gamma_{m})^{\mu\sigma}(\gamma^{m})^{\nu\rho} - (\gamma_{m})^{\mu\rho}(\gamma^{m})^{\nu\sigma}] \label{ap5}\\
(\gamma_{mnp})^{\mu\nu}(\gamma^{mnp})_{\rho\sigma} &=& 48(\delta^{\mu}_{\rho}\delta^{\nu}_{\sigma} - \delta^{\mu}_{\sigma}\delta^{\nu}_{\rho})\label{ap6}
\end{eqnarray}

\section{$D=11$ gamma matrix identities}\label{apB}
In $D=11$ dimensions, one has Majorana spinors and an antisymmetric tensor $C_{\alpha\beta}$ (and its inverse) which can be used to raise and lower spinor indices. The product of two spinors can be decomposed into the form
\begin{eqnarray}\label{app7}
\chi^{\alpha}\psi^{\beta} &=& -\frac{1}{32}C^{\alpha\beta}(\chi\psi) + \frac{1}{32}(\Gamma^{a})^{\alpha\beta}(\chi\Gamma_{a}\psi) - \frac{1}{2!.32}(\Gamma^{ab})^{\alpha\beta}(\chi\Gamma_{ab}\psi) + \frac{1}{3!.32}(\Gamma^{abc})^{\alpha\beta}(\chi\Gamma_{abc}\psi)\nonumber\\
&& - \frac{1}{4!.32}(\Gamma^{abcd})^{\alpha\beta}(\chi\Gamma_{abcd}\psi) + \frac{1}{5!.32}(\Gamma^{abcde})^{\alpha\beta}(\chi\Gamma_{abcde}\psi)
\end{eqnarray}
The 1-form, 2-form and 5-form are symmetric; and the 0-form, 3-form and 4-form are antisymmetric.

\vspace{2mm}
The crucial identity in eleven dimensions is
\begin{eqnarray}\label{ap7}
(\Gamma^{ab})_{(\alpha\beta}(\Gamma_{b})_{\delta\epsilon)} &=& 0
\end{eqnarray}

\vspace{2mm}
One can find analogous formulae to \eqref{ap1}-\eqref{ap6} for $D=11$ dimensions. However, they do not enter into any computations of this paper, therefore we will not list them.

\vspace{2mm}
From \eqref{ap7} and the pure spinor constraint, one can find several useful pure spinor identities. These were listed in Appendix \ref{apapB}.

\vspace{2mm}
If $a$ is a shift-symmetry index, there exists a very useful identity which states the following
\begin{eqnarray}\label{ap8}
(\bar{\lambda}\Gamma^{ab}\bar{\lambda})(\lambda\Gamma_{cb}\lambda) &=& \frac{1}{2}\delta^{a}_{c}\eta
\end{eqnarray}
This can be easily seen from the following argument. Eqn. \eqref{ap7} implies the relation
\begin{eqnarray*}
-(\bar{\lambda}\Gamma^{ab}\bar{\lambda})(\lambda\Gamma_{b}\Gamma_{c}\lambda) &=& 2(\bar{\lambda}\Gamma^{ab}\bar{\lambda})(\bar{\lambda}\Gamma_{b}\Gamma_{c}\lambda) + 2(\bar{\lambda}\Gamma^{ab}\Gamma_{c}\lambda)(\bar{\lambda}\Gamma_{c}\lambda)
\end{eqnarray*}
which can be rewritten in the more convenient form
\begin{eqnarray*}
-(\bar{\lambda}\Gamma^{ab}\bar{\lambda})(\lambda\Gamma_{b}\Gamma_{c}\lambda) &=& \lambda\Gamma^a\xi_c + 4\delta^a_c(\lambda\bar{\lambda})^2 - 4\delta^a_c (\bar{\lambda}\Gamma^b \lambda)(\bar{\lambda}\Gamma_b \lambda)
\end{eqnarray*}
where $\xi_c^\alpha$ is defined as follows
\begin{eqnarray*}
\xi_c^\alpha &=& -2(\bar{\lambda}\Gamma_c)^\alpha (\bar\lambda \lambda) - 2(\bar{\lambda}\Gamma^b)^\alpha (\bar{\lambda}\Gamma_b \Gamma_c \lambda) + 2(\bar{\lambda}\Gamma_{bc})^\alpha (\lambda\Gamma^b \bar{\lambda}) + 4\bar{\lambda}^\alpha (\lambda\Gamma_c \bar\lambda)
\end{eqnarray*}
The use of \eqref{app7} allows us to write
\begin{eqnarray*}
-(\lambda\bar{\lambda})^2 &=& -\frac{1}{64}\eta + \frac{1}{3840}(\lambda\Gamma^{abcde}\lambda)(\bar{\lambda}\Gamma_{abcde}\bar{\lambda})\\
(\bar{\lambda}\Gamma^a \lambda)(\bar{\lambda}\Gamma_a \lambda) &=& -\frac{7}{64}\eta - \frac{1}{3840}(\lambda\Gamma^{abcde}\lambda)(\bar{\lambda}\Gamma_{abcde}\bar{\lambda})
\end{eqnarray*}
Therefore,
\begin{eqnarray*}
(\bar{\lambda}\Gamma^{ab}\bar{\lambda})(\lambda\Gamma_{cb}\lambda) &=& \frac{1}{2}\delta^a_c \eta + \lambda\Gamma^a\xi_c
\end{eqnarray*}

\section{Pure spinor superfield formalism}\label{newapC}
In this appendix we give a summary of the systematic procedure used for constructing pure spinor master actions called the pure spinor superfield formalism \cite{Cederwall:2009ez,Cederwall:2010wf,Cederwall:2010tn,Cederwall:2011vy,Cederwall:2013vba}. We find it useful to first briefly review the standard antifield formalism.

\vspace{2mm}
The classical BV field theory was introduced as a generalization of the BRST method for quantizing interacting gauge systems. The standard procedure used to describe a theory in its BV version can be easily summarized as follows. For each gauge symmetry, one introduce a ghost field. If the system is reducible, one should also introduce a ghost for ghost for each generation of reducibility. All of these ghost fields together with the physical fields will form the field $\Phi^{I}$. Each of these fields $\Phi^{I}$ is supplemented by an antifield $\Phi_{I}^{\ast}$ with opposite statistics and a ghost number assignment satisfying $g(\Phi^{I}) + g(\Phi_{I}^{\ast}) = -1$, where $g(A)$ stands for the ghost number associated to the object $A$. Using these fields and antifields, a fermionic antibracket is defined
\begin{eqnarray}
(A\,,\,B) &=& \int\,d^{D} x \left(A\frac{\overleftarrow{\delta}}{\delta\Phi^{I}}\frac{\overrightarrow{\delta}}{\delta\Phi_{I}^{\ast}}B - A\frac{\overleftarrow{\delta}}{\delta\Phi_{I}^{\ast}}\frac{\overrightarrow{\delta}}{\delta\Phi^{I}}B \right)
\end{eqnarray}
Using this antibracket, one then defines the master action as a solution to the master equation defined as
\begin{eqnarray}
(S, S) &=& 0
\end{eqnarray}
In this sense, the action itself generates gauge transformations via the antibracket. The master action reduces to the original action for the physical fields after removing the ghosts and antifields.

\vspace{2mm}
We have seen that the pure spinor superfields in $D=10$ and $D=11$ dimensions possess a field-antifield structure similar to the BV prescription described above. However, unlike the usual case, one single pure spinor superfield contains both as fields and antifields. This suggests that the pure spinor superfield $\Psi$ should be self-conjugated with respect to a \emph{pure spinor antibracket}. A natural candidate for it is given by
\begin{eqnarray}
(A\,,\,B) &=& \int\,\left(A\frac{\overleftarrow{\delta}}{\delta\Psi}[dZ]\frac{\overrightarrow{\delta}}{\delta\Psi}B\right)
\end{eqnarray}
where $[dZ]$ was defined in previous section. Therefore, the pure spinor master action will be defined as a solution to the pure spinor master equation
\begin{eqnarray}\label{chapter5eq1}
(\mathcal{S},\mathcal{S}) &=& 0
\end{eqnarray}
The master action will generate gauge transformations which can be calculated from the relation
\begin{eqnarray}\label{chapter5eq2}
\delta \Psi &=& \frac{\overrightarrow{\delta}\mathcal{S}}{\delta\Psi}
\end{eqnarray}

\vspace{2mm}
One can easily see that the actions given in \eqref{eeeq102}, \eqref{eeeeq102} for $D=10$ super-Maxwell and $D=11$ linearized supergravity, respectively, satisfy eqn. \eqref{chapter5eq1} in a trivial way as a consequence of the nilpotency of $Q$. Moreover, the gauge transformations found there readily follow from eqn. \eqref{chapter5eq2}.

\vspace{2mm}
One can now use the pure spinor superfield framework for constructing pure spinor master actions for interacting theories like $D=10$ super Yang-Mills, $D=10$ abelian Born-Infeld and $D=11$ supergravity.

\subsection{$D=10$ super-Yang-Mills}
The $D=10$ super-Yang-Mills actions was defined in \eqref{eq222} to be
\begin{eqnarray}
\mathcal{S}_{SYM} &=& \int [dZ] \,Tr(\frac{1}{2}\Psi Q\Psi + \frac{g}{3}\Psi \Psi \Psi)
\end{eqnarray}
or equivalently,
\begin{eqnarray}\label{chapter5eq4}
\mathcal{S}_{SYM} &=& \int [dZ](\frac{1}{2}\Psi^{a}Q\Psi^{a} + \frac{g}{6}f^{abc}\Psi^{a}\Psi^{b}\Psi^{c})
\end{eqnarray}
This action can be readily shown to satisfy the master equation \eqref{chapter5eq1}:
\begin{eqnarray}
(\mathcal{S}_{SYM}, \mathcal{S}_{SYM}) &=& \int [dZ]\left(Q\Psi^{a} + gf^{abc}\Psi^{b}\Psi^{c}\right)\left(Q\Psi^{a} + gf^{ade}\Psi^{d}\Psi^{e}\right)
\end{eqnarray}
After integrating by parts the quadratic term in $Q\Psi^{a}$ and the mixed term, one finds they both identically vanish. The third term vanishes as a consequence of the Jacobi identity.

\vspace{2mm}
Furthermore, one can use eqn. \eqref{chapter5eq2} to calculate the BRST symmetry which leaves invariant \eqref{chapter5eq4}
\begin{eqnarray}
\delta\Psi^{a} &=& Q\Lambda^{a} + f^{abc}\Psi^{b}\Lambda^{c}
\end{eqnarray}
or in compact form
\begin{eqnarray}
\delta\Psi = Q\Lambda + [\Psi , \Lambda]
\end{eqnarray}
which coincides with eqn. \eqref{eq224}.

\subsection{$D=10$ Abelian Supersymmetric Born-Infeld}
The $D=10$ Abelian super-Born-Infeld action was defined in \eqref{eeq12} to be
\begin{eqnarray}\label{neweeq12}
\mathcal{S}_{SBI} &=& \int [dZ] \left[\frac{1}{2}\Psi Q\Psi + \frac{k}{4}\Psi(\lambda\gamma^{m}\hat{\chi}\Psi)(\lambda\gamma^{n}\hat{\chi}\Psi)\hat{F}_{mn}\Psi\right]
\end{eqnarray}
One can easily check that \eqref{neweeq12} is indeed a master action after rewriting \eqref{neweeq12} in terms of $\Delta_{m}$, given in eqn. \eqref{eeq5}, as follows
\begin{eqnarray}
\mathcal{S}_{SBI} &=& \int [dZ]\left(\frac{1}{2}\Psi Q\Psi + \frac{k}{32(\lambda\bar{\lambda})^{2}}(\bar{\lambda}\gamma^{mnp}r)\Psi\Delta_{m}\Psi\Delta_{n}\Psi\Delta_{p}\Psi\right)
\end{eqnarray}
The master equation then reads
\begin{eqnarray}
(\mathcal{S}_{SBI},\mathcal{S}_{SBI}) &=& \int (Q\Psi +\frac{k}{8(\lambda\bar{\lambda})^{2}}(\bar{\lambda}\gamma^{mnp}r)\Delta_{m}\Psi\Delta_{n}\Psi\Delta_{p}\Psi)( Q\Psi\nonumber\\
&& +\frac{k}{8(\lambda\bar{\lambda})^{2}}(\bar{\lambda}\gamma^{qrs}r)\Delta_{q}\Psi\Delta_{r}\Psi\Delta_{s}\Psi)
\end{eqnarray}
The first term vanishes because of the nilpotency of $Q$. The mixed term can be shown to vanish by using the result $Q(\frac{1}{(\lambda\bar{\lambda})^{2}}(\bar{\lambda}\gamma^{qrs}r)\Delta_{q}\Psi\Delta_{r}\Psi\Delta_{s}\Psi) = 0$ which is a direct consequence of the identity $[Q, \Delta_{m}] = \frac{1}{2(\lambda\bar{\lambda})}(\lambda\gamma^{n}\gamma_{m}r)\Delta_{n}$. The last term can be written in the form
\begin{eqnarray}
(\bar{\lambda}\gamma^{mnp}r)(\bar{\lambda}\gamma^{qrs}r)\Delta_{m}\Psi\Delta_{n}\Psi\Delta_{p}\Psi\Delta_{q}\Psi\Delta_{r}\Psi\Delta_{s}\Psi &=& \frac{3}{16}(r\gamma^{mnk}r)(\bar{\lambda}\gamma_{k}{}^{pqrs}\bar{\lambda})\Delta_{m}\Psi\Delta_{n}\Psi\nonumber\\
&& \Delta_{p}\Psi\Delta_{q}\Psi\Delta_{r}\Psi\Delta_{s}\Psi
\end{eqnarray}
which vanishes because of the pure spinor constraint and the antisymmetry in $(n,p,q,r,s,t)$. 

\vspace{2mm}
One can now use eqn. \eqref{chapter5eq2} to find that the BRST transformation which leaves invariant \eqref{neweeq12} is given by
\begin{eqnarray}
\delta\Psi &=& Q\Lambda + k(\lambda\gamma^{m}\hat{\chi}\Psi)(\lambda\gamma^{n}\hat{\chi}\Psi)\hat{F}_{mn}\Lambda + 2k(\lambda\gamma^{m}\hat{\chi}\Psi)(\lambda\gamma^{n}\hat{\chi}\Lambda)\hat{F}_{mn}\Psi
\end{eqnarray}
which is the same as the one displayed in eqn. \eqref{newequationnew1}.

\subsection{$D=11$ Supergravity}
We have seen that $D=11$ linearized supergravity can be obtained from the ghost number three cohomology of the eleven-dimensional pure spinor BRST charge. However, it is also possible to describe $D=11$ linearized supergravity through a ghost number one vector cohomology $\Phi^{a} = \lambda^{\alpha}h_{\alpha}{}^{a}$ subjected to the shift symmetry given by
\begin{eqnarray}
Q\Phi^{a} &=& (\lambda\Gamma^{a}\nu)
\end{eqnarray} 
where $\nu^{\alpha} = \lambda^{\beta}h_{\beta}{}^{\alpha}$. Unlike the ghost number three cohomology, $\Phi^{a}$ describes $D=11$ linearized supergravity through the graviton field, the gravitino field and the 4-form field strength. The lowest $\theta$-components read
\begin{eqnarray}
\Phi^{a} &=& (\Gamma_{b}\theta)^{\alpha}h^{ab} + (\Gamma^{b}\theta)_{\alpha}(\psi_{b}\Gamma^{a}\theta) + (\mathcal{D}^{cdef}\Gamma^{a}\theta)H_{cdef} + \ldots 
\end{eqnarray}
where 
\begin{eqnarray}
(\mathcal{D}^{cdef})_{\alpha}{}^{\mu} &=& \frac{1}{4}\delta^{\mu}_{\beta}[(\theta\mathcal{R}_{ab}{}^{cdef})_{\alpha}(\Gamma^{ab}\theta)^{\beta} + (\theta\Gamma^{a})_{\alpha}(\theta\mathcal{T}_{a}{}^{cdef})^{\beta}]\nonumber\\
\mathcal{R}_{ab}{}^{cdef} &=& \frac{1}{6}[\delta^{[c}_{a}\delta^{d}_{b}(\Gamma^{ef]}) + \frac{1}{24}\Gamma_{ab}{}^{cdef}]\nonumber\\
\mathcal{T}_{a}{}^{bcde} &=& -\frac{1}{36}[\delta_{a}^{[b}\Gamma^{cde]} + \frac{1}{8}\Gamma_{a}{}^{bcde}]
\end{eqnarray}
and $H_{abcd} = E_{a}{}^{M}E_{b}{}^{N}E_{c}{}^{P}E_{d}{}^{Q}\partial_{M}C_{NPQ}$.

\vspace{2mm}
Cederwall showed in \cite{Cederwall:2009ez} that it is possible relate these two superfields through a ghost number -2 operator satisfying 
\begin{eqnarray}
R^{a}\Psi &=& \Phi^{a} + \mbox{shift symmetry terms.}
\end{eqnarray}
with $[Q, R^{a}] = 0$. Its explicit form was displayed in eqn. \eqref{eeq13}. Using this vector superfield, one can then deform the action \eqref{eeeeq102} by adding a shift-shymmetry invariant 3-point coupling as follows
\begin{eqnarray}\label{neweq1new}
\mathcal{S}_{SG} &=& {1\over{\kappa^2}}\int [dZ][\frac{1}{2}\Psi Q\Psi + \frac{1}{6}(\lambda\Gamma_{ab}\lambda)\Psi R^{a}\Psi R^{b}\Psi]
\end{eqnarray}
It can be shown that it correctly reproduces the Chern-Simons term appearing in the $D=11$ supergravity action \cite{Cederwall:2009ez,Cederwall:2010tn}. However, \eqref{neweq1new} does not satisfy the full master equation, since the antibracket of $\mathcal{S}_{SG}$ with itself fails to vanish. In fact, one finds it takes the form
\begin{eqnarray}\label{neweq2new}
(\mathcal{S}_{SG},\mathcal{S}_{SG}) &=& \frac{1}{3}\int [dZ]\left((\lambda\Gamma_{ab}\lambda)R^{a}\Psi R^{b}\Psi\Psi R^{c}((\lambda\Gamma_{cd}\lambda)R^{d}\Psi)\right)
\end{eqnarray}
which is not zero. Remarkably, one can show that 
\begin{eqnarray}
R^{a}(\lambda\Gamma_{ab}\lambda)R^{b} = \frac{1}{2}(\lambda\Gamma_{ab}\lambda)[R^{a},R^{b}] = \frac{3}{2}\{Q,T\}
\end{eqnarray}
where $T$ is a fermionic operator of ghost number -3 defined in eqn. \eqref{eqq13}. In this way, eqn. \eqref{neweq2new} becomes
\begin{eqnarray}\label{neweq3new}
(\mathcal{S}_{SG},\mathcal{S}_{SG}) &=& \int [dZ]\left((\lambda\Gamma_{ab}\lambda)\Psi\{Q,T\}\Psi R^{a}\Psi R^{b}\Psi\right)
\end{eqnarray}
In order to cancel this contribution one needs add the term $(\lambda\Gamma_{ab}\lambda)\Psi T\Psi R^{a}\Psi R^{b}\Psi $ to \eqref{neweq1new}. The antibracket between this new term and the kinetic one kills \eqref{neweq3new}. Surprisingly the resulting action given in \eqref{eq105} identically satisfies the master equation.

\vspace{2mm}
One can again use eqn. \eqref{chapter5eq2} to show that the BRST transformation
\begin{eqnarray}
\delta\Psi &=& Q\Lambda + (\lambda\Gamma_{ab}\lambda)R^{a}\Psi R^{b}\Lambda + \frac{1}{2}\Psi\{Q,T\}\Lambda - \frac{1}{2}\Lambda\{Q,T\}\Psi - 2(\lambda\Gamma_{ab}\lambda)T\Psi R^{a}\Psi R^{b}\Lambda\nonumber\\
&& - (\lambda\Gamma_{ab}\lambda)(T\Lambda)R^{a}\Psi R^{b}\Psi
\end{eqnarray}
for any ghost number 2 pure spinor superfield $\Lambda$, leaves invariant the action \eqref{eq105}.

\section{Relation between $\Psi$ and $\Phi^{a}$}\label{apC}
At linearized level, there exists a simple relation between $\Psi$ and $\Phi^{a}$. To find this relation, define
\begin{eqnarray}\label{eqeqe18}
\hat {H}_{ABCD} &=& \hat{E}_{A}^{\hspace{2mm}M}\hat{E}_{B}^{\hspace{2mm}N}\hat{E}_{C}^{\hspace{2mm}P}\hat{E}_{D}^{\hspace{2mm}Q} {H}_{MNPQ}
\end{eqnarray}
as in \eqref{eqeq000}. Using the conventions
\begin{eqnarray}\label{eeq17}
H_{\alpha\beta\delta\gamma} = 0\hspace{2mm},\hspace{2mm}H_{a\alpha\beta\delta} = 0\hspace{2mm},\hspace{2mm}H_{ab\alpha\beta} = -\frac{1}{12}(\Gamma_{ab})_{\alpha\beta}\hspace{2mm},\hspace{2mm}H_{abc\alpha} = 0,
\end{eqnarray}
one finds that
\begin{eqnarray}
\lambda^\alpha \lambda^\beta\lambda^\gamma\hat{H}_{a\alpha\beta\gamma} &=& \lambda^\alpha \lambda^\beta\lambda^\gamma\hat E_a{}^M \hat E_\alpha{}^N \hat E_\beta{}^P \hat E_\gamma{}^Q E_M{}^A E_N{}^B E_P{}^C E_Q{}^D H_{ABCD}
\end{eqnarray}
\begin{eqnarray}
&=& 3\kappa \lambda^\alpha \lambda^\beta\lambda^\gamma \hat E_\alpha{}^N E_N^{(0)b} 
H_{ab \beta\gamma} + ...
\end{eqnarray}
\begin{eqnarray}
&=&  \frac{1}{4}\kappa \Phi^b \lambda^\beta\lambda^\gamma  
(\Gamma_{ab})_{\beta\gamma} + ...
\end{eqnarray}
where $...$ denotes terms of order $\kappa^2$ and $\Phi^b = -\lambda^\alpha \hat E_\alpha{}^N E_N^{(0)b}$.

Since 
$$\lambda^\alpha\lambda^\beta\lambda^\gamma\hat H_{a\alpha\beta\gamma} =
\kappa(\partial_a \Psi_0 - 3Q(\lambda^{\alpha}\lambda^\beta C_{a\alpha\beta})) ,$$
one obtains the relation
\begin{eqnarray}\label{eqeqe305}
\partial_{a}\Psi_{0} &=& \frac{1}{4}(\lambda\Gamma_{ab}\lambda)\Phi^{b} + 3Q(\lambda^{\alpha}\lambda^\beta C_{a\alpha\beta})
\end{eqnarray}

\vspace{2mm}
The use of equation \eqref{eqeqe305} and the linearized e.o.m
\begin{eqnarray}\label{eeq20}
D_{\alpha}\Psi_{0} + 3Q_{0}(C_{\alpha\beta\delta})\lambda^{\beta}\lambda^{\delta} = -6(\Gamma^{a}\lambda)_{\alpha}C_{a\beta\delta}\lambda^{\beta}\lambda^{\delta}
\end{eqnarray}
allows us to compute the action of $R^{a}$ on $\Psi_{0}$ in the form displayed in \eqref{eeq15}. To see this, it will be useful to express $R^{a}$ in the more convenient way \cite{Cederwall:2010tn}
\begin{eqnarray}
R^{a} &=& -8[\frac{1}{\eta}(\bar{\lambda}\Gamma^{ab}\bar{\lambda})\partial_{b} + \frac{1}{\eta^{2}}(\bar{\lambda}\Gamma^{ab}\bar{\lambda})(\bar{\lambda}\Gamma^{cd}r)(\lambda\Gamma_{bcd}D)\nonumber\\
&& - \{Q , \frac{1}{\eta^{2}}(\bar{\lambda}\Gamma^{ab}\bar{\lambda})(\bar{\lambda}\Gamma^{cd}r)\}(\lambda\Gamma_{bcd}w)]
\end{eqnarray} 
Therefore,
\begin{eqnarray}
R^{a}\Psi_{0} &=& -8\bigg(\frac{1}{\eta}(\bar{\lambda}\Gamma^{ab}\bar{\lambda})\partial_{b}\Psi_{0} + \frac{1}{\eta^{2}}(\bar{\lambda}\Gamma^{ab}\bar{\lambda})(\bar{\lambda}\Gamma^{cd}r)(\lambda\Gamma_{bcd}D\Psi_{0})\nonumber\\
&& + 3\{Q , \frac{1}{\eta^{2}}(\bar{\lambda}\Gamma^{ab}\bar{\lambda})(\bar{\lambda}\Gamma^{cd}r)\}(\lambda\Gamma_{bcd})^{\alpha}C_{\alpha\beta\delta}\lambda^{\beta}\lambda^{\delta}\bigg)\nonumber\\
&=& -8\bigg(\frac{1}{\eta}(\bar{\lambda}\Gamma^{ab}\bar{\lambda})\partial_{b}\Psi_{0} - \frac{3}{\eta^{2}}(\bar{\lambda}\Gamma^{ab}\bar{\lambda})(\bar{\lambda}\Gamma^{cd}r)(\lambda\Gamma_{bcd})^{\alpha} (Q C_{\alpha\beta\delta})\lambda^{\beta}\lambda^{\delta}\nonumber\\
&& - \frac{6}{\eta^{2}}(\bar{\lambda}\Gamma^{ab}\bar{\lambda})(\bar{\lambda}\Gamma^{cd}r)(\lambda\Gamma_{bcd}\Gamma^{e}\lambda)C_{e\alpha\beta}\lambda^{\alpha}\lambda^{\beta}\nonumber\\
&& + 3\{Q , \frac{1}{\eta^{2}}(\bar{\lambda}\Gamma^{ab}\bar{\lambda})(\bar{\lambda}\Gamma^{cd}r)\}(\lambda\Gamma_{bcd})^{\alpha}C_{\alpha\beta\delta}\lambda^{\beta}\lambda^{\delta}\bigg)\nonumber\\
&=& -8\bigg(\frac{1}{\eta}(\bar{\lambda}\Gamma^{ab}\bar{\lambda})\partial_{b}\Psi_{0} + Q\left[ \frac{3}{\eta^{2}}(\bar{\lambda}\Gamma^{ab}\bar{\lambda})(\bar{\lambda}\Gamma^{cd}r)(\lambda\Gamma_{bcd})^{\alpha}C_{\alpha\beta\delta}\lambda^{\beta}\lambda^{\delta}\right]\nonumber\\
&& - \frac{6}{\eta^{2}}(\bar{\lambda}\Gamma^{ab}\bar{\lambda})(\bar{\lambda}\Gamma^{cd}r)[-2(\lambda\Gamma_{bd}\lambda)\eta^{e}_{c} + (\lambda\Gamma_{cd}\lambda)\eta^{e}_{b}]C_{e\alpha\beta}\lambda^{\alpha}\lambda^{\beta}\bigg)\nonumber\\
&=& -8\bigg(\frac{1}{\eta}(\bar{\lambda}\Gamma^{ab}\bar{\lambda})\partial_{b}\Psi_{0} + Q\left[ \frac{3}{\eta^{2}}(\bar{\lambda}\Gamma^{ab}\bar{\lambda})(\bar{\lambda}\Gamma^{cd}r)(\lambda\Gamma_{bcd})^{\alpha}C_{\alpha\beta\delta}\lambda^{\beta}\lambda^{\delta}\right]\nonumber\\
&& + \frac{6}{\eta}(\bar{\lambda}\Gamma^{ab}r)C_{b\alpha\beta}\lambda^{\alpha}\lambda^{\beta} - \frac{6}{\eta^{2}}(\bar{\lambda}\Gamma^{ab}\bar{\lambda})
 (\bar{\lambda}\Gamma^{cd}r)(\lambda\Gamma_{cd}\lambda) C_{b\alpha\beta}\lambda^{\alpha}\lambda^{\beta}\bigg)\nonumber\\
&=& -8\bigg(\frac{1}{\eta}(\bar{\lambda}\Gamma^{ab}\bar{\lambda})\partial_{b}\Psi_{0} + Q\left[ \frac{3}{\eta^{2}}(\bar{\lambda}\Gamma^{ab}\bar{\lambda})(\bar{\lambda}\Gamma^{cd}r)(\lambda\Gamma_{bcd})^{\alpha}C_{\alpha\beta\delta}\lambda^{\beta}\lambda^{\delta}\right]\nonumber\\
&& + Q\bigg[\frac{3}{\eta}(\bar{\lambda}\Gamma^{ab}\bar{\lambda})C_{b\alpha\beta}\lambda^{\alpha}\lambda^{\beta}\bigg] - \frac{3}{\eta}(\bar{\lambda}\Gamma^{ab}\bar{\lambda})Q[C_{b\alpha\beta}\lambda^{\alpha}\lambda^{\beta}]\bigg)\nonumber\\
&=& -\frac{2}{\eta}(\bar{\lambda}\Gamma^{ab}\bar{\lambda})(\lambda\Gamma_{bc}\lambda)\Phi^{c} + Q\bigg[ -\frac{24}{\eta^{2}}(\bar{\lambda}\Gamma^{ab}\bar{\lambda})(\bar{\lambda}\Gamma^{cd}r)(\lambda\Gamma_{bcd})^{\alpha}C_{\alpha\beta\delta}\lambda^{\beta}\lambda^{\delta}\nonumber\\
&& - \frac{24}{\eta}(\bar{\lambda}\Gamma^{ab}\bar{\lambda})C_{b\alpha\beta}\lambda^{\alpha}\lambda^{\beta}\bigg]\nonumber\\
&=& \Phi^{a} + Q\left[ -\frac{24}{\eta^{2}}(\bar{\lambda}\Gamma^{ab}\bar{\lambda})(\bar{\lambda}\Gamma^{cd}r)(\lambda\Gamma_{bcd})^{\alpha}C_{\alpha\beta\delta}\lambda^{\beta}\lambda^{\delta} - \frac{24}{\eta}(\bar{\lambda}\Gamma^{ab}\bar{\lambda})C_{b\alpha\beta}\lambda^{\alpha}\lambda^{\beta}\right]\nonumber\\
\end{eqnarray}
Notice that in order for the normalization factor of $\Phi^{a}$ to be one after applying $R^{a}$ on $\Psi$, one should choose the conventions used for $R^{a}$ in \eqref{eeq13} and those displayed in \eqref{eeq17}.
\end{appendix}

\providecommand{\href}[2]{#2}\begingroup\raggedright\endgroup


\begin{thebibliography}{10}

\bibitem{Cremmer:1978km}
E.~Cremmer, B.~Julia, and J.~Scherk, ``{Supergravity Theory in
  Eleven-Dimensions},''
  \href{http://dx.doi.org/10.1016/0370-2693(78)90894-8}{{\em Phys. Lett.} {\bf
  B76} (1978)  409--412}.
[,25(1978)].

\bibitem{Hull:1994ys}
C.~M. Hull and P.~K. Townsend, ``{Unity of superstring dualities},''
  \href{http://dx.doi.org/10.1016/0550-3213(94)00559-W}{{\em Nucl. Phys.} {\bf
  B438} (1995)  109--137}, \href{http://arxiv.org/abs/hep-th/9410167}{{\tt
  arXiv:hep-th/9410167 [hep-th]}}.
[,236(1994)].

\bibitem{Townsend:1995kk}
P.~K. Townsend, ``{The eleven-dimensional supermembrane revisited},''
  \href{http://dx.doi.org/10.1016/0370-2693(95)00397-4}{{\em Phys. Lett.} {\bf
  B350} (1995)  184--187}, \href{http://arxiv.org/abs/hep-th/9501068}{{\tt
  arXiv:hep-th/9501068 [hep-th]}}.
[,265(1995)].

\bibitem{Witten:1995ex}
E.~Witten, ``{String theory dynamics in various dimensions},''
  \href{http://dx.doi.org/10.1016/0550-3213(95)00158-O}{{\em Nucl. Phys.} {\bf
  B443} (1995)  85--126}, \href{http://arxiv.org/abs/hep-th/9503124}{{\tt
  arXiv:hep-th/9503124 [hep-th]}}.
[,333(1995)].

\bibitem{Green:1999by}
M.~B. Green, M.~Gutperle, and H.~H. Kwon, ``{Light cone quantum mechanics of
  the eleven-dimensional superparticle},''
  \href{http://dx.doi.org/10.1088/1126-6708/1999/08/012}{{\em JHEP} {\bf 08}
  (1999)  012},
\href{http://arxiv.org/abs/hep-th/9907155}{{\tt arXiv:hep-th/9907155
  [hep-th]}}.

\bibitem{Siegel:1985xj}
W.~Siegel, ``{Classical Superstring Mechanics},''
\href{http://dx.doi.org/10.1016/0550-3213(86)90029-5}{{\em Nucl. Phys.} {\bf
  B263} (1986)  93--104}.

\bibitem{Mikovic:1988bj}
A.~R. Mikovic and W.~Siegel, ``{On-shell Equivalence of Superstrings},''
\href{http://dx.doi.org/10.1016/0370-2693(88)91827-8}{{\em Phys. Lett.} {\bf
  B209} (1988)  47--52}.

\bibitem{Essler:1990az}
F.~Essler, E.~Laenen, W.~Siegel, and J.~P. Yamron, ``{BRST operator for the
  first ilk superparticle},''
\href{http://dx.doi.org/10.1016/0370-2693(91)91176-V}{{\em Phys. Lett.} {\bf
  B254} (1991)  411--416}.

\bibitem{Essler:1990yq}
F.~Essler, M.~Hatsuda, E.~Laenen, W.~Siegel, J.~P. Yamron, T.~Kimura, and A.~R.
  Mikovic, ``{Covariant quantization of the first ilk superparticle},''
\href{http://dx.doi.org/10.1016/0550-3213(91)90578-L}{{\em Nucl. Phys.} {\bf
  B364} (1991)  67--84}.

\bibitem{Berkovits:1990yc}
N.~Berkovits, ``{A Supertwistor Description of the Massless Superparticle in
  Ten-dimensional Superspace},''
  \href{http://dx.doi.org/10.1016/0370-2693(90)91047-F,
  10.1016/0550-3213(91)90258-Y}{{\em Phys. Lett.} {\bf B247} (1990)  45--49}.
[Nucl. Phys.B350,193(1991)].

\bibitem{Nilsson:1985cm}
B.~E.~W. Nilsson, ``{Pure Spinors as Auxiliary Fields in the Ten-dimensional
  Supersymmetric {Yang-Mills} Theory},''
\href{http://dx.doi.org/10.1088/0264-9381/3/2/007}{{\em Class. Quant. Grav.}
  {\bf 3} (1986)  L41}.

\bibitem{HOWE1991141}
P.~Howe, ``Pure spinor lines in superspace and ten-dimensional supersymmetric
  theories,''
  \href{http://dx.doi.org/https://doi.org/10.1016/0370-2693(91)91221-G}{{\em
  Physics Letters B} {\bf 258} (1991) no.~1, 141 -- 144}.
  \url{http://www.sciencedirect.com/science/article/pii/037026939191221G}.

\bibitem{Howe:1991bx}
P.~S. Howe, ``{Pure spinors, function superspaces and supergravity theories in
  ten-dimensions and eleven-dimensions},''
\href{http://dx.doi.org/10.1016/0370-2693(91)90558-8}{{\em Phys. Lett.} {\bf
  B273} (1991)  90--94}.

\bibitem{Berkovits:2000fe}
N.~Berkovits, ``{Super Poincare covariant quantization of the superstring},''
  \href{http://dx.doi.org/10.1088/1126-6708/2000/04/018}{{\em JHEP} {\bf 04}
  (2000)  018},
\href{http://arxiv.org/abs/hep-th/0001035}{{\tt arXiv:hep-th/0001035
  [hep-th]}}.

\bibitem{Berkovits:2001rb}
N.~Berkovits, ``{Covariant quantization of the superparticle using pure
  spinors},'' \href{http://dx.doi.org/10.1088/1126-6708/2001/09/016}{{\em JHEP}
  {\bf 09} (2001)  016},
\href{http://arxiv.org/abs/hep-th/0105050}{{\tt arXiv:hep-th/0105050
  [hep-th]}}.

\bibitem{Berkovits:2002zk}
N.~Berkovits, ``{ICTP lectures on covariant quantization of the superstring},''
  in {\em {Superstrings and related matters. Proceedings, Spring School,
  Trieste, Italy, March 18-26, 2002}}, pp.~57--107.
\newblock 2002.
\newblock \href{http://arxiv.org/abs/hep-th/0209059}{{\tt arXiv:hep-th/0209059
  [hep-th]}}.
\newblock
\url{http://www.ictp.trieste.it/~pub_off/lectures/lns013/Berkovits/Berkovits.pdf}.
\newblock

\bibitem{Berkovits:2002uc}
N.~Berkovits, ``{Towards covariant quantization of the supermembrane},''
  \href{http://dx.doi.org/10.1088/1126-6708/2002/09/051}{{\em JHEP} {\bf 09}
  (2002)  051},
\href{http://arxiv.org/abs/hep-th/0201151}{{\tt arXiv:hep-th/0201151
  [hep-th]}}.

\bibitem{Bjornsson:2010wm}
J.~Bjornsson and M.~B. Green, ``{5 loops in 24/5 dimensions},''
  \href{http://dx.doi.org/10.1007/JHEP08(2010)132}{{\em JHEP} {\bf 08} (2010)
  132},
\href{http://arxiv.org/abs/1004.2692}{{\tt arXiv:1004.2692 [hep-th]}}.

\bibitem{Bjornsson:2010wu}
J.~Bjornsson, ``{Multi-loop amplitudes in maximally supersymmetric pure spinor
  field theory},'' \href{http://dx.doi.org/10.1007/JHEP01(2011)002}{{\em JHEP}
  {\bf 01} (2011)  002},
\href{http://arxiv.org/abs/1009.5906}{{\tt arXiv:1009.5906 [hep-th]}}.

\bibitem{Kugo:1979gm}
T.~Kugo and I.~Ojima, ``{Local Covariant Operator Formalism of Nonabelian Gauge
  Theories and Quark Confinement Problem},''
\href{http://dx.doi.org/10.1143/PTPS.66.1}{{\em Prog. Theor. Phys. Suppl.} {\bf
  66} (1979)  1--130}.

\bibitem{Cederwall:2009ez}
M.~Cederwall, ``{Towards a manifestly supersymmetric action for 11-dimensional
  supergravity},'' \href{http://dx.doi.org/10.1007/JHEP01(2010)117}{{\em JHEP}
  {\bf 01} (2010)  117},
\href{http://arxiv.org/abs/0912.1814}{{\tt arXiv:0912.1814 [hep-th]}}.

\bibitem{Cederwall:2010tn}
M.~Cederwall, ``{D=11 supergravity with manifest supersymmetry},''
  \href{http://dx.doi.org/10.1142/S0217732310034407}{{\em Mod. Phys. Lett.}
  {\bf A25} (2010)  3201--3212},
\href{http://arxiv.org/abs/1001.0112}{{\tt arXiv:1001.0112 [hep-th]}}.

\bibitem{Cederwall:2011vy}
M.~Cederwall and A.~Karlsson, ``{Pure spinor superfields and Born-Infeld
  theory},'' \href{http://dx.doi.org/10.1007/JHEP11(2011)134}{{\em JHEP} {\bf
  11} (2011)  134},
\href{http://arxiv.org/abs/1109.0809}{{\tt arXiv:1109.0809 [hep-th]}}.

\bibitem{Cederwall:2013vba}
M.~Cederwall, ``{Pure spinor superfields -- an overview},''
  \href{http://dx.doi.org/10.1007/978-3-319-03774-5_4}{{\em Springer Proc.
  Phys.} {\bf 153} (2014)  61--93},
\href{http://arxiv.org/abs/1307.1762}{{\tt arXiv:1307.1762 [hep-th]}}.

\bibitem{Anguelova:2004pg}
L.~Anguelova, P.~A. Grassi, and P.~Vanhove, ``{Covariant one-loop amplitudes in
  D=11},'' \href{http://dx.doi.org/10.1016/j.nuclphysb.2004.09.024}{{\em Nucl.
  Phys.} {\bf B702} (2004)  269--306},
\href{http://arxiv.org/abs/hep-th/0408171}{{\tt arXiv:hep-th/0408171
  [hep-th]}}.

\bibitem{Bandos:2007mi}
I.~A. Bandos, ``{Spinor moving frame, M0-brane covariant BRST quantization and
  intrinsic complexity of the pure spinor approach},''
  \href{http://dx.doi.org/10.1016/j.physletb.2007.10.048}{{\em Phys. Lett.}
  {\bf B659} (2008)  388--398},
\href{http://arxiv.org/abs/0707.2336}{{\tt arXiv:0707.2336 [hep-th]}}.

\bibitem{Bandos:2007wm}
I.~A. Bandos, ``{D=11 massless superparticle covariant quantization, pure
  spinor BRST charge and hidden symmetries},''
  \href{http://dx.doi.org/10.1016/j.nuclphysb.2007.12.019}{{\em Nucl. Phys.}
  {\bf B796} (2008)  360--401},
\href{http://arxiv.org/abs/0710.4342}{{\tt arXiv:0710.4342 [hep-th]}}.

\bibitem{Gran:2001yh}
U.~Gran, ``{GAMMA: A Mathematica package for performing gamma matrix algebra
  and Fierz transformations in arbitrary dimensions},''
\href{http://arxiv.org/abs/hep-th/0105086}{{\tt arXiv:hep-th/0105086
  [hep-th]}}.

\bibitem{Cederwall:2012es}
M.~Cederwall and A.~Karlsson, ``{Loop amplitudes in maximal supergravity with
  manifest supersymmetry},''
  \href{http://dx.doi.org/10.1007/JHEP03(2013)114}{{\em JHEP} {\bf 03} (2013)
  114},
\href{http://arxiv.org/abs/1212.5175}{{\tt arXiv:1212.5175 [hep-th]}}.

\bibitem{Karlsson:2014xva}
A.~Karlsson, ``{Ultraviolet divergences in maximal supergravity from a pure
  spinor point of view},''
  \href{http://dx.doi.org/10.1007/JHEP04(2015)165}{{\em JHEP} {\bf 04} (2015)
  165},
\href{http://arxiv.org/abs/1412.5983}{{\tt arXiv:1412.5983 [hep-th]}}.

\bibitem{Tsimpis:2004gq}
D.~Tsimpis, ``{Curved 11D supergeometry},''
  \href{http://dx.doi.org/10.1088/1126-6708/2004/11/087}{{\em JHEP} {\bf 11}
  (2004)  087},
\href{http://arxiv.org/abs/hep-th/0407244}{{\tt arXiv:hep-th/0407244
  [hep-th]}}.

\bibitem{Mason:2013sva}
L.~Mason and D.~Skinner, ``{Ambitwistor strings and the scattering
  equations},'' \href{http://dx.doi.org/10.1007/JHEP07(2014)048}{{\em JHEP}
  {\bf 07} (2014)  048},
\href{http://arxiv.org/abs/1311.2564}{{\tt arXiv:1311.2564 [hep-th]}}.

\bibitem{Berkovits:2013xba}
N.~Berkovits, ``{Infinite Tension Limit of the Pure Spinor Superstring},''
  \href{http://dx.doi.org/10.1007/JHEP03(2014)017}{{\em JHEP} {\bf 03} (2014)
  017},
\href{http://arxiv.org/abs/1311.4156}{{\tt arXiv:1311.4156 [hep-th]}}.

\bibitem{Guillen:2017mte}
M.~Guillen, ``{Equivalence of the 11D pure spinor and Brink-Schwarz-like
  superparticle cohomologies},''
  \href{http://dx.doi.org/10.1103/PhysRevD.97.066002}{{\em Phys. Rev.} {\bf
  D97} (2018) no.~6, 066002},
\href{http://arxiv.org/abs/1705.06316}{{\tt arXiv:1705.06316 [hep-th]}}.

\bibitem{Strassler:1992zr}
M.~J. Strassler, ``{Field theory without Feynman diagrams: One loop effective
  actions},'' \href{http://dx.doi.org/10.1016/0550-3213(92)90098-V}{{\em Nucl.
  Phys.} {\bf B385} (1992)  145--184},
\href{http://arxiv.org/abs/hep-ph/9205205}{{\tt arXiv:hep-ph/9205205
  [hep-ph]}}.

\bibitem{Berkovits:2018gbq}
N.~Berkovits and M.~Guillen, ``{Equations of motion from Cederwall's pure
  spinor superspace actions},''
  \href{http://dx.doi.org/10.1007/JHEP08(2018)033}{{\em JHEP} {\bf 08} (2018)
  033},
\href{http://arxiv.org/abs/1804.06979}{{\tt arXiv:1804.06979 [hep-th]}}.

\bibitem{Babalic:2008ga}
M.~Babalic and N.~Wyllard, ``{Towards relating the kappa-symmetric and
  pure-spinor versions of the supermembrane},''
  \href{http://dx.doi.org/10.1088/1126-6708/2008/10/059}{{\em JHEP} {\bf 10}
  (2008)  059},
\href{http://arxiv.org/abs/0808.3691}{{\tt arXiv:0808.3691 [hep-th]}}.

\bibitem{Berkovits:2005bt}
N.~Berkovits, ``{Pure spinor formalism as an N=2 topological string},''
  \href{http://dx.doi.org/10.1088/1126-6708/2005/10/089}{{\em JHEP} {\bf 10}
  (2005)  089},
\href{http://arxiv.org/abs/hep-th/0509120}{{\tt arXiv:hep-th/0509120
  [hep-th]}}.

\bibitem{Berkovits:2013pla}
N.~Berkovits, ``{Dynamical twisting and the b ghost in the pure spinor
  formalism},'' \href{http://dx.doi.org/10.1007/JHEP06(2013)091}{{\em JHEP}
  {\bf 06} (2013)  091},
\href{http://arxiv.org/abs/1305.0693}{{\tt arXiv:1305.0693 [hep-th]}}.

\bibitem{Brink:1981nb}
L.~Brink and J.~H. Schwarz, ``{Quantum Superspace},''
\href{http://dx.doi.org/10.1016/0370-2693(81)90093-9}{{\em Phys. Lett.} {\bf
  B100} (1981)  310--312}.

\bibitem{Bedoya:2009np}
O.~A. Bedoya and N.~Berkovits, ``{GGI Lectures on the Pure Spinor Formalism of
  the Superstring},'' in {\em {New Perspectives in String Theory Workshop
  Arcetri, Florence, Italy, April 6-June 19, 2009}}.
\newblock 2009.
\newblock \href{http://arxiv.org/abs/0910.2254}{{\tt arXiv:0910.2254
  [hep-th]}}.
\newblock
\url{https://inspirehep.net/record/833767/files/arXiv:0910.2254.pdf}.
\newblock

\bibitem{Jusinskas:2013yca}
R.~Lipinski~Jusinskas, ``{Nilpotency of the b ghost in the non-minimal pure
  spinor formalism},'' \href{http://dx.doi.org/10.1007/JHEP05(2013)048}{{\em
  JHEP} {\bf 05} (2013)  048},
\href{http://arxiv.org/abs/1303.3966}{{\tt arXiv:1303.3966 [hep-th]}}.

\bibitem{Berkovits:2014ama}
N.~Berkovits and O.~Chandia, ``{Simplified Pure Spinor b Ghost in a Curved
  Heterotic Superstring Background},''
  \href{http://dx.doi.org/10.1007/JHEP06(2014)001}{{\em JHEP} {\bf 06} (2014)
  001},
\href{http://arxiv.org/abs/1403.2429}{{\tt arXiv:1403.2429 [hep-th]}}.

\bibitem{Karlsson:2015qda}
A.~Karlsson, ``{Pure spinor indications of ultraviolet finiteness in D=4
  maximal supergravity},''
\href{http://arxiv.org/abs/1506.07505}{{\tt arXiv:1506.07505 [hep-th]}}.

\bibitem{Gomez:2013sla}
H.~Gomez and C.~R. Mafra, ``{The closed-string 3-loop amplitude and
  S-duality},'' \href{http://dx.doi.org/10.1007/JHEP10(2013)217}{{\em JHEP}
  {\bf 10} (2013)  217}, \href{http://arxiv.org/abs/1308.6567}{{\tt
  arXiv:1308.6567 [hep-th]}}.

\bibitem{Chang:2014nwa}
C.-M. Chang, Y.-H. Lin, Y.~Wang, and X.~Yin, ``{Deformations with Maximal
  Supersymmetries Part 2: Off-shell Formulation},''
  \href{http://dx.doi.org/10.1007/JHEP04(2016)171}{{\em JHEP} {\bf 04} (2016)
  171},
\href{http://arxiv.org/abs/1403.0709}{{\tt arXiv:1403.0709 [hep-th]}}.

\bibitem{Cederwall:2001td}
M.~Cederwall, B.~E.~W. Nilsson, and D.~Tsimpis, ``{D = 10 superYang-Mills at
  O(alpha-prime**2)},''
  \href{http://dx.doi.org/10.1088/1126-6708/2001/07/042}{{\em JHEP} {\bf 07}
  (2001)  042},
\href{http://arxiv.org/abs/hep-th/0104236}{{\tt arXiv:hep-th/0104236
  [hep-th]}}.

\bibitem{BRINK1981310}
L.~Brink and J.~H. Schwarz, ``Quantum superspace,''
  \href{http://dx.doi.org/https://doi.org/10.1016/0370-2693(81)90093-9}{{\em
  Physics Letters B} {\bf 100} (1981) no.~4, 310 -- 312}.
  \url{http://www.sciencedirect.com/science/article/pii/0370269381900939}.

\bibitem{Berkovits:2006vi}
N.~Berkovits and N.~Nekrasov, ``{Multiloop superstring amplitudes from
  non-minimal pure spinor formalism},''
  \href{http://dx.doi.org/10.1088/1126-6708/2006/12/029}{{\em JHEP} {\bf 12}
  (2006)  029},
\href{http://arxiv.org/abs/hep-th/0609012}{{\tt arXiv:hep-th/0609012
  [hep-th]}}.

\bibitem{BERGSHOEFF1987371}
E.~Bergshoeff, M.~Rakowski, and E.~Sezgin, ``Higher derivative super
  {Yang}-{Mills} theories,''
  \href{http://dx.doi.org/https://doi.org/10.1016/0370-2693(87)91017-3}{{\em
  Physics Letters B} {\bf 185} (1987) no.~3, 371 -- 376}.
  \url{http://www.sciencedirect.com/science/article/pii/0370269387910173}.

\bibitem{JAMESGATES1987172}
S.~J. Gates and S.~Vashakidze, ``On d = 10, n = 1 supersymmetry, superspace
  geometry and superstring effects (i),''
  \href{http://dx.doi.org/https://doi.org/10.1016/0550-3213(87)90470-6}{{\em
  Nuclear Physics B} {\bf 291} (1987)  172 -- 204}.
  \url{http://www.sciencedirect.com/science/article/pii/0550321387904706}.

\bibitem{Berkovits:2002ag}
N.~Berkovits and V.~Pershin, ``{Supersymmetric Born-Infeld from the pure spinor
  formalism of the open superstring},''
  \href{http://dx.doi.org/10.1088/1126-6708/2003/01/023}{{\em JHEP} {\bf 01}
  (2003)  023},
\href{http://arxiv.org/abs/hep-th/0205154}{{\tt arXiv:hep-th/0205154
  [hep-th]}}.

\bibitem{BRINK1980384}
L.~Brink and P.~Howe, ``Eleven-dimensional supergravity on the mass shell in
  superspace,''
  \href{http://dx.doi.org/https://doi.org/10.1016/0370-2693(80)91002-3}{{\em
  Physics Letters B} {\bf 91} (1980) no.~3, 384 -- 386}.
  \url{http://www.sciencedirect.com/science/article/pii/0370269380910023}.

\bibitem{Berkovits:2015yra}
N.~Berkovits, ``{Origin of the Pure Spinor and Green-Schwarz Formalisms},''
  \href{http://dx.doi.org/10.1007/JHEP07(2015)091}{{\em JHEP} {\bf 07} (2015)
  091},
\href{http://arxiv.org/abs/1503.03080}{{\tt arXiv:1503.03080 [hep-th]}}.

\bibitem{Aisaka:2009yp}
Y.~Aisaka and N.~Berkovits, ``{Pure Spinor Vertex Operators in Siegel Gauge and
  Loop Amplitude Regularization},''
  \href{http://dx.doi.org/10.1088/1126-6708/2009/07/062}{{\em JHEP} {\bf 07}
  (2009)  062},
\href{http://arxiv.org/abs/0903.3443}{{\tt arXiv:0903.3443 [hep-th]}}.

\bibitem{Oda:2003dz}
I.~Oda, ``{Covariant matrix model of superparticle in the pure spinor
  formalism},'' \href{http://dx.doi.org/10.1142/S0217732303010879}{{\em Mod.
  Phys. Lett.} {\bf A18} (2003)  1023--1036},
\href{http://arxiv.org/abs/hep-th/0302203}{{\tt arXiv:hep-th/0302203
  [hep-th]}}.

\bibitem{Baez:2001dm}
J.~C. Baez, ``{The Octonions},''
  \href{http://dx.doi.org/10.1090/S0273-0979-01-00934-X}{{\em Bull. Am. Math.
  Soc.} {\bf 39} (2002)  145--205},
\href{http://arxiv.org/abs/math/0105155}{{\tt arXiv:math/0105155 [math-ra]}}.

\bibitem{SIEGEL1983397}
W.~Siegel, ``Hidden local supersymmetry in the supersymmetric particle
  action,''
  \href{http://dx.doi.org/https://doi.org/10.1016/0370-2693(83)90924-3}{{\em
  Physics Letters B} {\bf 128} (1983) no.~6, 397 -- 399}.
  \url{http://www.sciencedirect.com/science/article/pii/0370269383909243}.

\bibitem{Berkovits:2004tw}
N.~Berkovits and D.~Z. Marchioro, ``{Relating the Green-Schwarz and pure spinor
  formalisms for the superstring},''
  \href{http://dx.doi.org/10.1088/1126-6708/2005/01/018}{{\em JHEP} {\bf 01}
  (2005)  018},
\href{http://arxiv.org/abs/hep-th/0412198}{{\tt arXiv:hep-th/0412198
  [hep-th]}}.

\bibitem{Mafra:2009wq}
C.~R. Mafra, {\em {Superstring Scattering Amplitudes with the Pure Spinor
  Formalism}}.
\newblock PhD thesis, Sao Paulo, IFT, 2008.
\newblock
\href{http://arxiv.org/abs/0902.1552}{{\tt arXiv:0902.1552 [hep-th]}}.
\newblock

\bibitem{Siegel:1978yi}
W.~Siegel, ``{Superfields in Higher Dimensional Space-time},''
\href{http://dx.doi.org/10.1016/0370-2693(79)90202-8}{{\em Phys. Lett.} {\bf
  80B} (1979)  220--223}.

\bibitem{Witten:1985nt}
E.~Witten, ``{Twistor - Like Transform in Ten-Dimensions},''
\href{http://dx.doi.org/10.1016/0550-3213(86)90090-8}{{\em Nucl. Phys.} {\bf
  B266} (1986)  245--264}.

\bibitem{Brink:1980az}
L.~Brink and P.~S. Howe, ``{Eleven-Dimensional Supergravity on the Mass-Shell
  in Superspace},''
\href{http://dx.doi.org/10.1016/0370-2693(80)91002-3}{{\em Phys. Lett.} {\bf
  91B} (1980)  384--386}.

\bibitem{Cederwall:2010wf}
M.~Cederwall, ``{From supergeometry to pure spinors},'' in {\em {Modern
  Mathematical Physics. Proceedings, 6th Summer School: Belgrade, Serbia,
  September 14-23, 2010}}, pp.~139--151.
\newblock 2011.
\newblock \href{http://arxiv.org/abs/1012.3334}{{\tt arXiv:1012.3334
  [hep-th]}}.
\newblock
\url{http://www.mphys6.ipb.ac.rs/proceedings6/13-Cederwall.pdf}.
\newblock

\end{thebibliography}

\end{document}